\documentclass[11pt,leqno]{article}
\usepackage{amssymb,amsfonts}
\usepackage{amsmath,amsthm,amsxtra}
\usepackage[all]{xy}

\usepackage{hyperref}

\hypersetup{
    bookmarks=true,         
    unicode=false,          
    pdftoolbar=true,        
    pdfmenubar=true,        
    pdffitwindow=false,     
    pdfstartview={FitH},    
    pdftitle={My title},    
    pdfauthor={Author},     
    pdfsubject={Subject},   
    pdfcreator={Creator},   
    pdfproducer={Producer}, 
    pdfkeywords={keywords}, 
    pdfnewwindow=true,      
    colorlinks=true,       
    linkcolor=blue,          
    citecolor=red, 
    filecolor=magenta,      
    urlcolor=cyan           
}

\usepackage{color}
\usepackage{verbatim}

\newcommand{\st}[1]{\ensuremath{^{\scriptstyle \textrm{#1}}}}

\newcommand\bigcheck[1]{#1 \raise1ex\hbox{$\hspace{-1ex}{}^\vee$}}
\newcommand\sucheck[1]{#1 \raise0.5ex\hbox{$\hspace{-1ex}{}^\vee$}}

\newcommand{\alphaparenlist}{
  \renewcommand{\theenumi}{\alph{enumi}}%
  \renewcommand{\labelenumi}{(\theenumi)}%
}

\newcommand{\romanparenlist}{
  \renewcommand{\theenumi}{\roman{enumi}}%
  \renewcommand{\labelenumi}{(\theenumi)}%
}

\newcommand{\ad}{\mathop{\rm ad}\,}

\newcommand{\Der}{{\rm Der}}

\newcommand{\End}{\mathop{\rm End }}

\newcommand{\im}{\mathop{\rm Im  \, }}
\newcommand{\Ker}{\mathop{\rm Ker \, }}

\newcommand{\ord}{\mathop{\rm ord \, }}

\newcommand{\Span}{\text{Span}}
\newcommand{\Hom}{\mathop{\rm Hom }}

\newcommand{\Dom}{\text{Dom}}

\renewcommand{\hat}{\widehat}

\newcommand{\fin}{\text{fin}}

\makeatletter
\renewcommand\section{\@startsection {section}{1}{\z@}%
                                   {-3.5ex \@plus -1ex \@minus -.2ex}%
                                   {2.3ex \@plus.2ex}%
                                   {\normalfont\large\bfseries}}
\renewcommand\subsection{\@startsection{subsection}{2}{\z@}%
                                     {-3.25ex\@plus -1ex \@minus -.2ex}%
                                     {0ex \@plus .0ex}%
                                     {\normalfont\normalsize\bfseries}}

\@addtoreset{equation}{section}
\makeatother

\newtheorem{theorem}{Theorem}[section]
\newtheorem{lemma}[theorem]{Lemma}
\newtheorem{corollary}[theorem]{Corollary}
\newtheorem{proposition}[theorem]{Proposition}
\newtheorem*{lemma*}{Lemma}

\theoremstyle{definition}
\newtheorem{definition}[theorem]{Definition}

\theoremstyle{remark}
\newtheorem{remark}[theorem]{Remark}
\newtheorem{example}[theorem]{Example}

\makeatletter
\def\@maketitle{\newpage
 \null
 \vskip 2em
 \begin{center}%
  \vskip 3em
  {\Large\bf \@title \par}%
  \vskip 1.5em
  {\normalsize
   \lineskip .5em
   \begin{tabular}[t]{c}\@author
   \end{tabular}\par}%
  \vskip 2em

 \end{center}%
 \par
 \vskip 2.5em}
\makeatother

\newcommand{\mc}[1]{{\mathcal #1}}
\newcommand{\mf}[1]{{\mathfrak #1}}
\newcommand{\mb}[1]{{\mathbb #1}}

\renewcommand{\epsilon}{\varepsilon}

\newcommand\tint{{\textstyle\int}}

\newcommand{\id}{{1 \mskip -5mu {\rm I}}}

\definecolor{light}{gray}{.9}

\begin{document}


\begin{center}
{\Large {\bf
Poisson vertex algebras in the theory of Hamiltonian equations
}
}

\vspace{20pt}

{\large
\begin{tabular}[t]{c}
$\mbox{Aliaa Barakat}^{1}\phantom{m} ,\phantom{mm} \mbox{Alberto De Sole}^{2}\phantom{m} ,
\phantom{mm}
\mbox{Victor G. Kac}^{3}$
\\
\end{tabular}
\par
}

\bigskip

{\small
\begin{tabular}[t]{ll}
{1} & Department of Mathematics, MIT \\
& 77 Massachusetts Avenue, Cambridge, MA 02139, USA \\
& E-mail: {\tt barakat@math.mit.edu} \\
{2} & {\it Dipartimento di Matematica, Universit\`a di Roma ``La Sapienza"} \\
&  Citt\`{a} Universitaria, 00185 Roma, Italy \\
& E-mail: {\tt desole@mat.uniroma1.it} \\
{3} & Department of Mathematics, MIT \\
& 77 Massachusetts Avenue, Cambridge, MA 02139, USA \\
& E-mail: {\tt kac@math.mit.edu}
\end{tabular}
}
\end{center}

\vspace{20pt}

\begin{abstract}
We lay down the foundations of the theory of Poisson vertex algebras aimed at its applications
to integrability of Hamiltonian partial differential equations.
Such an equation is called integrable if it can be included in an infinite hierarchy 
of compatible Hamiltonian equations, which admit an infinite sequence
of linearly independent integrals of motion in involution.
The construction of a hierarchy and its integrals of motion is achieved 
by making use of the so called Lenard scheme.
We find simple conditions which guarantee that the scheme produces an infinite sequence
of closed 1-forms $\omega_j,\,j\in\mb Z_+$, of the variational complex $\Omega$.
If these forms are exact, i.e.\ $\omega_j$ are variational derivatives of some
local functionals $\tint h_j$,
then the latter are integrals of motion in involution of the hierarchy formed 
by the corresponding Hamiltonian vector fields.
We show that the complex $\Omega$ is exact, provided that the algebra
of functions $\mc V$ is ``normal'';
in particular, for arbitrary $\mc V$, any closed form in $\Omega$ becomes exact
if we add to $\mc V$ a finite number of antiderivatives.
We demonstrate on the examples of the KdV, HD and CNW hierarchies how the 
Lenard scheme works.
We also discover a new integrable hierarchy, which we call the CNW hierarchy
of HD type.
Developing the ideas of Dorfman, we extend the Lenard scheme
to arbitrary Dirac structures, and demonstrate its applicability on the 
examples of the NLS, pKdV and KN hierarchies.
\end{abstract}

\smallskip


\vfill\eject

\tableofcontents

\vfill\eject

\setcounter{section}{-1}
\section{Introduction.}
\label{sec:intro}


An \emph{evolution equation} is
a system of partial differential equations of the form
\begin{equation}
  \label{eq:0.1}
  \frac{du_i}{dt}= P_i (u,u^\prime,u^{\prime\prime},\ldots)\, , \quad
     i\in I=\{1,\ldots ,\ell\}\qquad
     \text{ ($\ell$ may be infinite)}\,,
\end{equation}
where $u_i=u_i (t,x)$ are functions on a $1$-dimensional manifold
$M$, depending on time~$t$ and $x \in M$, and $P_i$ are
differentiable functions in $u = (u_i)_{i\in I}$ and a finite number of
its derivatives
$u^{\prime}= \big(\frac{\partial u_i}{\partial x}\big)_{i\in I},\,u^{\prime\prime}=\big(\frac{\partial^2u_i}{\partial x^2}\big)_{i\in I},\dots$. 

One usually views $u_i,u^{\prime}_i,u^{\prime\prime}_i, \ldots$ as generators of the
algebra of polynomials
\begin{equation}\label{eq:jan31}
  R=\mb{C} \big[u^{(n)}_i \,\big|\, i \in I,\, n  \in \mb{Z}_+ \big]\, ,
\end{equation}
equipped with the derivation $\partial$, defined by 
$\partial(u^{(n)}_i)=u^{(n+1)}_i$, $n \in \mb{Z}_+$.  
An \emph{algebra of differential functions} $\mc V$
is an extension of the algebra $R$,
endowed with commuting derivations $\frac{\partial}{\partial u_i^{(n)}},\,i\in I,\,n\in\mb Z_+$,
extending the usual partial derivatives in $R$,
and such that, given $f\in\mc V$, $\frac{\partial f}{\partial u_i^{(n)}}=0$
for all but finitely many $i\in I$ and $n\in\mb Z_+$.
Then $\partial$ extends to a derivation of $\mc V$ by
\begin{equation}\label{eq:v03}
\partial\,=\,\sum_{i\in I,n\in\mb Z_+}u_i^{(n+1)}\frac{\partial}{\partial u_i^{(n)}}\,.
\end{equation}

An element of $\mc V /\partial \mc V$ is
called a \emph{local functional}, and the image of $f \in \mc V$ in
$\mc V /\partial \mc V$ is denoted by $\tint f$, the reason being that in
the variational calculus local functionals are of the form
$\tint_M f(u,u^{\prime},\ldots)\, dx$, and taking $\mc V /\partial \mc V$ provides the
universal space for which integration by parts holds.
A local functional $\tint f$  is called an \emph{integral of
motion} of \eqref{eq:0.1}, or is said to be \emph{conserved by
the evolution equation} \eqref{eq:0.1}, if its evolution in time 
along \eqref{eq:0.1}
is constant, namely
\begin{equation}
  \label{eq:0.2}
  \int \frac{df}{dt} \,=\, 0 \, .
\end{equation}
The element $f$, defined modulo $\partial \mc V$, is then called a \emph{conserved density}.
By the chain rule, equation \eqref{eq:0.2} is equivalent to
\begin{equation}
  \label{eq:0.3}
  \tint X_P(f) \,=\, 0 \, ,
\end{equation}
where, for $P=(P_i)_{i\in I}\in\mc V^\ell$, $X_P$ is the following derivation of $\mc V$:
\begin{equation}
\label{eq:0.4}
   X_P = \sum_{\substack{n \in \mb{Z}_+\\ i \in I}} (\partial^n P_i)
          \frac{\partial}{\partial u_i^{(n)}}\, .
\end{equation}
Note that $[\partial ,X_P] =0$.  A derivation of the algebra $\mc V$
with this property is called an \emph{evolutionary vector field}.  
It is easy to see that all of these are of the form
\eqref{eq:0.4} for some $P \in \mc V^\ell$.

Here and further $\mc V^\ell$ denotes the space of all $\ell\times1$ column vectors with entries in $\mc V$.
Also, $\mc V^{\oplus\ell}$ will be used to denote the subspace of $\ell\times1$ column
vectors with only finitely many non-zero entries.
Though, in this paper we do not consider any example with infinite $\ell$,
we still use this distinction as a book-keeping device.

We have the usual pairing
$\mc V^{\oplus\ell}\times \mc V^\ell\to \mc V$, defined by $F\cdot P=\sum_{i\in I}F_i P_i$.
Note that integrating by parts transforms \eqref{eq:0.3} to
$$
\int \frac{\delta f}{\delta u} \cdot P
\,=\,0 \, , 
$$
where $\frac{\delta f}{\delta u} 
= \big( \frac{\delta f}{\delta  u_i} \big)_{i\in I} \in \mc V^{\oplus\ell}$ and
\begin{equation}  
\label{eq:0.6}  
\frac{\delta f}{\delta u_i} = \sum_{n\in\mb Z_+} (-\partial)^n \frac{\partial f}{\partial u^{(n)}_i}
\end{equation}
is the \emph{variational derivative} of $f$ by $u_i$.

Given a sequence of commuting evolutionary vector fields, i.e.\ a sequence 
$P^n\in \mc V^\ell,\,n=0,1,2,\dots$,
such that the corresponding evolutionary vector fields commute (compatibility condition):
\begin{equation}\label{dec20-1}
[X_{P^m},X_{P^n}]\,=\,0\,\,,\,\,\,\,\text{for all\,\,} m,n\in\mb Z_+\,,
\end{equation}
one considers a \emph{hierarchy of evolution equations}
\begin{equation}\label{v08}
  \frac{du}{dt_n} = P^n (u,u^{\prime},u^{\prime\prime},\ldots)\, , \,\,\, n\in\mb Z_+ \, ,
\end{equation}
where $\frac{du}{dt_n}=\big(\frac{du_i}{dt_n}\big)_{i\in I}\in \mc V^\ell$ 
and $u_i=u_i(t_1,t_2,\dots)$.
If the evolutionary vector fields $X_{P^n},\,n\in\mb Z_+$, span an infinite-dimensional vector space,
then each of the equations of the hierarchy \eqref{v08} 
(and the whole hierarchy) is called \emph{integrable}.
For example, the hierarchy of linear evolution equations
\begin{equation}\label{v09}
\frac{du_i}{dt_n}\,=\,u_i^{(n)}\,\,,\,\,\,\,i\in I,n\in\mb Z_+\,,
\end{equation}
is integrable.
Note that the compatibility condition \eqref{dec20-1} can be equivalently expressed
by the following identities:
$$
\frac{dP^m}{dt_n} \,=\, \frac{dP^n}{dt_m}\,,\,\,m,n\in\mb Z_+\,.
$$
Note also that the problem of classification of integrable 
evolution equations is equivalent to the problem
of classification of maximal infinite-dimensional abelian subalgebras
of the Lie algebra of evolutionary vector fields.

\begin{remark}
The above setup can be generalized to the case of any finite-dimensional manifold $M$,
replacing $\partial$ by partial derivatives $\partial_i =
\frac{\partial}{\partial x_i}$ and $\mc V/\partial \mc V$ by $ \mc V/\sum_i
(\partial_i \mc V)$.  However, we shall be concerned only with the
$1$-dimensional case.
\end{remark}

\medskip

A special case of an evolution equation is a system of equations of the
form
\begin{equation}
  \label{eq:0.8}
  \frac{du}{dt} = H(\partial) \frac{\delta h}{\delta u} \, ,
\end{equation}
where $H(\partial)=\big(H_{ij}(u,u^{\prime},\dots;\partial)\big)_{i,j\in I}$ is an $\ell \times \ell$ matrix,
whose entries are finite order differential operators in $\partial$
with coefficients in $\mc V$,
and $\tint h\in \mc V/\partial \mc V$ is a local functional (note that, 
since, in view of \eqref{eq:v03}, $\frac{\delta}{\delta u_i} \circ \partial =0\,\text{for all\,\,} i\in I$, the RHS is well defined).  In this case one
considers the ``local'' bracket \cite{FT} $(i,j \in I)$:
\begin{equation}
  \label{eq:0.9}
  \{ u_i (x) , u_j (y) \} = H_{ji} (u (y), u^{\prime}(y),\dots;\partial_y) 
      \delta (x-y) \, , \quad x,y \in M \, ,
\end{equation}
where $\delta (x-y)$ is the $\delta$-function:  $\tint_M f(x)\delta (x-y)\, dx = f(y)$.  
This bracket extends, by the Leibniz rule and bilinearity,
to arbitrary $f,g \in \mc V$:
\begin{equation}
  \label{eq:0.10}
  \{ f (x) , g (y) \} = \sum_{\substack{i,j \in I \\ m,n \in \mb{Z}_+}}
    \frac{\partial f (x)}{\partial u^{(m)}_i }\, 
    \frac{\partial g (y)}{\partial u^{(n)}_j} \partial ^m_x
      \partial^n_y \{ u_i (x) , u_j (y) \} \, .
\end{equation}
Using bilinearity and integration by parts, equation
\eqref{eq:0.10} suggests a way to define a bracket on the space 
of local functionals $\mc V/\partial \mc V$. Indeed we have
\begin{eqnarray*}
  \{ \tint_M f(x) dx \, , \, \tint_M g(y) dy \} 
  &=& \sum_{i,j \in I} \int_M \int_M \frac{\delta f(x)}{\delta u_i}
      \frac{\delta g(y)}{\delta u_j} \{u_i(x) , u_j(y)\}dxdy \nonumber \\
  &=& \sum_{i,j \in I} \int_M \frac{\delta g(y)}{\delta u_j}  
     H_{ji}(u(y),u^{\prime}(y),\dots,\partial_y) \frac{\delta f(y)}{\delta u_i} dy\,,
\end{eqnarray*}
where, for the last identity, we performed integration by parts.
Hence, given two local functionals $\tint f,\tint g\in \mc V/\partial \mc V$, we define their bracket
associated to the operator $H(\partial)$, as the local functional
\begin{equation}\label{dec20-2}
\big\{\tint f,\tint g\big\} \,=\, \int \sum_{i,j\in I} \frac{\delta g}{\delta u_j} 
H_{ji}(u,u^{\prime},\dots;\partial) \frac{\delta f}{\delta u_i}
\,=\, \int \frac{\delta g}{\delta u} \cdot \Big(H(\partial) \frac{\delta f}{\delta u}\Big) \,.
\end{equation}
Likewise, given a local functional $\tint f\in \mc V/\partial \mc V$ and a function $g\in \mc V$, 
we can define
their bracket, which will now be an element of $\mc V$, by the formula
\begin{equation}\label{dec20-3}
\big\{\tint f,g\big\} \,=\, \sum_{\substack{i,j\in I \\ n\in\mb Z_+}} 
\frac{\partial g}{\partial u_j^{(n)}} \partial^n H_{ji}(u,u^{\prime},\dots;\partial) \frac{\delta f}{\delta u_i}\,.
\end{equation}
Of course, by integration by parts, 
we get that the brackets \eqref{dec20-2} and \eqref{dec20-3}
are compatible in the sense that
$$
\tint \big\{\tint f,g\big\}\,=\,\big\{\tint f,\tint g\big\}\,.
$$
We can then rewrite the evolution equation  \eqref{eq:0.8}, using the above notation, 
in the \emph{Hamiltonian form}:
\begin{equation}
  \label{eq:0.13}
  \frac{du}{dt} = \{ \tint h , u \}\,.
\end{equation}
Here and further, for $u=(u_i)_{i\in I}\in \mc V^\ell$,
$\big\{\tint h,u\big\}$ stands for $\big(\{\tint h,u_i\}\big)_{i\in I}\in \mc V^\ell$.

\medskip

The bracket \eqref{eq:0.10} is called a \emph{Poisson bracket} if
the bracket \eqref{dec20-2} on $\mc V/\partial \mc V$ satisfies the Lie algebra
axioms (this is the basic integrability conditions 
of the evolution equation \eqref{eq:0.8}).
The skew-commutativity of the bracket \eqref{dec20-2}
simply means that the differential
operator $H(\partial)$ is skew-adjoint.
The Jacobi identity is more involved, but can be easily understood
using the language of  $\lambda$-brackets,
which, we believe, greatly simplifies the theory.  
The idea is to apply the Fourier
transform $F(x,y)\mapsto \tint_M \, dxe^{\lambda (x-y)}F(x,y)$ to both sides of
\eqref{eq:0.10}.  Denoting
\begin{displaymath}
  \{ f(y) _\lambda g(y) \} = \tint_M e^{\lambda (x-y)} \{ f(x) , g(y) \} dx \, ,
\end{displaymath}
a straightforward calculation gives the following important formula \cite{DK}:
\begin{equation}
  \label{eq:0.14}
  \{ f_\lambda g \} = \sum_{\substack{i,j \in I \\ m,n \in \mb{Z}_+}}
     \frac{\partial g}{\partial u^{(n)}_j} (\partial + \lambda)^n
       \{ {u_i}_{\partial +\lambda} u_j \}_\to (-\partial -\lambda)^m
          \frac{\partial f}{\partial u^{(m)}_i}\, , 
\end{equation}
where $\{ {u_i}_{\lambda} u_j \} = H_{ji}(u,u^{\prime},\dots;\lambda)$, 
and $\{ {u_i}_{\partial + \lambda} u_j \}_\to$ means that $\partial$ is moved to the right.

In terms of the $\lambda$-bracket \eqref{eq:0.14}, the bracket \eqref{dec20-2} on the space of local functionals $\mc V /\partial \mc V$ becomes
\begin{equation}\label{eq:jan30_1}
\{ \tint f, \tint g \} \,=\, \tint \{ f _\lambda g \} \big|_{\lambda =0} \, .
\end{equation}
Likewise the bracket \eqref{dec20-3} between a local functional $\tint f\in \mc V/\partial \mc V$ and a function
$g\in \mc V$ can be expressed as
$$
 \{ \tint f, g \} \,=\, \{ f _\lambda g \} \big|_{\lambda =0} \, ,
$$
so that the evolution equation \eqref{eq:0.8} becomes
\begin{equation}
  \label{eq:0.15}
  \frac{du}{dt} = \{ h_\lambda u \} \big|_{\lambda =0}\,.
\end{equation}

It is easy to see that the $\lambda$-bracket \eqref{eq:0.14}
satisfies the following \emph{sesquilinearity} properties
\begin{equation}
  \label{eq:0.17}
  \{ \partial f_\lambda g \} = -\lambda \{ f_\lambda g \} \, , 
  \, \{ f_\lambda \partial g \} = (\partial +\lambda) \{ f_\lambda g \} \, ,
\end{equation}
and the following \emph{left} and \emph{right Leibniz rules}:
\begin{equation}
  \label{eq:0.18}
   \{ f_\lambda gh \} = \{ f_\lambda g \} h 
      + g \{ f_\lambda h \} \, , \, \{ fg_\lambda h \} =
      \{ f_{\lambda + \partial} h \}_\to g
        + \{ g_{\lambda +\partial} h \}_\to f \, .
\end{equation}
Furthermore, the skew-commutativity of the bracket \eqref{dec20-2} is
equivalent to
\begin{equation}
  \label{eq:0.19}
  \{ f_\lambda g \} = -\{ g_{-\partial - \lambda} f \}\, ,
\end{equation}
where now $\partial$ is moved to the left, and the Jacobi identity is
equivalent to 
\begin{equation}
  \label{eq:0.20}
  \{ f_\lambda \{ g_\mu h \}\} - \{ g_\mu \{ f_\lambda h \}\}
     = \{\{ f_\lambda g \}_{\lambda +\mu}h\} \, .
\end{equation}

A commutative associative unital differential algebra $\mc{V}$, endowed with a
$\lambda$-bracket $\mc{V} \otimes \mc{V} \to \mb{C} [\lambda] \otimes \mc{V}$,
denoted by $a\otimes b\mapsto\{a_\lambda b\}$,
is called a \emph{Poisson vertex algebra} (PVA) if it satisfies all the
identities \eqref{eq:0.17}--\eqref{eq:0.20}.
If the $\lambda$-bracket \eqref{eq:0.14} defines a PVA structure on $\mc V$,
we say that $H(\partial)=\big(H_{ij}(u,u^{\prime},\dots;\partial)\big)_{i,j\in I}$ in \eqref{eq:0.9}
is a \emph{Hamiltonian operator},
and that equation \eqref{eq:0.8} (or, equivalently, \eqref{eq:0.15})
is a system of \emph{Hamiltonian equations}
associated to the Hamiltonian operator $H(\partial)$.

It follows from \eqref{eq:0.14} that 
$$
\big\{\tint h,\cdot\big\}
\,=\, \{h_\lambda \cdot \} \big|_{\lambda=0}
\,=\, X_{H(\partial)\delta h/\delta u} \,\,\,\,\,\, \text{ for } \tint h \in \mc V/ \partial \mc V \, .
$$
It is easy to check that,
provided that Jacobi identity \eqref{eq:0.20} holds, the map 
$\tint f \mapsto X_{H(\partial)\delta f/\delta u}$ defines a homorphism of the
Lie algebra $\mc V/\partial \mc V$ of local functionals 
(with the bracket \eqref{eq:jan30_1})
to the Lie algebra of evolutionary vector fields. 
Recall that, by definition, a local functional $\tint f$ is an integral of motion 
of the Hamiltonian equation \eqref{eq:0.13}  if 
\begin{equation}\label{dec29_1}
\int \frac {df}{dt} = \{\tint h \, ,\, \tint f\} = 0\,.
\end{equation}
By the above observation this implies that the corresponding evolutionary vector fields
$X_{H(\partial)\delta h/\delta u}$ and $X_{H(\partial)\delta f/\delta u}$ commute:
\begin{equation}\label{dec29_2}
[X_{H(\partial)\delta h/\delta u},X_{H(\partial)\delta f/\delta u}] = 0\,.
\end{equation}
In fact, we will see in Section \ref{sec:dec29}, that in many cases this is a necessary
and sufficient condition for the commutativity relation \eqref{dec29_1}.

\medskip

The basic problem in the theory of Hamiltonian equations is to establish ``integrability".
A system of Hamiltonian equations 
$du/dt=\{\tint h,u\}= H(\partial) \delta h/\delta u$,
is said to be \emph{integrable} if, first, 
$\tint h$ lies in an infinite-dimensional abelian subalgebra
of the Lie algebra $(\mc V/\partial \mc V,\{\,,\,\})$,
in other words, if there exists an infinite sequence of linearly independent local functionals
$\tint h_0=\tint h,\,\tint h_1,\,\tint h_2,\dots\in \mc V/\partial \mc V$,
commuting with respect to the Lie bracket \eqref{eq:jan30_1} on $\mc V/\partial \mc V$,
and, second, if the evolutionary vector fields $X_{H(\partial)\frac{\delta h_n}{\delta u}}$
span an infinite-dimensional vector space.
By the above observations, if such a sequence $\tint h_n$ exists, 
the corresponding Hamiltonian equations
$$
\frac{du}{dt_n} \,=\, \big\{\tint h_n,u\big\}
\,=\, H(\partial) \frac{\delta h_n}{\delta u}\,\,,\,\,\,\, n\in\mb Z_+\,,
$$
form a hierarchy of compatible Hamiltonian equations.
Thus, if the center of the Lie algebra $\mc V/\partial\mc V$ with bracket \eqref{eq:jan30_1}
is finite-dimensional, then we get a hierarchy of integrable evolution equations.
(Note that the hierarchy \eqref{v09} of all linear evolution equations is not Hamiltonian,
but its part with $n$ odd is.)

The problem of classification of integrable Hamiltonian equations
consists of two parts.
The first one is to classify all $\lambda$-brackets on $\mc V$, making it a PVA.
The second one is to classify all maximal infinite-dimensional abelian subalgebras
in the Lie algebra $\mc V/\partial \mc V$ with bracket \eqref{eq:jan30_1}.

What apparently complicates the problem is the possibility of having the same evolution 
equation \eqref{eq:0.15} in two different Hamiltonian forms:
\begin{equation}\label{v023b}
\frac{du}{dt} \,=\, \big\{{\tint h_1} _\lambda u\big\}_0\big|_{\lambda=0}
\,=\, \big\{{\tint h_0} _\lambda u\big\}_1\big|_{\lambda=0}\,,
\end{equation}
where $\{\cdot\,_\lambda\,\cdot\}_i,\,i=0,1$, are two linearly independent $\lambda$-brackets on $\mc V$
and $\tint h_i,\,i=0,1$, are some local functionals.
However, this disadvantage happens to be the main source of integrability in the Hamiltonian
approach, as can be seen by utilizing the so called Lenard scheme \cite{M}, \cite{O}, \cite{D}.
Under some mild conditions, this scheme works, provided that the two $\lambda$-brackets
form a bi-Hamiltonian pair, meaning that any their linear combination makes $\mc V$ a PVA.
Namely, the scheme generates a sequence of local functionals $\tint h_n,\,n\in\mb Z_+$, extending
the given first two terms,
such that for each $n\in\mb Z_+$ we have
\begin{equation}\label{v024b}
\big\{{\tint h_{n+1}} _\lambda u\big\}_0\big|_{\lambda=0}
\,=\, \big\{{\tint h_n} _\lambda u\big\}_1\big|_{\lambda=0}\,.
\end{equation}
In this case all $\tint h_n\in \mc V/\partial \mc V,\,n\in\mb Z_+$,
pairwise commute with respect to both brackets $\{\cdot\,,\,\cdot\}_i,\,i=0,1$,
on $\mc V/\partial \mc V$,
and hence they are integrals of motion in involution for the evolution equation \eqref{v023b}.

In Section \ref{sec:1half} we provide some conditions, which garantee
that the Lenard scheme works.
As applications, we discuss in detail the example of the KdV and the dispersionless KdV
hierarchies, based on the Gardner-Faddeev-Zakharov and the Virasoro-Magri PVA. 
In particular, we prove that the 
polynomials in $u$ form a maximal abelian subspace of integrals of motion
for the dispersionless KdV, and derive from this that the infinite-dimensional
space of integrals of motion for the KdV hierarchy, obtained via the Lenard
scheme, is maximal abelian as well.
In fact, due to the presence of a parameter in the Virasoro-Magri $\lambda$-bracket
(the central charge), we have a triple of compatible $\lambda$-brackets.
This allows one to choose another bi-Hamiltonian pair,
which leads to the so called HD integrable hierarchy,
which we discuss as well.
We also discuss the example of the coupled non-linear wave (CNW) system of Ito,
based on the PVA corresponding to the Virasoro Lie algebra acting on functions on the circle.
Here again, due to the presence of a parameter in the $\lambda$-brackets,
we can make another choice of a bi-Hamiltonian pair,
leading us to a new hierarchy,
which we call the CNW hierarchy of HD type.

\medskip

Another important class of Hamiltonian equations
is provided by symplectic operators.
Let $S(\partial)=\big(S_{ij}(u,u^{\prime},\dots;\partial)\big)_{i,j\in I}$
be a matrix differential operator with finitely many non-zero entries.
The operator $S(\partial)$ is called \emph{symplectic}
if it is skew-adjoint and it satisfies the following analogue of the Jacobi identity
($i,j,k\in I$):
\begin{equation}\label{v024}
\big\{u_i\, _\lambda\, S_{kj}(\mu)\big\}_B
-\big\{u_j\, _\mu\, S_{ki}(\lambda)\big\}_B
+\big\{S_{ji}(\lambda)\,_{\lambda+\mu}\,u_k\big\}_B\,=\,0\,.
\end{equation}
Here $\{\cdot \, _\lambda\, \cdot\}_B$ is called the Beltrami $\lambda$-bracket,
which we define by $\{{u_i}_\lambda u_j\}=\delta_{ij}$
and extended to $\mc V\otimes \mc V\to\mb C[\lambda]\otimes \mc V$ 
using \eqref{eq:0.17} and \eqref{eq:0.18}.

In the symplectic case there is again a Lie algebra bracket, similar to \eqref{eq:jan30_1},
but it is defined only on the following subspace of $\mc V/\partial \mc V$ of
\emph{Hamiltonian functionals}:
$$
\mc F_S\,=\,\Big\{\tint f\,\Big|\, \frac{\delta f}{\delta u}\in S(\partial)\mc V^\ell\Big\}\,.
$$
It is given by the following formula (cf.\ \eqref{dec20-3}):
\begin{equation}\label{v025}
\big\{\tint f,\tint g\big\}_S\,=\,\int \frac{\delta g}{\delta u}\cdot P\,,\,\,
\text{ where } \frac{\delta f}{\delta u}=S(\partial)P\,.
\end{equation}

Consider an evolution equation of the form
\begin{equation}\label{v023}
\frac{du}{dt}\,=\, P\,,
\end{equation}
where $P\in \mc V^\ell$ is such that
$S(\partial)P=\frac{\delta h}{\delta u}$ for some $\tint h\in \mc F_S$.
A local functional $\tint f\in\mc F_S$ is called an integral of motion
of the evolution equation \eqref{v023} if
$\big\{\tint h,\tint f\}_S=0$,
and equation \eqref{v023} is called integrable if it admits infinitely many 
linearly independent
commuting integrals of motion $\tint h_n, n \in \mb{Z}_+$,
commuting with $\tint h =\tint h_0$,
and the corresponding evolutionary vector fields $X_{P^n}$ commute 
and span an infinite-dimensional vector space.
Thus, classification of integrable Hamiltonian equations associated to the symplectic
operator $S(\partial)$
reduces to the classification of infinite-dimensional maximal abelian subspaces 
of the Lie algebra $\mc F_S$.
Based on the above definitions, the theory proceeds in the same way as in the Hamiltonian case,
including the Lenard scheme.
Following Dorfman \cite{D}, we establis in Section \ref{2006_sec-dirac}
integrability of the potential KdV equation and the Krichever-Novikov equation, using this scheme.

\medskip

In fact, there is a more general setup, in terms of Dirac structures, 
introduced by Dorfman \cite{D},
which embraces both the Hamiltonian and the symplectic setup.
A Dirac structure involves two matrix differential operators
$H(\partial)=\big(H_{ij}(u,u^{\prime},\dots;\partial)\big)_{i,j\in I}$ and 
$S(\partial)=\big(S_{ij}(u,u^{\prime},\dots;\partial)\big)_{i,j\in I}$,
and the corresponding \emph{Hamiltonian equations} are of the form
$\frac{du}{dt}=P$, where
$P\in \mc V^\ell$ is such that
\begin{equation}\label{v026}
S(\partial)P=H(\partial)\frac{\delta h}{\delta u}\,,
\end{equation}
for some $\tint h\in \mc F_{H,S}$, the corresponding 
Lie algebra of \emph{Hamiltonian functionals}, defined by
\begin{equation}\label{v027}
\mc F_{H,S}\,=\,\Big\{\tint f\in \mc V/\partial \mc V\,\Big|\, 
H(\partial)\frac{\delta f}{\delta u}\in S(\partial)\mc V^\ell\Big\}\,,
\end{equation}
with the Lie algebra bracket
\begin{equation}\label{v028}
\big\{\tint f,\tint g\big\}_S\,=\,\int \frac{\delta g}{\delta u}\cdot P\,,\,\,
\text{ where } H(\partial)\frac{\delta f}{\delta u}=S(\partial)P\,.
\end{equation}
Under a suitable integrability condition,
expressed in terms of the so-called Courant-Dorfman product,
\eqref{v028} is well defined and it satisfies the Lie algebra axioms.

We develop further the theory of the Lenard scheme for Dirac structures,
initiated by Dorfman \cite{D}, and, using this,
complete the proof of integrability of the non-linear Schr\"odinger system, 
sketched in \cite{D}.

Applying the Lenard scheme at the level of generality of an arbitrary algebra of 
differential functions $\mc V$
requires understanding of the exactness of the variational complex
$\Omega(\mc V)=\bigoplus_{k\in\mb Z_+}\Omega^k(\mc V)$ at $k=0$ and $1$.
We prove in Section \ref{2006_sec2} that, adding finitely many antiderivatives to 
$\mc V$,
one can make any closed $k$-cocycle exact for any $k\geq1$,
and, after adding a constant, for $k=0$
(which seems to be a new result).

The contents of the paper is as follows.
In Sections \ref{sec:1.0} and \ref{sec:1.1.5} we introduce the notion
of an algebra $\mc V$ of differential functions
and its normality property.
The main results here are Propositions \ref{prop:july19_1} and \ref{prop:july19_2}.
They provide for a normal $\mc V$
(in fact any $\mc V$ after adding a finite number of antiderivatives)
algorithms for computing,
for a given $f\in\mc V$ such that $\frac{\delta f}{\delta u}=0$,
an element $g\in\mc V$ such that $f=\partial g+$const.,
and for a given $F\in\mc V^{\oplus\ell}$ which is closed,
i.e.\ such that $D_F(\partial)=D_F^*(\partial)$,
where $D_F(\partial)$ is the Fr\'echet derivative (defined in Section \ref{sec:1.1.5}),
an element $f\in\mc V$ such that $\frac{\delta f}{\delta u}=F$.
These results are important for the proof of integrability of Hamiltonian equations,
discussed in Sections \ref{sec:1half} and \ref{2006_sec-dirac}.

In Sections \ref{sec:1.1} and \ref{sec:dic2008_1} we give several equivalent definitions 
of a Poisson vertex algebra (PVA) and we explain how to introduce a structure 
of a PVA in an algebra $\mc V$ of differential functions or its quotient
(Theorems \ref{prop:exten} and \ref{prop:quotient},
Propositions \ref{prop:09jan16} and \ref{prop:08dic22_1}).
In particular, we show that an equivalent notion is that of a Hamiltonian operator.
In Section \ref{sec:1.3} we show how to construct Hamiltonian equations
and their integrals of motion in the PVA language.
In Section \ref{sec:dec29} we explain how to compute the center of a PVA.
In Section \ref{sec:july19} we introduce the Beltrami $\lambda$-bracket
(which, unlike the Poisson $\lambda$-bracket,
is commutative, rather than skew-commutative),
and interpret the basic operators of the variational calculus,
like the variational derivative and the Fr\'echet derivative,
in terms of this $\lambda$-bracket.
Other applications of the Beltrami $\lambda$-bracket appear in Section \ref{2006_sec2}.

In Section \ref{sec:1.6} we introduce the notion of compatible PVA structures
on an algebra $\mc V$ of differential functions
and develop in this language the well-known Lenard scheme of integrability
of Hamiltonian evolution equations \cite{M}, \cite{O}, \cite{D}. 
The new results here are Propositions \ref{prop:feb28_2}, \ref{prop:feb28_3} 
and \ref{rem:mar19},
and Corollary \ref{cor:feb28},
which provide sufficient conditions for the Lenard scheme to work.

In Section \ref{sec:1.7}, using the Lenard scheme,
we discuss in detail the integrability of the KdV hierarchy,
which includes the classical KdV equation
$$
\frac{du}{dt}
\,=\,
3uu^\prime+cu^{\prime\prime\prime}\,\,,\,\,\,\,c\in\mb C\,.
$$
The new result here is Theorem \ref{prop:1.26} on the maximality of the sequence of integrals
of motion of the KdV hierarchy.
We discuss the integrability of the HD hierarchy, 
which includes the Harry Dim equation
$$
\frac{du}{dt}
\,=\,
\partial^3(u^{-1/2})+\alpha\partial(u^{-1/2})
\,\,,\,\,\,\,\alpha\in\mb C\,,
$$
in Section \ref{sec:hd},
and that of the CNW hierarchy, which includes the CNW system of Ito ($c\in\mb C$):
$$
\left\{\begin{array}{rcl}
\frac{du}{dt} &=& cu^{\prime\prime\prime}+3uu^\prime+vv^\prime \\
\frac{dv}{dt} &=& \partial(uv)
\end{array}\right.\,,
$$
in Section \ref{sec:n-wave}.
In Section \ref{sec:new-intsys} we prove integrability of the CNW hierarchy
of HD type, which seems to be new.
The simplest non-trivial equation of this hierarchy is ($c\in\mb C$):
$$
\left\{\begin{array}{l}
\frac{du}{dt} = \partial\big(\frac1v\big)+c\partial^3\big(\frac1v\big) \\
\frac{dv}{dt} = -\partial\big(\frac{u}{v^2}\big)
\end{array}\right.
$$

In Sections \ref{sec:abstractdr} and \ref{2006_sec2.1} we introduce,
following Gelfand and Dorfman \cite{GD2},
the variational complex $\Omega(\mc V)$
as the reduction of the de Rham $\mf g$-complex $\tilde\Omega(\mc V)$
by the action of $\partial$, i.e.\ $\Omega(\mc V)=\tilde\Omega(\mc V)/\partial\tilde\Omega(\mc V)$.
Our main new result on exactness of the complexes $\tilde\Omega(\mc V)$ and $\Omega(\mc V)$
provided that $\mc V$ is normal, is Theorem \ref{th:july23},
proved in Section \ref{sec:july23_1}.
Another version of the exactness theorem,
which is Theorem \ref{2006_th_cohom},
is well known (see e.g. \cite{D}, \cite{Di}).
However it is not always applicable, whereas Theorem \ref{th:july23}
works for any $\mc V$,
provided that we add to $\mc V$ a finite number of antiderivatives.

In Sections \ref{sec:july23_2} and \ref{sec:09jan2}, following \cite{DSK},
we link the variational complex $\Omega(\mc V)$
to the Lie conformal algebra cohomology, developed in \cite{BKV},
via the Beltrami $\lambda$-bracket,
which leads to an explicit construction of $\Omega(\mc V)$.
As an application, we give a classification of symplectic differential operators
and find simple formulas for the well-known 
Sokolov and Dorfman symplectic operators \cite{S}, \cite{D}, \cite{W}.
In Section \ref{sec:projop} we explore an alternative way of understanding the variational
complex, via the projection operators $\mc P_k$ of $\tilde\Omega^k$
onto a subspace complementary to $\partial\tilde\Omega^k$.

In Sections \ref{sec:derived}--\ref{sec:equations-dirac} we define the Dirac structures
and the corresponding evolution equations.
In Sections \ref{sec:bi-dirac} and \ref{2006_LenardScheme}
we introduce the notion of compatible Dirac structures
and discuss the corresponding Lenard scheme of integrability.
The exposition of these sections follows closely the book of Dorfman \cite{D},
except for Proposition \ref{prop:09jan22} and Corollary \ref{09jan26},
guaranteeing that the Lenard scheme works,
which seem to be new results.

In Section \ref{sec:09febmai} we prove integrability of the non-linear 
Schr\"odinger (NLS) system
$$
\left\{\begin{array}{rcl}
\frac{du}{dt} &=& v^{\prime\prime}+2v(u^2+v^2) \\
\frac{dv}{dt} &=& -u^{\prime\prime}-2u(u^2+v^2) 
\end{array}\right.\,,
$$
using a compatible pair of Dirac structures and the results of Section 
\ref{2006_LenardScheme}.

In Section \ref{sec:dirac-hamilt} we prove that the graph of a Hamiltonian differential operator
is a Dirac structure (see Theorem \ref{2006_th-may30}),
and a similar result on the bi-Hamltonian structure vs.\ a pair of compatible Dirac structures
(see Proposition \ref{2006_BiHamiltonian}),
which allows us to derive the key property of a bi-Hamiltonian structure,
stated in Theorem \ref{th:1.24} in Section \ref{sec:1.6}.

Likewise, in Section \ref{sec:dirac-sympl} we relate the symplectic differential operators
to Dirac structures
and prove Dorfman's criteria of compatibility of a pair of symplectic operators \cite{D}.
In Sections \ref{sec:pkdv} and \ref{sec:kn}
we derive the integrability of the potential KdV equation ($c\in\mb C$)
$$
\frac{du}{dt}
\,=\,
3(u^\prime)^2+cu^{\prime\prime\prime}\,,
$$
and of the Krichiver-Nivikov equation
$$
\frac{du}{dt}
\,=\,
u^{\prime\prime\prime}-\frac32 \frac{(u^{\prime\prime})^2}{u^\prime}\,.
$$
The latter uses the compatibility of the Sokolov and Dorfman symplectic operators.

In this paper we are considering only the translation invariant case,
when functions $f\in\mc V$ do not depend explicitly on $x$.
The general case amounts to adding $\frac{\partial}{\partial x}$ in the definition
\eqref{eq:v03} of $\partial$.

Many important integrable hierarchies, like the KP, the Toda lattice, and the Drinfeld-Sokolov,
are not treated in this paper. We are planning to fill in these gaps in future publications.

\section{Poisson vertex algebras and Hamiltonian operators.}
\label{sec:1}

\subsection{Algebras of differential functions.}~~
\label{sec:1.0}
By a differential algebra we shall mean a unital commutative associative algebra over $\mb C$
with a derivation $\partial$. Recall that $\partial1=0$.
One of the most important examples is the algebra of differential polynomials
in $\ell$ variables
$$
R\,=\,\mb C[u_i^{(n)}\,|\,i\in \{1,\dots,\ell\}=I,\,n\in\mb Z_+]\,,
$$
where $\partial$ is the derivation of the algebra $R$, defined by
$\partial(u_i^{(n)})=u_i^{(n+1)}$.
\begin{definition}\label{def:diff_alg}
An \emph{algebra of differential functions} $\mc V$ in a set of variables $\{u_i\}_{i\in I}$
is an extension of the algebra of polynomials
$R=\mb C[u_i^{(n)}\,|\,i\in I,n\in\mb Z_+]$,
endowed with linear maps
$\frac{\partial}{\partial u_i^{(n)}}\,:\,\,\mc V\to\mc V$, for all $i\in I$ and $n\in\mb Z_+$,
which are commuting derivations of the product in $\mc V$, extending the usual partial derivatives
in $R$, and such that, given $f\in\mc V$,
$\frac{\partial}{\partial u_i^{(n)}}f=0$ for all but finitely many $i\in I$ and $n\in\mb Z_+$.
Unless otherwise specified, we shall assume that $\mc V$ is an integral domain.
\end{definition}
Typical examples of algebras of differential functions
that we will consider are: 
the algebra of differential polynomials itself, $R=\mb C[u_i^{(n)}\,|\,i\in I,n\in\mb Z_+]$,
any localization of it by some element $f\in R$, or, more generally,
by some multiplicative subset $S\subset R$,
such as the whole field of fractions $Q=\mb C(u_i^{(n)}\,|\,i\in I,n\in\mb Z_+)$,
or any algebraic extension of the algebra $R$ or of the field $Q$ obtained by adding
a solution of certain polynomial equation.
An example of the latter type, which we will consider in Section \ref{sec:hd},
is $\mc V=\mb C[\sqrt{u}^{\pm1},u^{\prime},u^{\prime\prime},\dots]$,
obtained by starting from the algebra of differential 
polynomials $R=\mb C[u^{(n)}\,|\,n\in\mb Z_+]$,
adding the square root of the element $u$, and localizing by $\sqrt{u}$.
%

On any algebra of differential functions $\mc V$ we have 
a well defined derivation
$\partial:\, \mc V\to\mc V$, extending the usual derivation of $R$ (defined above), given by
\begin{equation}\label{partial}
\partial \,=\,\sum_{i\in I,n\in\mb Z_+} u_i^{(n+1)}\frac{\partial}{\partial u_i^{(n)}}\,.
\end{equation}
Moreover we have the usual commutation rule
\begin{equation}\label{eq:comm_frac}
\Big[\frac{\partial}{\partial u_i^{(n)}} , \partial\Big] = \frac{\partial}{\partial u_i^{(n-1)}}\,,
\end{equation}
where the RHS is considered to be zero if $n=0$.
These commutation relations imply the following lemma,
which will be used throughout the paper.
\begin{lemma}\label{2006_l-may31}
Let $D_i(z)=\sum_{n\in\mb Z_+}z^n\frac{\partial}{\partial u_i^{(n)}}$.
Then for every  
$h(\lambda)=\sum_{m=0}^Nh_m \lambda^m \in \mb C[\lambda]\otimes\mc V$ 
and $f\in \mc V$ the following identity holds:
\begin{equation}\label{2006_may31-8}
D_i(z)\big(h(\partial)f\big)
=
\big(D_i(z)\big(h(\partial)\big)\big)f
+h(z+\partial)\big(D_i(z)f\big)\,,
\end{equation}
where $D_i(z)\big(h(\partial)\big)$ is the differential operator obtained by applying
$D_i(z)$ to the coefficients of $h(\partial)$.
\end{lemma}
\begin{proof}
In the case $h(\lambda)=\lambda$, equation \eqref{2006_may31-8}
means \eqref{eq:comm_frac}.
It follows that
$$
D_i(z)\circ\partial^n=(z+\partial)^n\circ D_i(z)
\,\,\,\,\text{ for every }\,\, n\in\mb Z_+\,.
$$
The general case follows.
\end{proof}

As before, we denote by $\mc V^{\oplus r}\subset\mc V^r$ the subspace of all
$F=(F_i)_{i\in I}$ with finitely many non-zero entries ($r$ may be infinite),
and introduce a pairing $\mc V^r\times\mc V^{\oplus r}\to\mc V/\partial\mc V$
\begin{equation}\label{eq:may4_4}
(P,F)\,\mapsto\,\tint P\cdot F\,,
\end{equation}
where, as before, $\tint$ denotes the canonical map $\mc V\to\mc V/\partial\mc V$.

Recall the definition \eqref{eq:0.6} of the variational derivative 
$\frac\delta{\delta u}:\,\mc V\to\mc V^{\oplus\ell}$.
It follows immediately from \eqref{eq:comm_frac} that
\begin{equation}\label{eq:mar11_2}
\frac{\delta}{\delta u_i}\circ\partial = 0\,\,,\,\,\,\,i\in I\,,
\end{equation}
i.e.\ $\partial\mc V\subset\Ker \frac{\delta}{\delta u}$.
In fact, we are going to prove that, under some assumptions on $\mc V$,
apart from constant functions,
there is nothing else in $\Ker\frac\delta{\delta u}$.

First, we make some preliminary observations.
Given $f\in \mc V$,
we say that it has \emph{differential order} $n$,
and we write $\ord(f)=n$,
if $\frac{\partial f}{\partial u_i^{(n)}}\neq0$
for some $i\in I$ and $\frac{\partial f}{\partial u_j^{(m)}}=0$ for all $j\in I$ and $m> n$.
In other words,
the space of functions of differential order at most $n$ is
\begin{equation}\label{ord}
\mc V_n\,:=\,\Big\{ f\in\mc V\,\Big|\,\frac{\partial f}{\partial u_j^{(m)}}=0\,\,\text{for all\,\,} j\in I,\,m>n\Big\}\,.
\end{equation}
Clearly $\ord(f)=n$ if and only if $f\in\mc V_n\backslash\mc V_{n-1}$.
We also denote by 
$\mc C\subset\mc V$ the space of constant functions $f$,
namely such that $\frac{\partial f}{\partial u_i^{(m)}}=0$ for all $i\in I$ and $m\in\mb Z_+$,
or, equivalently, such that $\partial f=0$
(this equivalence is immediate by \eqref{partial}).
We will also let
$\mc C(u_i)\subset\mc V_0$ the subspace of functions $f$
depending only on $u_i$, 
namely such that $\frac{\partial f}{\partial u_j^{(n)}}=0$ unless $j=i$ and $n=0$.

We can refine filtration \eqref{ord} as follows.
For $i\in I$ and $n\in\mb Z_+$ we let 
\begin{equation}\label{eq:july21_1}
\mc V_{n,i}\,:=\,\Big\{ f\in\mc V_n\,\Big|\,\frac{\partial f}{\partial u_j^{(n)}}=0\,\,\text{for all\,\,} j>i\Big\}
\,\subset\,\mc V_n\,.
\end{equation}
In particular, $\mc V_{n,\ell}=\mc V_n$. We also let $\mc V_{n,0}=\mc V_{n-1}$ for $n\geq1$,
and $\mc V_{0,0}=\mc C$.

It is clear by the definition \eqref{partial} of $\partial$ that
if $f=\partial g\neq0$ and $g\in\mc V_{n,i}\backslash\mc V_{n,i-1}$,
then $f\in\mc V_{n+1,i}\backslash\mc V_{n+1,i-1}$, and in fact $f$ has the form
\begin{equation}\label{dec27_1_new}
f\,=\, \sum_{j\leq i} h_ju_j^{(n+1)}+\sum_{j>i}h_ju_j^{(n)}+r\,,
\end{equation}
where $h_j\in\mc V_{n,i}$ for all $j\in I,\, r\in\mc V_{n,i}$, 
and $h_i\neq0$.
In particular, it immediately follows that
\begin{equation}\label{eq:20081222_1}
\mc C\cap\partial\mc V\,=\,0\,.
\end{equation}
\begin{proposition}\label{prop:20081222}
\begin{enumerate}
\alphaparenlist
\item
The pairing \eqref{eq:may4_4} is non-degenerate, namely
$\tint P\cdot F=0$ for every $F\in\mc V^{\oplus r}$ if and only if $P=0$.
\item
Moreover, let $I$ be an ideal of $\mc V^{\oplus r}$ containing a non-zero 
element $\bar F$ 
which is not a zero divisor.
If $\tint P\cdot F=0$ for every $F$ in the ideal $I$, then $P=0$.
\end{enumerate}
\end{proposition}
\begin{proof}
We can assume, without loss of generality, that $r=1$.
Suppose that $P\neq0$ is such that $\tint P \cdot F=0$ for every $F\in\mc V$.
In this case, letting $F=1$, we have that $P\in\partial\mc V$ has the form \eqref{dec27_1_new}.
But then it is easy to see that 
$u_i^{(n+1)}P$ does not have this form, so that $\tint u_i^{(n+1)}P\neq0$.
This proves (a).
The argument for (b) is the same, by replacing $P$ by $P\bar F$.
\end{proof}

We  let $\mf g$ be the Lie algebra of all derivations of $\mc V$
of the form
\begin{equation}\label{2006_X}
X=\sum_{i\in I,n\in\mb{Z}_+} h_{i,n} \frac{\partial}{\partial u_i^{(n)}}\,\,,  \quad h_{i,n} \in \mc V\,.
\end{equation}
%
By \eqref{partial}, $\partial$ is an element of $\mf g$,
and we denote by $\mf g^\partial$ the centralizer of $\partial$ in $\mf g$.
Elements $X\in\mf g^\partial$ are called \emph{evolutionary vector fields}.
For $X\in\mf g^\partial$ we have
$X(u_i^{(n)})=X(\partial^nu_i)=\partial^nX(u_i)$,
so that $X=\sum_{i\in I,n\in\mb Z_+}X(u_i^{(n)})\frac{\partial}{\partial u_i^{(n)}}$ 
is completely determined by its values $X(u_i)=P_i,\,i\in I$.
We thus have a vector space isomorphism $\mc V^\ell\,\simeq\, \mf{g}^\partial$,
given by
\begin{equation}\label{2006_iso2}
\mc V^\ell\ni P=\big(P_i\big)_{i\in I}
\mapsto
X_P  = \sum_{i\in I,n\in\mb Z_+} (\partial^n P_i) \frac\partial{\partial u_i^{(n)}}\in\mf g^\partial\,.
\end{equation}
The $\ell$-tuple $P$ is called the \emph{characteristic} of the evolutionary vector field $X_P$.

\vspace{3ex}
\subsection{Normal algebras of differential functions.}~~
\label{sec:1.1.5}
\begin{definition}\label{def:july21}
We call an algebra of differential functions \emph{normal} if 
$\frac\partial{\partial u_i^{(n)}}\big(\mc V_{n,i}\big)=\mc V_{n,i}$ 
for all $i\in I,n\in\mb Z_+$.
Given $f\in \mc V_{n,i}$, we denote by $\tint du_i^{(n)} f\in\mc V_{n,i}$ a preimage of $f$
under the map $\frac{\partial}{\partial u_i^{(n)}}$.
This antiderivative is defined up to adding elements from $\mc V_{n,i-1}$.
\end{definition}
Clearly any algebra of differential functions can be embedded in a normal one.

We will discuss in Section \ref{2006_sec2}
the exactness of the variational complex $\Omega$.
In this language,
Propositions \ref{prop:july19_1} and \ref{prop:july19_2} below
provide algorithms to find,
for a ``closed" element in $\mc V$, a preimage of $\partial$ (up to adding a constant),
and, for a ``closed" element in $\Omega^1$, a preimage of $\frac\delta{\delta u}$.

\begin{proposition}\label{prop:july19_1}
Let $\mc V$ be a normal algebra of differential functions.
Then
\begin{equation}\label{eq:may4_1}
\Ker \frac\delta{\delta u} \,=\, \mc C\oplus\partial\mc V\,.
\end{equation}
In fact, given $f\in\mc V_{n,i}$ such that $\frac{\delta f}{\delta u}=0$,
we have $\frac{\partial f}{\partial u_i^{(n)}}\in\mc V_{n-1,i}$, and
letting
$$
g_{n-1,i}=\int du_i^{(n-1)}\frac{\partial f}{\partial u_i^{(n)}}\,\in\mc V_{n-1,i}\,,
$$
we have $f-\partial g_{n-1,i}\in\mc V_{n,i-1}$.
This allows one to compute an element $g\in\mc V$ such that
$f=\partial g+$const. by induction on the pair $(n,i)$ in the lexicographic order.
\end{proposition}
\begin{proof}
Equality \eqref{eq:may4_1} is a special case of exactness of the variational complex,
which will be established in Theorem \ref{th:july23}.
Hence, if $f\in\Ker\frac{\delta }{\delta u}\cap\mc V_{n,i}$, 
we have $f=\partial g+c$ for some $g\in\mc V_{n-1,i}$ and $c\in\mc C$.
By equation \eqref{dec27_1_new},
$\frac{\partial f}{\partial u_i^{(n)}}\in\mc V_{n-1,i}$  for every $i\in I$.
The rest immediately follows by the definition \eqref{partial} of $\partial$.
\end{proof}
\begin{example}\label{ex:may4}
In general \eqref{eq:may4_1} does not hold. For example, in the non-normal algebra
$\mc V=\mb C[u^{\pm1},u^{\prime},u^{\prime\prime},\dots]$ the element $\frac{u^{\prime}}u$
is in $\Ker\frac\delta{\delta u}$, but not in $\partial\mc V$.
But in a normal extension of $\mc V$ this element can be written as $\partial\log u$.
\end{example}

In order to state the next proposition we need to introduce the Fr\'echet derivative.
\begin{definition}\label{def:july19}
The \emph{Fr\'echet derivative} $D_f$ of $f\in\mc V$ is defined as the following
differential operator from $\mc V^\ell$ to $\mc V$:
$$
{D_f(\partial)}P
\,=\,
X_P(f)
\,=\,
\sum_{i\in I,\,n\in\mb Z_+}\frac{\partial f}{\partial u_i^{(n)}}\partial^nP_i\,.
$$
\end{definition}
One can think of $D_f(\partial)P$
as the differential of the function $f(u)$
in the usual sense, that is, up to higher order in $P$,
$$
f(u+P)-f(u) \approx \sum_{j\in I,n\in\mb Z_+}
\frac{\partial f(u)}{\partial u_i^{(n)}} P_i^{(n)} 
= D_f(\partial)P\,,
$$
hence the name Fr\'echet derivative.

More generally, for any collection $F=(f_\alpha)_{\alpha\in \mc A}$ of elements of $\mc V$
(where $\mc A$ is an index set),
the corresponding Fr\'echet derivative is 
the map $D_F:\,\mc V^\ell\to\mc V^{\mc A}$ given by 
\begin{equation}\label{eq:july18_5}
(D_F(P))_\alpha=D_{f_\alpha}(\partial)P\,\Big(=X_P(f_\alpha)\Big),\,\alpha\in \mc A\,.
\end{equation}
Its adjoint with respect to the pairing \eqref{eq:may4_4} 
is the linear map $D^*_F(\partial):\, \mc V^{\oplus \mc A}\to \mc V^{\oplus\ell}$, given by
\begin{equation}\label{eq:may4_7}
\big({D^*_F(\partial)G}\big)_{i}=\sum_{\alpha\in \mc A,\,n\in\mb{Z}_+}(-\partial)^n 
\Big( 
\frac{\partial F_\alpha}{\partial u_i^{(n)}}G_\alpha
\Big) 
\,\,,\,\,\,\,
i\in I\,.
\end{equation}
On the other hand, if $F$ has only finitely many non-zero entries, namely 
$F\in\mc V^{\oplus \mc A}\subset\mc V^{\mc A}$, it is clear from the definition 
\eqref{eq:may4_7} that $D_F^*$ can be extended to a map $\mc V^{\mc A}\to\mc V^{\oplus\ell}$.
In particular, for $F\in\mc V^{\oplus\ell}$, both $D_F(\partial)$ and $D^*_F(\partial)$
are maps from $\mc V^\ell$ to $\mc V^{\oplus\ell}$.

The following lemma will be useful in computations with Fr\'echet derivatives.
\begin{lemma}\label{lem:20090323}
\begin{enumerate}
\alphaparenlist
\item Both the Fr\'echet derivative $D_F(\partial)$ and the adjoint operator $D_F^*(\partial)$
are linear in $F\in\mc V^{\mc A}$.
\item Let $M$ be an $\mc A\times\mc B$ matrix with entries in $\mc V$.
We have, for $F\in\mc V^{\mc B},\,G\in\mc V^{\oplus A}$ and $P\in\mc V^{\ell}$,
$$
\begin{array}{c}
\displaystyle{
D_{MF}(\partial)P
=
\sum_{\beta\in\mc B} F_\beta \big(D_{M_{\alpha\beta}}(\partial)P\big)
+\sum_{\beta\in\mc B} M_{\alpha\beta} \big(D_{F_\beta}(\partial)P\big) 
}\\
\displaystyle{
= 
\big(D_{M}(\partial)P\big)\cdot F + M\cdot\big(D_{F}(\partial)P\big)\,,
}
\end{array}
$$
and
$$
\begin{array}{c}
\displaystyle{
D^*_{MF}(\partial)G
=
\sum_{\alpha\in\mc A,\beta\in\mc B} D^*_{M_{\alpha\beta}}(\partial) \big(F_\beta G_\alpha\big)
+\sum_{\alpha\in\mc A,\beta\in\mc B} D^*_{F_\beta}(\partial) \big(M_{\alpha\beta} G_\alpha\big) 
}\\
\displaystyle{
=
D^*_{M}(\partial) \big(G^T\otimes F\big) + D^*_{F}(\partial) \big(G^TM\big)\,.
}
\end{array}
$$
\item 
Let $F\in\mc V^{\mc A},\,G\in\mc V^{\oplus\mc A}$ and $P\in\mc V^\ell$.
We have
$$
D_{\partial F}(\partial)P=\partial \big(D_{F}(\partial)P\big)
\,\,,\,\,\,\,
D_{\partial F}^*(\partial)G=-D_{F}^*(\partial)(\partial G)\,.
$$
\end{enumerate}
\end{lemma}
\begin{proof}
Part (a) is obvious. For part (b) the first equation follows immediately from the definition
\eqref{eq:july18_5} of the Fr\'echet derivative,
while the second equation can be derived taking the adjoint operator.
Part (c) can be easily proved using equation \eqref{eq:comm_frac}.
\end{proof}

We have the following formula for the commutator of evolutionary vector fields in terms
of Fr\'echet derivatives:
\begin{equation}\label{eq:09jan21_1}
[X_P,X_Q]
\,=\,
X_{D_Q(\partial)P-D_P(\partial)Q}\,.
\end{equation}

Elements of the form $\frac{\delta f}{\delta u}\in\mc V^{\oplus\ell}$ are called \emph{exact}.
An element $F\in\mc V^{\oplus\ell}$ is called \emph{closed}
if its Fr\'echet derivative is a self-adjoint differential operator:
\begin{equation}\label{eq:july18_5_b}
{D_F^*(\partial)}\,=\, {D_F(\partial)}\,.
\end{equation}
It is well known and not hard to check, applying Lemma \ref{2006_l-may31} twice,
that any exact element in $\mc V^{\oplus\ell}$ is closed:
\begin{equation}\label{eq:july18_5_c}
{D_{\frac{\delta f}{\delta u}}^*(\partial)}\,=\, {D_{\frac{\delta f}{\delta u}}(\partial)}
\,\,\,\,\,\,\,
\text{ for every } \tint f\in\mc V/\partial\mc V\,. 
\end{equation}

If $\mc V$ is normal, the converse holds. In fact,
the next proposition provides an algorithm for constructing 
$f\in\mc V$ such that $\frac{\delta f}{\delta u}=F$, provided that $F$ is closed,
by induction on the following filtration of $\mc V^{\oplus\ell}$.
Given a triple $(n,i,j)$ with $n\in\mb Z_+$ and $i,j\in I$,
the subspace
$\big(\mc V^{\oplus\ell}\big)_{n,i,j}\subset\mc V^{\oplus\ell}$
consists of elements $F\in\mc V^{\oplus\ell}$ such that
$F_k\in\mc V_{n,i}$ for every $k\in I$ and $j$ is the maximal index such that 
$F_j\in\mc V_{n,i}\backslash\mc V_{n,i-1}$.
We also let $\big(\mc V^{\oplus\ell}\big)_{n,0,j}=\big(\mc V^{\oplus\ell}\big)_{n-1,\ell,j}$
and $\big(\mc V^{\oplus\ell}\big)_{n,i,0}=\big(\mc V^{\oplus\ell}\big)_{n,i-1,\ell}$.
\begin{proposition}\label{prop:july19_2}
Let $\mc V$ be a normal algebra of differential functions.
Then $F\in\mc V^{\oplus\ell}$ is closed if and only if
it is exact, i.e.\ $F=\frac{\delta f}{\delta u}$ for some $f\in\mc V$.
In fact, if $F\in\big(\mc V^{\oplus\ell}\big)_{n,i,j}$ is closed,
then there are two cases:
\begin{enumerate}
\alphaparenlist
\item if $n$ is even, we have $j\leq i$ and $\frac{\partial F_j}{\partial u_i^{(n)}}\in\mc V_{n/2,j}$, and
letting
\begin{equation}\label{eq:A}
f_{n,i,j} = (-1)^{n/2}\int du_i^{(n/2)}\int du_j^{(n/2)}\frac{\partial F_j}{\partial u_i^{(n)}}\,\in\,\mc V_{n/2,i}\,,
\end{equation}
we have $F-\frac{\delta f_{n,i,j}}{\delta u}\in\big(\mc V^{\oplus\ell}\big)_{n,i,j-1}$; 
\item if $n$ is odd, we have $j<i$ and $\frac{\partial F_j}{\partial u_i^{(n)}}\in\mc V_{(n-1)/2,i}$,
and letting
\begin{equation}\label{eq:B}
f_{n,i,j} = (-1)^{(n+1)/2}\int du_j^{((n+1)/2)}\int du_i^{((n-1)/2)}\frac{\partial F_j}{\partial u_i^{(n)}}\,\in\,\mc V_{(n+1)/2,j}\,,
\end{equation}
we have $F-\frac{\delta f_{n,i,j}}{\delta u}\in\big(\mc V^{\oplus\ell}\big)_{n,i,j-1}$. 
\end{enumerate}
(Note that $f_{n,i,j}$ in \eqref{eq:A} is defined up to adding an arbitrary
element $g\in\mc V_{n/2,i}$ such that $\frac{\partial g}{\partial u_i^{(n/2)}}\in\mc V_{n/2,j-1}$,
and similarly for $f_{n,i,j}$ in \eqref{eq:B}.)
Equations \eqref{eq:A} and \eqref{eq:B} allow one to compute an element $f\in\mc V$ such that
$F=\frac{\delta f}{\delta u}$ by induction on the triple $(n,i,j)$ in the lexicographic order.
\end{proposition}
\begin{proof}
The first statement of the proposition is a special case of the exactness of the complex
of variational calculus, which will be established in Theorem \ref{th:july23}.
Hence there exists $f\in\mc V$, defined up to adding elements from $\partial\mc V$,
such that $F=\frac{\delta f}{\delta u}$.
\begin{lemma}\label{lem:july22}
\begin{enumerate}
\alphaparenlist
\item Up to adding elements from $\partial\mc V$, $f$ can be chosen such that
$f\in\mc V_{m,k}\backslash\mc V_{m,k-1}$ 
and $\frac{\partial f}{\partial u_k^{(m)}}\in\mc V_{m,k}\backslash\mc V_{m-1,k}$ 
for some $m\in\mb Z_+$ and $k\in I$.
\item If $f$ is chosen as in part (a), 
let $(n,i,j)$ be the maximal triple (with respect to the lexicographic order)
such that
\begin{equation}\label{eq:july22_2}
\frac{\partial^2 f}{\partial u_i^{(n-p)}\partial u_j^{(p)}}\,\neq\,0\,\,\,\,
\text{ for some } p\in\mb Z_+\,.
\end{equation}
Then, there are two possibilities:
\begin{enumerate}
\item[A.] $n=2m,\, j\leq i\leq k$;
\item[B.] $n=2m-1,\,i>j=k$.
\end{enumerate}
In both cases the only $p\in\mb Z_+$ for which \eqref{eq:july22_2} holds is $p=m$.
\item $\frac{\delta f}{\delta u}\in\big(\mc V^{\oplus\ell}\big)_{n,i,j}\backslash\big(\mc V^{\oplus\ell}\big)_{n,i,j-1}$, 
where $(n,i,j)$ is as in (b).
\end{enumerate}
\end{lemma}
\begin{proof}
For (a) it suffices to notice that, if $f\in\mc V_{m,k}\backslash\mc V_{m,k-1}$ is such that
$\frac{\partial f}{\partial u_k^{(m)}}\in\mc V_{m-1,k}$,
then
$$
f-\partial\int du_k^{(m-1)} \frac{\partial f}{\partial u_k^{(m)}}\,\in\,\mc V_{m,k-1}\,,
$$
and after finitely many such steps we arrive at the $f$ that we want.

Let $(n,i,j)$ be as in (b). Notice that, since condition \eqref{eq:july22_2} is symmetric in $i$ and $j$,
we necessarily have $i\geq j$.
Since $f\in\mc V_{m,k}$, it is clear that $(n,i,j)\leq(2m,k,k)$.
Moreover, since, by (a), $\frac{\partial f}{\partial u_k^{(m)}}\in\mc V_{m,k}\backslash\mc V_{m-1,k}$,
it easily follows that $(n,i,j)\geq(2m-1,k+1,k)$ in the lexicographic order.
These two inequalities immediately imply that either A.\ or B.\ occurs.
In order to prove (c) we consider the two cases A. and B. separately.
In case A. condition \eqref{eq:july22_2} exactly means that:
\begin{enumerate}
\item $\frac{\partial f}{\partial u_h^{(m)}}\in\mc V_{m,i-1}$ for every $h>j$,
so that $\frac{\delta f}{\delta u_h}\in\mc V_{2m,i-1}$; 
\item $\frac{\partial f}{\partial u_j^{(m)}}\in\mc V_{m,i}\backslash\mc V_{m,i-1}$,
so that $\frac{\delta f}{\delta u_j}\in\mc V_{2m,i}\backslash\mc V_{2m,i-1}$;
\item $\frac{\partial f}{\partial u_h^{(m)}}\in\mc V_{m,i}$ for every $h<j$,
so that $\frac{\delta f}{\delta u_h}\in\mc V_{2m,i}$.
\end{enumerate}
By definition, the above conditions imply that $\frac{\delta f}{\delta u}\in\big(\mc V^{\oplus\ell}\big)_{2m,i,j}$.
Case B. is treated in a similar way.
\end{proof}
Returning to the proof of Proposition \ref{prop:july19_2}, let $F=\frac{\delta f}{\delta u}$, where
$f\in\mc V$ is chosen as in Lemma \ref{lem:july22}(a).
We then have
$$
\frac{\partial F_j}{\partial u_i^{(n)}}
=\frac{\partial}{\partial u_i^{(n)}}\frac{\delta f}{\delta u_j}
=\frac{\partial}{\partial u_i^{(n)}}\Big((-\partial)^m\frac{\partial f}{\partial u_j^{(m)}}
+(-\partial)^{m-1}\frac{\partial f}{\partial u_j^{(m-1)}}+\cdots\Big)
=(-1)^m\frac{\partial^2 f}{\partial u_i^{(n-m)}\partial u_j^{(m)}}\,,
$$
where, for the last equality, we used the commutation relation \eqref{eq:comm_frac}
and Lemma \ref{lem:july22}(b)-(c).
The rest follows easily.
\end{proof}

In Section \ref{2006_sec2} we shall prove exactness of the whole complex $\Omega$ 
in two different contexts.
First, under the assumption that $\mc V$ is normal,
and second,
under an assumption involving the \emph{degree evolutionary vector field}:
\begin{equation}\label{2006_degop}
\Delta=\sum_{\substack{1\leq i\leq\ell \\ n\in\mb{Z}_+}} u_i^{(n)} \tfrac{\partial}{\partial u_i^{(n)}}\,,
\end{equation}
for which we are going to need the following commutation rules:
\begin{equation}\label{eq:mar11_3}
\partial\,\Delta\,=\,\Delta\,\partial\,\,,\,\,\,\,
\frac{\partial}{\partial u_i^{(n)}}\,\Delta\,=\,(\Delta+1)\,\frac{\partial}{\partial u_i^{(n)}}\,\,,\,\,\,\,
\frac{\delta}{\delta u_i}\,\Delta\,=\,(\Delta+1)\,\frac{\delta}{\delta u_i}\,.
\end{equation}
Using this, one finds a simpler formula (if applicable) 
for the preimages of the
variational derivative, given by the following well-known result.
\begin{proposition}\label{prop:july22}
Let $\mc V$ be an algebra of differential functions
and suppose that $F\in\mc V^{\oplus\ell}$ is closed, namely \eqref{eq:july18_5_b} holds.
Then, if $f\in\Delta^{-1}(u\cdot F)$, we have
\begin{equation}\label{eq:may3_3_new}
\frac{\delta f}{\delta u_i}-F_i\,\in\,\Ker(\Delta+1)\,,\,\,\text{for all\,\,} i\in I\,,
\end{equation}
where $u\cdot F=\sum_{i\in I}u_iF_i$.
\end{proposition}
\begin{proof}
We have, for $i\in I$,
\begin{eqnarray*}
\frac{\delta}{\delta u_i}(u\cdot F)
&=& \sum_{j\in I,n\in\mb Z_+}(-\partial)^n\frac{\partial}{\partial u_i^{(n)}}\big(u_j F_j\big)
= F_i+ \sum_{j\in I,n\in\mb Z_+}(-\partial)^n \Big(\frac{\partial F_j}{\partial u_i^{(n)}} u_j\Big) \nonumber\\
&=& F_i+ \sum_{j\in I,n\in\mb Z_+} \frac{\partial F_i}{\partial u_j^{(n)}} \partial^n u_j
= (\Delta+1)F_i\,.
\end{eqnarray*}
In the third equality we used the assumption that $F$ is closed,
i.e.\ it satisfies \eqref{eq:july18_5_b},
while in the last equality we used definition \eqref{2006_degop}
of the degree evolutionary vector field.
If we then substitute $u\cdot F=\Delta f$ in the LHS and use the last equation in \eqref{eq:mar11_3},
we immediately get \eqref{eq:may3_3_new}.
\end{proof}
\begin{example}\label{eq:july22}
Consider the closed element 
$F=(u^\prime_3,\,-u_2^{\prime\prime},\,-u_1^{\prime})\in R^3_{2,2,2}$,
where $R$ is the algebra of differential polynomials in $u_1,\,u_2,\,u_3$.
Since $\Ker(\Delta+1)=0$ in $R$, we can apply Proposition \ref{prop:july22}
to find that $f=\Delta^{-1}(u_1u_3^\prime-u_2u_2^{\prime\prime}-u_3u_1^{\prime})=\frac12(u_1u_3^{\prime}-u_2u_2^{\prime\prime}-u_3u_1^{\prime})$
satisfies $F=\frac{\delta f}{\delta u}$.

On the other hand, consider the localization $\mc V$ by $u_4$ of the algebra of differential
polynomials in $u_1,\,u_2,\,u_3,\,u_4$ and the following closed element
$F\in(\mc V^4)_{2,2,2}$:
$$
F_1=u_4^{-2}u_3^{\prime}\,,\,\,F_2=2u_4^{-3}u_2^{\prime}u_4^{\prime}-u_4^{-2}u_2^{\prime\prime}\,,\,\,
F_3=2u_4^{-3}u_1u_4^{\prime}-u_4^{-2}u_1^{\prime}\,,\,\,
F_4=-u_4^{-3}u_1u_3^{\prime}-u_4^{-3}(u_2^{\prime})^2)\,.
$$
In this case $\Delta(u\cdot F)=0$, hence  formula \eqref{eq:may3_3_new} is not applicable,
but we still can use the algorithm provided by Proposition \ref{prop:july19_2},
to find $f=f_{2,2,2}+f_{1,4,3}$, where
$f_{2,2,2}=\frac12 u_4^{-1}(u_2^{\prime})^2$ and $f_{1,4,3}=u_4^{-2}u_1u_3^{\prime}$.
\end{example}

\vspace{3ex}
\subsection{Poisson vertex algebras.}~~
\label{sec:1.1}
\begin{definition}\label{def:v11}
Let $\mc V$ be a $\mb C[\partial]$-module.
A $\lambda$-\emph{bracket} on $\mc V$ is a $\mb C$-linear map
$\mc V\otimes\mc V\to\mb C[\lambda]\otimes\mc V$,
denoted by $f\otimes g\mapsto\{f_\lambda g\}$,
which is \emph{sesquilinear}, namely one has ($f,g,h\in\mc V$)
\begin{equation}\label{sesq}
  \{ \partial f_\lambda g \} =- \lambda \{ f_\lambda g \} \, , \, 
      \{  f_\lambda \partial g \} = (\partial + \lambda) \{ f_\lambda g \} \, .
\end{equation}
If, moreover, $\mc V$ is a 
commutative associative unital differential algebra with a derivation 
$\partial$,
a $\lambda$-\emph{bracket} on $\mc V$ 
is defined to obey, in addition, the \emph{left Leibniz rule}
\begin{equation}
  \label{eq:1.7}
  \{ f_\lambda gh \} = \{ f_\lambda g \} h+\{ f_\lambda h\} g \, ,
\end{equation}
and the \emph{right Leibniz rule}
\begin{equation}
  \label{eq:1.8}
  \{ fg_\lambda h \} = \{ f_{\lambda +\partial} h \}_{\to} g + 
      \{ g_{\lambda +\partial} h\}_{\to} f \, .
\end{equation}
\end{definition}

One writes $\{f_\lambda g\}=\sum_{n\in\mb Z_+}\frac{\lambda^n}{n!}(f_{(n)}g)$,
where the $\mb C$-bilinear products $f_{(n)}g$ are called the $n$-th products on $\mc V$ 
($n\in\mb Z_+$), and $f_{(n)}g=0$ for $n$ sufficiently large.
In \eqref{eq:1.8} and further on, the arrow on the right means that $\lambda +\partial$ 
should be moved to the right (for example, 
$\{ f_{\lambda +\partial} h\}_{\to} g 
= \sum_{n \in  \mb{Z}_+} (f_{(n)}h) \frac{(\lambda +\partial)^n}{n!} g$).  

An important property of a $\lambda$-bracket $\{.\,_\lambda\,.\}$ is 
\emph{commutativity} (resp. \emph{skew-commutativity}):
\begin{equation}
  \label{eq:1.6}
  \{ g_\lambda f \} =\, _\leftarrow\!\{ f_{-\lambda-\partial} g \}\,\, (\hbox{resp.  } 
     = - \,_\leftarrow\!\{ f_{-\lambda -\partial}g \} )\,.
\end{equation}
Here and further the arrow on the left means that $-\lambda-\partial$ is moved to the left, i.e.\ 
$_\leftarrow\!\{ f_{-\lambda-\partial} g \}
=\sum_{n \in  \mb{Z}_+} \frac{(-\lambda-\partial)^n}{n!} (f_{(n)}g)
=e^{\partial\frac d{d\lambda}}\{ f_{-\lambda} g \}$. 
In case there is no arrow, it is assumed to be to the left.

Another important property of a $\lambda$-bracket 
$\{\cdot\,_\lambda\,\cdot\}$ is the \emph{Jacobi identity}:
\begin{equation}
  \label{eq:1.10}
  \{ f_\lambda \{ g_\mu h \}\} - \{g_\mu \{f_\lambda h \}\}
     = \{\{f_\lambda g \}_{\lambda + \mu}h \} \, .
\end{equation}

\begin{definition}
  \label{def:1.10}
A \emph{Poisson vertex algebra} (PVA) is a differential algebra $\mc V$
with a $\lambda$-bracket 
$\{\cdot\,_\lambda\,\cdot\}:\,\mc V\otimes\mc V\to\mb C[\lambda]\otimes\mc V$
(cf.\ Definition \ref{def:v11}) satisfying 
skew-commutativity \eqref{eq:1.6} and Jacobi identity \eqref{eq:1.10}.
A \emph{Lie conformal algebra} is a $\mb C[\partial]$-module, 
endowed with a $\lambda$-bracket, satisfying the same two properties \cite{K}.
\end{definition}

We next want to explain how to extend an arbitrary ``non-linear" $\lambda$-bracket 
on a set of variables $\{u_i\}_{i\in I}$
with value in some algebra $\mc V$ of differential functions,
to a Poisson vertex algebra structure on $\mc V$.
\begin{theorem}\label{prop:exten}
Let $\mc V$ be an algebra of differential functions,
which is an extension of the algebra of 
differential polynomials $R=\mb C[u_i^{(n)}\,|\,i\in I,n\in\mb Z_+]$.
For each pair $i,j\in I$, choose $\{{u_i}_\lambda u_j\}\in\mb C[\lambda]\otimes\mc V$.
\begin{enumerate}
\alphaparenlist
\item Formula
\begin{equation}\label{eq:1.9}
\{f_\lambda g\}\,=\,
\sum_{\substack{i,j\in I \\ m,n\in\mb Z_+}}
\frac{\partial g}{\partial u_j^{(n)}}(\lambda+\partial)^n
\{{u_i}_{\lambda+\partial}u_j\}_\to (-\lambda-\partial)^m \frac{\partial f}{\partial u_i^{(m)}}\,.
\end{equation}
defines a $\lambda$-bracket on $\mc V$ (cf.\ Definition \ref{def:v11}),
which extends the given $\lambda$-brackets on the generators $u_i,\,i\in I$.
\item The $\lambda$-bracket \eqref{eq:1.9} on $\mc V$ satisfies the commutativity 
(resp. skew-commutativity) condition \eqref{eq:1.6},
provided that the same holds on generators:
\begin{equation}\label{eq:feb5_1}
\{{u_i}_\lambda u_j\}\,=\,\pm \,_\leftarrow\!\{ {u_j}_{-\lambda -\partial}u_i \}\,,\,\,\text{for all\,\,} i,j\in I\,.
\end{equation}
\item Assuming that  the skew-commutativity condition \eqref{eq:feb5_1} holds,
the $\lambda$-bracket \eqref{eq:1.9} satisfies the Jacobi identity \eqref{eq:1.10}
(thus making $\mc V$ a PVA),
provided that the Jacobi identity holds on any triple of generators:
\begin{equation}\label{eq:feb5_2}
\{{u_i}_\lambda \{{u_j}_\mu u_k\}\}
- \{{u_j}_\mu \{{u_i}_\lambda u_k\}\}
= \{{\{{u_i}_\lambda {u_j}\}}_{\lambda+\mu} u_k\}\,,\,\,\text{for all\,\,} i,j,k\in I\,.
\end{equation}
\end{enumerate}
\end{theorem}
\begin{proof}
First, notice that the sum in equation \eqref{eq:1.9} is finite (cf.\ Definition \ref{def:diff_alg}),
so that  \eqref{eq:1.9} gives a well defined $\mb C$-linear
map $\{\cdot\,_\lambda\,\cdot\}:\,\mc V\otimes\mc V\to\mb C[\lambda]\otimes\mc V$.
Moreover, for $f=u_i,\,g=u_j,\,i,j\in I$, such map clearly reduces to the
given polynomials $\{{u_i}_\lambda u_j\}\in\mb C[\lambda]\otimes\mc V$.
More generally, for $f=u_i^{(m)},\,g=u_j^{(n)}$,
equation \eqref{eq:1.9} reduces to
\begin{equation}\label{eq:feb5_5}
\{{u_i^{(m)}}_\lambda u_j^{(n)}\}
=(-\lambda)^m(\lambda+\partial)^n \{{u_i}_\lambda u_j\}\,.
\end{equation}
It is also useful to rewrite equation \eqref{eq:1.9} in the following equivalent forms, 
which can be checked directly:
\begin{equation}\label{eq:1.9_cons}
\{f_\lambda g\} 
= \sum_{j\in I, n\in\mb Z_+}
\frac{\partial g}{\partial u_j^{(n)}}
(\lambda+\partial)^n \{f\, _\lambda\, u_j\}
= \sum_{i\in I,m\in\mb Z_+}
\{{u_i}\, _{\lambda+\partial}\, g\}_\to (-\lambda-\partial)^m \frac{\partial f}{\partial u_i^{(m)}} \,.
\end{equation}
or, using \eqref{eq:feb5_5}, in the following further form:
\begin{equation}\label{eq:feb5_6}
\{f_\lambda g\}\,=\,
\sum_{\substack{i,j\in I \\ m,n\in\mb Z_+}}
\Big(e^{\partial\frac{d}{d\lambda}} \frac{\partial f}{\partial u_i^{(m)}}\Big)
\{{u_i^{(m)}}\, _{\lambda}\, u_j^{(n)}\} 
\frac{\partial g}{\partial u_j^{(n)}}\,,
\end{equation}
where the parentheses indicate that $\partial$ in the exponent is acting only on 
$\frac{\partial f}{\partial u_i^{(m)}}$,
but $\frac{d}{d\lambda}$ acts on $\lambda$.
Here and further we use the following identity for $f(\lambda)\in\mc V[\lambda]$,
coming from Taylor's formula:
$$
f(\partial+\lambda)u=\big(e^{\partial\frac{d}{d\lambda}}u\big)f(\lambda)\,.
$$

We start by proving the sesquilinearity relations \eqref{sesq}.
For the first one we have, by the second identity in \eqref{eq:1.9_cons} and 
the commutativity relation \eqref{eq:comm_frac}:
\begin{eqnarray*}
\{\partial f\, _\lambda\, g\} 
&=& \sum_{i\in I,m\in\mb Z_+}
\{{u_i}\, _{\lambda+\partial}\, g\}_\to 
(-\lambda-\partial)^m \Big(\frac{\partial }{\partial u_i^{(m)}}\partial f\Big) \\
&=& \sum_{i\in I,m\in\mb Z_+}
\{{u_i}\, _{\lambda+\partial}\, g\}_\to 
(-\lambda-\partial)^m \Big(\frac{\partial f}{\partial u_i^{(m-1)}} 
+ \partial \frac{\partial f}{\partial u_i^{(m)}} \Big) \\
&=& -\lambda \sum_{i\in I,m\in\mb Z_+}
\{{u_i}\, _{\lambda+\partial}\, g\}_\to 
(-\lambda-\partial)^m
\frac{\partial f}{\partial u_i^{(m)}} 
\,=\, -\lambda \{f\, _\lambda\, g\}\,,
\end{eqnarray*}
where, as in \eqref{eq:comm_frac}, 
we replace $\frac{\partial f}{\partial u_i^{(m-1)}}$ by zero for $m=0$.
Similarly, for the first sesquilinearity condition, we can use the first identity in \eqref{eq:1.9_cons}
and \eqref{eq:comm_frac} to get:
\begin{eqnarray*}
\{f_\lambda \partial g\} 
&=& \sum_{j\in I, n\in\mb Z_+}
\Big(\frac{\partial }{\partial u_j^{(n)}}\partial g\Big)
(\lambda+\partial)^n \{f\, _\lambda\, u_j\} \\
&=& \sum_{j\in I, n\in\mb Z_+}
\Big(\frac{\partial g}{\partial u_j^{(n-1)}} + \partial \frac{\partial g}{\partial u_j^{(n)}}\Big)
(\lambda+\partial)^n \{f\, _\lambda\, u_j\} \\
&=& (\lambda+\partial) \sum_{j\in I, n\in\mb Z_+}
\frac{\partial g}{\partial u_j^{(n)}}
(\lambda+\partial)^n \{f\, _\lambda\, u_j\} 
\,=\, (\lambda+\partial) \{f_\lambda g\} \,.
\end{eqnarray*}
In order to complete the proof of part (a) we are left to prove the left and right Leibniz rules
\eqref{eq:1.7} and \eqref{eq:1.8}.
For both we use that the partial derivatives $\frac\partial{\partial u_i^{(m)}}$
are derivations of the product in $\mc V$.
For the left Leibniz rule, we use the first identity in \eqref{eq:1.9_cons} to get:
$$
\{f_\lambda gh\} 
= \!\!\! \sum_{j\in I, n\in\mb Z_+}\!\!\! 
\Big(h\frac{\partial g}{\partial u_j^{(n)}}+g\frac{\partial h}{\partial u_j^{(n)}}\Big)
(\lambda+\partial)^n \{f\, _\lambda\, u_j\}
= h \{f_\lambda g\} + g \{f_\lambda h\} \,,
$$
and similarly, for the right Leibniz rule, we can use the second identity in \eqref{eq:1.9_cons} to get:
$$
\{fg _\lambda h\} 
= \!\!\! \!\!\! \sum_{i\in I,m\in\mb Z_+}\!\!\! \!\!\! 
\{{u_i}\, _{\lambda+\partial}\, h\}_\to (-\lambda-\partial)^m 
\Big(\frac{\partial f}{\partial u_i^{(m)}}g + \frac{\partial g}{\partial u_i^{(m)}}f\Big)
= \{f _{\lambda+\partial} h\}_\to g + \{g _{\lambda+\partial} h\}_\to f\,.
$$
We next prove part (b). With the notation introduced in equation \eqref{eq:1.6}
we have, by \eqref{eq:1.9}:
\begin{equation}\label{eq:feb5_3}
_\leftarrow\!\{ g_{-\lambda-\partial} f \}
= e^{\partial\frac d{d\lambda}}\{ g_{-\lambda} f \}
= e^{\partial\frac d{d\lambda}} 
\sum_{\substack{i,j\in I \\ m,n\in\mb Z_+}}
\frac{\partial f}{\partial u_i^{(m)}}(-\lambda+\partial)^m
\{{u_j} _{-\lambda+\partial} u_i\}_\to (\lambda-\partial)^n \frac{\partial g}{\partial u_j^{(n)}}\,.
\end{equation}
It immediately follows from \eqref{eq:feb5_1} that
$\{{u_j}_{-\lambda+\partial}u_i\}_\to F=
\pm e^{- \partial\frac d{d\lambda}}  \Big(\{{u_i} _\lambda u_j\} F\Big)$,
for arbitrary $F\in\mc V$.
Equation \eqref{eq:feb5_3} then gives
\begin{eqnarray*}
_\leftarrow\!\{ g_{-\lambda-\partial} f \}
&=& \pm \sum_{\substack{i,j\in I \\ m,n\in\mb Z_+}}
\frac{\partial g}{\partial u_j^{(n)}} (\lambda+\partial)^n
\{{u_i} _{\lambda+\partial} u_j\}_\to
(-\lambda-\partial)^m
\frac{\partial f}{\partial u_i^{(m)}} = \pm \{ f_{\lambda} g \} \,,
\end{eqnarray*}
thus proving (skew-)commutativity.

We are left to prove part (c). 
First, observe that, since, by \eqref{eq:feb5_2}, Jacobi identity holds on any 
triple of generators $u_i,u_j,u_k$,
applying $(-\lambda)^m(-\mu)^n(\lambda+\mu+\partial)^p$ to both sides of \eqref{eq:feb5_2}
and using sesquilinearity, Jacobi identity holds on any triple of elements of type
$u_i^{(m)}$ for $i\in I$ and $m\in\mb Z_+$:
\begin{equation}\label{eq:feb5_7}
\{{u_i^{(m)}}\, _\lambda \{{u_j^{(n)}}\ _\mu\, u_k^{(p)}\}\}
- \{{u_j^{(n)}}\, _\mu \{{u_i^{(m)}}\, _\lambda\, u_k^{(p)}\}\}
= \{{\{{u_i^{(m)}}\, _\lambda\, {u_j^{(n)}}\}}_{\lambda+\mu} u_k^{(p)}\}\,.
\end{equation}
We have to prove that for every $f,g,h\in\mc V$
Jacobi identity \eqref{eq:1.10} holds as well.
We will study separately each term in equation \eqref{eq:1.10}.
For the first term in the LHS of \eqref{eq:1.10} we have, by the first identity in \eqref{eq:1.9_cons},
combined with sesquilinearity \eqref{sesq} and the left Leibniz rule \eqref{eq:1.7}:
\begin{eqnarray}\label{eq:feb6_1}
&\displaystyle{
\{f_\lambda \{g_\mu h\}\}
\,=\, \sum_{k\in I,p\in\mb Z_+} 
\Big\{f\, _\lambda \frac{\partial h}{\partial u_k^{(p)}}\{g\, _\mu\, u_k^{(p)}\}\Big\} 
}\nonumber\\
&\displaystyle{
= \sum_{k\in I,p\in\mb Z_+} 
\Big\{f\, _\lambda \frac{\partial h}{\partial u_k^{(p)}}\Big\} 
\{g\, _\mu\, u_k^{(p)}\}
+ \sum_{k\in I,p\in\mb Z_+} 
\frac{\partial h}{\partial u_k^{(p)}}
\{f\, _\lambda \{g\, _\mu\, u_k^{(p)}\}\}
}
\end{eqnarray}
The fist term in the RHS of \eqref{eq:feb6_1} can be rewritten,
using again the first identity in \eqref{eq:1.9_cons}, as
\begin{equation}\label{eq:feb6_2}
\sum_{\substack{k,l\in I \\ p,q\in\mb Z_+}} 
\frac{\partial^2 h}{\partial u_l^{(q)} \partial u_k^{(p)}}
\{f\, _\lambda u_l^{(q)}\}  \{g\, _\mu\, u_k^{(p)}\}\,.
\end{equation}
Notice that this expression is unchanged if we switch $f$ with $g$ and $\lambda$ with $\mu$,
therefore it does not give any contribution to the LHS of \eqref{eq:1.10}.
For the second term in the RHS of \eqref{eq:feb6_1},
we apply twice the second identity in \eqref{eq:1.9_cons}, combined 
with sesquilinearity \eqref{sesq}
and the left Leibniz rule \eqref{eq:1.7}, to get:
\begin{eqnarray}\label{eq:feb6_3}
&&
\sum_{k\in I,p\in\mb Z_+} 
\frac{\partial h}{\partial u_k^{(p)}}
\{f\, _\lambda \{g\, _\mu\, u_k^{(p)}\}\} 
\nonumber\\
&&\,\,\,\,\,\,\,\,\,\,\,\,
\,=\, \sum_{\substack{i,j,k\in I \\ m,n,p\in\mb Z_+}} 
\frac{\partial h}{\partial u_k^{(p)}}
\Big(e^{\partial\frac{d}{d\lambda}} \frac{\partial f}{\partial u_i^{(m)}}\Big)
\Big\{{u_i^{(m)}}\, _\lambda 
\Big(e^{\partial\frac{d}{d\mu}} \frac{\partial g}{\partial u_j^{(n)}}\Big)
\{{u_j^{(n)}}\, _\mu\, u_k^{(p)}\}
\Big\}
\nonumber \\
&&\,\,\,\,\,\,\,\,\,\,\,\,
=\, \sum_{\substack{i,j,k\in I \\ m,n,p\in\mb Z_+}} 
\frac{\partial h}{\partial u_k^{(p)}}
\Big(e^{\partial\frac{d}{d\lambda}} \frac{\partial f}{\partial u_i^{(m)}}\Big)
\Big(e^{\partial\frac{d}{d\mu}} \frac{\partial g}{\partial u_j^{(n)}}\Big)
\{{u_i^{(m)}}\, _\lambda \{{u_j^{(n)}}\, _\mu\, u_k^{(p)}\}\}
\\
&&\,\,\,\,\,\,\,\,\,\,\,\,\,\,\,\,\,\,\,\,\,
+\, \sum_{\substack{i,j,k\in I \\ m,n,p\in\mb Z_+}} 
\frac{\partial h}{\partial u_k^{(p)}}
\Big(e^{\partial\frac{d}{d\lambda}} \frac{\partial f}{\partial u_i^{(m)}}\Big)
\Big\{{u_i^{(m)}}\, _\lambda 
\Big(e^{\partial\frac{d}{d\mu}} \frac{\partial g}{\partial u_j^{(n)}}\Big)\Big\}
\{{u_j^{(n)}}\, _\mu\, u_k^{(p)}\}\,.
\nonumber
\end{eqnarray}
Furthermore, if we use again both identities \eqref{eq:1.9_cons} and sesquilinearity \eqref{sesq},
we get the following identity for
the last term in the RHS of \eqref{eq:feb6_3},
\begin{equation}\label{eq:feb6_4}
\sum_{\substack{i,j,k\in I \\ m,n,p\in\mb Z_+}}\!\!\! \!\!\! 
\frac{\partial h}{\partial u_k^{(p)}}
\!\Big(\! e^{\partial\frac{d}{d\lambda}} \frac{\partial f}{\partial u_i^{(m)}} \!\Big)\!
\!\Big\{\! {u_i^{(m)}} _\lambda 
\!\Big(\! e^{\partial\frac{d}{d\mu}} \frac{\partial g}{\partial u_j^{(n)}} \!\Big)\! \!\Big\}\!
\{{u_j^{(n)}} _\mu u_k^{(p)}\}
\!=
\!\! \!\!\! \!\!\! \sum_{j\in I,n\in\mb Z_+} \!\!\! \!\!\!
\{{u_j^{(n)}} _{\lambda+\mu+\partial} h\}_\to
\!\Big\{\! f _\lambda \frac{\partial g}{\partial u_j^{(n)}} \!\Big\}.
\end{equation}
The second term in the LHS of \eqref{eq:1.10} is the same as the first term, 
after exchanging $f$ with $g$ and $\lambda$ with $\mu$.
Therefore, combining the results from equations \eqref{eq:feb6_1}-\eqref{eq:feb6_4},
we get that the LHS of \eqref{eq:1.10} is
\begin{eqnarray}\label{eq:feb6_5}
&& \displaystyle{
\{f_\lambda \{g_\mu h\}\}-\{g_\mu \{f_\lambda h\}\} 
}\nonumber \\
&& \displaystyle{
=\!\!\! \!\! \sum_{\substack{i,j,k\in I \\ m,n,p\in\mb Z_+}} \!\!\! \!
\frac{\partial h}{\partial u_k^{(p)}}
\Big(\! e^{\partial\frac{d}{d\lambda}} \frac{\partial f}{\partial u_i^{(m)}} \!\Big)\!
\Big(\! e^{\partial\frac{d}{d\mu}} \frac{\partial g}{\partial u_j^{(n)}} \!\Big)\!
\Big(\!
\{{u_i^{(m)}}\, _\lambda \{{u_j^{(n)}}\, _\mu\, u_k^{(p)}\}\}
- \{{u_j^{(n)}}\, _\mu \{{u_i^{(m)}}\, _\lambda\, u_k^{(p)}\}\}
\!\Big)
} \nonumber \\
&& \displaystyle{
+\, \sum_{j\in I,n\in\mb Z_+} 
\{{u_j^{(n)}}\, _{\lambda+\mu+\partial}\, h\}_\to
\Big\{f\, _\lambda \frac{\partial g}{\partial u_j^{(n)}}\Big\}
- \sum_{i\in I,m\in\mb Z_+} 
\{{u_i^{(m)}}\, _{\lambda+\mu+\partial}\, h\}_\to
\Big\{g\, _\mu \frac{\partial f}{\partial u_i^{(m)}}\Big\}\,.
} 
\end{eqnarray}
We finally use \eqref{eq:feb5_7},
and then the first equation in \eqref{eq:1.9_cons},
to rewrite the first term in the RHS of \eqref{eq:feb6_5} to get:
\begin{eqnarray}\label{eq:feb7_1}
& \displaystyle{
\{f_\lambda \{g_\mu h\}\}-\{g_\mu \{f_\lambda h\}\} 
=\! \sum_{\substack{i,j\in I \\ m,n\in\mb Z_+}} \!
\Big(e^{\partial\frac{d}{d\lambda}} \frac{\partial f}{\partial u_i^{(m)}}\Big)
\Big(e^{\partial\frac{d}{d\mu}} \frac{\partial g}{\partial u_j^{(n)}}\Big)
\{\{{u_i^{(m)}}\, _\lambda {u_j^{(n)}}\}\, _{\lambda+\mu}\, h\}
} \nonumber \\
& \displaystyle{
+\, \sum_{j\in I,n\in\mb Z_+} 
\{{u_j^{(n)}}\, _{\lambda+\mu+\partial}\, h\}_\to
\Big\{f\, _\lambda \frac{\partial g}{\partial u_j^{(n)}}\Big\}
- \sum_{i\in I,m\in\mb Z_+} 
\{{u_i^{(m)}}\, _{\lambda+\mu+\partial}\, h\}_\to
\Big\{g\, _\mu \frac{\partial f}{\partial u_i^{(m)}}\Big\}\,.
} 
\end{eqnarray}
We next look at the RHS of equation \eqref{eq:1.10} with $\nu$ in place of $\lambda+\mu$.
By the first identity in \eqref{eq:1.9_cons}, combined with sesquilinearity \eqref{sesq},
and using the right Leibniz rule \eqref{eq:1.8}, we get:
\begin{eqnarray}\label{eq:feb7_2}
&\displaystyle{
\{\{f_\lambda g\}_\nu h\}
= \sum_{j\in I,n\in\mb Z_+} \Big\{\frac{\partial g}{\partial u_j^{(n)}} 
\{f\, _\lambda\, u_j^{(n)}\}\, _\nu\, h\Big\}
= \sum_{j\in I,n\in\mb Z_+} 
\Big\{\frac{\partial g}{\partial u_j^{(n)}}\, _{\nu+\partial}\, h \Big\}_\to
\{f\, _\lambda\, u_j^{(n)}\}
}\nonumber \\
&\displaystyle{
+ \sum_{j\in I,n\in\mb Z_+} 
\Big(e^{\partial\frac{d}{d\nu}}\frac{\partial g}{\partial u_j^{(n)}} \Big)
\{\{f\, _\lambda\, u_j^{(n)}\}\, _\nu\, h\}\,.
}
\end{eqnarray}
We then expand $\{f_\lambda u_j^{(n)}\}$ using the second identity in \eqref{eq:1.9_cons}
combined with sesquilinearity \eqref{sesq}, and then apply the right Leibniz rule \eqref{eq:1.8},
to rewrite the last term in the RHS of \eqref{eq:feb7_2} as
\begin{eqnarray}\label{eq:feb7_3}
&&
\sum_{j\in I,n\in\mb Z_+} 
\Big(e^{\partial\frac{d}{d\nu}}\frac{\partial g}{\partial u_j^{(n)}} \Big)
\{\{f\, _\lambda\, u_j^{(n)}\}\, _\nu\, h\} 
\nonumber\\
&&\,\,\,\,\,\,\,\,\,\,\,\,
=\sum_{\substack{i,j\in I \\ m,n\in\mb Z_+}}
\Big(e^{\partial\frac{d}{d\nu}}\frac{\partial g}{\partial u_j^{(n)}} \Big)
\Big\{\Big(e^{\partial\frac{d}{d\lambda}}\frac{\partial f}{\partial u_i^{(m)}} \Big)
\{{u_i^{(m)}}\, _\lambda\, u_j^{(n)}\}\, _\nu\, h\Big\}
\nonumber \\
&&\,\,\,\,\,\,\,\,\,\,\,\,
= \sum_{\substack{i,j\in I \\ m,n\in\mb Z_+}}
\Big(e^{\partial\frac{d}{d\nu}}\frac{\partial g}{\partial u_j^{(n)}} \Big)
\Big(e^{\partial\frac{d}{d\nu}}e^{\partial\frac{d}{d\lambda}}\frac{\partial f}{\partial u_i^{(m)}} \Big)
\{\{{u_i^{(m)}}\, _\lambda\, u_j^{(n)}\}\, _\nu\, h\}
\\
&&\,\,\,\,\,\,\,\,\,\,\,\, \,\,\,\,\,\,
+ \sum_{\substack{i,j\in I \\ m,n\in\mb Z_+}}
\Big(e^{\partial\frac{d}{d\nu}}\frac{\partial g}{\partial u_j^{(n)}} \Big)
\Big\{\Big(e^{\partial\frac{d}{d\lambda}}\frac{\partial f}{\partial u_i^{(m)}} \Big)\, _{\nu+\partial}\, h \Big\}_\to
\{{u_i^{(m)}}\, _\lambda\, u_j^{(n)}\} 
\,.
\nonumber
\end{eqnarray}
Furthermore, the last term in the RHS of equation \eqref{eq:feb7_3}
can be rewritten, using equation \eqref{eq:1.9_cons} combined with 
sesquilinearity \eqref{sesq}, as
\begin{eqnarray}\label{eq:feb7_4}
&&\displaystyle{
\sum_{\substack{i,j\in I \\ m,n\in\mb Z_+}}
\Big(e^{\partial\frac{d}{d\nu}}\frac{\partial g}{\partial u_j^{(n)}} \Big)
\Big\{\Big(e^{\partial\frac{d}{d\lambda}}\frac{\partial f}{\partial u_i^{(m)}} \Big)\, _{\nu+\partial}\, h \Big\}_\to
\{{u_i^{(m)}}\, _\lambda\, u_j^{(n)}\} 
}\nonumber \\
&&\displaystyle{
=\!\!\! \!\!\! \sum_{\substack{i,j\in I \\ m,n\in\mb Z_+}}\!\!\! \!\!\! 
\Big\{\!\Big(\! e^{\partial\frac{d}{d\lambda}}\frac{\partial f}{\partial u_i^{(m)}} \!\Big)
_{\nu+\partial} h \!\Big\}\!_\to
\frac{\partial g}{\partial u_j^{(n)}} \{{u_i^{(m)}} _\lambda u_j^{(n)}\} 
=\!\!\! \!\!\!  \sum_{i\in I,n\in\mb Z_+}\!\!\! \!\!\! 
\Big\{\!\Big(\! e^{\partial\frac{d}{d\lambda}}\frac{\partial f}{\partial u_i^{(m)}} \!\Big)
_{\nu+\partial} h \!\Big\}\!_\to
\{{u_i^{(m)}}\, _\lambda\, g\} 
}\nonumber \\
&&\displaystyle{
=\!\!\! \sum_{i\in I,n\in\mb Z_+}\!\!\! 
\Big\{\frac{\partial f}{\partial u_i^{(m)}} \, _{\nu+\partial}\, h \Big\}_\to
\,_\leftarrow\!\{{u_i^{(m)}}\, _{\lambda-\nu-\partial}\, g\} 
= - \!\!\! \sum_{i\in I,n\in\mb Z_+}\!\!\! 
\Big\{\frac{\partial f}{\partial u_i^{(m)}} \, _{\nu+\partial}\, h \Big\}_\to
\{g\,_{\nu-\lambda}\, {u_i^{(m)}}\}\,.
}
\end{eqnarray}
For the last equality, we used that, by part (b), the $\lambda$-bracket \eqref{eq:1.9} satisfies
the skew-commutativity \eqref{eq:1.6}.
We can then put together equations \eqref{eq:feb7_2}-\eqref{eq:feb7_4} with $\nu=\lambda+\mu$, 
to get the following
expression for the RHS of equation \eqref{eq:1.10},
\begin{eqnarray}\label{eq:feb7_5}
&&\displaystyle{
\{\{f_\lambda g\}_{\lambda+\mu} h\}
= \sum_{\substack{i,j\in I \\ m,n\in\mb Z_+}}
\Big(e^{\partial\frac{d}{d\mu}}\frac{\partial g}{\partial u_j^{(n)}} \Big)
\Big(e^{\partial\frac{d}{d\lambda}}\frac{\partial f}{\partial u_i^{(m)}} \Big)
\{\{{u_i^{(m)}}\, _\lambda\, u_j^{(n)}\}\, _{\lambda+\mu}\, h\}
}\nonumber \\
&&\displaystyle{
+\!\!\! \sum_{j\in I,n\in\mb Z_+}\!\!\!
\Big\{\frac{\partial g}{\partial u_j^{(n)}}\, _{\lambda+\mu+\partial}\, h \Big\}_\to
\{f\, _\lambda\, u_j^{(n)}\}
- \!\!\! \sum_{i\in I,n\in\mb Z_+}\!\!\! 
\Big\{\frac{\partial f}{\partial u_i^{(m)}} \, _{\lambda+\mu+\partial}\, h \Big\}_\to
\{g\,_{\mu}\, {u_i^{(m)}}\}\,.
}
\end{eqnarray}
To conclude, we notice that the first term in the RHS of \eqref{eq:feb7_1}
coincides with the first term in the RHS of \eqref{eq:feb7_5}.
Moreover, it is not hard to check, using \eqref{eq:1.9_cons} and 
the fact that $\frac{\partial}{\partial u_i^{(m)}}$ and $\frac{\partial}{\partial u_j^{(n)}}$ commute, 
that the second and third terms
in the RHS of \eqref{eq:feb7_1} coincide, respectively, with the second and third terms
in the RHS of \eqref{eq:feb7_5}.
\end{proof}

Theorem \ref{prop:exten}(a) says that, in order to define a $\lambda$-bracket
on an algebra of differential functions $\mc V$ 
extending $R=\mb C[u_i^{(n)}\,|\,i\in I,\,n\in\mb Z_+]$,
one only needs to define for any pair $i,j \in I$ the $\lambda$-bracket 
\begin{equation}\label{eq:mar19}
\{ u_{i_\lambda} u_j \} = H_{ji} (\lambda) \in \mc V[\lambda]\,.
\end{equation}
In particular $\lambda$-brackets on $\mc V$ are in one-to-one correspondence
with $\ell\times\ell$-matrices $H(\lambda)=\big(H_{ij}(\lambda)\big)_{i,j\in I}$
with entries $H_{ij}(\lambda)=\sum_{n=0}^NH_{ij;n}\lambda^n$ in $\mc V[\lambda]$,
or, equivalently, with the corresponding matrix valued differential operators
$H(\partial)=\big(H_{ij}(\partial)\big)_{i,j\in I}:\,\mc V^{\oplus\ell}\to\mc V^\ell$.
We shall denote by $\{\cdot\,_\lambda\,\cdot\}_H$ the $\lambda$-bracket 
on $\mc V$ corresponding to the operator $H(\partial)$ via equation 
\eqref{eq:mar19}.
\begin{proposition}\label{prop:09jan16}
Let $H(\partial)=\big(H_{ij}(\partial)\big)_{i,j\in I}$ be an $\ell\times\ell$ matrix valued
differential operator.
\begin{enumerate}
\alphaparenlist
\item
The $\lambda$-bracket $\{\cdot\,_\lambda\,\cdot\}_H$
satisfies the (skew-)commutativity condition \eqref{eq:1.6}
if and only if the differential operator $H(\partial)$ is self (skew-)adjoint, meaning that
\begin{equation}\label{eq:apr23_skadj}
H^*_{ij}(\partial)\,:=\,\sum_{n=0}^N(-\partial)^n \circ H_{ji;n} \,=\, \pm H_{ij}(\partial)\,.
\end{equation}
\item
If $H(\partial)$ is skew-adjoint,
the following conditions are equivalent:
\begin{enumerate}
\item[(i)]
the $\lambda$-bracket $\{\,\cdot_\lambda\cdot\,\}_H$
defines a PVA structure on $\mc V$,
\item[(ii)]
the following identity holds for every $i,j,k\in I$:
\begin{equation}\label{2006_may31-14_second}
\begin{array}{c}
\displaystyle{
\sum_{h\in I,\, n\in \mb{Z}_+}
\Bigg(
\frac{\partial H_{kj}(\mu)}{\partial u_h^{(n)}}
(\lambda+\partial)^n H_{hi}(\lambda)
- 
\frac{\partial H_{ki}(\lambda)}{\partial u_h^{(n)}} 
(\mu+\partial)^n H_{hj}(\mu)
\Bigg)
} \\
=
\displaystyle{
\sum_{h\in I,\, n\in\mb{Z}_+}
H_{kh}(\lambda+\mu+\partial) (-\lambda-\mu-\partial)^n 
\frac{\partial H_{ji}(\lambda)}{\partial u_h^{(n)}} \,,
}
\end{array}
\end{equation}
\item[(iii)]
the following identity holds for every $F,G\in\mc V^{\oplus\ell}$:
\begin{eqnarray}\label{eq:09jan16}
&
\vphantom{\Big(}
H(\partial)D_G(\partial)H(\partial)F
+H(\partial)D^*_{H(\partial)F}(\partial)G
-H(\partial)D_F(\partial)H(\partial)G
\nonumber\\
&
+H(\partial)D^*_F(\partial)H(\partial)G 
\,=\,
D_{H(\partial)G}(\partial)H(\partial)F
-D_{H(\partial)F}(\partial)H(\partial)G\,.
\end{eqnarray}
\end{enumerate}
\end{enumerate}
\end{proposition}
\begin{proof}
Equation \eqref{eq:feb5_1} is the same as equation \eqref{eq:apr23_skadj},
if we use the identification \eqref{eq:mar19}.
Hence part (a) follows from Theorem \ref{prop:exten}(b).
Next, equation \eqref{eq:feb5_2} is easily translated, using the identification \eqref{eq:mar19}
and formula \eqref{eq:1.9},
into equation \eqref{2006_may31-14_second}.
Hence, the equivalence of (i) and (ii) follows immediately from Theorem \ref{prop:exten}(c).
We are thus left to prove that, if $H(\partial)$ is skew-adjoint, then conditions (ii) and (iii) are equivalent.
Written out explicitly, 
using formulas for the Fr\'echet derivatives in Section \ref{sec:1.1.5},
equation \eqref{eq:09jan16} becomes  ($k\in I$):
\begin{equation}\label{2006_may31-5}
\begin{array}{c}
\displaystyle{
\sum_{\substack{i,j,h\in I \\ n\in\mb{Z}_+}}
H_{ki}(\partial)\Bigg(
\frac{\partial G_i}{\partial u_j^{(n)}} \Big(\partial^n H_{jh}(\partial)F_h \Big)
+  (-\partial)^n\bigg(\frac{\partial \big(H_{jh}(\partial)F_h\big)}{\partial u_i^{(n)}} G_j\bigg)
}\\
\displaystyle{
- \frac{\partial F_i}{\partial u_j^{(n)}} \Big(\partial^n H_{jh}(\partial)G_h \Big)
+ (-\partial)^n\bigg(\frac{\partial F_j}{\partial u_i^{(n)}} \Big(H_{jh}(\partial)G_h\Big)\bigg)\Bigg)
}\\
\displaystyle{
\,=\!\!
\sum_{\substack{i,j,h\in I \\ n\in \mb{Z}_+}} 
\Bigg(
\Big(\partial^n H_{ij}(\partial)F_j\Big)\frac{\partial (H_{kh}(\partial)G_h)}{\partial u_i^{(n)}} 
- \Big(\partial^n H_{ij}(\partial)G_j\Big)\frac{\partial (H_{kh}(\partial)F_h)}{\partial u_i^{(n)}}
\Bigg)\,.
}
\end{array}
\end{equation}
We then use Lemma \ref{2006_l-may31} to check that 
the second term in the LHS of \eqref{2006_may31-5} is
\begin{equation}\label{2006_may31-9}
\sum_{\substack{i,j,h\in I \\ n\in\mb{Z}_+}}
H_{ki}(\partial) (-\partial)^n \Bigg(
\bigg( \frac{\partial H_{jh}(\partial)}{\partial u_i^{(n)}}F_h\bigg) G_j
+
\frac{\partial F_h}{\partial u_i^{(n)}} 
\Big(
H^*_{hj}(\partial) G_j\Big)
\Bigg)\,,
\end{equation}
the first term in the RHS of \eqref{2006_may31-5} is
\begin{equation}\label{2006_may31-10}
\sum_{\substack{i,j,h\in I \\ n\in \mb{Z}_+}}
\Big(\partial^n H_{ij}(\partial)F_j\Big)
\Bigg(
\frac{\partial H_{kh}(\partial)}{\partial u_i^{(n)}} G_h 
\Bigg)
+ \sum_{\substack{i,j,h\in I \\ n\in \mb{Z}_+}}
H_{kh}(\partial)\Bigg(
\frac{\partial G_h}{\partial u_i^{(n)}} \Big(\partial^n H_{ij}(\partial)F_j\Big)
\Bigg)\,,
\end{equation}
and the second term in the RHS of \eqref{2006_may31-5} is
\begin{equation}\label{2006_may31-11}
- \sum_{\substack{i,j,h\in I \\ n\in \mb{Z}_+}} 
\Big(\partial^n H_{ij}(\partial)G_j\Big)
\Bigg(
\frac{\partial H_{kh}(\partial)}{\partial u_i^{(n)}} F_h
\Bigg)
- \sum_{\substack{i,j,h\in I \\ n\in \mb{Z}_+}} 
H_{kh}(\partial) \Bigg(
\frac{\partial F_h}{\partial u_i^{(n)}} \Big(\partial^n H_{ij}(\partial)G_j\Big)
\Bigg)\,.
\end{equation}
We then notice that the second term in \eqref{2006_may31-10} cancels with the
first term in the LHS of \eqref{2006_may31-5},
the second term in \eqref{2006_may31-11} cancels with the
third term in the LHS of \eqref{2006_may31-5}
and, finally, the second term in \eqref{2006_may31-9} cancels with the
last term in the LHS of \eqref{2006_may31-5}, since $H(\partial)$ is skew-adjoint.
In conclusion equation \eqref{2006_may31-5} becomes
\begin{equation}\label{2006_may31-14}
\begin{array}{c}
\displaystyle{
\sum_{\substack{i,j,h\in I \\ n\in \mb{Z}_+}}\!\!
\Bigg(
\Big(\partial^n H_{ij}(\partial)F_j\Big)\!
\bigg(
\frac{\partial H_{kh}(\partial)}{\partial u_i^{(n)}} G_h 
\bigg)\!
- 
\! \Big(\partial^n H_{ij}(\partial)G_j\Big)\!
\bigg(
\frac{\partial H_{kh}(\partial)}{\partial u_i^{(n)}} F_h
\bigg)\!\!
\Bigg)
} \\
=
\displaystyle{
\sum_{\substack{i,j,h\in I \\ n\in\mb{Z}_+}}
H_{ki}(\partial) (-\partial)^n \Bigg(
G_j \bigg( \frac{\partial H_{jh}(\partial)}{\partial u_i^{(n)}}F_h\bigg) 
\Bigg)\,.
}
\end{array}
\end{equation}
Since the above equation holds for every $F,G\in \mc V^{\oplus\ell}$,
we can replace $\partial$ acting on $F_i$ by $\lambda$
and $\partial$ acting on $G_j$ by $\mu$ and write it as an identity
between polynomials in $\lambda$ and $\mu$.
Hence \eqref{2006_may31-14} is equivalent to \eqref{2006_may31-14_second}.
\end{proof}
\begin{definition}\label{def:09jan19}
A matrix valued differential operator $H(\partial)=\big(H_{ij}(\partial)\big)_{i,j\in I}$
which is skew-adjoint and satisfies one of the three equivalent conditions (i)--(iii) of Proposition \ref{prop:09jan16}(b),
is called a \emph{Hamiltonian operator}.
\end{definition}

\begin{example}
  \label{ex:1.3}
The Gardner--Faddeev--Zakharov (GFZ) PVA structure
on $R=\mb C[u,u^{\prime},u^{\prime\prime},\dots]$ is defined by
\begin{equation}
  \label{eq:1.2}
  \{ u_\lambda u \} = \lambda 1
\end{equation}
(further on we shall usually drop 1).
In fact, one can replace $\lambda$ in the RHS of \eqref{eq:1.2} by any 
odd polynomial $P (\lambda)\in\mb C[\lambda]$, and still get a PVA structure on $R$.  
Indeed, the bracket
\eqref{eq:1.2} is skew-commutative and it satisfies the Jacobi
identity for the triple $u,u,u$, since each triple commutator in the Jacobi identity is zero.
\end{example}

\begin{example}
  \label{ex:1.4}
The \emph{Virasoro-Magri} PVA on $R=\mb C[u,u^{\prime},u^{\prime\prime},\dots]$,  with central charge $c \in\mb{C}$,
is defined by
\begin{equation}
  \label{eq:1.3}
  \{ u_\lambda u \} = (\partial +2\lambda) u + \lambda^3 c \, .
\end{equation}
It is easily seen that the bracket \eqref{eq:1.3} is skew-commutative
and it satisfies the Jacobi identity for the triple $u,u,u$.
\end{example}

\begin{example}
  \label{ex:1.5}
Let $\mf{g}$ be a Lie algebra with a symmetric invariant bilinear form $(\,. \, | \, . \, )$, 
and let $p$ be an element of $\mf{g}$.
Let $\{u_i\}_{i\in I}$ be a basis for $\mf g$.
The affine PVA associated to the triple $\big(\mf g,(\cdot\,|\,\cdot),p\big)$
is the algebra $R=\mb C[u_i^{(n)}\,|\,i\in I,n\in\mb Z_+]$ together with the following 
$\lambda$-bracket
\begin{equation}
  \label{eq:1.4}
 \{ a_\lambda b \} = [a,b] + (p | [a,b]) + \lambda (a|b)\, , \, 
    a,b \in \mf{g} \, .
\end{equation}
Note that, taking in the RHS of \eqref{eq:1.4} any of the three summands, 
or, more generally, any linear combination of them,
endows $R$ with a PVA structure.
\end{example}

Note that Example~\ref{ex:1.3} is a special case of
Example~\ref{ex:1.5}, when $\mf{g} =\mb C u$ is the 1-dimensional abelian Lie algebra
and $(u|u)=1$.

The following theorem further generalizes the results from Theorem \ref{prop:exten},
as it allows us to consider not only extensions of $R$,
but also quotients of such extensions by ideals.
\begin{theorem}\label{prop:quotient}
Let $\mc V$ be an algebra of differential functions,
which is an extension of the algebra of 
differential polynomials $R=\mb C[u_i^{(n)}\,|\,i\in I,n\in\mb Z_+]$.
For each pair $i,j\in I$, let $\{{u_i}_\lambda u_j\}=H_{ji}(\lambda)\in\mb C[\lambda]\otimes\mc V$,
and consider the induced $\lambda$-bracket on $\mc V$ defined by formula \eqref{eq:1.9}.
Suppose that $J\subset\mc V$ is a subspace such that
$\partial J\subset J,\,J\cdot\mc V\subset J,\,
\{J_\lambda\mc V\}\subset\mb C[\lambda]\otimes J,\,\{\mc V_\lambda J\}\subset\mb C[\lambda]\otimes J$,
and consider the quotient space $\mc V/J$
with the induced action of $\partial$, the induced commutative associative product
and the induced $\lambda$-bracket.
\begin{enumerate}
\alphaparenlist
\item The $\lambda$-bracket on $\mc V/J$ satisfies the commutativity 
(resp. skew-commutativity) condition \eqref{eq:1.6},
provided that
\begin{equation}\label{eq:feb18_1}
\{{u_i}_\lambda u_j\} \mp \,_\leftarrow\!\{ {u_j}_{-\lambda -\partial}u_i \}\,\in\,\mb C[\lambda]\otimes J\,
,\,\,\text{for all\,\,} i,j\in I\,.
\end{equation}
\item Furthermore, assuming that  the skew-commutativity condition \eqref{eq:feb18_1} holds,
the $\lambda$-bracket on $\mc V/J$ satisfies the Jacobi identity \eqref{eq:1.10},
thus making $\mc V/J$ a PVA,
provided that
\begin{equation}\label{eq:feb18_2}
\{{u_i}_\lambda \{{u_j}_\mu u_k\}\}
- \{{u_j}_\mu \{{u_i}_\lambda u_k\}\}
- \{{\{{u_i}_\lambda {u_j}\}}_{\lambda+\mu} u_k\}\,\in\,\mb C[\lambda,\mu]\otimes J\,
\,,\,\,\text{for all\,\,} i,j,k\in I\,.
\end{equation}
\end{enumerate}
\end{theorem}
\begin{proof}
By the assumptions on $J$, it is clear that
the action of $\partial$, the commutative associative product
and the $\lambda$-bracket are well defined on the quotient space $\mc V/J$.
It is not hard to check, using equation \eqref{eq:feb5_3} and the assumption 
\eqref{eq:feb18_1}, that
$\{f_\lambda g\}
\mp _\leftarrow\!\{ g_{-\lambda-\partial} f \} \in\mb C[\lambda] J$
for all $f,g\in\mc V$, thus proving part (a).
Similarly, if we compare equations \eqref{eq:feb6_5}
and \eqref{eq:feb7_5} and use the assumptions \eqref{eq:feb18_1} and \eqref{eq:feb18_2},
we get that 
$\{f_\lambda \{g_\mu h\}\} - \{g_\mu \{f_\lambda h\}\} 
- \{\{f_\lambda g\}_{\lambda+\mu} h\}\in\mb C[\lambda,\mu]J$
for all $f,g,h\in\mc V$, thus proving part (b).
\end{proof}

\begin{example}\label{ex:final}
Let $A$ be a Lie conformal algebra with a central element $K$ such that $\partial K=0$.
Then $\mc V^k(A)=S(A)/(K-k1),\,k\in\mb C$, 
carries the usual structure of a unital commutative
associative differential algebra endowed with the $\lambda$-bracket,
extending that from $A$ by the left and right Leibniz rules,
making it a PVA.
This generalization of Examples \ref{ex:1.3}, \ref{ex:1.4}, \ref{ex:1.5},
may be viewed as a PVA analogue of the Lie-Kirillov-Kostant Poisson algebra
$S(\mf g)$, associated to a Lie algebra $\mf g$.
\end{example}

\vspace{3ex}
\subsection{The Beltrami $\lambda$-bracket.}~~
\label{sec:july19}
All the examples of $\lambda$-brackets in the previous section were skew-commutative.
In this section we introduce a commutative $\lambda$-bracket,
called the Beltrami $\lambda$-bracket,
which is important in the calculus of variations.
\begin{definition}
  \label{ex:1.6}
The \emph{Beltrami $\lambda$-bracket} is defined on the generators by,
\begin{equation}
  \label{eq:1.5}
  \{ u_{i_\lambda} u_j \}_B = \delta_{ij}\, , \quad i,j \in I \, .
\end{equation}
By Theorem \ref{prop:exten}, this extends to a well defined 
commutative $\lambda$-bracket on any algebra of differential functions $\mc V$,
given by the following formula,
\begin{equation}\label{eq:july19_2}
  \{ f_\lambda g \}_B = \sum_{\substack{i \in I\\ m,n \in \mb{Z}_+}}
  (-1)^m \frac{\partial g}{\partial u_i ^{(n)}} (\lambda +\partial)^{m+n}
    \frac{\partial f}{\partial u_i ^{(m)}} \, .
\end{equation}
\end{definition}


Both the variational derivative $\frac\delta{\delta u}$ and the Fr\'echet derivative $D_F(\partial)$
have nice interpretation in terms of this $\lambda$-bracket.
Indeed we have
\begin{equation}\label{eq:july19_1}
\frac{\delta f}{\delta u_i}
\,=\, \big\{f_\lambda u_i\big\}_B\big|_{\lambda=0}
\,\,,\,\,\,\,
{D_F(\lambda)}_{ij}
\,=\, \big\{{u_j}_\lambda F_i\big\}_B\,.
\end{equation}
Moreover the whole $\lambda$-bracket $\big\{f_\lambda u_i\big\}_B$ can be interpreted in terms
of the well-known higher order Euler operators 
$E_i^{(m)}=\sum_{n\in\mb Z_+}\binom{n}{m}(-1)^n\partial^{n-m}\frac{\partial}{\partial u_i^{(n)}}$:
$$
\big\{f_\lambda u_i\big\}_B
\,=\, \sum_{n\in\mb Z_+} \lambda^n E_i^{(n)}f\,.
$$

The Beltrami $\lambda$-bracket does not 
satisfy the Jacobi identity \eqref{eq:1.10}.  Indeed, when
restricted to the subalgebra of functions, depending only on $u_i, i \in I$ 
, the Beltrami $\lambda$-bracket becomes the bracket:
$\{ f,g \}_B = \sum_{i \in I} \frac{\partial f}{\partial u_i}\, \frac{\partial 
g}{\partial u_i}$,
which does not satisfy the Jacobi identity. The latter bracket
is one of the well-known Beltrami operations in classical differential 
geometry, hence our name ``Beltrami $\lambda$-bracket''.

On the other hand we have the following two identities for the triple $\lambda$-brackets,
which can be easily checked using \eqref{eq:july19_2} together with the sesquilinearity
and the Leibniz rules:
\begin{eqnarray}
{\{ {u_i} _\lambda \{f _\mu g\}_B\}_B} - {\{ f _\mu \{{u_i} _\lambda g\}_B\}_B}
&=& {\{ {\{{u_i} _\lambda f\}_B} _{\lambda+\mu} g\}_B} \,,\label{eq:july19_3}\\
{\{ f _\lambda \{g _\mu u_i\}_B\}_B} + {\{ g _\mu \{f _\lambda u_i\}_B\}_B}
&=& {\{ {\{f _\lambda g\}_B} _{\lambda+\mu} u_i\}_B} \,.\label{eq:july19_4}
\end{eqnarray}
Both Propositions \ref{prop:july19_1} and \ref{prop:july19_2}
can be nicely reformulated in terms of the Beltrami $\lambda$-bracket.
Namely, equation \eqref{eq:may4_1} states that
$f\in\partial\mc V+\mc C$ if and only if $\{f_\lambda u_i\}_B\big|_{\lambda=0}=0,\,\text{for all\,\,} i\in I$,
where the only if part is immediate from sesquilinearity \eqref{sesq}.
Similarly, Proposition \ref{prop:july19_2} states that
$F=\{f_\lambda u_i\}_B\big|_{\lambda=0}$ for some $f\in \mc V$
if and only if $\{{u_j}_\lambda F_i\}_B=\{{F_j}_\lambda u_i\}_B,\,\text{for all\,\,} i,j\in I$.
Again the only if part follows immediately by identity \eqref{eq:july19_4}
with $g=u_j$ and $\lambda=0$.

Even though the Beltrami $\lambda$-bracket does not make $\mc V$ into a PVA,
it has the following nice interpretation 
in the theory of Lie conformal algebras.
Consider the Lie conformal algebra $A=\oplus_{i\in I}\mb C[\partial]u_i$
with the zero $\lambda$-bracket.
Then $\mc V$ is a representation of the Lie conformal algebra $A$
given by 
\begin{equation}\label{eq:july24_8}
{u_i}_\lambda f=\{{u_i}_\lambda f\}_B\,.
\end{equation}
This is indeed a representation due to equation \eqref{eq:july19_3} with $f=u_j,\,g\in\mc V$.
We will see in Section \ref{2006_sec2} that, in fact,
the whole variational complex
can be nicely interpreted as the Lie conformal algebra cohomology complex
for the Lie conformal algebra $A$ and its module $\mc V$.

\vspace{3ex}
\subsection{Hamiltonian operators and Hamiltonian equations.}~~
\label{sec:1.3}
Lie conformal algebras, and, in particular, Poisson vertex
algebras, provide a very convenient framework for 
systems of Hamiltonian equations associated to a Hamiltonian operator. 
This is based on the following observation:
\begin{proposition}\label{prop:08dic22}
Let $\mc V$ be a $\mb C[\partial]$-module
endowed with a $\lambda$-bracket $\{\,_\lambda\,\}:\,\mc V\otimes\mc V\to\mb C[\lambda]\otimes\mc V$,
and consider the bracket on $\mc V$ obtained by setting $\lambda=0$:
\begin{equation}
  \label{eq:1.12}
  \{f, g \} = \{ f_\lambda g \} |_{\lambda =0}, \,\, f,g \in \mc V\,.
\end{equation}
\begin{enumerate}
\alphaparenlist
\item 
The bracket \eqref{eq:1.12} induces a well defined bracket 
on the quotient space $\mc V/\partial\mc V$.
\item
If the $\lambda$-bracket on $\mc V$ satisfies the Lie conformal algebra axioms \eqref{eq:1.6} and \eqref{eq:1.10},
then the bracket \eqref{eq:1.12}
induces a structure of a Lie algebra on $\mc V/\partial \mc V$,
and a structure of left $\mc V/\partial \mc V$-module on $\mc V$.
\item
If $\mc V$ is a PVA, 
the corresponding Lie algebra $\mc{V} /\partial\mc{V}$ acts on $\mc{V}$ (via \eqref{eq:1.12}) 
by derivations of the commutative associative product on $\mc V$,
commuting with the derivation~$\partial$, and this defines a Lie algebra homomorphism
from $\mc{V}/\partial \mc{V}$ to the Lie algebra of derivations of $\mc V$.
\end{enumerate}
\end{proposition}
\begin{proof}
Part (a) is clear from the sesquilinearity conditions \eqref{sesq} for the $\lambda$-bracket.
Clearly,
the bracket \eqref{eq:1.12} on $\mc V$ satisfies the Jacobi identity for 
Lie algebras
(due to the Jacobi identity \eqref{eq:1.10} for Lie conformal algebras with $\lambda=\mu=0$).
Moreover, the skew-commutativity of the bracket on $\mc V/\partial\mc V$
induced by \eqref{eq:1.12}
follows from the skew-commutativity \eqref{eq:1.6} of the $\lambda$-bracket.
Furthermore, it is immediate to check that, letting 
$\{\tint h , u\} = \{h_\lambda u \} |_{\lambda =0}$ 
for $\tint h=h+\partial \mc V\in \mc V/\partial \mc V$ and $u\in \mc V$,
we get a well-defined left $\mc V/\partial \mc V$-module structure on $\mc V$, proving (b). 
Claim (c) follows from the left Leibniz rule and the Jacobi identity with 
$\lambda = \mu =0$.
\end{proof}

Proposition \ref{prop:08dic22}
motivates the following definition and justifies the subsequent remarks.
\begin{definition}
  \label{def:1.14}
\alphaparenlist
\begin{enumerate}
\item
Elements of $\mc V/\partial \mc V$ are called \emph{local functionals}.  Given $f\in \mc V$, 
its image in $\mc V/\partial \mc V$ is denoted by $\tint f$.
\item
Given a local functional $\tint h\in \mc V/\partial \mc V$, the corresponding 
\emph{Hamiltonian equation} is
\begin{equation}
  \label{eq:1.14}
  \frac{du}{dt}= \{ h_\lambda u \} |_{\lambda =0}=\{\tint h,u\}\, 
\end{equation}
and $\{\tint h,\cdot\}$
is the corresponding \emph{Hamiltonian vector field}.
\item
A local functional $\tint f\in \mc V/\partial \mc V$ is called an \emph{integral of motion}
of equation \eqref{eq:1.14} if $\frac{df}{dt}=0 \mod \partial \mc V$,
or, equivalently, if
$$
  \big\{ \tint h ,\tint f \big\}=0 \, .
$$
\item
The local functionals $\tint h_n,\,n\in\mb Z_+$ are \emph{in  involution} 
if $\{ \tint h_m,\tint h_n \} =0$ for all $m,n\in\mb Z_+$.  The
corresponding \emph{hierarchy of Hamiltonian equations} is 
$$
  \frac{du}{dt_n} = \{ {h_n}_\lambda u\} |_{\lambda =0}=:\{\tint h_n,u\}\, , \quad n\in\mb Z_+\, .
$$
(Then all $\tint h_n$'s are  integrals of motion of each of the
equations of the hierarchy.)
\end{enumerate}
\end{definition}

From now on we restrict ourselves to the case in which the PVA $\mc V$ is an algebra
of differential functions in the variables $\{u_i\}_{i\in I}$ (cf.\ Definition \ref{def:diff_alg}).
In this case, by Theorem \ref{prop:exten}, the $\lambda$-bracket $\{\cdot\,_\lambda\,\cdot\}_H$
is uniquely defined by the corresponding Hamiltonian operator
$H(\partial)=\big(H_{ij}(\partial)\big)_{i,j\in I}$ (cf.\ equation \eqref{eq:mar19}),
and the $\lambda$-bracket of any two elements $f,g\in\mc V$ can be computed using
formula \eqref{eq:1.9}.
In this case the Hamiltonian vector field $\big\{\tint h,\cdot\big\}$ is equal to the evolutionary
vector field $X_{H(\partial)\frac{\delta h}{\delta u}}$ 
(defined by \eqref{eq:0.4})
and the Hamiltonian equation \eqref{eq:1.14} has the form
\begin{equation}\label{k17}
  \frac{du}{dt} = H(\partial) \frac{\delta h}{\delta u} \,,
\end{equation}
where $\frac{\delta h}{\delta u} =\big( \frac{\delta h}{\delta u_i} \big)_{i\in I}\in {\mc V}^{\oplus\ell}$
is the variational derivative of $h$ (defined by \eqref{eq:0.6}).
Moreover,
the corresponding Lie algebra structure of $\mc V/\partial\mc V$ is given by
\begin{equation}\label{k18}
 \big\{ \tint f,\tint g \big\} 
 = \int \frac{\delta g}{\delta u}\cdot \Big(H(\partial) \frac{\delta f}{\delta u}\Big)
 = \sum_{i,j\in I} \int \frac{\delta g}{\delta u_j} H_{ji}(\partial) \frac{\delta f}{\delta u_i}  \, .
\end{equation}
\begin{remark}\label{rem:vicapr09}
Since $\tint h\mapsto X_{H(\partial)\frac{\delta h}{\delta u}}$
is a Lie algebra homomorphism,
local functionals in involution correspond to commuting evolutionary vector fields. 
If a sequence $\tint h_n\in\mc V/\partial\mc V$ is such that
$\frac{\delta h_n}{\delta u}\in\mc V^{\oplus\ell}$ span infinite-dimensional subspace
and $\dim\Ker H(\partial)<\infty$,
then the vector fields $X_{H(\partial)\frac{\delta h_n}{\delta u}}$
span an infinite-dimensional space as well.
\end{remark}

\begin{definition}\label{def:vicapr09}
The Hamiltonian equation \eqref{eq:1.14} is called \emph{integrable}
if there exists an infinite sequence of  local functionals $\tint h_n$,
including $\tint h$, which span an infinite-dimensional abelian subspace
in the Lie algebra $\mc V/\partial\mc V$ (with Lie bracket \eqref{eq:1.12}),
and such that the evolutionary vector fields $X_{H(\partial)\frac{\delta h_n}{\delta u}}$
span an infinite-dimensional space
(they commute by Remark \ref{rem:vicapr09}).
\end{definition}

\vspace{3ex}
\subsection{Center of the Lie algebra $\mc V/\partial \mc V$.}~~
\label{sec:dec29}
Consider a Hamiltonian operator 
$H(\partial)=\big(H_{ij}(\partial)\big)_{i,j\in I}:\, {\mc V}^{\oplus\ell}\to {\mc V}^\ell$,
where $\mc V$ is an algebra of differential functions extending
$R=\mb C[u_i^{(n)}\,|\,i\in I,n\in\mb Z_+]$,
and consider the corresponding PVA structure on $\mc V$,
defined by equations \eqref{eq:1.9} and \eqref{eq:mar19}.
Recall
that two local functionals $\tint f$ and $\tint g$ are said to be in \emph{involution} 
if $\big\{\tint f,\tint g\big\}=0$.
We have, in fact, three different ``compatibility conditions" for the local functionals
$\tint f,\tint g\in\mc V/\partial\mc V$ or the corresponding Hamiltonian vector fields
$X_{H(\partial)\frac{\delta f}{\delta u}},\,X_{H(\partial)\frac{\delta g}{\delta u}}$,
namely:
\begin{enumerate}
\item[(i)] $\big\{\tint f,\tint g\big\}=0$,
\item[(ii)] $\big[X_{H(\partial)\frac{\delta f}{\delta u}},
X_{H(\partial)\frac{\delta g}{\delta u}}\big]=0$,
\item[(iii)] $\tint\big[X_{H(\partial)\frac{\delta f}{\delta u}},
X_{H(\partial)\frac{\delta g}{\delta u}}\big](h)=0$
for every $h\in \mc V$.
\end{enumerate}
By definition the action of the Hamiltonian vector field 
$X_{H(\partial)\frac{\delta f}{\delta u}}$ on $\mc V$ is given by
$$
X_{H(\partial)\frac{\delta f}{\delta u}}(h)\,=\,
\sum_{i,j\in I}\Big(\partial^nH_{ij}(\partial)\frac{\delta f}{\delta u_j}\Big)\frac{\partial h}{\partial u_i^{(n)}}
\,=\,\big\{\tint f,h\big\}\,,
$$
hence condition (ii) above is equivalent to saying that $\big\{\tint f,\tint g\big\}$
is in the kernel of the representation of the Lie algebra $\mc V/\partial \mc V$ 
on the space $\mc V$ (see Proposition \ref{prop:08dic22}(b)),
namely to the equation
$H(\partial)\frac{\delta}{\delta u}\big\{\tint f,\tint g\big\}=0$,
while condition (iii) is equivalent to
saying that $\big\{\tint f,\tint g\big\}$ lies in the center of the Lie algebra $\mc V/\partial \mc V$.
In particular, we always have the implications (i) $\implies$ (ii) $\implies$ (iii).
The opposite implications, in general, are not true.
To understand when they hold, we need to study
the kernel of the representation of the Lie algebra $\mc V/\partial\mc V$ on the space $\mc V$
and the center of the Lie algebra $\mc V/\partial \mc V$.
Clearly the former is always contained in the latter,
and we want to understand when they are both zero.
\begin{lemma}\label{lem:sstef_3}
Let $i\in I$ and let $f\in \mc V$ be a non-zero element
such that
\begin{equation}\label{dec27_2}
u_i^n f\in\partial \mc V
\quad,\quad
\text{ for every } n\in \mb Z_+\,.
\end{equation}
Then $\ord(f)=1$, and in fact $f=\partial\alpha(u_i)$, for 
some $\alpha(u_i)\in\mc C(u_i)\subset \mc V$.
In particular, if $\ell\geq2$ and condition \eqref{dec27_2} holds for every $i\in I$ and $n\in\mb Z_+$,
then $f=0$.
\end{lemma}
\begin{proof}
Let $f\in \mc V\backslash\{0\}$ satisfy \eqref{dec27_2} for some $i\in I$.
If we let $n=0$ in \eqref{dec27_2} we get that $f=\partial g$ for some $g\in \mc V$.
Let $N=\ord(g)$ (so that $\ord(f)=N+1$).

Let us consider first the case $N=0$, namely $g\in\mc V_0$.
For $\ell=1$ we have $i=1$ and $f=\partial g$ with $g\in\mc V_0=\mc C(u_1)$,
which is what we wanted.
For $\ell\geq2$, we have from condition \eqref{dec27_2} that, for every $n\in\mb Z_+$,
there exists $g_n\in \mc V_0$ such that
$$
u_i^n\partial g\,=\,\partial g_n\,,
$$
and, if we apply $\frac{\partial}{\partial u_j^\prime},\,j\in I$, to both sides of the above equation,
we get, using \eqref{eq:comm_frac}, that
\begin{equation}\label{jan2_3}
u_i^n\frac{\partial g}{\partial u_j}\,=\,\frac{\partial g_n}{\partial u_j}\,,\,\,j\in I\,.
\end{equation}
Using that 
$\frac{\partial^2 g_n}{\partial u_i\partial u_j}=\frac{\partial^2 g_n}{\partial u_j\partial u_i},\,i,j\in I$, 
we get from \eqref{jan2_3} that
$$
\frac\partial{\partial u_j}\Big(u_i^n\frac{\partial g}{\partial u_i}\Big)
\,=\,\frac\partial{\partial u_i}\Big(u_i^n\frac{\partial g}{\partial u_j}\Big)\,,
$$
which gives $\frac{\partial g}{\partial u_j}=0$ for every $j\in I$ different from $i$.
Hence $g\in\mc C(u_i)$, as we wanted.

Next, we will prove that, for $N\geq1$, no element $f\in\mc V$ 
of differential order $N+1$,
satisfying condition \eqref{dec27_2} for every $n\in\mb Z_+$, can exist.
By assumption \eqref{dec27_2}, for every $n\geq1$ there exists
an element $g_n\in\mc V$ of differential order $N$ such that
\begin{equation}\label{jan2_4}
u_i^n\partial g\,=\,\partial g_n\,.
\end{equation}
If we apply $\frac{\partial}{\partial u_j^{(N+1)}}$ to both sides of the above equaiton
we get, using \eqref{eq:comm_frac}, that
$$
\frac{\partial g_n}{\partial u_j^{(N)}}\,=\,u_i^n \frac{\partial g}{\partial u_j^{(N)}}\,,\,\,n\geq1\,,
$$
which is equivalent to saying, after performing integration in $u_j^{(N)},\,j\in I$, 
that the elements $g_n\in R$ are of the form
$$
g_n\,=\,u_i^ng+r_n\,,
$$
for some $r_n\in\mc V_{N-1}$.
Substituting back in equation \eqref{jan2_4} we thus get
$$
u_i^n\partial g\,=\,\partial \big(u_i^ng+r_n\big)\,,
$$
which immediately gives
\begin{equation}\label{dec27_4}
u_i^{n}gu_i^{\prime}=-\frac1{n+1}\partial r_{n+1}\,\in\partial \mc V\,,
\end{equation}
for every $n\in\mb Z_+$.
Suppose first that $N=1$. Then $\ord(g)=1$ and
equation \eqref{dec27_4} with $n=0$ implies that $gu^{\prime}_i\in\partial \mc V$.
But this is is not possible since $gu^{\prime}_i$, as function of $u^{\prime}_1,\dots,u^{\prime}_\ell$, 
is not linear, namely $gu^{\prime}_i$ does not have the form \eqref{dec27_1_new} with $N=0$.
Next, suppose, by induction, that no element in $\mc V_N$
exists which satisfies condition \eqref{dec27_2} for every $n\in\mb Z_+$,
and let $f\in \mc V_{N+1}\backslash\mc V_N$, be as above.
Then equation \eqref{dec27_4} exactly says that $gu^{\prime}_i$, which has differential order $N$,
also satisfies assumption \eqref{dec27_2}, thus contradicting the inductive assumption.
\end{proof}
\begin{remark}\label{rem:feb18_1}
Consider the pairing $\mc V\times \mc V\to \mc V/\partial \mc V$ 
given by \eqref{eq:may4_4} with $r=1$.
Lemma \ref{lem:sstef_3} can then be stated in terms of the following orthogonality condition:
\begin{equation}\label{eq:feb18_3}
\big(\mb C[u_i]\big)^\perp\,\subset\,\partial\mc C(u_i)\,
\end{equation}
and the above inclusion becomes an equality under the additional assumption
that $\frac\partial{\partial u_i}:\,\mc C(u_i)\to\mc C(u_i)$ is surjective.
\end{remark}

Lemma \ref{lem:sstef_3} will be useful in Section \ref{sec:1.7} to prove
that the integrals of motion of the KdV hierarchy span a maximal abelian
subalgebra of the Lie algebra $R/\partial R$,
where $R=\mb C[u^{(n)}\,|\,n\in\mb Z_+]$, with Lie bracket \eqref{k18} with $H(\partial)=\partial$
(cf.\ Theorem \ref{prop:1.26}).
An analogous result will be needed to prove that the integrals of motion of the
``linear" KdV hierarchy span a maximal abelian subalgebra.
This is given by the following
\begin{lemma}\label{lem:feb18}
Let $i\in I$ and let $\partial f\in \partial \mc V$ be an element 
such that
\begin{equation}\label{eq:feb18_4}
u_i^{(2n)}\partial f \,\in\, \partial \mc V
\quad,\quad
\text{ for every } n\in\mb Z_+\,.
\end{equation}
Then $\partial f$ is a $\mc C$-linear combination of the monomials $u_i^{(2n+1)},\,n\in\mb Z_+$.
In particular, if $\ell\geq2$ and condition \eqref{eq:feb18_4} holds for every $i\in I$,
then $\partial f=0$.
\end{lemma}
\begin{proof}
We start by observing that, under the assumption \eqref{eq:feb18_4},
$f$ must be a function only of $u_i$ and its derivatives,
namely $\frac{\partial f}{\partial u_j^{(n)}}=0,\,\text{for all\,\,} j\neq i,n\in\mb Z_+$.
In other words, we reduce the statement to the case $\ell=1$.
Indeed suppose, by contradiction, that, for some $j\neq i$ and some $N\geq0$,
we have $\frac{\partial f}{\partial u_j^{(N)}}\neq0$ 
and $\frac{\partial f}{\partial u_j^{(n)}}=0\,\text{for all\,\,} n>N$.
As a consequence of \eqref{eq:feb18_4} we get that
$$
\frac{\delta}{\delta u_j} \big(u_i^{(2n)} \partial f\big)
\,=\, - \sum_{p=0}^N (-\partial)^p \Big(\frac{\partial f}{\partial u_j^{(p)}} u_i^{(2n+1)}\Big) \,=\, 0\,.
$$
Here we are using the fact that, in any algebra of differential functions $\mc V$,
$\partial\mc V\subset\Ker\frac{\delta}{\delta u}$.
If we then take $n$ large enough (such that $2n>\ord(f)\geq N$),
by looking at the terms of highest differential order, 
the above equation has the form
$$
(-1)^{N+1}\frac{\partial f}{\partial u_j^{(N)}} u_i^{(2n+N+1)} + r\,=\,0\,,
$$
where $r\in\mc V_{2n+N}$.
If we then apply $\frac{\partial}{\partial u_i^{(2n+N+1)}}$ to both sides,
we get $\frac{\partial f}{\partial u_j^{(N)}}=0$,
contradicting our original assumption.

Next, we consider the case $\ell=1$. 
We need to prove that if $f\in\mc V$ is such that
\begin{equation}\label{eq:feb19_1}
f u^{(2n+1)}\in\partial \mc V\,\,,\,\,\,\,\text{for all\,\,} n\in\mb Z_+\,,
\end{equation}
then $\partial f\in\Span_{\mc C}\{u^{(2m+1)},\,m\in\mb Z_+\}$.

It is clear from \eqref{eq:feb19_1}, that $f$ has even differential order.
Indeed, if $\ord(f)=2n+1$ is odd, $u^{(2n+1)}f$ is not of the form \eqref{dec27_1_new}
and hence it cannot be in $\partial \mc V$, contradicting \eqref{eq:feb19_1}.
We thus let $\ord(f)=2N$ and we will prove the statement by induction on $N\geq0$.

Let us consider first the case $N=0$, namely $f=f(u)\in\mc V_0=\mc C(u)$.
If we let $n=1$ in \eqref{eq:feb19_1},
we have $u^{\prime\prime\prime}f(u)\in\partial \mc V$, so that
$$
\frac{\delta}{\delta u} \big(u^{\prime\prime\prime} f\big)
\,=\, \frac{df}{du}u^{\prime\prime\prime}-\partial^3 \big(f(u)\big)
\,=\, 0\,.
$$
Expanding $\partial^3\big(f(u)\big)$ 
and looking at the coefficient of $u^{\prime\prime}$, it follows that $\frac{d^2f}{du^2}=0$,
namely $\partial f=\alpha u^{\prime}$ for some $\alpha\in\mc C$, as we wanted.

Next, let $N\geq1$. Condition \eqref{eq:feb19_1} says that $u^{\prime}f\in\partial \mc V$,
hence, by \eqref{dec27_1_new}, $f$ must have the form
\begin{equation}\label{eq:feb19_2}
f=a u^{(2N)}+b\,,
\end{equation}
where $a,b\in \mc V_{2N-1}$. 
We want to prove that, necessarily, $a\in\mc C$ and $b\in\mc V_{2N-2}$
(so that, using the inductive assumption, we can conclude that 
$\partial f\in\Span_{\mc C}\{u^{\prime},u^{(3)},\dots,u^{(2N+1)}\}$).
Condition \eqref{eq:feb19_1} with $n=N$ gives $u^{(2N+1)}f\in\partial \mc V$.
Using \eqref{eq:feb19_2} and integrating by parts we have
\begin{equation}\label{eq:feb19_3}
\frac12 (\partial a) \big(u^{(2N)}\big)^2 +(\partial b) u^{(2N)}\,\in\,\partial \mc V\,.
\end{equation}
In particular the LHS of \eqref{eq:feb19_3} must have the form \eqref{dec27_1_new}.
It follows that the coefficients of $\big(u^{(2N)}\big)^3$ and $\big(u^{(2N)}\big)^2$ in \eqref{eq:feb19_3} 
must vanish, namely
\begin{equation}\label{eq:feb19_4}
\frac{\partial a}{\partial u^{(2N-1)}}=0\,,\,\,\frac12 \partial a+\frac{\partial b}{\partial u^{(2N-1)}}=0\,.
\end{equation}
The first of the above equations says that $a\in\mc V_{2N-2}$.
It follows that $\partial a$, viewed as a function of $u^{(2N-1)}$, is linear,
and therefore, by the second equation in \eqref{eq:feb19_4}, it follows that $b$,
viewed as a function of $u^{(2N-1)}$, has degree at most 2, namely
\begin{equation}\label{eq:feb19_5}
b=\alpha+\beta u^{(2N-1)}+\frac12 \gamma\big(u^{(2N-1)}\big)^2\,,\,\,
-\frac12\partial a=\beta + \gamma u^{(2N-1)}\,,
\end{equation}
where $\alpha,\beta,\gamma\in \mc V_{2N-2}$.
Next, let us apply condition \eqref{eq:feb19_1} with $n=N-1$, namely $u^{(2N-1)}f\in\partial\mc V$.
By equation \eqref{eq:feb19_2}  and integration by parts, this condition gives
$$
-\frac12 (\partial a) \big(u^{(2N-1)}\big)^2+bu^{(2N-1)}\,\in\,\partial \mc V\,,
$$
which, by \eqref{eq:feb19_5}, is equivalent to
\begin{equation}\label{eq:feb19_6}
\alpha u^{(2N-1)} + 2\beta \big(u^{(2N-1)}\big)^2 + \frac32 \gamma \big(u^{(2N-1)}\big)^3
\,\in\,\partial \mc V\,.
\end{equation}
To conclude we observe that the LHS of \eqref{eq:feb19_6},
being in $\partial \mc V$, must have the form \eqref{dec27_1_new},
and this is possible only if $\beta=\gamma=0$, 
which is exactly what we wanted.
\end{proof}
\begin{remark}\label{rem:feb18_2}
Lemma \ref{lem:feb18} can be stated, equivalently, in terms of the pairing 
$\mc V\times \mc V\to \mc V/\partial \mc V$ given by $f,g\mapsto\tint fg$,
as an orthogonality condition
similar to \eqref{eq:feb18_3}. Let $E=\Span_{\mc C}\{1\,;\, u_i^{(2n)},\,n\in\mb Z_+\}$.
Then:
\begin{equation}\label{eq:feb18_5}
E^\perp\,=\,\partial E
\end{equation}
\end{remark}

We next use Lemma \ref{lem:sstef_3} and Lemma \ref{lem:feb18} 
to study the center of the Lie algebra $\mc V/\partial \mc V$.
\begin{corollary}\label{cor:jan2}
\begin{enumerate}
\alphaparenlist
\item If $\ell\geq2$, the center of the Lie algebra $\mc V/\partial \mc V$ with Lie bracket \eqref{k18}
coincides with the kernel of the Lie algebra representation of $\mc V/\partial \mc V$ on $\mc V$
defined in Proposition \ref{prop:08dic22}(b),
namely it consists of the local functionals $\tint f\in \mc V/\partial \mc V$ such that
\begin{equation}\label{jan2_5}
H(\partial)\frac{\delta f}{\delta u}\,=\,0\,.
\end{equation}
In particular, the ``compatibility conditions" (ii) and (iii) above are equivalent.
\item  If $\ell=1$, the center of the Lie algebra $\mc V/\partial \mc V$ with Lie bracket \eqref{k18}
consists of the local functionals $\tint f\in \mc V/\partial \mc V$ such that
\begin{equation}\label{jan2_6}
H(\partial)\frac{\delta f}{\delta u}\,=\,\alpha u^{\prime}\,\,,\,\,\,\,\text{ where } \alpha\in\mc C\,.
\end{equation} 
\end{enumerate}
\end{corollary}
\begin{proof}
Suppose that the local functional $\tint f$ is in the center of the Lie algebra $\mc V/\partial \mc V$.
Then necessarily, for every $i\in I$ and $n\in\mb Z_+$,
\begin{eqnarray*}
\int u_i^n\Big(\sum_{j\in I}H_{ij}(\partial)\frac{\delta f}{\delta u_j}\Big)
&=& \frac1{n+1}\big\{\tint f,\tint u_i^{n+1}\big\}\,=\,0\,, \\
\int u_i^{(2n)}\Big(\sum_{j\in I}H_{ij}(\partial)\frac{\delta f}{\delta u_j}\Big)
&=& (-1)^n\frac12 \big\{\tint f,\tint \big(u_i^{(n)}\big)^2\big\}\,=\,0\,. 
\end{eqnarray*}
We can thus apply Lemma \ref{lem:sstef_3} and Lemma \ref{lem:feb18} to deduce that,
if $\ell\geq2$, then \eqref{jan2_5} holds,
while, if $\ell=1$, then 
\begin{equation}\label{jan2_7}
H(\partial)\frac{\delta f}{\delta u}\,\in\,
\partial\mc C(u)\cap\Span_{\mc C}\big\{u^{(2n+1)}\,|\,n\in\mb Z_+\big\}
\,=\, \{\alpha u^{\prime}\,|\,\alpha\in\mc C\}\,.
\end{equation}
To conclude we just notice that, integrating by parts,
$\tint\frac{\delta g}{\delta u}u^{\prime}=\tint\partial g=0$ for every $g\in \mc V$,
so that, if condition \eqref{jan2_6} holds, 
then $\tint f$ lies in the center of the Lie algebra $\mc V/\partial\mc V$.
\end{proof}

In conclusion of this section, we discuss two examples:
in Proposition \ref{cor:jan2_2} we consider 
the GFZ PVA from Example \ref{ex:1.3},
for which the center of the Lie algebra $R/\partial R$ is strictly bigger than the kernel
of the representation of $R/\partial R$ on $R$,
and in Example \ref{ex:heis}, we discuss a simple case in which the
``compatibility conditions" (ii) and (iii) hold, but (i) fails.
\begin{proposition}\label{cor:jan2_2}
Consider the Hamiltonian operator $H(\partial)=\partial$ on $R=\mb C[u^{(n)}\,|\,n\in\mb Z_+]$,
so that the corresponding Lie algebra structure on $R/\partial R$ 
is given by the bracket
$$
\big\{\tint f,\tint g\big\}\,=\,\int \frac{\delta g}{\delta u}\partial\frac{\delta f}{\delta u}\,,
$$
and the representation of the Lie algebra $R/\partial $R on the space $R$
is given by
$$
\big\{\tint f,g\big\}\,=\,X_{\partial\frac{\delta f}{\delta u}}(g)
\,=\, \sum_{n\in\mb Z_+}\Big(\partial ^{n+1} \frac{\delta f}{\delta u}\Big)\frac{\partial g}{\partial u^{(n)}}\,.
$$
\begin{enumerate}
\alphaparenlist
\item The kernel of the Lie algebra representation of $R/\partial R$ on $R$ is two-dimensional,
spanned by
$$
\tint 1\,,\,\,\tint u\,\in\,R/\partial R\,.
$$
\item The center of the Lie algebra $R/\partial R$ is three-dimensional,
spanned by
$$
\tint 1\,,\,\,\tint u\,,\,\,\tint u^2\,\in\,R/\partial R\,.
$$
\item The ``compatibility" conditions (i), (ii) and (iii) 
for the local functionals $\tint f,\,\tint g\in R/\partial R$
are equivalent.
\end{enumerate}
\end{proposition}
\begin{proof}
A local functional $\tint f$ is in the kernel of the representation of $R/\partial R$ on $R$
if and only if $\partial\frac{\delta f}{\delta u}=0$,
which is equivalent to saying that $\tint f\in\Span_{\mb C}\big\{\tint 1,\tint u\big\}$.
For the last assertion we use the fact that $\Ker\frac\delta{\delta u}=\mb C+\partial R$
(see Proposition \ref{prop:july19_1}).
Similarly, by Corollary \ref{cor:jan2}(b), a local functional $\tint f\in R/\partial R$
is in the center of the Lie bracket if and only if $\partial \frac{\delta f}{\delta u}\in\mb Cu^{\prime}$,
which, by the same argument as before, is equivalent to saying that
$\tint f\in\Span_{\mb C}\big\{\tint 1,\tint u,\tint u^2\big\}$.

We are left to prove part (c).
We already remarked that (i) implies (ii), which implies (iii),
and we only need to prove that (iii) implies (i).
Recall that condition (iii) is equivalent to saying that
$\big\{\tint f,\tint g\big\}$ lies in the center of the Lie algebra $R/\partial R$.
Hence, by part (b) we have
\begin{equation}\label{jan2_11}
\int\frac{\delta g}{\delta u}\partial\frac{\delta f}{\delta u}
\,\in\, \Span_{\mb C}\big\{\tint 1,\tint u,\tint u^2\big\}\subset R/\partial R\,.
\end{equation}
On the other hand, $R$ admits the decomposition 
\begin{equation}\label{eq:2009july3}
R=\mb C[u]\oplus R^\partial\,,
\end{equation}
where $R^\partial\subset R$ is the ideal (of the commutative associative product)
generated by the elements $u^{(n)},\,n\geq1$,
and clearly $\partial R\subset R^\partial$.
Hence, by \eqref{jan2_11}, we get
$$
\frac{\delta g}{\delta u}\partial\frac{\delta f}{\delta u}
\,\in\,R^\partial\cap\big(\mb C[u]+\partial R\big)=\partial R\,,
$$
which implies $\tint \frac{\delta g}{\delta u}\partial\frac{\delta f}{\delta u}=0$,
as we wanted.
\end{proof}

\begin{remark}\label{rem:jan15}
By the same argument as in the proof of Proposition \ref{cor:jan2_2},
it is not hard to show that, if $\mc V=R$ and 
$H(\partial)=\big(H_{ij}(\partial)\big)_{ij\in I}:\,R^{\oplus\ell}\to R^\ell$
is an injective Hamiltonian operator
(or, more generally, if $\Ker H(\partial)\cap \im \frac{\delta}{\delta u}=0$),
then the commutativity condition of Hamiltonian vector fields
$[X_{H(\partial)\frac{\delta f}{\delta u}},X_{H(\partial)\frac{\delta g}{\delta u}}]=0$,
is equivalent to the following condition
for the corresponding local functionals $\tint f,\,\tint g\in R/\partial R$:
\begin{equation}\label{jan15}
\big\{\tint f,\tint g\big\}\,\in\,\mb C\,\tint 1\,.
\end{equation}
Moreover equation \eqref{jan15} is clearly equivalent to $\big\{\tint f,\tint g\big\}=0$
if the operator $H(\partial)$ does not have a constant term
(by taking the image of both sides of equation \eqref{jan15}
via the quotient map $R/\partial R\twoheadrightarrow\tint\mb C$).
\end{remark}

\begin{example}\label{ex:heis}
Take $R=\mb C[p^{(n)},q^{(n)},z^{(n)}\,|\,n\in\mb Z_+]$,
and define the $\lambda$-bracket on $R$ by letting
$z$ to be central and
$$
\{p_\lambda q\}=z=-\{q_\lambda p\}
\,\,,\,\,\,\,
\{p_\lambda p\}=-\{q_\lambda q\}=0\,.
$$
Let $H(\partial)$ be the corresponding Hamiltonian operator.
In this example we have $\ell=3$, therefore by Corollary \ref{cor:jan2}(a) 
we know that the kernel of the representation of $R/\partial R$ on $R$
and the center of the Lie algebra $R/\partial R$ coincide.
In particular conditions (ii) and (iii) above are equivalent.
Next, take the local functionals $\tint p,\tint q\in R/\partial R$.
By the above definition of the $\lambda$-bracket on $R$ we have
$\{\tint p,\tint q\}=\tint z\neq 0$, namely condition (i) fails,
but $\tint z$ is in the center of the Lie algebra $R/\partial R$, namely condition (iii) holds.
\end{example}

\vspace{3ex}
\subsection{Poisson vertex algebras and Lie algebras.}~~
\label{sec:dic2008_1}
In this section we shall prove a converse statement to Proposition \ref{prop:08dic22}(b).
In order to do so,
we will need the following simple lemma:
\begin{lemma}\label{cor:08dic22}
Let $\mc V$ be an algebra of differential functions in the variables $\{u_i\}_{i\in I}$.
\begin{enumerate}
\alphaparenlist
\item
If $P\in\mc V^\ell$ is such that $\tint P\frac{\delta f}{\delta u}=0$ for every $\tint f\in\mc V/\partial\mc V$,
then $P=0$ if $\ell\geq2$, and $P\in\mb Cu_1^\prime$ if $\ell=1$.
\item
If $H(\partial)=\big(H_{ij}(\partial)\big)_{i,j\in I}$ and $n\in\mb Z_+$ are such that 
$H(\partial)\frac{\delta f}{\delta u}\in{\mc V_{n}}^{\oplus\ell}$
for every $\tint f\in\mc V/\partial\mc V$, then $H=0$.
\end{enumerate}
\end{lemma}
\begin{proof}
Note that $\frac{\delta }{\delta u_j}\frac{u_i^{n+1}}{n+1}=\delta_{i,j}u_i^n$,
and $\frac{\delta }{\delta u_j}\big(\frac{(-1)^n}2 \big(u_i^{(n)}\big)^2\big)=\delta_{i,j}u_i^{(2n)}$.
Hence, part (a) is an immediate corollary of Lemmas \ref{lem:sstef_3} and \ref{lem:feb18}.
For part (b),
suppose that $H_{ij}(\partial)=\sum_{n=0}^Nh_{ijn}\partial^n$ is a differential operator
such that $h_{ijN}\neq0$.
Let then $h=\frac{(-1)^M}{2}\big(u_j^{(M)}\big)^2$, so that $\frac{\delta h}{\delta u_k}
=\delta_{k,j}u_j^{(2M)}$.
If $M$ is sufficiently large, we have $\frac{\partial}{\partial u_j^{(M+N)}}\big(H(\partial)\frac{\delta h}{\delta u}\big)_i=h_{ijN}\neq0$.
\end{proof}
\begin{proposition}\label{prop:08dic22_1}
Let $\mc V$ be an algebra of differential functions in the variables $\{u_i\}_{i\in I}$,
endowed with a $\lambda$-bracket $\{\cdot\,_\lambda\,\cdot\}$.
Suppose that the bracket on $\mc V/\partial\mc V$ induced by \eqref{eq:1.12}
is a Lie algebra bracket.
Then the $\lambda$-bracket satisfies skew-commutativity \eqref{eq:1.6} and Jacobi identity \eqref{eq:1.10},
thus making $\mc V$ into a PVA.
\end{proposition}
\begin{proof}
Recall that the $\lambda$-bracket on $\mc V$ is given by \eqref{eq:1.9},
hence, the induced bracket on $\mc V/\partial\mc V$
is given by \eqref{k18} where $H_{ji}(\partial)=\{{u_i}_\partial u_j\}_\to$.
Integrating by parts,
the skew-symmetry of this bracket reads
$$
\int\frac{\delta g}{\delta u}\big(H(\partial)+H^*(\partial)\big)\frac{\delta f}{\delta u}\,=\,0\,,
$$
for every $\tint f,\tint g\in\mc V/\partial\mc V$.
By Lemma \ref{cor:08dic22}(a), it follows that 
$\big(H(\partial)+H^*(\partial)\big)\frac{\delta f}{\delta u}\in\mc V_1^{\oplus\ell}$ 
for every $\tint f\in\mc V/\partial\mc V$.
Hence, by Lemma \ref{cor:08dic22}(b), we have  $H^*(\partial)=-H(\partial)$.
Recalling \eqref{eq:apr23_skadj},
this identity can be equivalently written in terms of the $\lambda$-bracket 
as $\{{u_i}_\lambda{u_j}\}=-\{{u_j}_{-\lambda-\partial}{u_i}\}$,
which implies the skew-symmetry of the $\lambda$-bracket by Theorem \ref{prop:exten}(b).

Next, let us prove the Jacobi identity for the $\lambda$-bracket,
which, by Theorem \ref{prop:exten}, reduces to checking \eqref{eq:feb5_2}.
We will use the Jacobi identity for the Lie bracket:
\begin{equation}\label{eq:dic23_3}
\big\{\tint f,\{\tint g,\tint h\}\big\}-\big\{\tint g,\{\tint f,\tint h\}\big\}
=\big\{\{\tint f,\tint g\},\tint h\big\}\,.
\end{equation}
Recalling that the bracket \eqref{k18} is the same as \eqref{eq:1.9} with $\lambda=0$, we have
\begin{eqnarray}\label{eq:dic23_1}
&\displaystyle{
\big\{\tint f,\{\tint g,\tint h\}\big\}
\,=\,
\sum_{\substack{i,j,p,q\in I\\m,n\in\mb Z_+}}\int
\Big(
\frac{\partial}{\partial u_j^{(m)}}
\Big(
\frac{\partial h}{\partial u_q^{(n)}}\partial^n H_{qp}(\partial)\frac{\delta g}{\delta u_p}
\Big)\Big)
\Big(
\partial^m H_{ji}(\partial)\frac{\delta f}{\delta u_i}
\Big) 
}\nonumber\\
&\displaystyle{
=\,
\sum_{\substack{i,j,p,q\in I\\m,n\in\mb Z_+}}\int
\frac{\partial^2 h}{\partial u_j^{(m)}\partial u_q^{(n)}}
\Big(
\partial^n H_{qp}(\partial)\frac{\delta g}{\delta u_p}
\Big)
\Big(
\partial^m H_{ji}(\partial)\frac{\delta f}{\delta u_i}
\Big)
}\\
&\displaystyle{
+
\sum_{\substack{i,j,p,q\in I\\m,n\in\mb Z_+}}\int
\frac{\partial h}{\partial u_q^{(n)}}
\Big(
\frac{\partial}{\partial u_j^{(m)}}
\partial^n H_{qp}(\partial)\frac{\delta g}{\delta u_p}
\Big)
\Big(
\partial^m H_{ji}(\partial)\frac{\delta f}{\delta u_i}
\Big) \,.
}\nonumber
\end{eqnarray}
Note that the first term in the RHS is symmetric in $f$ and $g$,
hence it cancels out in the LHS of \eqref{eq:dic23_3}.
By Lemma \ref{2006_l-may31},
the last term in the RHS of \eqref{eq:dic23_1} is
\begin{eqnarray}\label{eq:dic23_2}
&& \sum_{\substack{i,j,p,q\in I\\m,n\in\mb Z_+}}\int
\frac{\partial h}{\partial u_q^{(n)}}\partial^n
\Bigg(
\bigg(
\frac{\partial H_{qp}(\partial)}{\partial u_j^{(m)}}
\frac{\delta g}{\delta u_p}
\bigg)
\bigg(
\partial^m H_{ji}(\partial)\frac{\delta f}{\delta u_i}
\bigg) \\
&& \,\,\,\,\,\,\,\,\,\, +
H_{qp}(\partial)
\bigg(
\bigg(
\frac{\partial}{\partial u_j^{(m)}}\frac{\delta g}{\delta u_p}
\bigg)
\bigg(
\partial^m H_{ji}(\partial)\frac{\delta f}{\delta u_i}
\bigg)
\bigg)
\Bigg) \,.
\nonumber
\end{eqnarray}
Integrating by parts, we can replace $\sum_{n\in\mb Z_+}\frac{\partial h}{\partial u_q^{(n)}}\partial^n$
by the variational derivative $\frac{\delta h}{\delta u_q}$.
Hence the LHS of \eqref{eq:dic23_3} reads,
\begin{eqnarray}\label{eq:dic23_4}
&\displaystyle{
\sum_{\substack{i,j,p,q\in I\\m\in\mb Z_+}}\int
\frac{\delta h}{\delta u_q}
\Bigg(
\bigg(
\frac{\partial H_{qp}(\partial)}{\partial u_j^{(m)}}
\frac{\delta g}{\delta u_p}
\bigg)
\bigg(
\partial^m H_{ji}(\partial)\frac{\delta f}{\delta u_i}
\bigg) 
-
\bigg(
\frac{\partial H_{qi}(\partial)}{\partial u_j^{(m)}}
\frac{\delta f}{\delta u_i}
\bigg)
\bigg(
\partial^m H_{jp}(\partial)\frac{\delta g}{\delta u_p}
\bigg) 
} \\
&\displaystyle{ 
+
H_{qp}(\partial)
\bigg(
\bigg(
\frac{\partial}{\partial u_j^{(m)}}\frac{\delta g}{\delta u_p}
\bigg)
\bigg(
\partial^m H_{ji}(\partial)\frac{\delta f}{\delta u_i}
\bigg)
\bigg)
-
H_{qi}(\partial)
\bigg(
\bigg(
\frac{\partial}{\partial u_j^{(m)}}\frac{\delta f}{\delta u_i}
\bigg)
\bigg(
\partial^m H_{jp}(\partial)\frac{\delta g}{\delta u_p}
\bigg)
\bigg)
\Bigg) \,.
}\nonumber
\end{eqnarray}
Similarly, the RHS of \eqref{eq:dic23_3} reads
\begin{eqnarray}\label{eq:dic23_5}
&& \sum_{\substack{i,j,p,q\in I\\n\in\mb Z_+}}\int
\frac{\delta h}{\delta u_q}
H_{qp}(\partial)(-\partial)^n
\Bigg(
\bigg(
\frac{\partial}{\partial u_p^{(n)}}
\frac{\delta g}{\delta u_j}
\bigg)
\bigg(
H_{ji}(\partial)\frac{\delta f}{\delta u_i}
\bigg) \nonumber\\
&& \,\,\,\,\,\,\,\,\,\, +
\frac{\delta g}{\delta u_j}
\bigg(
\frac{\partial H_{ji}(\partial)}{\partial u_p^{(n)}}
\frac{\delta f}{\delta u_i}
\bigg) 
+
\bigg(
\frac{\partial}{\partial u_p^{(n)}}
\frac{\delta f}{\delta u_i}
\bigg)
\bigg(
H^*_{ij}(\partial)\frac{\delta g}{\delta u_j}
\bigg) 
\Bigg) \,,
\end{eqnarray}
where in the last term we used Lemma \ref{2006_l-may31}.
The third term in \eqref{eq:dic23_4} is equal to the first term in \eqref{eq:dic23_5},
due to \eqref{eq:july18_5_c}.
Similarly, the last term in \eqref{eq:dic23_4} is equal to the last term in \eqref{eq:dic23_5},
due to skew-adjointness of $H(\partial)$.
Hence, the Jacobi identity \eqref{eq:dic23_3} is equivalent to the following relation
\begin{eqnarray*}
&\displaystyle{
\sum_{\substack{i,j,p,q\in I\\n\in\mb Z_+}}\int
\frac{\delta h}{\delta u_q}
\Bigg(
\bigg(
\frac{\partial H_{qp}(\partial)}{\partial u_j^{(n)}}
\frac{\delta g}{\delta u_p}
\bigg)
\bigg(
\partial^n H_{ji}(\partial)\frac{\delta f}{\delta u_i}
\bigg) 
-
\bigg(
\frac{\partial H_{qi}(\partial)}{\partial u_j^{(n)}}
\frac{\delta f}{\delta u_i}
\bigg)
\bigg(
\partial^n H_{jp}(\partial)\frac{\delta g}{\delta u_p}
\bigg) 
}\nonumber \\
&\displaystyle{
- 
H_{qj}(\partial)(-\partial)^n
\Bigg(
\frac{\delta g}{\delta u_p}
\bigg(
\frac{\partial H_{pi}(\partial)}{\partial u_j^{(n)}}
\frac{\delta f}{\delta u_i}
\bigg) 
\Bigg)
\Bigg) \,=\,0
\,.
}
\end{eqnarray*}
Since this equation holds for every $\tint f,\tint g,\tint h\in\mc V/\partial\mc V$,
we conclude, using Lemma \ref{cor:08dic22}, that
\begin{eqnarray*}
&&\displaystyle{
\sum_{j\in I,n\in\mb Z_+}
\Bigg(
\frac{\partial H_{qp}(\mu)}{\partial u_j^{(n)}}
(\lambda+\partial)^n H_{ji}(\lambda)
-
\frac{\partial H_{qi}(\lambda)}{\partial u_j^{(n)}}
(\mu+\partial)^n H_{jp}(\mu)
}\nonumber \\
&&\,\,\,\,\,\,\,\,\,\displaystyle{
- 
H_{qj}(\lambda+\mu+\partial)(-\lambda-\mu-\partial)^n
\frac{\partial H_{pi}(\lambda)}{\partial u_j^{(n)}}
\Bigg)
\,=\,0
\,\,,\,\,\,\,
\text{for all\,\,} i,p,q\in I\,,
}
\end{eqnarray*}
where we replaced $\partial$ acting on $\frac{\delta f}{\delta u}$ by $\lambda$,
and $\partial$ acting on $\frac{\delta g}{\delta u}$ by $\mu$.
The latter equation is the same as \eqref{eq:feb5_2}, due to equation \eqref{eq:1.9}.
\end{proof}
A more difficult proof of this proposition, in terms of Hamiltonian operators,
is given in \cite{O}.

\section{Compatible PVA structures and the Lenard scheme of integrability.}\label{sec:1half}
%

\subsection{Compatible PVA structures.}~~
\label{sec:1.6}
In this section we discuss a method, known as the \emph{Lenard scheme of integrability},
based on a \emph{bi-Hamiltonian pair} $(H,K)$,
to construct an infinite sequence of local functionals $\tint h_n\in\mc V/\partial\mc V,\,n\in\mb Z_+$,
which are pairwise in involution with respect to both Hamiltonian operators 
$H(\partial)$ and $K(\partial)$ (cf.\ Definition \ref{def:1.14}(d)).
This scheme will produce an infinite hierarchy of integrable Hamiltonian equations
\begin{equation}\label{eq:feb28_4}
\frac{du}{dt_n} \,=\, \{{h_n}_\lambda u\}_H\big|_{\lambda=0}
\,\,\Big(\,=\, \{{h_{n+1}}_\lambda u\}_K\big|_{\lambda=0}\Big)
\,\,,\,\,\,\,\,n\in\mb Z_+\,,
\end{equation}
such that the local functionals $\tint h_n,\,n\in\mb Z_+$,
are integrals of motion of every evolution equation in the hierarchy.

\begin{definition}
  \label{def:1.22}
Several $\lambda$-brackets $ \{ \, . \, _\lambda \, . \, \}_n,\,n=1,2,\dots,N$, 
on a differential algebra
$\mc{V}$ are called \emph{compatible} if any $\mb C$-linear
combination $\{ \, . \, _\lambda \, . \, \} = \sum_{n=0}^N c_n\{ \, . \,_\lambda \, . \, \}_n $ 
of them makes it a PVA.
If $\mc V$ is an algebra of differential functions,
and $H_n(\partial),\,n=1,2,\dots,N$, are the corresponding
to the compatible $\lambda$-brackets Hamiltonian operators,
defined by \eqref{eq:mar19},
we say that they are \emph{compatible} as well.
A \emph{bi-Hamiltonian pair} $(H,K)$ is a pair of compatible Hamiltonian operators
$H(\partial),K(\partial)$.
\end{definition}

\begin{example}
  \label{ex:1.23} 
(cf.\ Examples \ref{ex:1.3} and \ref{ex:1.4})  Let $R = \mb{C} [u,u^{\prime},u^{\prime\prime},\ldots]$.  The
$\lambda$-brackets
\begin{displaymath}
  \{ u_\lambda u \}_1 = (\partial +2 \lambda) u \, , \quad
     \{ u_\lambda u \}_2 = \lambda \,,\quad
     \{ u_\lambda u \}_3 = \lambda^3 \,,
\end{displaymath}
are compatible.  The corresponding Hamiltonian operators, defined by \eqref{eq:mar19}, are:
$$
  H_1(\partial) = u^{\prime} + 2u \partial\, , \quad H_2(\partial) = \partial\, , \quad H_3(\partial) = \partial^3 \, .
$$
\end{example}
\begin{example}
  \label{ex:1.24}
(cf.\ Example \ref{ex:1.5})  Let $\mf{g}$ be a Lie algebra with a
non-degenerate symmetric bilinear form $(\, . \, | \,
. \, )$, let $\{ u_i \}_{i \in I}$ be an orthonormal basis of
$\mf{g}$ and $[u_i,u_j] = \sum_k c^k_{ij} u_k$.  Let $R = \mb{C}
[u^{(n)}_i | i \in I\, , \, n \in \mb{Z} \}$.
The following $\lambda$-brackets on $R$ are compatible:
\begin{displaymath}
  \{ u_{i \lambda} u_j \}^{\prime} = \sum_k c^k_{ij} u_k \, , \quad
  \{ u_{i_\lambda} u_j \}^{\prime\prime} = \lambda\delta_{ij}\, , \quad
\{ u_{i_\lambda} u_j \}_k = c^k_{ij}\,,\,\, k \in I \, .
\end{displaymath}
The corresponding Hamiltonian operators, $H^{\prime},H^{\prime\prime}$ and $H^k,\,k\in I$, are given by:
\begin{displaymath}
H^{\prime}_{ij}(\partial) =-\sum_{k\in I} c^k_{ij} u_k\, , \quad
H^{\prime\prime}_{ij}(\partial) = \delta_{ij}\partial\, , \quad 
H^k_{ij}(\partial) = -c^k_{ij}\, .
\end{displaymath}
\end{example}

\begin{definition}\label{def:feb28}
Let $\mc V$ be an algebra of differential functions
and let $H(\partial)=\big(H_{ij}(\partial)\big)_{i,j\in I}$, $K(\partial)=\big(K_{ij}(\partial)\big)_{i,j\in I}$, 
be any two differential operators on $\mc V^{\oplus\ell}$.
An $(H,K)$-\emph{sequence} is a collection $\{F^n\}_{0\leq n\leq N}\subset\mc V^{\oplus\ell}$
such that
\begin{equation}\label{eq:1.27}
  K(\partial)F^{n+1}\,=\,H(\partial)F^{n}\,\,,\,\,\,\, 0\leq n\leq N-1\,.
\end{equation}
If $N=\infty$, we say that $\{F^n\}_{n\in\mb Z_+}$ is an \emph{infinite} $(H,K)$-\emph{sequence}.
\end{definition}
\begin{remark}\label{rem:july18_1}
Equation \eqref{eq:1.27} for an infinite $(H,K)$-sequence can be rewritten using 
the generating series
$F(z)=\sum_{n\in\mb Z_+} F^nz^{-n}$, as follows:
\begin{equation}\label{eq:july18_1}
\big(H(\partial)-zK(\partial)\big)F(z)=-zK(\partial)F^0\,.
\end{equation}
Assuming that
\begin{equation}\label{eq:july18_2}
K(\partial)F^0=0\,,
\end{equation}
and taking the pairing of both sides of \eqref{eq:july18_1} with $F(z)$,
we get the equation
\begin{equation}\label{eq:july18_3}
F(z)\cdot \big(H(\partial)-zK(\partial)\big)F(z)=0\,.
\end{equation}
Note that for a skew-adjoint differential operator $M(\partial)$,
we have $\tint F\cdot M(\partial)F=0$ for every $F\in\mc V^{\oplus\ell}$. 
Therefore,
if both operators $H(\partial)$ and $K(\partial)$ are skew-adjoint,
the LHS of the equation \eqref{eq:july18_3} is a total derivative,
hence we can reduce its order.
If, in addition, $\ell=1$ and $K(\partial)$ is a differential operator of order 1,
this produces a recursive formula for the sequence $F^{n},\, n\geq1$.
Explicitly, for $K(\partial)=\partial(k)+2k\partial$, where $k\in\mc V$, 
equation \eqref{eq:july18_3} becomes
\begin{equation}\label{eq:july18_4}
\partial \big(k F^0 F^{n+1}\big)
=\frac12 \sum_{m=0}^n F^{n-m}H(\partial)F^m
-\frac12 \sum_{m=1}^n\partial \big(k F^{n+1-m}F^m\big)\,.
\end{equation}
Notice that the RHS is a total derivative (since $H(\partial)$ is skew-adjoint), 
so that, given $F^0$ satisfying \eqref{eq:july18_2},
the above equation defines recursively $F^{n+1}$ up to adding a constant multiple of $(kF^0)^{-1}$.
\end{remark}
Note that in the special case in which $F^n=\frac{\delta h_n}{\delta u}$ 
for some $\tint h_n\in\mc V/\partial\mc V$, equation \eqref{eq:1.27}
can be written in terms of the $\lambda$-brackets associated to the operators $H(\partial)$
and $K(\partial)$ (cf.\ \eqref{eq:mar19}):
$$
\{{h_{n+1}}\,_\lambda\, u\}_K\big|_{\lambda=0}\,=\,\{{h_{n}}\,_\lambda\, u\}_H\big|_{\lambda=0}\,.
$$
\begin{lemma}\label{lem:feb28}
Suppose that the operators $H(\partial),\,K(\partial)$, acting on $\mc V^{\oplus\ell}$,
are skew-adjoint.
Then any $(H,K)$-sequence $\{F^n\}_{0\leq n\leq N}$ satisfies the orthogonality relations:
\begin{equation}\label{eq:feb28_6}
\tint F^m \cdot H(\partial)F^n \,=\, \tint F^m \cdot K(\partial)F^n \,=\, 0\,\,,
\,\,\,\,\ 0\leq m,n\leq N\,.
\end{equation}
\end{lemma}
\begin{proof}
If $m=n$, \eqref{eq:feb28_6} clearly holds, since both $H(\partial)$ and $K(\partial)$ 
are skew-adjoint operators.
We may assume, without loss of generality, that $m<n$ (hence $n\geq1$),
and we prove equation \eqref{eq:feb28_6} by induction on $n-m$.
By \eqref{eq:1.27} and the inductive assumption we have
$$
\tint F^m \cdot K(\partial)F^n = \tint F^m \cdot H(\partial)F^{n-1} = 0\,,
$$
and similarly, since $H(\partial)$ is skew-adjoint,
$$
\tint F^m \cdot H(\partial)F^n = -\tint F^n \cdot H(\partial)F^m = -\tint F^n \cdot K(\partial)F^{m+1} = 0\,.
$$
\end{proof}
The key result for the Lenard scheme of integrability is contained in the following theorem,
which will be proved in Section~\ref{2006_sec-dirac}.
\begin{theorem}\label{th:1.24}
Let $\mc V$ be an algebra of differential functions in the variables $\{u_i\}_{i\in I}$,
and suppose that $H(\partial)=\big(H_{ij}(\partial)\big)_{i,j\in I}$
and $K(\partial)=\big(K_{ij}(\partial)\big)_{i,j\in I}$ 
form a bi-Hamiltonian pair.
Assume moreover that $K(\partial)$ is non-degenerate,
in the sense that $K(\partial)M(\partial)K(\partial)=0$
implies $M(\partial)=0$ for any skew-adjoint differential operator
$M(\partial)$ of the form
$D_F(\partial)-D^*_F(\partial)$, $F\in\mc V^{\oplus\ell}$.
Let $\{F^n\}_{n=0,\cdots,N}\subset\mc V^{\oplus\ell}$ ($N\geq1$, and it can be infinite) 
be an $(H,K)$-sequence,
and assume that $F^0=\frac{\delta h_0}{\delta u},\,F^1=\frac{\delta h_1}{\delta u}$,
for two local functionals $\tint h_0,\,\tint h_1\in\mc V/\partial\mc V$.
Then all the elements of the sequence $F^n,\,n=0,\cdots,N$,
are closed, i.e.\ \eqref{eq:july18_5_b} holds for $F=F^n$.
\end{theorem}
\begin{proof}
This theorem is a corollary of the results in Section \ref{2006_sec-dirac}, see Remark \ref{rem:09jan21}.
\end{proof}


\begin{remark}\label{rem:jan2}
If $\ell=1$, the non-degeneracy condition for the differential operator 
$K(\partial)$
in Theorem \ref{th:1.24},
i.e.\ that $K(\partial)M(\partial)K(\partial)=0$ implies $M(\partial)=0$,
holds automatically unless $K(\partial)=0$.
It is because the ring of differential operators has no zero divisors when 
$\ell=1$. For general $\ell$ a sufficient condition of non-degenaracy
of $K(\partial)$ is 
non-vanishing of the determinant of its principal symbol.
\end{remark}

Theorem \ref{th:1.24} provides a way to construct an infinite hierarchy
of Hamiltonian equations, $\frac{du}{dt_n}=\{\tint h_n,u\}_H$,
and the associated infinite sequence of integrals of motion $\tint h_n,\,n\in\mb Z_+$.
In order to do this, we need to solve two problems.
First, given a bi-Hamiltonian pair $H(\partial),K(\partial)$ acting on $\mc V^{\oplus\ell}$,
we need to find an infinite $(H,K)$-sequence $\{F^n\}_{n\in\mb Z_+}$,
such that $F^0,\,F^1\in\im\frac\delta{\delta u}$.
Second, given an element $F\in\mc V^{\oplus\ell}$ which is closed,
we want to prove, that it is also ``exact", 
i.e.\ that $F=\frac{\delta h}{\delta u}$ for some local functional 
$\tint h\in\mc V/\partial\mc V$,
and we want to find an explicit formula for the local functional $\tint h\in\mc V/\partial\mc V$.
Then, if we have $F^n=\frac{\delta h_n}{\delta u}$ for all $n\in\mb Z_+$,
by Lemma \ref{lem:feb28},
the corresponding local functionals $\tint h_n$
are pairwise in involution with respect to the Lie brackets associated to both
Hamiltonian operators $H(\partial)$ and $K(\partial)$:
\begin{equation}\label{eq:feb28_3}
\big\{\tint h_m,\tint h_n\big\}_H\,=\,\big\{\tint h_m,\tint h_n\big\}_K\,=\,0\,\,,\,\,\,\,\text{for all\,\,} m,n\in\mb Z_+\,.
\end{equation}
The solution to the second problem is given by Propositions \ref{prop:july19_2} and \ref{prop:july22}.
We shall discuss, in Propositions \ref{prop:feb28_2} and \ref{prop:feb28_3},
a solution to the first problem:
we find sufficient conditions for the existence of an infinite 
$(H,K)$-sequence $\{F^n\}_{n\in\mb Z_+}$,
and we describe a practical method to construct inductively the whole sequence.

Let $\mc V$ be an algebra of differential functions.
Given a subset $U\subset \mc V^{\oplus\ell}$, let
$$
U^\perp \,:=\, \big\{P\in\mc V^\ell\,\big|\, \tint F\cdot P=0\,,\,\,\text{for all\,\,} F\in U\big\} 
\,\subset\, \mc V^\ell\,,
$$
and similarly, given a subset $\tilde U\subset \mc V^\ell$, let
$$
{\tilde U}^\perp \,:=\, \big\{F\in\mc V^{\oplus\ell}\,\big|\, \tint F\cdot P=0\,,\,\,\text{for all\,\,} P\in \tilde U\big\} 
\,\subset\, \mc V^{\oplus\ell}\,.
$$
\begin{proposition}\label{prop:feb28_2}
Let   $H(\partial)=\big(H_{ij}(\partial)\big)_{i,j\in I},\,K(\partial)=\big(K_{ij}(\partial)\big)_{i,j\in I}$,
be skew-adjoint operators on $\mc V^{\oplus\ell}$,
and let $\{F^n\}_{0\leq n\leq N}$ be a finite $(H,K)$-sequence satisfying the following condition:
\begin{equation}\label{eq:feb28_7}
\big(\Span_{\mb C}\{F^n\}_{0\leq n\leq N}\big)^\perp\subset \im K(\partial)\,.
\end{equation}
Then it can be extended 
to an infinite $(H,K)$-sequence $\{F^n\}_{n\in\mb Z_+}$.

In particular, letting $U=\Span_{\mb C}\big\{F^n\,\big|\,n\in\mb Z_+\big\}\subset\mc V^{\oplus\ell}$
and $U^\prime=\Span_{\mb C}\big\{F^n\,\big|\,n\geq1\big\}\subset\mc V^{\oplus\ell}$,
we have:
\begin{equation}\label{eq:feb28_10}
U^\prime\subset U\,\,,\,\,\,\,
H(\partial)U\,=\,K(\partial)U^\prime\,\,,\,\,\,\,
U^\perp\subset\im K(\partial)\,.
\end{equation}
\end{proposition}
\begin{proof}
We prove, by induction on $n\geq N+1$, that $F^n\in\mc V^{\oplus\ell}$
exists such that $K(\partial)F^n=H(\partial)F^{n-1}$.
Indeed, by the inductive assumption $\{F^k\}_{0\leq k\leq n-1}$ is an $(H,K)$-sequence,
therefore by Lemma \ref{lem:feb28} $H(\partial)F^{n-1}\perp F^m$ for every $m\in\{0,\dots,N\}$,
and hence by assumption \eqref{eq:feb28_7} $H(\partial)F^{n-1}\in\im K(\partial)$,
as we wanted.
The second part of the proposition is obvious.
\end{proof}

\begin{proposition}\label{prop:feb28_3}
Let $\mc V$ be an algebra of differential functions in the variables $\{u_i\}_{i\in I}$
and let $H(\partial),\,K(\partial)$, be skew-adjoint operators on $\mc V^{\oplus\ell}$.
Suppose that $U,U^\prime\subset\mc V^{\oplus\ell}$ are subspaces
satisfying conditions \eqref{eq:feb28_10}.
Then:
\begin{enumerate}
\alphaparenlist
\item $\Ker K(\partial)\subset \big(H(\partial)U\big)^\perp$,
\item for every $F^0\in \big(H(\partial)U\big)^\perp$ there exists
an infinite $(H,K)$-sequence $\{F^n\}_{n\in\mb Z_+}$,
and this $(H,K)$-sequence is such that 
$\Span_{\mb C}\{F^n\}_{n\in\mb Z_+}\subset \big(H(\partial)U\big)^\perp$,
\item if $F^0\in\Ker K(\partial)$ and $\{F^n\}_{n\in\mb Z_+},\,\{G^n\}_{n\in\mb Z_+}$
are two infinite $(H,K)$-sequences, then they are ``compatible", in the sense that
\begin{equation}\label{eq:feb28_11}
\tint F^m \cdot H(\partial)G^n \,=\, \tint F^m \cdot K(\partial)G^n \,=\, 0\,\,,
\,\,\,\,\text{for all\,\,}\,  m,n\in\mb Z_+\,.
\end{equation}
\end{enumerate}
\end{proposition}
\begin{proof}
Since $K(\partial)$ is skew-adjoint, we clearly have 
$\im K(\partial)\subset\big(\Ker K(\partial)\big)^\perp$.
Hence, from 
$H(\partial)U=K(\partial)U^\prime\subset\im K(\partial)\subset\big(\Ker K(\partial)\big)^\perp$,
it follows that
$\Ker K(\partial)\subset\big(H(\partial)U\big)^\perp$.
For part (b), 
we first observe that, if $\{F^n\}_{0\leq n\leq N}$ is any $(H,K)$-sequence starting 
at $F^0\in \big(H(\partial)U\big)^\perp$, then $F^n\in \big(H(\partial)U\big)^\perp$ for every 
$n\in\{0,\dots,N\}$.
Indeed, let $1\leq n\leq N$ and suppose by induction that $F^{n-1}\in \big(H(\partial)U\big)^\perp$.
If $G\in U$, by assumption \eqref{eq:feb28_10}
we have that $H(\partial)G=K(\partial)G_1$ for some $G_1\in U^\prime\subset U$.
Hence
\begin{eqnarray*}
\tint F^n\cdot H(\partial)G &=& \tint F^n\cdot K(\partial)G_1
= - \tint G_1\cdot K(\partial) F^n \\
&=& - \tint G_1\cdot H(\partial) F^{n-1}
= \tint F^{n-1}\cdot H(\partial)G_1 = 0\,,
\end{eqnarray*}
hence $F^n\in\big(H(\partial)U\big)^\perp$, as claimed.
Next, we need to prove the existence of an infinite $(H,K)$-sequence
starting at $F^0$.
Suppose, by induction on $N\geq0$,
that there exists an $(H,K)$-sequence $\{F^n\}_{0\leq n\leq N}$.
By the above observation, $F^N\in\big(H(\partial)U\big)^\perp$, 
or equivalently $H(\partial)F^N\in U^\perp$, so that, 
by the third condition in \eqref{eq:feb28_10},
there exists $F^{N+1}\in\mc V^{\oplus\ell}$ such that $H(\partial)F^N=K(\partial)F^{N+1}$,
as we wanted.
We are left to prove part (c).
We will prove equation \eqref{eq:feb28_11} by induction on $m\in\mb Z_+$.
For $m=0$ we have $\tint F^0\cdot K(\partial)G^n = - \tint G^n\cdot K(\partial)F^0 = 0$,
and $\tint F^0\cdot H(\partial)G^n = \tint F^0\cdot K(\partial)G^{n+1} = 0$, for every $n\in\mb Z_+$.
We then let $m\geq1$ and we use inductive assumption to get, for every $n\in\mb Z_+$,
$$
\tint F^m\cdot K(\partial)G^n = - \tint G^n\cdot K(\partial)F^m 
= - \tint G^n\cdot H(\partial)F^{m-1}
=  \tint F^{m-1}\cdot H(\partial)G^n
= 0\,,
$$
and similarly
$$
\tint F^m\cdot H(\partial)G^n
= \tint F^m\cdot K(\partial)G^{n+1} = 0\,.
$$
\end{proof}

\begin{remark}\label{rem:feb28_2}
If $H(\partial)$ and $K(\partial)$ are pseudodifferential operators on $\mc V^\ell$,
i.e.\ the corresponding $\lambda$-brackets on $\mc V$ are ``non-local'',
then the statements of Propositions \ref{prop:feb28_2} and \ref{prop:feb28_3}
should be modified to make sure that the elements $F^n,\,n\geq0$,
belong to the domains of the operators $H(\partial)$ and $K(\partial)$.
In Proposition \ref{prop:feb28_2} we then need the further assumption
\begin{equation}\label{eq:feb28_8}
\big(\Span_{\mb C}\{H(\partial)F^n\}_{0\leq n\leq N}\big)^\perp\subset \Dom H(\partial)\,.
\end{equation}
Similarly, in Proposition \ref{prop:feb28_3} we need the further assumption
\begin{equation}\label{eq:feb28_9}
\big(H(\partial)U)^\perp\subset \Dom H(\partial)\,.
\end{equation}
\end{remark}

By Proposition \ref{prop:feb28_2}, a sufficient condition for the existence of an
infinite $(H,K)$-sequence $\{F^n\}_{n\in\mb Z_+}$, is the existence of a finite 
$(H,K)$-sequence $\{F^n\}_{0\leq n\leq N}$ satisfying condition \eqref{eq:feb28_7}.
By Proposition \ref{prop:feb28_3},
if a finite $(H,K)$-sequence satisfying condition \eqref{eq:feb28_7} \emph{exists},
then any element $F^0\in\Ker K(\partial)$ can be extended to
an infinite $(H,K)$-sequence $\{F^n\}_{n\in\mb Z_+}$.
Therefore, in some sense,
the best place to look for the starting point of an infinite $(H,K)$-sequences is $\Ker K(\partial)$.

Moreover, by part (c) of Proposition \ref{prop:feb28_3},
the infinite $(H,K)$-sequences starting in $\Ker K(\partial)$
are especially nice since they are compatible with any other infinite $(H,K)$-sequence.
Notice that the space $\Ker K(\partial)$ is related to the center of the
Lie algebra $\mc V/\partial\mc V$ with Lie bracket $\{\cdot\,,\,\cdot\}_K$.
Indeed, if $F^0=\frac{\delta h_0}{\delta u}$ belongs to $\Ker K(\partial)$,
then the local functional $\tint h_0\in\mc V/\partial\mc V$ is central
with respect to the Lie bracket $\{\cdot\,,\,\cdot\}_K$,
and the converse is almost true (cf.\ Corollary \ref{cor:jan2}).
Proposition \ref{prop:feb28_3}(c) says that,
if $F^0\in\Ker K(\partial)$, then the corresponding infinite $(H,K)$-sequence 
$\{F^n\}_{n\in\mb Z_+}$ is ``central" among all the other infinite $(H,K)$-sequences,
in the sense of the compatibility condition \eqref{eq:feb28_11}.

We conclude this section with
the following result, which incorporates the results
of Theorem \ref{th:1.24} and Propositions \ref{prop:july19_2}, 
\ref{prop:feb28_2} and \ref{prop:feb28_3}.
It will be used in the following Sections \ref{sec:1.7} and \ref{sec:hd}
to implement the Lenard scheme of integrability in some explicit examples.
\begin{corollary}\label{cor:feb28}
Let $\mc V$ be a normal algebra of differential functions.
Let $H(\partial),\,K(\partial)$ be a bi-Hamiltonian pair acting on $\mc V^{\oplus\ell}$,
with $K(\partial)$ non-degenerate.
Let $\tint h_0,\, \tint h_1\in \mc V/\partial\mc V$ satisfy the following conditions:
\begin{enumerate}
\romanparenlist
\item $H(\partial)\frac{\delta h_0}{\delta u}=K(\partial)\frac{\delta h_1}{\delta u}$,
\item $\Big(\Span_{\mb C}\big\{\frac{\delta h_0}{\delta u},\frac{\delta h_1}{\delta u}\big\}\Big)^\perp
\subset\im K(\partial)$.
\end{enumerate}
Then there exists an infinite sequence of local functionals
$\tint h_n,\,n\in\mb Z_+$, which are pairwise in involution
with respect to the Lie brackets associated to both $H(\partial)$ and $K(\partial)$,
i.e.\ \eqref{eq:feb28_3} holds. 
Such hierarchy of compatible Hamiltonian functionals can be constructed,
recursively, using Proposition \ref{prop:july19_2} and the following equations:
\begin{equation}\label{eq:feb28_12}
H(\partial)\frac{\delta h_{n}}{\delta u}=K(\partial)\frac{\delta h_{n+1}}{\delta u}\,\,,\,\,\,\, n\in\mb Z_+\,.
\end{equation}
\end{corollary}
\begin{proof}
By the conditions (i) and (ii), the elements 
$\big\{\frac{\delta h_0}{\delta u},\frac{\delta h_1}{\delta u}\big\}$
form a finite $(H,K)$-sequence
satisfying condition \eqref{eq:feb28_7}.
Hence, by Proposition \ref{prop:feb28_2}, 
they can be extended to an infinite $(H,K)$-sequence $\{F^n\}_{n\in\mb Z_+}$
such that $F^0=\frac{\delta h_0}{\delta u}$ and $F^1=\frac{\delta h_1}{\delta u}$.
Since, by assumption, $(H,K)$ form a bi-Hamiltonian pair and $K$ is non-degenerate,
we can apply Theorem \ref{th:1.24} to deduce that every element $F^n,\,n\in\mb Z_+$,
is closed.
Hence we can apply Proposition \ref{prop:july19_2}
to conclude that $F^n$ is exact, namely there is $h_n\in\mc V$
such that $F^n=\frac{\delta h_n}{\delta u}$.
\end{proof}
\begin{proposition}\label{rem:mar19}
Let $\mc V$ be a normal algebra of differential polynomials in one variable $u$.
Let $K(\partial)=\partial$ and let $H(\partial)$ be a Hamiltonian operator
forming a bi-Hamiltonian pair with $\partial$.
Then there exists $\tint h_1\in\mc V/\partial\mc V$ such that
\begin{equation}\label{eq:july18_6}
H(\partial)1=\partial\frac{\delta h_1}{\delta u}\,.
\end{equation}
Furthermore, we can extend $\tint h_0=\tint u$ and $\tint h_1$ 
to an infinite sequence $\tint h_n,\,n\in\mb Z_+$,
satisfying \eqref{eq:feb28_12}.
\end{proposition}
\begin{proof}
First, notice that, since $H(\partial)$ is a skew-adjoint operator, $\tint H(\partial)1=0$,
namely there exists $F\in\mc V$ such that $H(\partial)1=\partial F$.
Next, the condition that the Hamiltonian operators $H(\partial)$ and $K=\partial$ 
form a bi-Hamiltonian pair reads
$$
\big\{u_\lambda H(\mu)\big\}_K-\big\{u_\mu H(\lambda)\big\}_K
=\big\{H(\lambda)_{\lambda+\mu} u\big\}_K\,.
$$
Using equation \eqref{eq:1.9} and letting $\lambda=0$ we then get
$$
\sum_{n\in\mb Z_+} \frac{\partial H(0)}{\partial u^{(n)}}\mu^{n+1}
-\sum_{n\in\mb Z_+} (-\mu-\partial)^n\frac{\partial H(0)}{\partial u^{(n)}} = 0\,.
$$
Since $H(0)=\partial F$, using the commutation rule \eqref{eq:comm_frac},
we obtain, after simple algebraic manipulations, that
$\mu(\partial+\mu)\big(D_F(\mu)-D_F(-\mu-\partial)\big)=0$.
This implies that $D_F(\partial)$ is a self-adjoint differential operator,
i.e.\ $F\in\mc V$ is closed.
By Proposition \ref{prop:july19_2}, $F$ is a variational derivative.
Thus, all the assumptions of Corollary \ref{cor:feb28} hold,
which proves the proposition.
\end{proof}

\vspace{3ex}
\subsection{The KdV hierarchy.}~~
\label{sec:1.7}
Let $R=\mb{C} [u,u^{\prime},u^{\prime\prime},\dots]$ with the following two compatible
$\lambda$-brackets (see ~Example~\ref{ex:1.23}):
\begin{eqnarray}\label{sstef_1}
\{ u_\lambda u \}_H 
&=& \{ u_\lambda u \}_1 + c\{ u_\lambda u \}_3
= (\partial+2\lambda)u+c\lambda^3\,, \nonumber\\
\{ u_\lambda u \}_K 
&=& \{ u_\lambda u \}_2 = \lambda\, ,
\end{eqnarray}
where $ c \in \mb{C}$, so that the corresponding Hamiltonian operators are:
\begin{equation}\label{eq:1.30}
H(\partial)= u^{\prime} + 2u \partial + c \partial^3\,\,,\,\,\,\, K(\partial)=\partial\, .
\end{equation}
We shall use the Lenard scheme discussed in the previous Section,
to construct an infinite hierarchy of integrable Hamiltonian equations,
which includes the classical \emph{KdV-equation}
\begin{equation}\label{eq:kdveq}
\frac{du}{dt}\,=\,3uu^{\prime}+cu^{\prime\prime\prime}\,.
\end{equation}

By Proposition \ref{rem:mar19},
we conclude that
there exists a sequence of local functionals $\tint h_n\in R/\partial R,\,n\in\mb Z_+$,
where $h_0=u$,
which are in involution with respect to both Lie brackets 
$\{\cdot\,,\,\cdot\}_H$ and $\{\cdot\,,\,\cdot\}_K$,
and such that the corresponding variational derivatives 
$F^n=\frac{\delta h_n}{\delta u},\,n\in\mb Z_+$
form an infinite $(H,K)$-sequence (cf.\ Definition \ref{def:feb28}).

In fact, we can compute the whole hierarchy of Hamiltonian functionals $\tint h_n,\,n\in\mb Z_+$,
iteratively as follows:
assuming we computed $\tint h_{n},\,n\in\mb Z_+$,
we can find $\tint h_{n+1}$ by solving equation \eqref{eq:feb28_12}.
Clearly $F^n=\frac{\delta h_n}{\delta u}$ is defined by equation \eqref{eq:feb28_12}
uniquely up to an additive constant.
To compute it explicitly we have to find a preimage of $H(\partial)F^{n-1}$
under the operator $\partial$,
which is easy to do, inductively on the differential order of $H(\partial)F^{n-1}$.
Similarly, $\tint h_n$ is defined uniquely up to adding arbitrary 
linear combinations of $\tint 1$ and $\tint u$,
and to compute it explicitly we can use Proposition \ref{prop:july22},
since $\Ker(\Delta+1)=0$.
For example, for the first few terms of the hierarchy we have:
\begin{eqnarray}\label{sstef_2}
&&  {F^0} = 1\,\,,\,\,\,\, h_0 = u\,\,;\,\,\,\,
  {F^1} = u \,\, , \,\,\,\, h_1 = \frac12 u^2\,\,;\,\,\,\,
  {F^2} = \frac{3}{2} u^2+cu^{\prime\prime} \,\, , \,\,\,\, h_2 = \frac{1}{2} u^3 + \frac12 cuu^{\prime\prime}\,\,;\nonumber\\
&&  {F^3} = \frac{5}{2} u^3 + 5cuu^{\prime\prime} + \frac52 c(u^{\prime})^2 + c^2 u^{(4)}\,\,,\,\,\,\,
h_3 = \frac58 u^4 + \frac53c u^2u^{\prime\prime} + \frac56 cuu^{\prime 2} + \frac12 c^2uu^{(4)}\,.
\end{eqnarray}

By Corollary \ref{cor:feb28}  the local functionals $\tint h_n,\,n\in\mb Z_+$, 
are in involution with respect to both Lie brackets associated to $H(\partial)$ and $K(\partial)$,
i.e.\ \eqref{eq:feb28_3} holds.
Hence, if we let $\tint h_{-1}=\tint 1$,
we have an increasing filtration of abelian subalgebras of $R/\partial R$
with Lie bracket $\{\cdot\,,\,\cdot\}_K$ (or $\{\cdot\,,\,\cdot\}_H$),
$\mf H^{-1}_c\subset \mf H^0_c\subset \mf H^1_c\subset\mf H^2_c\subset\dots$, 
where 
$\mf H^n_c \,=\, \Span_{\mb C} \big\{\tint h_k\}_{k=-1}^n$.
Notice that the space $\mf H^n_c$
does not depend on the choice of the local functionals $\tint h_n,\,n\in\mb Z_+$,
solving equation \eqref{eq:feb28_12}. We also denote
$\mf H_c=\bigcup_{n\in\mb Z_+}\mf H^n_c=\Span_{\mb C}\{\tint h_n\}_{n=-1}^\infty$.

The local functionals $\tint h_n,\,n\geq0$ are integrals of motion of the hierarchy of evolution
equations \eqref{eq:feb28_4}.
The 0\st{th} and the 1\st{st} equations are $\frac{du}{dt_0}=0$
and $\frac{du}{dt_1} = \frac{\partial u}{\partial x}$, and they say that 
$u(x,t_0,t_1,\dots ) = u (x+t_1, t_2,\dots)$,
the 2\st{nd} equation is the classical KdV equation \eqref{eq:kdveq}
provided that $c \neq 0$ (then $c$ can be made equal 1 by rescaling),
and the 3\st{rd} equation is the simplest higher KdV equation:
$$
  \frac{du}{dt_3} = \frac{15}{2} u^2u^{\prime} + 10cu^{\prime}u^{\prime\prime} +
     5cuu^{\prime\prime\prime} + c^2 u^{(5)}\ ,
$$
provided again that $c \neq 0$ (it can be made equal to 1 by rescaling).

\begin{remark}
  \label{rem:1.25}
Equation \eqref{eq:july18_4} for $k=1/2$ gives
an explicit recursion formula for the sequence $F^0=1,\,F^1,\,F^2,\dots$,
which in this case reads
\begin{equation}\label{sstef_7}
F^{n+1}\,=\,
\sum_{m=0}^n\Big(cF^{n-m}(\partial^2 F^m)-\frac c2 (\partial F^{n-m})(\partial F^m)
+u F^{n-m}F^m\Big)-\frac12 \sum_{m=1}^n F^{n+1-m}F^m\,,
\end{equation}
up to adding an arbitrary constant.
This gives an alternative way to compute the $F^n$'s.
\end{remark}


If we put $c=0$ in the KdV hierarchy \eqref{sstef_2} we get the so called 
\emph{dispersionless KdV hierarchy}.
We denote in this case $\overline{F}^n$ and $\tint \overline{h}_n$
in place of $F^n$ and $\tint h_n$.
The recursion relation \eqref{eq:feb28_12} for the elements 
$\overline{F}^n=\frac{\delta h_n}{\delta u},\,n\in\mb Z_+$, is
$$
\partial \overline{F}^{n+1} \,=\, (u^{\prime}+2u\partial) \overline{F}^{n}\,,\,\,n\in\mb Z_+\,,
$$
so if we put, as before, $\overline{F}^0=1$, it is immediate to find an explicit solution 
by induction:
\begin{equation}\label{sstef_8}
\overline{F}^n\,=\,\frac{(2n-1)!!}{n!}u^n
\,\,,\,\,\,\,
\tint \overline{h}_n=\frac{(2n-1)!!}{(n+1)!} \tint u^{n+1}, \, n\in\mb Z_+\,,
\end{equation}
where $(2n-1)!!:=1\cdot 3\cdots(2n-1)$.
The corresponding dispersionless KdV hierarchy is
$$
\frac{du}{dt_n}\,=\,\frac{(2n-1)!!}{n!} \partial (u^n)\,\,,\,\,\,\, n\in\mb Z_+\,.
$$
As before, we denote $\tint \overline h_{-1}=\tint 1$,
and we have a uniquely defined increasing filtration of abelian subalgebras,
with respect to the Lie bracket $\{\cdot\,,\,\cdot\}_K$
(or $\{\cdot\,,\,\cdot\}_H$ with $c=0$):
$\mf H^{-1}_0\subset \mf H^0_0\subset\mf H^1_0\subset\dots\subset\mf H_0$,
where 
$\mf H^n_0=\Span_{\mb C}\big\{\tint \overline h_k\big\}_{k=-1}^n$
and $\mf H_0=\bigcup_{n=-1}^\infty \mf H^n_0$.


Finally, as $c\to\infty$ the KdV hierarchy turns into the \emph{linear KdV hierarchy},
corresponding to the pair $H(\partial)=\partial^3,\,K(\partial)=\partial$.
We denote in this case $\underline{F}^n$ and $\tint \underline{h}_n$
in place of $F^n$ and $\tint h_n$.
The Lenard recursion relation \eqref{eq:1.27} becomes
$\partial \underline F^{n+1}=\partial^3 \underline F^n,\,n\geq1$,
and a solution of such recursion relation, 
such that all $\underline F^n$ are in the image of $\frac\delta{\delta u}$,
is obtained starting with $\underline F^1=u$.
In this case, $\underline F^{n+1}$ is defined by $\underline F^n$
uniquely up to adding an arbitrary constant,
while $\tint \underline h_{n+1}$ is defined uniquely
up to adding an arbitrary linear combination of $\tint 1$ and $\tint u$.
An explicit solution is given by 
\begin{equation}\label{eq:v151a}
\underline F^{n+1}=u^{(2n)}\,\,,\,\,\,\,
\tint \underline h_{n+1}=\frac12 \tint uu^{(2n)}=\frac{(-1)^n}2 \tint (u^{(n)})^2,\,n\in\mb Z_+\,,
\end{equation}
giving the linear KdV hierarchy
$$
\frac{du}{dt_n}=u^{(2n+1)}\,\,,\,\,\,\,n\in\mb Z_+\,.
$$
By Theorem \ref{th:1.24}, the local functionals $\tint \underline h_n,\,n\geq1$,
are in involution with respect to both the Lie brackets $\{\cdot\,,\,\cdot\}_H$ 
and $\{\cdot\,,\,\cdot\}_K$.
If we let $\tint \underline h_{-1}=\tint 1$ and $\tint \underline h_0=\tint u$,
we thus get, as before, a uniquely defined increasing
filtration of abelian subalgebras,
with respect to both $\{\cdot\,,\,\cdot\}_K$ and $\{\cdot\,,\,\cdot\}_H$:
$\mf H^{-1}_\infty\subset \mf H^0_\infty\subset\mf H^1_\infty\subset\dots\subset\mf H_\infty$,
where 
$\mf H^n_\infty=\Span_{\mb C}\big\{\tint \underline h_k\big\}_{k=-1}^n$
and $\mf H_\infty=\bigcup_{n=-1}^\infty \mf H^n_\infty$.

\begin{theorem}
  \label{prop:1.26}
The spaces $\mf H_c=\Span_{\mb C} \big\{\tint h_n\big\}_{n=-1}^\infty,\,c\neq0$,
$\mf H_0=\Span_{\mb C} \big\{\tint \overline h_n\big\}_{n=-1}^\infty$, 
and $\mf H_\infty=\Span_{\mb C} \big\{\tint \underline h_n\big\}_{n=-1}^\infty$,
defined by \eqref{sstef_2}, \eqref{sstef_8} and \eqref{eq:v151a} respectively,
are maximal abelian subalgebras of the Lie algebra $R/\partial R$ with the
bracket
\begin{equation}
  \label{eq:1.35}
  \{ \tint f,\tint g \} \,=\, \int \frac{\delta g}{\delta u} \partial 
  \frac{\delta f}{\delta u} \qquad  \big(\text{cf.\ }\, (1.60)\big)\,.
\end{equation}
Moreover, the local functionals $\tint h_n,\,n\geq-1$
(resp. $\tint \overline h_n,\,n\geq-1$, and $\tint \underline h_n,\,n\geq-1$),
form a maximal linearly independent set of integrals of motion of
the KdV (resp. dispersionless KdV, or linear KdV) hierarchy.
\end{theorem}
\begin{proof}
We consider first the case of the dispersionless KdV hierarchy $\tint \overline h_n,\,n\geq 1$.
Notice that, by \eqref{sstef_8}, we have $\mf H_0=\mb C[u]\subset R/\partial R$.
Let $\tint f\in R/\partial R$ be a local functional commuting 
with all elements $\tint \overline h_n,\,n\geq 1$.
We have
$$
\int u^n\partial\frac{\delta f}{\delta u}\,=\,\frac1{n+1} \big\{\tint f,\tint u^{n+1}\big\}\,=\,0
\,\,,\,\,\text{ for every }\,\,\,\, n\in\mb Z_+\,.
$$
Hence by Lemma \ref{lem:sstef_3} we get that $\partial\frac{\delta f}{\delta u}\in\partial\mb C[u]$,
namely (since $\Ker(\partial)=\mb C$) $\frac{\delta f}{\delta u}\in\mb C[u]$.
On the other hand $\mb C[u]=\frac\delta{\delta u}(\mb C[u])$,
so that we can find $\alpha(u)\in\mb C[u]$ such that 
$\frac{\delta f}{\delta u}=\frac{\delta\alpha(u)}{\delta u}$.
Equivalently, $f-\alpha(u)\in\Ker\big(\frac\delta{\delta u}\big)=\mb C+\partial R$
(see Proposition \ref{prop:july19_1}), so that $f\in \partial R+\mb C[u]$, as we wanted.

Next, we consider the linear KdV hierarchy $\tint \underline h_n,\,n\geq-1$.
Let $\tint f\in R/\partial R$ be a local functional commuting with 
all elements $\tint \underline h_n,\,n\geq-1$, namely, 
since $\frac{\delta h_{n+1}}{\delta u}=u^{(2n)}$,
such that
$$
u^{(2n)}\partial \frac{\delta f}{\delta u}\,\in\,\partial R
\,\,,\,\,\text{ for every } n\in\mb Z_+\,.
$$
By Lemma \ref{lem:feb18} this is possible if and only if 
$\partial \frac{\delta f}{\delta u}\in\partial\, \Span_{\mb C}\{u^{(2n)}\,,\,n\in\mb Z_+\}$,
which is equivalent to saying that $\tint f\in\Span_{\mb C}\{\tint 1,\tint u,\tint uu^{(2n)}\}$,
as we wanted.

We finally prove the general case, for arbitrary value of the central charge $c\in\mb C$, 
by reducing it to the case $c=0$.
According to the decomposition \eqref{eq:2009july3},
%
%
the elements $F^n,\,n\in\mb Z_+$,
defined by the recurrence formula \eqref{sstef_7},
can be written, for arbitrary $c\in \mb C$, in the form
\begin{equation}\label{sstef_9}
F^n\,=\,\overline{F}^n+G^n\,,
\end{equation}
where $\overline{F}^n\in\mb C[u]$ is given by \eqref{sstef_8} and $G^n\in R^\partial$.
Likewise, the local functionals $\tint h_n\in R/\partial R$ in \eqref{sstef_2} 
admit a decomposition
\begin{equation}\label{sstef_10}
h_n\,=\,\overline{h}_n + r_n\,,
\end{equation}
where $\overline{h}_n\in \mb C[u]$, is given by \eqref{sstef_8}, 
and $r_n\in R^\partial$.
Indeed,
let $\pi_0:\,R\twoheadrightarrow \mb C[u]$ be the projection
defined by the decomposition \eqref{eq:2009july3}. We need to prove 
that $\pi_0(F^n)=\overline{F}^n$
and $\pi_0(h_n)=\overline h_n$ for every $n\in\mb Z_+$.
If we take the image under $\pi_0$ of both sides of equation \eqref{sstef_7},
we get the following recurrence relation for the elements $\pi_0(F^n)\in\mb C[u]$:
$$
\pi_0(F^{n+1})\,=\,
\sum_{m=0}^n u \pi_0(F^{n-m})\pi_0(F^m)-\frac12 \sum_{m=1}^n \pi_0(F^{n+1-m})\pi_0(F^m)\,.
$$
Since in the above equation the central charge $c$ does not appear anywhere, 
its solution $\pi_0(F^n)\in \mb C[u]$ is the same for every value of $c$, 
in particular it is the same as the solution for $c=0$, i.e.\ $\pi_0(F^n)=\overline{F}^n$.
Furthermore, by Proposition \ref{prop:july22},
a representative for the local functional $\tint h_n\in R/\partial R$ can be obtained by taking
$h_n=\Delta^{-1}(uF^n)$,
where by $\Delta^{-1}$ we mean the inverse of $\Delta$, restricted to the space of differential
polynomials with zero constant term.
Therefore such $h_n$ clearly has a decomposition as in \eqref{sstef_10} 
since $\Delta^{-1}(u\overline{F}^n)=\overline{h}_n\in\mb C[u]$,
while $\Delta^{-1}(u G)\in R^\partial$ for every $G\in R^\partial$.

The basic idea to reduce to the case $c=0$ is to consider the change of variable 
$x\mapsto x/\epsilon$ in $u=u(x)$ and take the limit $\epsilon\to0$.
In order to formalize this idea, 
we define the map $\pi_\epsilon:\, \mb C[\lambda]\otimes R\to \mb C[\epsilon,\lambda]\otimes R$ 
given by
$$
\pi_\epsilon\big(F(\lambda;u,u^{\prime},u^{\prime\prime},\dots)\big)\,=\,F(\epsilon\lambda;u,\epsilon u^{\prime},\epsilon^2 u^{\prime\prime},\dots)\,.
$$
Clearly, we have induced maps $\pi_\epsilon$ both on the space of differential polynomials,
$\pi_\epsilon:\,R\to\mb C[\epsilon]\otimes R$, 
defined by the natural embedding $R\subset \mb C[\lambda]\otimes R$,
and on the space of local functionals 
$\pi_\epsilon:\,R/\partial R\to\mb C[\epsilon]\otimes R/\partial R
\simeq \big(\mb C[\epsilon]\otimes R\big)/\partial \big(\mb C[\epsilon]\otimes R\big)$,
defined by $\pi_\epsilon\big(\tint f\big):=\tint \pi_\epsilon f$.
The decomposition \eqref{sstef_10}
can then be restated by saying that
\begin{equation}\label{eq:mar8_2}
\pi_\epsilon(h_n)=\bar{h}_n+\epsilon r_{\epsilon,n}\,,
\end{equation}
where $\bar{h}_n$ is given by \eqref{sstef_8} and $r_{\epsilon,n}\in \mb C[\epsilon]\otimes R$.

It is clear by the definition of $\pi_\epsilon$ that
$\frac{\partial}{\partial u^{(n)}}\pi_\epsilon=\epsilon^n\pi_\epsilon\frac{\partial}{\partial u^{(n)}}$,
so that $\frac{\delta}{\delta u}\pi_\epsilon=\pi_\epsilon\frac{\delta}{\delta u}$.
In particular, recalling the definition \eqref{eq:1.35}
of the Lie bracket on $R/\partial R$, we have
\begin{equation}\label{eq:mar8_4}
\big\{\tint \pi_\epsilon f , \tint \pi_\epsilon g\big\}
\,=\,
\epsilon^{-1}\pi_\epsilon\big\{\tint f , \tint g\big\}\,,
\end{equation}
for every $f,g\in R$.
Let then $\tint f\in R/\partial R$, for some $f\in R$, 
be a local functional commuting with all $\tint h_n,\,n\in\mb Z_+$,
and assume, by contradiction, that $\tint f\not\in\Span_{\mb C}\big\{\tint h_n\big\}_{n=-1}^\infty$.
Notice that, by the decomposition \eqref{sstef_10} and by equation \eqref{sstef_8},
we can subtract from $f$ an appropriate
linear combination of the elements  $h_n$, so that,
without loss of generality, $f\in R^\partial$ and $\tint f\neq0$.
In other words, we have
\begin{equation}\label{eq:mar8_3}
\pi_\epsilon f \,=\, \epsilon^k\big(\overline f + \epsilon \, r_\epsilon\big)\,.
\end{equation}
where $k\geq1$ and $\overline f\in R^\partial$ is such that $\tint \overline f\neq0$.
By equation \eqref{eq:mar8_4} and by the assumption on $\tint f$,
we have $\big\{\tint \pi_\epsilon f , \tint \pi_\epsilon h_n\big\}=0$
for all $n\in\mb Z_+$, namely
\begin{equation}\label{eq:mar8_5}
\big\{\pi_\epsilon f \,_\lambda\, \pi_\epsilon h_n\big\}\big|_{\lambda=0}
\,\in\,\mb C[\epsilon]\otimes\partial R\,,\,\,\text{for all\,\,}\,n\in\mb Z_+\,.
\end{equation}
On the other hand, by the decompositions \eqref{eq:mar8_2} and \eqref{eq:mar8_3},
the minimal power of $\epsilon$ appearing in the LHS of \eqref{eq:mar8_5}
is $\epsilon^k$, and looking at the coefficient of $\epsilon^k$ we get that
$$
\big\{\bar f _\lambda \bar h_n\big\}\big|_{\lambda=0}\,\in\,\partial R\,,\,\,\text{for all\,\,}\,n\in\mb Z_+\,,
$$
namely $\big\{\tint\bar f,\tint\bar h_n\big\}=0$ for all $n\in\mb Z_+$.
By the maximality of $\mf H_0$,
it follows that $\tint \overline f\in\mf H_0$, namely $\overline{f}\in\mb C[u]+\partial R$,
contradicting the assumption that
$\overline{f}\in R^\partial\backslash\partial R$.

Finally, all local functionals $\tint h_n$ are linearly independent.
Indeed, since $\frac{\delta h_n}{\delta u}=F^n$, it suffices to show that the elements 
$F^n,\,n\in\mb Z_+$ are linearly independent in $R=\mb C[u^{(n)}\,|\,n\in\mb Z_+]$.
For $c=0$ (respectively $c=\infty$) this is clear from the explicit expression of $\overline F^n$
in \eqref{sstef_8} (resp. $\underline F^n$ in \eqref{eq:v151a}).
For $c\in\mb C\backslash\{0\}$, it follows by the decomposition \eqref{eq:2009july3}
and by the fact that $\pi_0(F^n)=\overline F^n\in\mb C[u],\,n\in\mb Z_+$.
\end{proof}
In view of Remark 1.26, the proof of Theorem \ref{prop:1.26} 
shows that the KdV hierarchy is integrable.

The following result shows that the only hierarchy that can be obtained
using the bi-Hamiltonian pair $(H=\partial^3,K=\partial)$ is $\tint\underline h_n$
defined in \eqref{eq:v151a},
for any choice of $\tint h_0,\,\tint h_1$.
We believe that the same should be true for all values of $c$.
\begin{proposition}\label{prop:feb20}
All the solutions $\tint f,\,\tint g\in R/\partial R$ of the following equation
\begin{equation}\label{eq:feb20}
\frac{\delta f}{\delta u}=\partial^2\frac{\delta g}{\delta u}\,.
\end{equation}
are obtained by taking linear combinations of the pairs $\tint h_{n+1},\,\tint h_n,\,n\geq0$,
defined in \eqref{eq:v151a}.
\end{proposition}
\begin{proof}
Let $N=\max\{\ord(f),\ord(g)+1\}$, so that
$f=f(u,u^{\prime},\dots,u^{(N)}),\,g=g(u,u^{\prime},\dots,u^{(N-1)})$.
By induction, we only need to show that
\begin{equation}\label{eq:feb20_2}
f\equiv\gamma\big(u^{(N)}\big)^2+f_1\,\,,
\,\,\,\,g\equiv-\gamma\big(u^{(N-1)}\big)^2+g_1\,\mod\partial R\,,
\end{equation}
for some $\gamma\in\mb C$ and $f_1,g_1\in R$
such that $\ord(f_1)\leq N-1,\,\ord(g_1)\leq N-2$.
We start by observing that, if $\phi\in R$ has $\ord(\phi)\leq N-1$,
then there exists $\psi\in R$ with $\ord(\psi)\leq N-1$
such that $\phi u^{(N)}\equiv\psi\mod\partial R$.
Indeed, in general we have $\phi=\sum_{k\geq0}\phi_k \big(u^{(N-1)}\big)^k$
with $\ord(\phi_k)\leq N-2$,
and we can then take $\psi=-\sum_{k\geq 0} \frac1{k+1}(\partial \phi_k)\big(u^{(N-1)}\big)^{k+1}$.
Therefore, since $f$ and $g$ are defined modulo $\partial R$,
we can assume, without loss of generality,
that $f$, as a polynomial in $u^{(N)}$, and $g$, as a polynomial in $u^{(N-1)}$,
do not have any linear term.
Under this assumption, claim \eqref{eq:feb20_2} reduces to saying that
\begin{equation}\label{eq:feb21_1a}
\frac{\partial^2 f}{\partial {u^{(N)}}^2} \,=\, - \frac{\partial^2 g}{\partial {u^{(N-1)}}^2} \,\in\, \mb C\,.
\end{equation}
By assumption $\ord(f)\leq N$, hence we can use the commutation rule \eqref{eq:comm_frac}
and the definition \eqref{eq:0.6} of the variational derivative to get
\begin{eqnarray}\label{eq:feb21_1b}
\frac{\partial}{\partial u^{(2N)}}\frac{\delta f}{\delta u}
&=& \sum_{n=0}^N (-1)^n \frac{\partial}{\partial u^{(2N)}} \partial^n \frac{\partial f}{\partial u^{(n)}}
= (-1)^N \frac{\partial^2 f}{\partial {u^{(N)}}^2}\,, \\
\frac{\partial}{\partial u^{(2N-1)}}\frac{\delta f}{\delta u}
&=& \sum_{n=0}^N (-1)^n \frac{\partial}{\partial u^{(2N-1)}} \partial^n \frac{\partial f}{\partial u^{(n)}} 
= (-1)^N N \partial \frac{\partial^2 f}{\partial {u^{(N)}}^2}\,,
\nonumber
\end{eqnarray}
and similarly, since $\ord(g)\leq N-1$, we have
\begin{eqnarray}\label{eq:feb21_2}
\frac{\partial}{\partial u^{(2N)}} \partial^2 \frac{\delta g}{\delta u}
&=& \sum_{n=0}^{N-1} (-1)^n \frac{\partial}{\partial u^{(2N)}} \partial^{n+2} \frac{\partial f}{\partial u^{(n)}}
= (-1)^{N-1} \frac{\partial^2 g}{\partial {u^{(N-1)}}^2}\,, \\
\frac{\partial}{\partial u^{(2N-1)}} \partial^2 \frac{\delta g}{\delta u}
&=& \sum_{n=0}^{N-1} (-1)^n \frac{\partial}{\partial u^{(2N-1)}} 
\partial^{n+2} \frac{\partial g}{\partial u^{(n)}} 
= (-1)^{N-1} (N+1) \partial \frac{\partial^2 g}{\partial {u^{(N-1)}}^2}\,. \nonumber
\end{eqnarray}
The first identity in \eqref{eq:feb21_1b} and the first identity in \eqref{eq:feb21_2},
combined with assumption \eqref{eq:feb20}, give
\begin{equation}\label{eq:feb21_3}
\frac{\partial^2 f}{\partial {u^{(N)}}^2} = - \frac{\partial^2 g}{\partial {u^{(N-1)}}^2}\,,
\end{equation}
while the second identities in \eqref{eq:feb21_1b} and \eqref{eq:feb21_2} give
\begin{equation}\label{eq:feb21_4}
N \partial \frac{\partial^2 f}{\partial {u^{(N)}}^2} 
= - (N+1) \partial \frac{\partial^2 g}{\partial {u^{(N-1)}}^2}\,.
\end{equation}
Clearly, equations \eqref{eq:feb21_3} and \eqref{eq:feb21_4} imply \eqref{eq:feb21_1a},
thus completing the proof.
\end{proof}

\vspace{3ex}
\subsection{The Harry Dym (HD) hierarchy.}~~
\label{sec:hd}
We can generalize the arguments of Section \ref{sec:1.7} to the case in which we take 
$\{\,_\lambda\,\}_H$ and $\{\,_\lambda\,\}_K$ to be two arbitrary
linearly independent linear combinations
of the $\lambda$-brackets $\{\,_\lambda\,\}_n,\,n=1,2,3$, in Example \ref{ex:1.23}.
Namely, in terms of Hamiltonian operators, we take
\begin{equation}\label{jan2_14}
H(\partial)\,=\,\alpha_1\big(u^{\prime}+2u\partial\big)+\alpha_2\partial+\alpha_3\partial^3\,\,,\,\,\,\,
K(\partial)\,=\,\beta_1\big(u^{\prime}+2u\partial\big)+\beta_2\partial+\beta_3\partial^3\,,
\end{equation}
acting on $\mc V$, a certain algebra of differential functions
extending $R=\mb C[u^{(n)}\,|\,n\in\mb Z_+]$, which will be specified later.

Clearly, if we rescale $H(\partial)$ or $K(\partial)$ by a constant factor $\gamma\in\mb C\backslash\{0\}$,
we still get a bi-Hamiltonian pair and the theory of integrability does not change:
any $(H,K)$ sequence $\{F^n\}_{0\leq n\leq N}$ (cf.\ Definition \ref{def:feb28})
can be rescaled to the $(\gamma H,K)$-sequence $\{\gamma^n F^n\}_{0\leq n\leq N}$,
or to the $(H,\gamma K)$-sequence $\{\gamma^{-n} F^n\}_{0\leq n\leq N}$.
The corresponding local functionals $\tint h_n,\,n\in\mb Z_+$,
satisfying $F^n=\frac{\delta h_n}{\delta u}$, can be rescaled accordingly.
Furthermore, if $(H,K)$ is a bi-Hamiltonian pair, so is $(H+\alpha K,K)$
for any $\alpha\in\mb C$.
It is immediate to check that,
if $\{F^n\}_{0\leq n\leq N}$ is an $(H,K)$-sequence,
then 
$$
F^n_\alpha = \sum_{k=0}^n\binom{n}{k}\alpha^k F^{n-k}\,,\,\,0\leq n\leq N\,,
$$
defines an $(H+\alpha K,K)$-sequence.
In particular, if $\{\tint h_n\}_{n\in\mb Z_+}$ is an infinite hierarchy of local functionals
satisfying equation \eqref{eq:feb28_12},
hence in involution with respect to both Lie brackets $\{\cdot\,,\cdot\}_H$ and $\{\cdot\,,\cdot\}_K$
(or any linear combination of them),
then $\tint h_{\alpha,n}=\sum_{k=0}^n\binom{n}{k}\alpha^k\tint h_{n-k},\,n\in\mb Z_+$,
defines an infinite hierarchy of local functionals
satisfying \eqref{eq:feb28_12} with $H$ replaced by $H+\alpha K$,
hence still in involution with respect to both the Lie brackets $\{\cdot\,,\cdot\}_H$ 
and $\{\cdot\,,\cdot\}_K$.
The corresponding increasing filtration of abelian subalgebras,
$\mf H^0\subset\mf H^1\subset\mf H^2\subset\dots\mc V/\partial\mc V$,
given by $\mf H^N=\Span_{\mb C}\big\{\tint h_{\alpha,n}\big\}_{0\leq n\leq N}$,
is independent of $\alpha$.
In conclusion, from the point of view of integrability and of the Lenard scheme as discussed
in Section \ref{sec:1.6},
nothing changes if we replace the bi-Hamiltonian pair $(H,K)$
by any pair of type $(\alpha K+\beta H,\gamma K)$,
with $\alpha,\beta,\gamma\in\mb C$ and $\beta,\gamma\neq0$.
Namely the integrability depends only on 
the flag of Hamiltonian operators $\mb CK\subset\mb CH+\mb CK$.

We next observe that, in order to apply successfully the Lenard scheme of integrability,
we must have $\beta_3=0$.
Indeed, suppose by contradiction that $\beta_3\neq0$.
By the above observations we can assume, 
after replacing $H$ with $H-\frac{\alpha_3}{\beta_3}K$, that $\alpha_3=0$.
In order to apply the Lenard scheme, we need 
to find a sequence $F^n,\,n\in\mb Z_+$, solving equations \eqref{eq:1.27}.
By looking at the differential order of the various terms in equation \eqref{eq:1.27},
unless $F^n\in\mb C$, we get that $\ord(F^n)+3=\ord(F^{n-1})+1$,
namely the differential order of $F^n$ decreases by 2 at each step, which is impossible 
since differential orders are non-negative integers,
unless $F^n\in\mb C$ for $n>>0$, i.e.\ the Lenard scheme fails.

Let then $K(\partial)=\beta_1(u^{\prime}+2u\partial)+\beta_2\partial$.
If $\beta_1=0$, we may assume that $K(\partial)=\partial$
and $H(\partial)=\alpha_1(u^{\prime}+2u\partial)+\alpha_3\partial^3$,
which gives the KdV, the dispersionless KdV or the linear KdV hierarchies,
if $\alpha_1\alpha_3\neq0,\,\alpha_1\neq0$ and $\alpha_3=0$,
or $\alpha_1=0$ and $\alpha_3\neq0$ respectively.
These cases were studied in detail Section \ref{sec:1.7}.
In this section we consider the remaining case, when $\beta_1\neq0$.
Then we may assume that 
$H(\partial)=\alpha\partial+\beta\partial^3,\, K(\partial)=u^{\prime}+2u\partial+\gamma\partial$,
for $\alpha,\beta,\gamma\in\mb C$, and $\alpha,\beta$ not both zero.

As a last observation, we note that the case $\gamma\neq0$
can be reduced to the case $\gamma=0$ with the change of variable
\begin{equation}\label{eq:mar12_1}
u\mapsto u-\frac\gamma2\,\,,\,\,\,\,u^{(n)}\mapsto u^{(n)}\,\,,\,\,\,\,n\geq1\,,
\end{equation}
in the polynomial algebra $R=\mb C[u^{(n)}\,|\,n\in\mb Z_+]$,
or in an algebra of differential functions $\mc V$ extending $R$.
Indeed, it is immediate to check that \eqref{eq:mar12_1}
defines an algebra automorphism $\varphi$ of $R$ commuting with the action of $\partial$,
hence the operator $H(\partial)$ commutes with $\varphi$, 
while $\varphi K(\partial)\varphi^{-1}=u^{\prime}+2u\partial$.

We thus arrive at the following one-parameter family (parametrized by $\mb P^1$)
of bi-Hamiltonian pairs:
\begin{equation}\label{eq:mar12_2}
H(\partial)=\alpha\partial+\beta\partial^3\,\,,\,\,\,\,K(\partial)=u^{\prime}+2u\partial\,,
\end{equation}
where $\alpha,\beta\in\mb C$ are not both zero.
We want to apply the Lenard scheme of integrability, namely Corollary \ref{cor:feb28},
to construct an infinite hierarchy of local functionals $\tint h_n,\,n\in\mb Z_+$,
which are pairwise in involution with respect to the Lie brackets $\{\cdot\,,\,\cdot\}_H$
and $\{\cdot\,,\,\cdot\}_K$
(thus defining an infinite hierarchy of integrable Hamiltonian equations \eqref{eq:feb28_4}).

As we pointed out in Section \ref{sec:1.6}, the best place to look for
the starting point $F^0$ of an infinite $(H,K)$-sequence $\{F^n\}_{n\in\mb Z_+}$
is the kernel of the operator $K(\partial)$.
Namely we want to find solutions of the equation
\begin{equation}\label{eq:mar12_3}
K(\partial)F^0=(u^{\prime}+2u\partial)F^0=0\,.
\end{equation}
If we write $K(\partial)=2u^{1/2}\partial\circ u^{1/2}$,
we immediately see that, up to a constant factor,
the only solution of equation \eqref{eq:mar12_3} is $F^0=1/\sqrt u$.
Therefore, if we want to have a non-zero kernel of the operator $K(\partial)$,
we are forced to enlarge the algebra $R=\mb C[u^{(n)}\,|\,n\in\mb Z_+]$
to a bigger algebra of differential functions $\mc V$, which includes $1/\sqrt u$.
The smallest such algebra (which is invariant by the action of $\partial$) is
\begin{equation}\label{eq:mar12_4}
\mc V=\mb C[u^{\pm\frac12},u^{\prime},u^{\prime\prime},\dots]\supset R\,.
\end{equation}
%

The degree evolutionary vector field $\Delta$ is diagonalizable on the algebra $\mc V$ 
with eigenvalues in $\frac12\mb Z$.
If we let $\mc U=u^{\frac12}\mb C[u^{\pm1},u^{\prime},u^{\prime\prime},\dots]\subset\mc V$,
the operator $\Delta$ is diagonalizable on $\mc U$ with eigenvalues in $\mb Z+\frac12$,
so that both $\Delta$ and $\Delta+1$ are invertible.
We also have $K(\partial)^{-1}\big(H(\partial)\mc U\big)\subset\mc U$.
We then let 
\begin{equation}\label{eq:247}
F^0\,=\,u^{-\frac12}\,,
\end{equation}
and we look for $F^1\in\mc V$
such that $K(\partial)F^1=H(\partial)F^0$.
By the definition \eqref{eq:mar12_2} of the operators $H(\partial)$ and $K(\partial)$,
the equation for $F^1$ is
$$
2u^{\frac12}\partial\big(u^{\frac12}F^1\big)
\,=\, (\alpha\partial+\beta\partial^3) u^{-\frac12}\,,
$$
which, by the identities
$u^{-\frac12}\partial u^{-\frac12}=\frac12\partial u^{-1}$
and $u^{-\frac12}\partial^3 u^{-\frac12}=2\partial \big(u^{-\frac34}\partial^2 u^{-\frac14}\big)$,
gives, up to adding elements of $\mb Cu^{-\frac12}$,
\begin{equation}\label{eq:mar12_9}
F^1 \,=\, 
\frac14 \alpha u^{-\frac32} + \beta u^{-\frac54}\partial^2 u^{-\frac14}\,.
\end{equation}
Clearly $F^0,\,F^1\in\mc U$, and it is not hard to show that both 
$F^0$ and $F^1$ are in the image of the variational
derivative $\frac\delta{\delta u}$.
In fact we can use Proposition \ref{prop:july22} to compute preimages 
$h_0,\,h_1\in\mc U$ explicitly.
Since $F^0$ (respectively $F^1$), is homogeneous of degree $-\frac12$ (resp. $-\frac32$),
we have $F^n=\frac{\delta h_n}{\delta u},\,n=0,1$,
where 
\begin{equation}\label{eq:mar12_10}
h_0 = 2 u^{\frac12}\,\,,\,\,\,\,
h_1 = -\frac12\alpha u^{-\frac12} -2\beta u^{-\frac14}\partial^2 u^{-\frac14}\,.
\end{equation}
We finally observe that the orthogonality condition (ii) of Corollary \ref{cor:feb28}
holds. Indeed we have
$$
\big(\Span_{\mb C}\{F^0,F^1\}\big)^\perp\subset \big(\mb Cu^{-\frac12}\big)^\perp=
u^{\frac12}\partial \mc V=\im K(\partial)\,.
$$
We can therefore apply Corollary \ref{cor:feb28} to conclude that
there exists an infinite sequence of local functionals 
$\tint h_n,\,n\in\mb Z_+$, with $h_n$ in some normal extension of $\mc V$,
which are in involution with respect to both Lie brackets 
$\{\cdot\,,\,\cdot\}_H$ and $\{\cdot\,,\,\cdot\}_K$,
and such that the corresponding variational derivatives 
$F^n=\frac{\delta h_n}{\delta u},\,n\in\mb Z_+$
form an $(H,K)$-sequence.
In fact, we can compute the whole hierarchy of Hamiltonian functionals $\tint h_n,\,n\in\mb Z_+$,
iteratively by solving equation \eqref{eq:feb28_12}.
Since $K(\partial)^{-1}\big(H(\partial)\mc U\big)\subset\mc U$,
we can choose all elements $h_n$ in $\mc U\subset\mc V$.

In the special case $\alpha=1,\beta=0$,
we have a closed formula for the whole sequence $F^n,\,n\in\mb Z_+$,
and the corresponding local functionals $\tint h_n,\,n\in\mb Z_+$,
which can be easily proved by induction:
\begin{equation}\label{eq:sprbr}
F^n = \frac{(2n-1)!!}{2^n\,(2n)!!} u^{-n-\frac12}\,\,,\,\,\,\,
h_n = - \frac{(2n-3)!!}{2^{n-1}\,(2n)!!} u^{-n+\frac12}\,\,,
\,\,\,\,n\in\mb Z_+\,.
\end{equation}
For arbitrary $\alpha,\beta\in\mb C$, 
the first two terms of the sequence are given by \eqref{eq:247}, \eqref{eq:mar12_9} 
and \eqref{eq:mar12_10},
but already the explicit computation of the next term, namely $F^2$, 
and the corresponding local functional $\tint h_2$,
is quite challenging if we try to use the equation $K(\partial)F^2=H(\partial)F^1$.

\begin{remark}
  \label{rem:mar12}
Equation \eqref{eq:july18_4} for $k=u$ gives
an explicit recursion formula for the sequence 
$F^0=u^{-\frac12},\,F^1,\,F^2,\dots$,
which in this case reads
\begin{eqnarray}\label{eq:mar12_14}
F^{n+1}&=&
\frac12 u^{-\frac12} \sum_{m=0}^n\Big(
\frac12 \alpha F^{n-m}F^m
+\beta F^{n-m}\big(\partial^2F^m\big) 
- \frac12\beta \big(\partial F^{n-m}\big)\big(\partial F^m\big)
\Big) \nonumber\\
&& -\frac12 u^{\frac12} \sum_{m=1}^n F^{n+1-m}F^m\,,
\end{eqnarray}
up to adding a constant multiple of $u^{-\frac12}$.
It is clear from this recursive formula that $F^n$ has degree $-\frac{2n+1}2$ for every $n\in\mb Z_+$,
so that, by Proposition \ref{prop:july22}, we have $F^n=\frac{\delta h_n}{\delta u}$,
where
\begin{equation}\label{eq:mar12_17}
h_n\,=\, -\frac{2}{2n-1}uF^n\,,\,\,n\in\mb Z_+\,.
\end{equation}
\end{remark}

We can use equations \eqref{eq:mar12_14} and \eqref{eq:mar12_17} 
to compute $F^2$ and $h_2$. Recalling that $F^0$
and $F^1$ are given by \eqref{eq:247} and \eqref{eq:mar12_9} respectively, 
we have, after some lengthy algebraic manipulations,
\begin{eqnarray}\label{eq:mar12_16}
F^2&=&
\frac3{32} \alpha^2 u^{-\frac52}
+\frac5{12} \alpha\beta u^{-\frac74} \partial^2 u^{-\frac34}
+\frac16 \beta^2 u^{-\frac74} \partial^4 u^{-\frac34}\,, \\
h_2&=&
-\frac1{16} \alpha^2 u^{-\frac32}
-\frac5{18} \alpha\beta u^{-\frac34} \partial^2 u^{-\frac34}
-\frac19 \beta^2 u^{-\frac34} \partial^4 u^{-\frac34}\,.\nonumber
\end{eqnarray}
The local functionals $\tint h_n,\,n\geq0$, are integrals of motion of the integrable 
hierarchy of evolution equations, called the Harry Dym hierarchy:
\begin{equation}\label{eq:254}
\frac{du}{dt_n}\,=\,(\alpha\partial+\beta\partial^3)F^n\,\,,\,\,\,\,n\in\mb Z_+\,.
\end{equation}
The first three equations of the hierarchy are then obtained by  
\eqref{eq:247}, \eqref{eq:mar12_9} and \eqref{eq:mar12_16} respectively.
The first equation for $\alpha=0,\,\beta=1$ is the classical Harry Dym equation:
$$
\frac{du}{dt}\,=\,\partial^3\big(u^{-\frac12}\big)\,.
$$

\begin{remark}\label{rem:mar19_1}
As in the case of the KdV hierarchy, we point out that the abelian subalgebras
$\mf H_{\alpha,\beta}=\big\{\tint h_n,\,n\in\mb Z_+\big\}\subset\mc V/\partial\mc V$
are infinite-dimensional for every $\alpha,\beta\in\mb C$.
Indeed, since $\frac{\delta h_n}{\delta u}=F^n$,
it suffices to show that $F^n,\,n\in\mb Z_+$, are linearly independent in $\mc V$.
Since the operators $H(\partial)$ and $K(\partial)$
in \eqref{eq:mar12_2} have degree 0 and 1 respectively,
then, by the recursive formula \eqref{eq:1.27}, $F^n$ has degree $-n-\frac12$.
Therefore, to prove that they are linearly independent it suffices to show that they are non-zero.
It is clear from the definition \eqref{eq:mar12_2} and some obvious differential order consideration, 
that $\Ker H(\partial)=\mb C$.
In particular $F^n$ is in $\Ker H(\partial)$ if and only if $F^n=0$.
But then we use equation \eqref{eq:1.27} and induction
to deduce that, if $F^{n-1}\neq0$,
then $F^{n-1}\not\in\Ker H(\partial)$, so that $F^n\neq0$.
Hence, in view of Remark 1.26, the Harry Dim hierarchy is integrable.
\end{remark}

\vspace{3ex}
\subsection{The coupled non-linear wave (CNW) system.}~~
\label{sec:n-wave}
Let $\mc V=\mb C[u^{(n)}, v^{(n)}\,|\,n\in\mb Z_+]$
be the algebra of differential polynomials in the indeterminates $(u,v)$.
Then the formulas
\begin{eqnarray}\label{eq:252}
\{{u}_\lambda u\}_H=(\partial+2\lambda)u+c\lambda^3
\,\,,\,\,\,\,
\{{v}_\lambda v\}_H=0\,,\\
\{{u}_\lambda v\}_H=(\partial+\lambda)v
\,\,,\,\,\,\,
\{{v}_\lambda u\}_H=\lambda v\,,\nonumber
\end{eqnarray}
and 
\begin{equation}\label{eq:253}
\{{u}_\lambda u\}_K=\{{v}_\lambda v\}_K=\lambda
\,\,,\,\,\,\,
\{{u}_\lambda v\}_K=\{{v}_\lambda u\}_K=0\,,
\end{equation}
define a compatible pair of $\lambda$-brackets on $\mc V$ for any $c \in \mb C$.
Indeed, formulas \eqref{eq:252} define a structure of a Lie conformal algebra 
on $\mb C[\partial]u\oplus\mb C[\partial]v\oplus\mb C$,
corresponding to the formal distribution Lie algebra,
which is the semi-direct sum of the Virasoro algebra and its abelian ideal,
defined by the action of the former on Laurent polynomials $\mb C[t,t^{-1}]$.
Moreover, adding a constant multiple of $\lambda$ in the first line of \eqref{eq:252}
corresponds to adding a trivial 2-cocycle \cite{K}.
The claim now follows from Theorem \ref{prop:exten}(c) or Example \ref{ex:final}.

The Hamiltonian operators (on $\mc V^{\oplus 2}$)
corresponding to the $\lambda$-brackets 
$\{\cdot\,_\lambda\,\cdot\}_H$ and $\{\cdot\,_\lambda\,\cdot\}_K$ are
\begin{equation}\label{eq:09may1_2}
H(\partial)=
\left(\begin{array}{cc}
c\partial^3+2u\partial+u^\prime & v\partial \\
v\partial+v^\prime & 0
\end{array}\right)
\quad,\qquad
K(\partial)=
\left(\begin{array}{cc}
\partial & 0 \\
0 & \partial
\end{array}\right)\,.
\end{equation}
Let $h_0=v,\, h_1=u\in\mc V$.
We have
$F^0=
\left(\begin{array}{c} 
\frac{\delta h_0}{\delta u} \\ 
\frac{\delta h_0}{\delta v} 
\end{array}\right)
=\left(\begin{array}{c} 0 \\ 1 \end{array}\right)$,
$F^1
=\left(\begin{array}{c} 
\frac{\delta h_1}{\delta u} \\ 
\frac{\delta h_1}{\delta v} 
\end{array}\right)
=\left(\begin{array}{c} 1 \\ 0 \end{array}\right)$,
and
$H(\partial)F^0=K(\partial)F^1=
\left(\begin{array}{c} 0 \\ 0 \end{array}\right)$.
Moreover, obviously $\big(\mb CF^0\oplus\mb CF^1\big)^\perp=\im(\partial)$.
Hence conditions (i) and (ii) of Corollary \ref{cor:feb28} hold,
and the sequence $\tint h_0,\,\tint h_1$ can be extended to an infinite sequence
$\tint h_n\in\mc V/\partial\mc V,\,n\in\mb Z_+$,
satisfying \eqref{eq:feb28_12}.
Thus, we get an infinite hierarchy of Hamiltonian evolution equations ($n\in\mb Z_+$):
$$
\frac d{dt}
\left(\begin{array}{c}
u \\ v
\end{array}\right)
=
H(\partial)
\left(\begin{array}{c}
\frac{\delta h_{n}}{\delta u} \\ \frac{\delta h_{n}}{\delta v}
\end{array}\right)
=
K(\partial)
\left(\begin{array}{c}
\frac{\delta h_{n+1}}{\delta u} \\ \frac{\delta h_{n+1}}{\delta v}
\end{array}\right)\,,
$$
for which all $\tint h_n$ are integrals of motion
for both brackets $\{\cdot\,,\,\cdot\}_H$ and $\{\cdot\,,\,\cdot\}_K$ on $\mc V/\partial\mc V$,
defined as in \eqref{eq:1.12}.
One easily computes the whole hieararchy of Hamiltonian functionals $\tint h_n$
and their variational derivatives $F^n\in\mc V^{\oplus 2}$
by induction on $n$,
starting with the given $\tint h_0$ and $\tint h_1$,
as we did for KdV and HD in Sections \ref{sec:1.7} and \ref{sec:hd}.
For example, we have:
\begin{eqnarray*}
&\displaystyle{
F^0=
\left(\begin{array}{c}
0 \\ 1
\end{array}\right)
\,\,,\,\,\,\,
h_0=v
\,\,;\,\,\,\,
F^1=
\left(\begin{array}{c}
1 \\ 0
\end{array}\right)
\,\,,\,\,\,\,
h_1=u
\,;}\\
&\displaystyle{
F^2=
\left(\begin{array}{c}
u \\ v
\end{array}\right)
\,\,,\,\,\,\,
h_2=\frac12(u^2+v^2)
\,;}\\
&\displaystyle{
F^3=
\left(\begin{array}{c}
cu^{\prime\prime}+\frac32 u^2 +\frac12 v^2 \\ uv
\end{array}\right)
\,\,,\,\,\,\,
h_3=\frac12(cuu^{\prime\prime}+u^3+uv^2)
\,,}
\end{eqnarray*}
and the first four equations of the hierarchy are:
\begin{eqnarray*}
& \displaystyle{
\frac{d}{dt_0}
\left(\begin{array}{c}
u \\ v
\end{array}\right)
=0
\,\,,\,\,\,\,
\frac{d}{dt_1}
\left(\begin{array}{c}
u \\ v
\end{array}\right)
=0
\,\,,\,\,\,\,
\frac{d}{dt_2}
\left(\begin{array}{c}
u \\ v
\end{array}\right)
= \left(\begin{array}{c}
u^\prime \\ v^\prime
\end{array}\right)
}\,,\\
& \displaystyle{
\frac{d}{dt_3}
\left(\begin{array}{c}
u \\ v
\end{array}\right)
= \left(\begin{array}{c}
cu^{\prime\prime\prime}+3uu^\prime+vv^\prime \\ \partial(uv)
\end{array}\right)\,.
}
\end{eqnarray*}
The last equation is called the CNW system \cite{I}, \cite{D}.
It is easy to see by induction that the first coordinate of $F^n\in\mc V^{\oplus2}$
has total degree $n-1$ if $n\geq1$.
It follows that the $F^n$, and hence the $\tint h_n$, are linearly independent for all $n\in\mb Z_+$.
Thus, in view of Remark 1.26, the CNW hierarchy is integrable.

\vspace{3ex}
\subsection{The CNW of HD type hierarchy.}~~
\label{sec:new-intsys}
Recall that the three compatible Hamiltonian operators $H_n(\partial),\,n=1,2,3$,
from Example \ref{ex:1.23}
were used to construct, in Section \ref{sec:1.7}, the KdV hierarchy,
and, in Section \ref{sec:hd}, the HD hierarchy.
Here we describe an analogous picture for an algebra of differential functions $\mc V$ 
in two variables $u$ and $v$.

We have three compatible $\lambda$-brackets in two variables,
$\{\cdot\,_\lambda\,\cdot\}_n,\,n=1,2,3$, defined as follows:
\begin{eqnarray*}
&& \{{u}_\lambda u\}_1=(\partial+2\lambda)u
\,\,,\,\,\,\,
\{{v}_\lambda v\}_1=0
\,\,,\,\,\,\,
\{{u}_\lambda v\}_1=(\partial+\lambda)v
\,\,,\,\,\,\,
\{{v}_\lambda u\}_1=\lambda v\,, \\
&& \{{u}_\lambda u\}_2=\{{v}_\lambda v\}_2=\lambda
\,\,,\,\,\,\,
\{{u}_\lambda v\}_2=\{{v}_\lambda u\}_2=0\,, \\
&& \{{u}_\lambda u\}_3=\lambda^3
\,\,,\,\,\,\,
\{{v}_\lambda v\}_3=\{{u}_\lambda v\}_3=\{{v}_\lambda u\}_3=0\,.
\end{eqnarray*}
The corresponding Hamiltonian operators $H_n(\partial):\,\mc V^{\oplus 2}\to\mc V^2,\,n=1,2,3$, 
are given by the following differential operators:
$$
H_1(\partial)=
\left(\begin{array}{cc}
u^\prime+2u\partial & v\partial \\
v^\prime+v\partial & 0
\end{array}\right)
\quad,\qquad
H_2(\partial)=
\left(\begin{array}{cc}
\partial & 0 \\
0 & \partial
\end{array}\right)
\quad,\qquad
H_3(\partial)=
\left(\begin{array}{cc}
\partial^3 & 0 \\
0 & 0
\end{array}\right)\,.
$$
It is clear from the discussion at the beginning of Section \ref{sec:n-wave}
that the above operators are all compatible,
namely any linear combination of them is again a Hamiltonian operator.
The CNW system is constructed using the bi-Hamilonian pair 
$H=H_1+cH_3,\,K=H_2$,
and it can be viewed as the analogue, in two variables, 
of the KdV hierarchy.
It is therefore natural to look for a two-variable analogue of the HD hierarchy,
which should be associated to the 1-parameter family (parametrized by $\mb P^1$) 
of bi-Hamiltonian pairs
$H=\alpha H_2+\beta H_3,\,K=H_1$, namely
\begin{equation}\label{eq:09may1_1}
H(\partial)
=\left(\begin{array}{cc}
\alpha\partial+\beta\partial^3 & 0 \\
0 & \alpha\partial
\end{array}\right)
\quad,\qquad
K(\partial)
=\left(\begin{array}{cc}
u^\prime+2u\partial & v\partial \\
v^\prime+v\partial & 0
\end{array}\right)\,,
\end{equation}
with $\alpha,\beta\in\mb C$.
Note that, using arguments similar to those at the beginning Section \ref{sec:hd},
any other bi-Hamiltonian pair $(H,K)$ obtained with linear combinations of the operators
$H_1,H_2,H_3$ can be reduced, from the point of view of finding integrable systems,
either to \eqref{eq:09may1_2} or to \eqref{eq:09may1_1}.

We next apply the Lenard scheme of integrability
to construct an infinite hierarchy of integrable Hamiltonian equations associated
to the Hamiltonian operators $H(\partial)$ and $K(\partial)$ in  \eqref{eq:09may1_1},
which we can call the ``CNW of HD type'' hierarchy.

First, by Remark \ref{rem:jan2},
the operator $K(\partial)$ is non-degenerate.
According to  Proposition \ref{prop:feb28_3} and the subsequent discussion,
the best place to look for the starting point of an integrable hierarchy is 
the kernel of $K(\partial)$.
All the solutions of $K(\partial)F=0$ 
are linear combinations of the following two elements:
\begin{equation}\label{eq:09may1_4}
F^0=
\left(\begin{array}{c}0 \\ 1\end{array}\right)
\quad,\qquad
F^1=
\left(\begin{array}{c}\frac1v \\ -\frac u{v^2}\end{array}\right)\,\in\mc V^{\oplus 2}\,.
\end{equation}
To make sense of them, we choose an algebra of differential functions 
$\mc V$ which contains $\frac1v$, for example we can take
\begin{equation}\label{eq:09may1_5}
\mc V=\mb C[u,v^{\pm1},u^\prime,v^\prime,u^{\prime\prime},v^{\prime\prime},\dots]\,.
\end{equation}
Note that $F^0$ and $F^1$ are the variational derivatives, 
respectively, of
\begin{equation}\label{eq:09may1_3}
\tint h_0=\tint v
\quad,\qquad
\tint h_1=\tint \frac uv\,\in\,\mc V/\partial\mc V\,.
\end{equation}
Moreover, we clearly have $H(\partial)F^0=0$, so that condition (i) 
of Corollary \ref{cor:feb28} holds trivially.

Next we want to check condition (ii) of Corollary \ref{cor:feb28},
namely
\begin{equation}\label{eq:09may1_6}
\big(\Span_{\mb C}\{F^0,F^1\}\big)^\perp\subset\im K(\partial)\,.
\end{equation}
Let $P=\left(\begin{array}{c}p \\ q\end{array}\right)\in\big(\Span_{\mb C}\{F^0,F^1\}\big)^\perp$.
Since $\tint P\cdot F^0=0$, we have that $q=\partial(vf)$ for some $f\in\mc V$.
Let $r=\frac1v\big(p-u^\prime f-2u\partial f\big)$, so that
$$
P=\left(\begin{array}{c}u^\prime f+2u\partial f +vr \\ \partial(vf) \end{array}\right)\,.
$$
In order to prove that 
$P\in\im K(\partial)$ 
we only have to check that 
$r\in\partial\mc V$.
Since $P\perp F^1$, we have
$$
0=\tint P\cdot F^1
=\tint \frac1v(u^\prime a+2u\partial a+vr)-\frac{u}{v^2}\partial(va)
=\int \Big(r+\partial\big(\frac{u}{v}a\big)\Big)\,,
$$
so that $r\in\partial\mc V$, as we wanted. This proves \eqref{eq:09may1_6}.

According to the above observations all the assumption of Corollary \ref{cor:feb28} hold.
Hence there exists an infinite sequence of local functionals
$\tint h_n\in\mc V/\partial\mc V,\,n\in\mb Z_+$, with $h_n$ in some normal extension of $\mc V$,
which are in involution with respect 
to both Lie brackets $\{\cdot\,,\,\cdot\}_H$ and $\{\cdot\,,\,\cdot\}_K$,
and such that the corresponding variational derivatives $F^n\in\mc V^{\oplus2},\,n\in\mb Z_+$,
form an $(H,K)$-sequence.
In fact, we can compute the whole sequence
of Hamiltonian functionals $\tint h_n,\,n\in\mb Z_+$,
iteratively by solving equation \eqref{eq:feb28_12}.
The corresponding hierarchy of evolution equations is given by \eqref{eq:feb28_4},
namely
\begin{equation}\label{eq:09may1_9}
\left\{\begin{array}{l}
\frac{du}{dt_n} \,=\, (\alpha\partial+\beta\partial^3) \frac{\delta h_n}{\delta u} \\
\frac{dv}{dt_n} \,=\, \alpha\partial \frac{\delta h_n}{\delta v}
\end{array}\right.\,.
\end{equation}
Is the above hierarchy integrable?
According to Definition \ref{def:vicapr09}, in order to establish integrability,
it suffices to prove that the elements $K(\partial)F^n,\,n\in\mb Z_+$
span an infinite-dimensional subspace of $\mc V^2$,
and since $\dim\Ker K(\partial)=2<\infty$, it suffices to prove that
the $F^n$'s are linearly independent.

As for the HD hierarchy, we observe that $H(\partial)$ and $K(\partial)$
in \eqref{eq:09may1_1} are homogenous of degree 0 and 1 respectively.
Hence, by the recursive formula \eqref{eq:feb28_12},
we can choose $F^n\in\mc V^{\oplus2},\,n\in\mb Z_+$, 
homogeneous of total degree $-n$.
Two claims follow from this observation.
First, we can choose the $\tint h_n$ in $\mc V$,
due to Proposition \ref{prop:july22}.
Second, to prove linear independence, we only have to check that 
the $F^n$'s are not zero.

We consider separately the two cases $\alpha=0$ and $\alpha\neq0$.
For $\alpha=0$ we can compute explicitly the whole sequence:
$F^0$ and $F^1$ are as in \eqref{eq:09may1_4},
$F^2=
\left(\begin{array}{c}
0 \\ 
\frac32 \beta \frac{(v^\prime)^2}{v^4}-\beta\frac{v^{\prime\prime}}{v^3}
\end{array}\right)$,
and 
$F^n=0$ for all $n\geq3$.
In particular the $F^n$'s span a finite-dimensional space and 
the Lenard scheme fails.

If $\alpha\neq0$ we may rescale so that $\alpha$ is replaced by 1 
and we let $c=\frac\beta\alpha$.
We want to prove that, in this case,  $F^n\neq 0$ for every $n\geq0$,
namely integrability holds.
Notice first that $\Ker (\partial) = \Ker(\partial+c\partial^3)=\mb C$.
In particular, since each component of $F^n$ has degree $-n$, if, for $n>0$, 
$F^n_i\neq0$, then necessarily $(H(\partial)F^n)_i\neq0,\,i=1,2$.
Moreover, we have by the expression \eqref{eq:09may1_1} of $K(\partial)$,
that, for $n\in\mb Z_+$,
\begin{equation}\label{eq:09may1_7}
(K(\partial)F^n)_2=\partial(vF^n_1)\,,
\end{equation}
while, if $F^n_1=0$, then
\begin{equation}\label{eq:09may1_8}
(K(\partial)F^n)_1=v\partial F^n_2\,.
\end{equation}
Suppose that $F^n_2\neq0$ for some $n\geq1$.
It follows that $(H(\partial)F^n)_2\neq0$
and hence, by the recursive formula \eqref{eq:1.27} and equation \eqref{eq:09may1_7},
that $F^{n+1}_1\neq0$.
On the other hand, if $F^n_2=0$ and $F^n_1\neq0$ for some $n\geq1$,
namely $(H(\partial)F^n)_2=0$ and $(H(\partial)F^n)_1\neq0$,
we have, by the recursive formula \eqref{eq:1.27} and equations 
\eqref{eq:09may1_7} and \eqref{eq:09may1_8},
that $F^{n+1}_1=0$ and $F^{n+1}_2\neq0$.
In conclusion, $F^n\neq0$ for every $n\in\mb Z_+$,
thus, in view of Remark 1.26, proving integrability of the hierarchy \eqref{eq:09may1_9}.

We can compute explicitly the first few terms of the hierarchy.
The first two Hamiltonian functionals $\tint h_0,\,\tint h_1$ are given by \eqref{eq:09may1_3},
the corresponding variational derivatives $F^0,\,F^1$ are in \eqref{eq:09may1_4},
the 0-th evolution equation is trivial and the 1-st is
$$
\frac{d}{dt_1}\left(\begin{array}{c} u \\ v \end{array}\right)
=
\left(\begin{array}{c} 
(\partial+c\partial^3)\big(\frac1v\big) \\ 
-\partial\big(\frac{u}{v^2}\big)
 \end{array}\right)\,.
$$
We call this equation the CNW system of HD type.
We next compute $F^2$ by solving the equation $H(\partial)F^1=K(\partial)F^2$.
The result is
$$
F^2
=
\left(\begin{array}{c} 
-\frac{u}{v^3} \\ 
\frac32 \frac{u^2}{v^4}-\frac12 \frac1{v^2} 
+\frac32 c \frac{(v^\prime)^2}{v^4} - c\frac{v^{\prime\prime}}{v^3}
 \end{array}\right)\,,
$$
which is the variational derivative of the Hamiltonian functional
$$
\tint h_2
=
\int -\frac12 \frac{u^2}{v^3} + \frac12 \frac1v +\frac12 c\frac{(v^\prime)^2}{v^3}\,.
$$
The corresponding evolution equation is
$$
\frac{d}{dt_2}\left(\begin{array}{c} u \\ v \end{array}\right)
=
\left(\begin{array}{c} 
-(\partial+c\partial^3)\frac{u}{v^3} \\ 
\partial\big(\frac32 \frac{u^2}{v^4}-\frac12 \frac1{v^2} 
+\frac32 c\frac{(v^\prime)^2}{v^4} - c\frac{v^{\prime\prime}}{v^3}\big)
\end{array}\right)\,.
$$

\section{The variational complex.}\label{2006_sec2}

\subsection{$\mf g$-complex and its reduction.}~~
\label{sec:abstractdr}
Given a Lie algebra $\mf g$, we can extend it to a Lie superalgebra $\hat{\mf g}$
containing $\mf g$ as even subalgebra as follows:
\begin{equation}\label{2006_hatg}
\hat{\mf{g}} 
= \mf g[\xi]\rtimes \mb{C}\partial_\xi
 = \mf{g}\oplus\mf{g}\xi\oplus\mb{C}\partial_\xi\,,
\end{equation}
where $\xi$ is an odd indeterminate such that $\xi^2=0$.
A $\mf{g}$-\emph{complex} is a $\mb{Z}_+$-graded vector space
$\Omega=\bigoplus_{n\in\mb{Z}_+}\Omega^n$,
with a representation of $\hat{\mf{g}}$ on $\Omega$, $\pi:\,\hat{\mf{g}}\to\End\Omega$,
such that, for $a\in\mf{g}$, the endomorphisms $\pi(a),\,\pi(a\xi),\,\pi(\partial_\xi)$ have
degrees $0$, $-1$ and $+1$, respectively.

Given an element $\partial\in\mf g$,
we can consider its centralizer $\mf g^\partial=\{a\in\mf g\,|\,[\partial,a]=0\}\subset\mf g$,
and the corresponding superalgebra
$\hat{\mf{g}}^\partial = \mf{g}^\partial \oplus\mf{g}^\partial\xi\oplus\mb{C}\partial_\xi
\subset\hat{\mf g}$.
Notice that $\hat{\mf g}^\partial=\hat{\mf g^\partial}$.
Let $\tilde\Omega$ be a $\mf g$-complex.
Clearly $\pi(\partial)\tilde\Omega=\bigoplus_{n\in\mb{Z}_+}\pi(\partial)\tilde\Omega^n
\subset\tilde\Omega$ is a graded submodule
with respect to the action of $\hat{\mf{g}}^\partial$.
We can thus take the corresponding quotient $\hat{\mf g}^\partial$-module
$\Omega=\bigoplus_{n\in\mb{Z}_+}\Omega^n$,
where $\Omega^n=\tilde\Omega^n/\pi(\partial)\tilde\Omega^n$,
which is called the \emph{reduced} (by $\partial$) $\mf g^\partial$-complex.

\vspace{3ex}
\subsection{The basic and the reduced de Rham complexes $\tilde\Omega$ and $\Omega$
over $\mc V$.}~~
\label{2006_sec2.1}
As in Section \ref{sec:1}, we let $R=\mb{C}[u_i^{(n)} \,|\, i\in I, n\in \mb{Z}_+]$ be the 
$\mb{C}$-algebra of differential polynomials in the variables $u_i,\,i\in\{1,\dots,\ell\}=I$,
and we let $\mc V$ be an algebra of differential functions extending $R$.
Again, as in Section \ref{sec:1}, we denote by $\mf g$ 
the Lie algebra of all vector fields of the form \eqref{2006_X}.

The \emph{basic de Rham complex} $\tilde\Omega=\tilde\Omega(\mc V)$
is defined as the
free commutative superalgebra over $\mc V$
with odd generators $\delta u_i^{(n)},\,i\in I,n\in\mb{Z}_+$.
In other words $\tilde\Omega$ consists of finite sums of the form
\begin{equation}\label{eq:apr24_1}
\tilde\omega=\sum_{
\substack{i_1,\cdots,i_k\in I \\ m_1,\cdots,m_k\in\mb{Z}_+}
}
f^{m_1\cdots m_k}_{i_1\cdots i_k}\, 
\delta u_{i_1}^{(m_1)}\wedge\cdots\wedge\delta u_{i_k}^{(m_k)}
\,\,,  \quad f^{m_1\cdots m_k}_{i_1\cdots i_k} \in \mc V\,,
\end{equation}
and it has a (super)commutative product given by the wedge product $\wedge$.
We have a natural $\mb Z_+$-grading $\tilde\Omega=\bigoplus_{k\in\mb Z_+}\tilde\Omega^k$
defined by letting elements in $\mc V$ have degree 0,
while the generators $\delta u_i^{(n)}$ have degree 1.
The space $\tilde\Omega^k$ is a free module over $\mc V$ with basis given by the elements
$\delta u_{i_1}^{(m_1)}\wedge\cdots\wedge\delta u_{i_k}^{(m_k)}$,
with $(m_1,i_1)>\dots>(m_k,i_k)$ (with respect to the lexicographic order).
In particular $\tilde\Omega^0=\mc V$ 
and $\tilde\Omega^1=\bigoplus_{i\in I,n\in\mb Z_+}\mc V\delta u_i^{(n)}$.
Notice that there is a $\mc V$-linear pairing $\tilde\Omega^1\times\mf g\to\mc V$
defined on generators by 
$\big(\delta u_i^{(m)},\frac{\partial}{\partial u_j^{(n)}}\big)=\delta_{i,j}\delta_{m,n}$,
and extended to $\tilde\Omega^1\times\mf g$ by $\mc V$-bilinearity.

Furthermore, we want to define a representation $\pi:\,\hat{\mf g}\to\Der\,\tilde\Omega$ 
of the Lie algebra $\hat{\mf g}$ on the space $\tilde\Omega$ given by derivations
of the wedge product in $\tilde\Omega$,
which makes $\tilde\Omega$ into a $\mf g$-complex.
We let $\delta$ be an odd derivation of degree 1 of $\tilde\Omega$,
such that $\delta f=\sum_{i\in I,\,n\in\mb Z_+}\frac{\partial f}{\partial u_i^{(n)}}\delta u_i^{(n)}$
for $f\in\mc V$, and $\delta(\delta u_i^{(n)})=0$.
It is immediate to check that $\delta^2=0$ and that,
for $\tilde\omega\in\tilde\Omega^k$ as in \eqref{eq:apr24_1}, we have
\begin{equation}\label{eq:apr24_2}
\delta(\tilde\omega)
\,=\,
\sum_{j\in I,n\in\mb Z_+}\sum_{
\substack{i_1,\cdots,i_k\in I \\ m_1,\cdots,m_k\in\mb{Z}_+}
}
\frac{\partial f^{m_1\cdots m_k}_{i_1\cdots i_k}}{\partial u_j^{(n)}}\, 
\delta u_j^{(n)}\wedge \delta u_{i_1}^{(m_1)}\wedge\cdots\wedge\delta u_{i_k}^{(m_k)}\,.
\end{equation}
For $X\in\mf g$ we define the \emph{contraction operator}
$\iota_X:\,\tilde\Omega\to\tilde\Omega$,
as an odd derivation of $\tilde\Omega$ of degree -1,
such that $\iota_X(f)=0$ for $f\in\mc V$,
and $\iota_X(\delta u_i^{(n)})=X(u_i^{(n)})$.
If $X\in\mf g$ is as in \eqref{2006_X}
and $\tilde\omega\in\tilde\Omega^k$ is as in \eqref{eq:apr24_1}, we have
\begin{equation}\label{eq:apr24_3}
\iota_X(\tilde\omega)
\,=\,
\sum_{
\substack{i_1,\cdots,i_k\in I \\ m_1,\cdots,m_k\in\mb{Z}_+}
} 
\sum_{r=1}^k (-1)^{r+1}
f^{m_1\cdots m_k}_{i_1\cdots i_k} h_{i_r,m_r}\, 
\delta u_{i_1}^{(m_1)}\wedge\stackrel{r}{\check{\cdots}}\wedge\delta u_{i_k}^{(m_k)}\,.
\end{equation}
In particular, for $f\in\mc V$ we have
\begin{equation}\label{eq:apr24_8}
\iota_X(\delta f)\,=\,X(f)\,.
\end{equation}
It is easy to check that the operators $\iota_X,\,X\in\mf g$, form an abelian
(purely odd) subalgebra of the Lie superalgebra $\Der\,\tilde\Omega$, namely
\begin{equation}\label{eq:apr24_5}
[\iota_X,\iota_Y]=\iota_X\circ\iota_Y+\iota_Y\circ\iota_X=0\,.
\end{equation}
The \emph{Lie derivative} along $X\in\mf g$ is the even derivation $L_X$ of $\tilde\Omega$ of degree 0,
defined by \emph{Cartan's formula}:
\begin{equation}\label{eq:apr24_6}
L_X\,=\,[\delta,\iota_X] = \delta\circ\iota_X+\iota_X\circ\delta\,.
\end{equation}
In particular, for $f\in\mc V=\tilde\Omega^0$ we have, by \eqref{eq:apr24_8},
$L_X(f)=\iota_X(\delta f)=X(f)$.
Moreover, it immediately follows by the fact that $\delta^2=0$ that
\begin{equation}\label{eq:apr24_7}
[\delta,L_X] = \delta\circ L_X-L_X\circ \delta = 0\,.
\end{equation}
We next want to prove the following:
\begin{equation}\label{eq:apr24_9}
[\iota_X,L_Y] = \iota_X\circ L_Y-L_Y\circ \iota_X = \iota_{[X,Y]}\,.
\end{equation}
It is clear by degree considerations that both sides of \eqref{eq:apr24_9}
act as zero on $\tilde\Omega^0=\mc V$.
Moreover, it follows by \eqref{eq:apr24_8} that
$[\iota_X,L_Y](\delta f)=\iota_X \delta \iota_Y \delta f -\iota_Y\delta \iota_X\delta f
=X(Y(f))-Y(X(f))=[X,Y](f)=\iota_{[X,Y]}(\delta f)$ for every $f\in\mc V$.
Equation \eqref{eq:apr24_9} then follows by the fact that both sides are even derivations
of the wedge product in $\tilde\Omega$.
Finally, as an immediate consequence of equations \eqref{eq:apr24_7} and \eqref{eq:apr24_9}, we get that
\begin{equation}\label{eq:apr24_10}
[L_X,L_Y] = L_X\circ L_Y-L_Y\circ L_X = L_{[X,Y]}\,.
\end{equation}
We then let $\pi(\partial_\xi)=\delta,\,\pi(X)=L_X,\,\pi(X\xi)=\iota_X$, for $X\in\mf g$,
and equations \eqref{eq:apr24_5}, \eqref{eq:apr24_6}, \eqref{eq:apr24_7},
\eqref{eq:apr24_9} and \eqref{eq:apr24_10} exactly mean that
$\pi:\,\hat{\mf g}\to\Der\,\tilde\Omega$ is a Lie superalgebra homomorphism.
Hence the basic  de Rham complex $\tilde\Omega$ is a $\mf g$-complex.

\begin{remark}\label{rem:apr24}
To each differential $k$-form $\tilde\omega\in\tilde\Omega^k$
we can associate a $\mc V$-multilinear skew-symmetric map
$\tilde\omega:\underbrace{\mf{g}\times\ldots\times\mf{g}}_k\to \mc V$ 
by letting
$$
\delta u_{i_1}^{(m_1)}\wedge\cdots\wedge\delta u_{i_k}^{(m_k)}
\Big(\frac{\partial}{\partial u_{j_1}^{(n_1)}},\dots,\frac{\partial}{\partial u_{j_k}^{(n_k)}}\Big)
\,=\,\sum_{\sigma\in S_k}\text{sign}(\sigma)
\prod_{r=1}^k\delta_{i_r,j_{\sigma_r}}\delta_{m_r,n_{\sigma_r}}\,,
$$
and extending it by $\mc V$-linearity in each argument.
We can then identify the complex $\tilde\Omega^k$ 
with the space of such maps $\Hom_{\mc V}(\bigwedge{\!}^k_{\mc V} \mf{g},\mc V)$.
The action of the differential $\delta$ on $\tilde\omega\in\tilde\Omega^k$ is then given by the 
usual formula for the Lie algebra cohomology differential:
\begin{eqnarray}\label{2006_diff}
 \delta\tilde\omega (X_1,\ldots,X_{k+1}) =  \sum_{1\le i \le k+1} (-1)^{i+1}
  X_i\big(\tilde\omega(X_1,\stackrel{i}{\check{\cdots}},X_{k+1})\big) \\
   + \sum_{i<j} (-1)^{i+j} \tilde\omega([X_i,X_j],
   X_1,\stackrel{i}{\check{\cdots}}\stackrel{j}{\check{\cdots}},X_{k+1})\,, \nonumber
\end{eqnarray}
the corresponding formula for the  contraction operator $\iota_X$ is
\begin{equation}\label{2006_contr}
  (\iota_X\tilde\omega)(X_1,\ldots,X_{k-1}) = \tilde\omega(X,X_1,\ldots,X_{k-1})\,\,\,\text{ if } k\geq 1\,,
\end{equation}
and that for the Lie derivative $L_X$ is
\begin{equation}\label{2006_lieder}
(L_X\tilde\omega)(X_1,\ldots,X_k) = X(\tilde\omega(X_1,\ldots,X_k)) 
-\sum_{r=1}^k \tilde\omega(X_1,\cdots,[X,X_r],\cdots,X_k)\,.
\end{equation}
\end{remark}

As in Section \ref{sec:1},
we denote by 
$\mf g^\partial$ the subalgebra of $\mf g$ consisting of evolutionary vector fields,
namely the centralizer of $\partial\in\mf g$, given by \eqref{partial}.
By an abuse of notation we also denote by $\partial$ the Lie derivative 
$L_\partial:\,\tilde\Omega^k\to\tilde\Omega^k$.
Since $\partial$ commutes with $\delta$,
we may consider the corresponding reduced $\mf g^\partial$-complex 
$\Omega=\tilde\Omega/\partial\tilde\Omega=\bigoplus_{k\in\mb Z_+}\Omega^k$.
This is known as the \emph{reduced de Rham complex} $\Omega=\Omega(\mc V)$,
or the \emph{complex of variational calculus}, or the \emph{variational complex}.

Elements $\omega\in\Omega^k=\tilde\Omega^k/\partial\tilde\Omega^k$
are called \emph{local} $k$-\emph{forms}.
In particular $\Omega^0=\mc V/\partial\mc V$ is the space of \emph{local functionals}
(cf.\ Definition \ref{def:1.14}).
By an abuse of notation, we denote by $\delta$ and, for $X\in\mf g^\partial$, by $\iota_X,\,L_X$,
the maps induced on the quotient space $\Omega^k$ by the corresponding maps 
on $\tilde\Omega^k$.
In particular $L_\partial$ acts as zero on the whole complex $\Omega$.
The contraction operators induce an $\Omega^0$-valued pairing between $\Omega^1$ 
and $\mf g^\partial$:
\begin{equation}\label{eq:may4_3}
(X,\omega)\,=\,(\omega,X)\,=\,\iota_X(\omega)
\,\,,\,\,\,\,
\text{ for } X\in\mf g^\partial,\,\omega\in\Omega^1\,.
\end{equation}

We have the following
identity, valid in any $\mf g$-complex, which can be easily checked using
equations \eqref{eq:apr24_6} and \eqref{eq:apr24_9}:
$$
\iota_{[X,Y]}
=L_X\iota_Y-L_Y\iota_X+\delta\iota_Y\iota_X-\iota_Y\iota_X\delta\,.
$$
In particular, for $\omega\in\Omega^1$ and $X,Y\in\mf g^\partial$,
we get the following identity, which will be used later:
\begin{equation}\label{eq:09jan14_1}
(\omega,[X,Y])
=
L_X(\omega,Y)-L_Y(\omega,X)-\iota_Y\iota_X\delta\omega\,.
\end{equation}
Another useful form of the above identity is, for $\omega\in\Omega^1$,
\begin{equation}\label{eq:09jan14_1_new}
(\omega,[X,Y])
=
L_X(\omega,Y)-(L_X\omega,Y)
\,.
\end{equation}

\vspace{3ex}
\subsection{Cohomology of $\tilde\Omega$ and $\Omega$.}~~
\label{sec:july23_1}
In this section we compute the cohomology of the complexes $\tilde\Omega$ and $\Omega$
under two different assumptions on the algebra of differential functions $\mc V$.
In Theorem \ref{th:july23} we assume that $\mc V$ is normal,
while in Theorem \ref{2006_th_cohom} we assume that the degree evolutionary vector field 
$\Delta$
is diagonalizable on $\mc V$.

We first need to extend filtration \eqref{eq:july21_1} of $\mc V$ to the whole
complex $\tilde\Omega$.
For $n\in\mb Z_+$ and $i\in I$, we let
\begin{equation}\label{eq:july24_1}
\tilde\Omega^k_{n,i}
\,=\,\Big\{\tilde\omega\in\tilde\Omega^k\,\Big|\,
\iota_{\frac{\partial}{\partial u_j^{(m)}}}\tilde\omega=L_{\frac{\partial}{\partial u_j^{(m)}}}\tilde\omega=0\,,
\,\,\text{for all\,\,} (m,j)>(n,i)\Big\}\,,
\end{equation}
where the inequality is understood in the lexicographic order.
We let $\tilde\Omega^k_{n,0}=\tilde\Omega^k_{n-1,\ell}$ for $n\geq1$, 
and $\tilde\Omega^k_{0,0}=\delta_{k,0}\mc C$.
Clearly $\tilde\Omega^0_{n,i}=\mc V_{n,i}$.
Note that, $\tilde\Omega^k_{n,i}$ consists of finite sums of the form
\begin{equation}\label{eq:july24_2}
\tilde\omega=\sum_{
\substack{
i_1,\cdots,i_k\in I,\, m_1,\cdots,m_k\in\mb{Z}_+
\\
(n,i)\geq (m_1,i_1)>\dots>(m_k,i_k)
}
}
f^{m_1\cdots m_k}_{i_1\cdots i_k}\, \delta u_{i_1}^{(m_1)}\wedge\cdots\wedge\delta u_{i_k}^{(m_k)}
\,\,,  \quad f^{m_1\cdots m_k}_{i_1\cdots i_k} \in \mc V_{n,i}\,.
\end{equation}

Given $i\in I,\,n\in\mb Z_+$, we have 
$\iota_{\frac{\partial}{\partial u_i^{(n)}}}\big(\tilde\Omega^k_{n,i}\big)\subset\tilde\Omega^{k-1}_{n,i},\,
L_{\frac{\partial}{\partial u_i^{(n)}}}\big(\tilde\Omega^k_{n,i}\big)\subset\tilde\Omega^{k}_{n,i}$
and $\delta\big(\tilde\Omega^k_{n,i}\big)\subset\tilde\Omega^{k+1}_{n,i}$.
Explicitly, if $\tilde\omega$ is as in \eqref{eq:july24_2}, we have
\begin{eqnarray}\label{eq:july24_3}
&& \iota_{\frac{\partial}{\partial u_i^{(n)}}}(\tilde\omega)
\,=\, \sum_{\substack{
i_1,\cdots,i_k\in I,\, m_1,\cdots,m_k\in\mb{Z}_+
\\
(n,i)= (m_1,i_1)>\dots>(m_k,i_k)
}}
f^{m_1\cdots m_k}_{i_1\cdots i_k}\, 
\delta u_{i_2}^{(m_2)}\wedge\cdots\wedge\delta u_{i_k}^{(m_k)}\,\in\tilde\Omega^{k-1}_{n,i}
\,,\nonumber\\
&& L_{\frac{\partial}{\partial u_i^{(n)}}}(\tilde\omega)
\,=\, \sum_{\substack{
i_1,\cdots,i_k\in I,\, m_1,\cdots,m_k\in\mb{Z}_+
\\
(n,i)\geq (m_1,i_1)>\dots>(m_k,i_k)
}}
\frac{\partial f^{m_1\cdots m_k}_{i_1\cdots i_k}}{\partial u_i^{(n)}}\, 
\delta u_{i_1}^{(m_1)}\wedge\cdots\wedge\delta u_{i_k}^{(m_k)}\,\in\tilde\Omega^{k}_{n,i}\,,\\
&& \delta(\tilde\omega)
\,=\, \sum_{\substack{
i_1,\cdots,i_k\in I,\, m_1,\cdots,m_k\in\mb{Z}_+
\\
(n,i)\geq (m_1,i_1)>\dots>(m_k,i_k)
}}
\sum_{\substack{j\in I, m\in\mb{Z}_+\\(n,i)\geq (m,j)}}
\frac{\partial f^{m_1\cdots m_k}_{i_1\cdots i_k}}{\partial u_j^{(m)}}\, 
\delta u_j^{(m)}\wedge\delta u_{i_1}^{(m_1)}\wedge\cdots\wedge\delta u_{i_k}^{(m_k)}\,
\in\tilde\Omega^{k+1}_{n,i}\,.\nonumber
\end{eqnarray}
Suppose next that the algebra of differential functions $\mc V$ is normal.
Recall that this means that, 
for $f\in\mc V_{n,i}$,  
there exists $g\in\mc V_{n,i}$, denoted by $\tint du_i^{(n)}f$,
such that $\frac{\partial g}{\partial u_i^{(n)}}=f$;
the antiderivative $\tint du_i^{(n)}f$ is defined up to adding elements from 
$\Ker\frac{\partial }{\partial u_i^{(n)}}=\mc V_{n,i-1}$.
Now we introduce ``local'' homotopy operators 
$h_{n,i}:\,\tilde\Omega^k_{n,i}\to\tilde\Omega^{k-1}_{n,i}$
by the following formula
\begin{equation}\label{eq:july24_4}
h_{n,i}(\tilde\omega)
\,=\, \sum_{\substack{
i_1,\cdots,i_k\in I,\, m_1,\cdots,m_k\in\mb{Z}_+
\\
(n,i)= (m_1,i_1)>\dots>(m_k,i_k)
}}
\big(\tint du_i^{(n)} f^{m_1\cdots m_k}_{i_1\cdots i_k}\big)\, 
\delta u_{i_2}^{(m_2)}\wedge\cdots\wedge\delta u_{i_k}^{(m_k)}\,\in\tilde\Omega^{k-1}_{n,i}\,,
\end{equation}
where $\tilde\omega$ is as in \eqref{eq:july24_2}.
The element \eqref{eq:july24_4} 
is defined up to adding elements from $\tilde\Omega^{k-1}_{n,i-1}$.
It is easy to check that
\begin{equation}\label{eq:2009july3_2}
\iota_{\frac{\partial}{\partial u_i^{(n)}}}\big(h_{n,i}(\tilde\omega)\big)=0
\,\,\,\,\,\,,\,\,\,\,\,\,\,\,
L_{\frac{\partial}{\partial u_i^{(n)}}}\big(h_{n,i}(\tilde\omega)\big)
=\iota_{\frac{\partial}{\partial u_i^{(n)}}}(\tilde\omega)\,.
\end{equation}
\begin{theorem}\label{th:july23}
Let $\mc V$ be a normal algebra of differential functions.
\begin{enumerate}
\alphaparenlist
\item For $\tilde\omega\in\tilde\Omega^k_{n,i}$, we have
$h_{n,i}(\delta\tilde\omega)+\delta h_{n,i}(\tilde\omega)-\tilde\omega\,\in\,\tilde\Omega^k_{n,i-1}$.
\item $H^k(\tilde\Omega,\delta)=\delta_{k,0}\mc C$,
\item $H^k(\Omega,\delta)=\delta_{k,0}\mc C$.
\end{enumerate}
\end{theorem}
\begin{proof}
Clearly all three elements $h_{n,i}(\delta\tilde\omega),\,\delta h_{n,i}(\tilde\omega)$ and $\tilde\omega$
lie in $\tilde\Omega^k_{n,i}$.
Hence, to prove (a) we just have to check the following two identities:
\begin{eqnarray*}
\iota_{\frac{\partial}{\partial u_i^{(n)}}}\big(h_{n,i}(\delta\tilde\omega)+\delta h_{n,i}(\tilde\omega)-\tilde\omega\,\in\,\tilde\Omega^k_{n,i-1}) &=& 0\,,\\
L_{\frac{\partial}{\partial u_i^{(n)}}}\big(h_{n,i}(\delta\tilde\omega)+\delta h_{n,i}(\tilde\omega)-\tilde\omega\,\in\,\tilde\Omega^k_{n,i-1}) &=& 0\,.
\end{eqnarray*}
This is a straightforward computation using \eqref{eq:july24_3}, \eqref{eq:july24_4}
and \eqref{eq:2009july3_2}.
Part (b) immediately follows from (a) by induction on the lexicographic order.
We next claim that 
\begin{equation}\label{eq:july24_5}
H^k(\partial\tilde\Omega)=0 \,\,,\,\,\,\,\text{for all\,\,} k\geq0\,.
\end{equation}
Indeed, if $\delta(\partial\tilde\omega)=0$, then $\delta\tilde\omega=0$ 
since $\delta$ and $\partial$ commute and $\Ker\partial|_{\tilde\Omega^{k+1}}=0$ for $k\geq0$.
Hence, by (b), $\tilde\omega=\delta\tilde\eta$, where $\tilde\eta\in\tilde\Omega^{k-1}$, if $k\geq1$,
and $\tilde\omega\in\mc C$ if $k=0$.
In both cases we have $\partial\tilde\omega\in\delta(\partial\tilde\Omega)$, 
thus proving \eqref{eq:july24_5}.
Next, consider the following short exact sequence of complexes:
$$
0\,\to\,\partial\tilde\Omega\,\to\,\tilde\Omega\,\to\,\Omega=\tilde\Omega/\partial\tilde\Omega\,\to\,0\,.
$$
This leads to the long exact sequence in cohomology
$$
\cdots\,\to
H^k(\partial\tilde\Omega)\,\to\,H^k(\tilde\Omega)\,\to\,H^k(\Omega)\,\to\,
H^{k+1}(\partial\tilde\Omega)\,\to\,
\cdots
$$
By part (b) and equation \eqref{eq:july24_5}, 
we get $H^k(\Omega,\delta)\simeq H^k(\tilde\Omega,\delta),\,\text{for all\,\,} k\in\mb Z_+$, proving (c).
\end{proof}

\begin{remark}\label{rem:09jan4}
It follows from the proof of Theorem  \ref{th:july23} that,
for any algebra of differential functions $\mc V$,
given a closed $k$-cocycle,
we can write it as $\delta \omega$, where $\omega$ is a $k-1$-cocyle
with coefficients in an extension $\tilde{\mc V}$ of $\mc V$
obtained by adding finitely many antiderivatives.
\end{remark}

\begin{remark}\label{rem:09jan2}
Theorem \ref{th:july23} holds in the more general setup of \cite{DSK}
where an algebra of differential functions is defined as a differential algebra $\mc V$
with commuting partial derivatives $\frac{\partial}{\partial u_i^{(n)}},\,i\in I,n\in\mb Z_+$,
such that all but finitely many of them annihilate any given element of $\mc V$,
and the commutation relations \eqref{eq:comm_frac} hold.
In this case, in parts (b) and (c) of this theorem, 
one should replace $\mc C$ by $\mc C/(\mc C\cap\partial\mc V)$.
\end{remark}

We next construct a ``global'' homotopy operator, different from \eqref{eq:july24_4},
based on the assumption that the degree evolutionary vector field $\Delta$, defined by \eqref{2006_degop},
is diagonalizable on $\mc V$.

Let $\mc V=\bigoplus_{\alpha}\mc V[\alpha]$ be the eigenspace decomposition of $\mc V$
with respect to the diagonalizable operator $\Delta$.
Then the Lie derivative $L_\Delta$ is diagonalizable 
on $\tilde\Omega^k$ for every $k\in\mb Z_+$.
In fact, 
we have the eigenspace decomposition
$\tilde\Omega^k=\bigoplus_\alpha\tilde\Omega^k[\alpha+k]$,
where $\tilde\Omega^k[\alpha+k]$ is the linear span of elements 
of the form $f\delta u_{i_1}^{(m_1)}\wedge\cdots\wedge\delta u_{i_k}^{(m_k)}$ 
with $f\in\mc V[\alpha]$.
We define the operator $L_{\Delta^{-1}}:\,\tilde\Omega^k\to\tilde\Omega^k$, 
letting it act as zero 
on the subspace $\tilde\Omega^k[0]=\Ker L_\Delta\big|_{\tilde\Omega^k}$,
and as the inverse of $L_\Delta$ 
on $\tilde\Omega^k_{\neq0}=\bigoplus_{\alpha\neq-k}\tilde\Omega^k[\alpha+k]$.
Equivalently, $L_{\Delta^{-1}}$ is uniquely defined by the following equations:
\begin{equation}\label{eq:may4_notte1}
L_\Delta\circ L_{\Delta^{-1}}\,=\,L_{\Delta^{-1}}\circ L_\Delta\,=\,\id-\pi_0\,,
\end{equation}
where $\pi_0:\,\tilde\Omega^k\to\tilde\Omega^k$ denotes the projection 
onto $\tilde\Omega^k[0]$.
We also notice that $\delta$ preserves the degree of a differential 
form $\tilde\omega$, hence it commutes with both $L_\Delta$ and $L_{\Delta^{-1}}$.
Moreover, since $\Delta$ is an evolutionary vector field, the total derivative $\partial$
commutes with $\iota_\Delta,\,L_\Delta$ and $L_{\Delta^{-1}}$.
In particular, we have an induced eigenspace decomposition for $\Delta$ on the reduced complex,
$\Omega^k=\bigoplus_\alpha\Omega^k[\alpha+k]$.
\begin{theorem}\label{2006_th_cohom} 
Let $\mc V$ be an algebra of differential functions,
and assume that the degree evolutionary vector field $\Delta$ is diagonalizable on $\mc V$.
Consider the map 
$\tilde h=L_{\Delta^{-1}}\circ\iota_\Delta:\,\tilde\Omega^k\to\tilde\Omega^{k-1},\,k\geq0$.
It satisfies the following equation:
\begin{equation}\label{2006_24nov16}
\tilde h(\delta\tilde\omega) + \delta\tilde h(\tilde\omega) = \tilde\omega-\pi_0\tilde\omega\,\,,
\,\,\,\,\tilde\omega\in\tilde\Omega\,.
\end{equation}
Hence
$$
H^k(\Omega) \simeq \Ker\big(\delta:\,\Omega^k[0]\to\Omega^{k+1}[0]\big)
\big/\delta\Omega^{k-1}[0]
\,\,,\,\,\,\,\text{for all\,\,} k\in\mb Z_+\,.
$$
In particular, if $\mc V[0]=\mc C$ and $\Delta$ does not have
negative integer eigenvalues on $\mc V$, we have $H^k(\Omega)\simeq\delta_{k,0}\mc C$.
\end{theorem}
\begin{proof}
Equation \eqref{2006_24nov16} is immediate from the definition of $\tilde h$,
the fact that $\delta$ commutes with $L_{\Delta^{-1}}$ 
and equations \eqref{eq:apr24_6} and \eqref{eq:may4_notte1}.
For the second statement of the theorem, we notice that, since $\delta$ and $L_\Delta$ commute,
we have the corresponding $\Delta$-eigenspace decomposition in the cohomology:
$$
H^k(\Omega)=\bigoplus_\alpha H^k(\Omega[\alpha+k])\,,
$$
and we only need to show that $H^k(\Omega[\alpha+k])=0$ when $\alpha+k\neq0$.
For this, let $\omega\in\Omega^k[\alpha+k]$, with $\alpha+k\neq0$, be such that $\delta\omega=0$,
and let $\tilde\omega\in\tilde\Omega^k[\alpha+k]$ be a representative of $\omega$,
so that $\delta\tilde\omega=\partial\tilde\eta$ for some $\tilde\eta\in\tilde\Omega^{k+1}[\alpha+k]$.
Since $\partial$ commutes with $\tilde h$, by \eqref{2006_24nov16} we have 
$$
\tilde\omega=\tilde h(\delta\tilde\omega) + \delta\tilde h(\tilde\omega)
=\partial\tilde h(\tilde\eta)+\delta\tilde h(\tilde\omega)
\equiv \delta\tilde h(\tilde\omega) \,\mod\partial\tilde\Omega\,.
$$
Hence $\omega\in\im\delta$, as we wanted.
\end{proof}

\vspace{3ex}
\subsection{The variational complex as a Lie conformal algebra cohomology complex.}~~
\label{sec:july23_2}
Let us review, following \cite{BKV}, the definition of the basic 
and reduced cohomology complexes
associated to a Lie conformal algebra $A$ and an $A$-module $M$.
A $k$-\emph{cochain} of $A$ with coefficients in $M$ is a $\mb C$-linear map
$$
\tilde\gamma:\, A^{\otimes k}\to M[\lambda_1,\dots,\lambda_k]\,\,,\,\,\,\,
a_1\otimes\cdots\otimes a_k\mapsto \tilde\gamma_{\lambda_1,\dots,\lambda_k}(a_1,\dots,a_k)\,,
$$
satisfying the following two conditions:
\begin{enumerate}
\item $\tilde\gamma_{\lambda_1,\dots,\lambda_k}(a_1,\dots,\partial a_i,\dots,a_k)=
-\lambda_i\tilde\gamma_{\lambda_1,\dots,\lambda_k}(a_1,\dots,a_k)$ for all $i$,
\item $\tilde\gamma$ is skew-symmetric with respect to simultaneous permutations
of the $a_i$'s and the $\lambda_i$'s.
\end{enumerate}
We let $\tilde\Gamma^k(A,M)$ be the space of all $k$-cochains, 
and $\tilde\Gamma(A,M)=\bigoplus_{k\geq0}\tilde\Gamma^k(A,M)$.
The differential $\delta$ of a $k$-cochain $\tilde\gamma$ is defined by the following formula
(cf.\ \eqref{2006_diff}):
\begin{eqnarray}\label{eq:july24_7}
(\delta\tilde\gamma)_{\lambda_1,\dots,\lambda_{k+1}}(a_1,\dots,a_{k+1})
=\sum_{1\le i \le k+1} (-1)^{i+1} {a_i}_{\lambda_i}
\tilde\gamma_{\lambda_1,\stackrel{i}{\check{\cdots}},\lambda_{k+1}}
(a_1,\stackrel{i}{\check{\cdots}},a_{k+1}) \\
+ \sum_{i<j} (-1)^{i+j} 
\tilde\gamma_{\lambda_i+\lambda_j,
\lambda_1,\stackrel{i}{\check{\cdots}}\stackrel{j}{\check{\cdots}},\lambda_{k+1}}
([{a_i}_{\lambda_i} a_j],a_1,\stackrel{i}{\check{\cdots}}\stackrel{j}{\check{\cdots}},a_{k+1})\,. \nonumber
\end{eqnarray}
One checks that $\delta$ maps $\tilde\Gamma^k(A,M)$ to $\tilde\Gamma^{k+1}(A,M)$,
and that $\delta^2=0$.
The space $\tilde\Gamma(A,M)$ with the differential $\delta$ 
is called the \emph{basic cohomology complex} associated to $A$ and $M$.

Define the structure of a $\mb C[\partial]$-module on $\tilde\Gamma(A,M)$ by letting
\begin{equation}\label{eq:july24_8_b}
(\partial\tilde\gamma)_{\lambda_1,\dots,\lambda_{k}}(a_1,\dots,a_{k})
=(\partial+\lambda_1+\cdots+\lambda_k)\tilde\gamma_{\lambda_1,\dots,\lambda_{k}}(a_1,\dots,a_{k})\,.
\end{equation}
One checks that $\delta$ and $\partial$ commute, and therefore 
$\partial\tilde\Gamma(A,M)\subset\tilde\Gamma(A,M)$ is a subcomplex,
and we can consider the \emph{reduced cohomology complex}
$\Gamma(A,M)=\tilde\Gamma(A,M)/\partial\tilde\Gamma(A,M)$.

For us, the interesting example is $A=\bigoplus_{i\in I}\mb C[\partial]u_i$,
with the zero $\lambda$-bracket,
and $M=\mc V$, where $\mc V$ is an algebra of differential functions
in the variables $u_i,\,i\in I$,
and the representation of $A$ on $M$ is given by the Beltrami $\lambda$-bracket, 
as in \eqref{eq:july24_8}.
In this case we consider the subspace 
$\tilde\Gamma_\fin=\bigoplus_{k\in\mb Z_+}\tilde\Gamma^k_\fin\subset \tilde\Gamma(A,\mc V)$,
where $\tilde\Gamma^k_\fin$ consists of $k$-cochains $\tilde\gamma$ with finite support,
namely such that
$\tilde\gamma_{\lambda_1,\cdots,\lambda_k}(u_{i_1},\cdots,u_{i_k})=0$
for all but finitely many choices of the indices $i_i,\dots,i_k\in I$.
Clearly, $\tilde\Gamma_\fin$ is a subcomplex of $\tilde\Gamma(A,\mc V)$,
and it is a $\mb C[\partial]$-submodule.
Hence, we can consider the associated reduced complex 
$\Gamma_\fin=\tilde\Gamma_\fin/\partial\tilde\Gamma_\fin$.
The following theorem was proved in \cite{DSK} (the first part of Theorem 4.6),
in the case when $I$ is finite.
However the proof is the same for arbitrary $I$.
\begin{theorem}\label{th:july24}
The map $\phi:\,\tilde\Gamma_\fin\to\tilde\Omega$, given by
\begin{equation}\label{eq:july24_9}
\phi(\tilde\gamma)\,=\,
\frac1{k!} \sum_{
\substack{
i_1,\cdots,i_k\in I \\ m_1,\cdots,m_k\in\mb{Z}_+
}
}
f^{m_1\cdots m_k}_{i_1\cdots i_k}\, \delta u_{i_1}^{(m_1)}\wedge\cdots\wedge\delta u_{i_k}^{(m_k)}\,,
\end{equation}
where $f^{m_1\cdots m_k}_{i_1\cdots i_k}\in\mc V$ is the coefficient 
of $\lambda_1^{m_1}\cdots\lambda_k^{m_k}$ 
in $\tilde\gamma_{\lambda_1,\dots,\lambda_k}(u_{i_1},\dots,u_{i_k})$,
is bijective, and it commutes with the actions of $\delta$ and $\partial$.
Hence it induces isomorphisms of complexes 
$\tilde\Omega\stackrel{\sim}{\longrightarrow}\tilde\Gamma_\fin$
and
$\Omega\stackrel{\sim}{\longrightarrow}\Gamma_\fin$.
\end{theorem}

\vspace{3ex}
\subsection{The variational complex as a complex of poly-$\lambda$-brackets
or poly-differential operators.}~~
\label{sec:09jan2}
Consider, as in Section \ref{sec:july23_2}, the Lie conformal algebra
$A=\bigoplus_{i\in I}\mb C[\partial]u_i$, with trivial $\lambda$-bracket,
and recall the Beltrami $\lambda$-bracket \eqref{eq:july19_2}
on an algebra of differential functions $\mc V$.
Recall from \cite{DSK} that, 
for $k\geq1$, a \emph{$k$-$\lambda$-bracket} on $A$ with coefficients in $\mc V$ 
is, by definition, a $\mb C$-linear map
$c:\,A^{\otimes k}\to\mb C[\lambda_1,\dots,\lambda_{k-1}]\otimes\mc V$,
denoted by
$$
a_1\otimes\cdots\otimes a_k\,\mapsto\,\{{a_1}_{\lambda_1}\cdots {a_{k-1}}_{\lambda_{k-1}} a_k\}_c\,,
$$
satisfying the following conditions:
\begin{enumerate}
\item[B1.] $\{{a_1}_{\lambda_1}\cdots (\partial a_i)_{\lambda_i}\cdots {a_{k-1}}_{\lambda_{k-1}} a_k\}_c
=-\lambda_i\{{a_1}_{\lambda_1}\cdots {a_{k-1}}_{\lambda_{k-1}} a_k\}_c$,
for $1\leq i\leq k-1$;
\item[B2.] $\{{a_1}_{\lambda_1}\cdots {a_{k-1}}_{\lambda_{k-1}} (\partial a_k)\}_c
=(\lambda_1+\cdots+\lambda_{k-1}+\partial)
\{{a_1}_{\lambda_1}\cdots {a_{k-1}}_{\lambda_{k-1}} a_k\}_c$;
\item[B3.] $c$ is skew-symmetric with respect to simultaneous permutations
of the $a_i$'s and the $\lambda_i$'s in the sense that,
for every permutation $\sigma$ of the indices $\{1,\dots,k\}$, we have:
$$
\{{a_1}_{\lambda_1}\cdots {a_{k-1}}_{\lambda_{k-1}} a_k\}_c
=\text{sign}(\sigma)
\{{a_{\sigma(1)}}_{\lambda_{\sigma(1)}}\cdots {a_{\sigma(k-1)}}_{\lambda_{\sigma(k-1)}} 
a_{\sigma(k)}\}_c\,\Big|_{\lambda_k\mapsto\lambda_k^\dagger}\,.
$$
The notation in the RHS means that $\lambda_k$ is replaced 
by $\lambda_k^\dagger=-\sum_{j=1}^{k-1}\lambda_j-\partial$,
if it occurs, and $\partial$ is moved to the left.
\end{enumerate}

We let $C^0=\mc V/\partial\mc V$ and, for $k\geq1$,
we denote by $C^k$ the space of all $k$-$\lambda$-brackets
on $A$ with coefficients in $\mc V$.
For example, $C^1$ is the space of all $\mb C[\partial]$-module homomorphisms 
$c:\,A\to \mc V$
(which in turn is isomorphic to $\mc V^\ell$).
We let $C=\bigoplus_{k\in\mb Z_+}C^k$, 
the space of all \emph{poly} $\lambda$-\emph{brackets}.

We also consider the subspace $C_\fin=\bigoplus_{k\in\mb Z_+}C^k_\fin\subset C$,
where $C^0_\fin=C^0$ and, for $k\geq1$,
$C^k_\fin$ consists of $k$-$\lambda$-brackets $c$ with finite support,
namely such that
$\{{u_{i_1}}_{\lambda_1}\cdots {u_{i_{k-1}}}_{\lambda_{k-1}} u_{i_k}\}_c=0$
for all but finitely many choices of the indices $i_i,\dots,i_k\in I$.
For example, $C^1_\fin\simeq\mc V^{\oplus\ell}$.

We next define, following \cite{DSK}, 
a differential $d$ on the space $C$ of poly $\lambda$-brackets
such that $d(C^k)\subset C^{k+1}$ and $d^2=0$,
thus making $C$ a cohomology complex,
and such that $C_\fin\subset C$ is a subcomplex.
The difference with \cite{DSK} is that here we use the Beltrami $\lambda$-bracket
in place of the equivalent language of left modules defined by the Beltrami $\lambda$-bracket.

For $\tint f\in C^0=C^0_\fin=\mc V/\partial\mc V$, 
we let $d\tint f\in C^1$ be the following
$\mb C[\partial]$-module homomorphism:
\begin{equation}\label{eq:d0}
\big(d\tint f\big)(a)\,
\Big(=\{a\}_{d\tint f}\Big)
\,:=\,
\{a_{-\partial}f\}_B\,.
\end{equation}
For $c\in C^k$, with $k\geq1$, we let
$dc\in C^{k+1}$ be the following $k+1$-$\lambda$-bracket:
\begin{eqnarray}\label{eq:d>}
&\displaystyle{
\{{a_1}_{\lambda_1}\cdots {a_{k}}_{\lambda_{k}} a_{k+1}\}_{dc} 
\,:=\,
\sum_{i=1}^k (-1)^{i+1} 
\Big\{
{a_i}_{\lambda_i}
\big\{{a_1}_{\lambda_1}\stackrel{i}{\check{\cdots}}{a_{k}}_{\lambda_{k}} a_{k+1}\big\}_{c} 
\Big\}_B
}\nonumber\\
&\displaystyle{
+(-1)^k 
\Big\{
{\big\{{a_1}_{\lambda_1}\cdots{a_{k-1}}_{\lambda_{k-1}} a_{k}\big\}_{c}}\,\,
_{\lambda_1+\cdots+\lambda_k}
{a_{k+1}}
\Big\}_B
\,.
}
\end{eqnarray}

For example, for a 1-cocycle, namely 
a $\mb C[\partial]$-module homomorphism $c:\,A\to \mc V$, we have
\begin{equation}\label{eq:july29_1}
\{a_\lambda b\}_{dc}
=\{a_\lambda c(b)\}_B-\{c(a)_{\lambda}b\}_B\,.
\end{equation}

We define, for $k\geq1$, 
a $\mb C$-linear map $\psi^k:\,{\tilde\Gamma}^k\to C^k$, as follows.
Given $\tilde\gamma\in\tilde\Gamma^k$, we define 
$\psi^k(\tilde\gamma):\,A^{\otimes k}\to\mb C[\lambda_1,\dots,\lambda_{k-1}]\otimes\mc V$, by:
\begin{equation}\label{eq:5}
\{{a_1}_{\lambda_1}\cdots{a_{k-1}}_{\lambda_{k-1}}a_k\}_{\psi^k(\tilde\gamma)}
=
\tilde\gamma_{\lambda_1,\cdots,\lambda_{k-1},\lambda_k^\dagger}(a_1,\cdots,a_{k})\,,
\end{equation}
where, as before,
$\lambda_k^\dagger=-\sum_{j=1}^{k-1}\lambda_j-\partial$,
and $\partial$ is moved to the left.
The following theorem is a special case of Theorem 1.5 from \cite{DSK}.
\begin{theorem}\label{th:red}
The identity map on $\mc V/\partial\mc V$ and the maps $\psi^k,\,k\geq1$, 
induce isomorphisms of cohomology complexes $\Gamma\stackrel{\sim}{\to} C$
and $\Gamma_\fin\stackrel{\sim}{\to} C_\fin$.
\end{theorem}
Due to Theorems \ref{th:july24} and \ref{th:red},
we can identify the variational complex $\Omega$
with the complex of poly-$\lambda$-brackets $C_\fin$.
Explicitly, we have:
\begin{corollary}\label{cor:09jan02}
Let $k\geq1$ and let $\tilde\omega\in\tilde\Omega^k$ be as in \eqref{eq:apr24_1},
where we assume that the coefficients $f^{m_1\cdots m_k}_{i_1\cdots i_k}$
are skew-symmetric with respect to simultaneous permutations 
of the lower and upper indices.
Let $\rho^k(\tilde\omega)$ be the $k$-$\lambda$-bracket
defined by 
\begin{equation}\label{eq:09jan02_1}
\{{u_{i_1}}_{\lambda_1}\cdots{u_{i_{k-1}}}_{\lambda_{k-1}}u_{i_k}\}_{\rho^k(\tilde\omega)}
=
k! 
\!\!\!\!
\sum_{m_1,\cdots,m_k\in\mb Z_+}
\!\!\!
\lambda_1^{m_1}\cdots\lambda_{k-1}^{m_{k-1}}(-\lambda_1-\cdots-\lambda_{k-1}-\partial)^{m_k}
f^{m_1\cdots m_k}_{i_1\cdots i_k}\,,
\end{equation}
and extended to $A^{\otimes k}$ by sesquilinearity.
Then,
the identity map on $\mc V/\partial\mc V$ and the maps 
$\rho^k:\,\tilde\Omega^k\to C^k,\,k\geq1$ 
factor through an isomorphism of complexes:
$\rho:\,\Omega\stackrel{\sim}{\longrightarrow}C_\fin$.
\end{corollary}
\begin{proof}
It follows immediately from the definitions \eqref{eq:july24_9} of $\phi$ 
and \eqref{eq:5} of $\psi^k$.
\end{proof}
Next, for $X_P\in\mf g^\partial,\,P\in\mc V^{\ell}$, we define, 
the contraction operator $\iota_{X_P}:\, C^k_\fin\to C^{k-1}_\fin,\,k\geq1$,
making $\rho$ an isomorphism of $\mf g^\partial$-complexes.
For $c\in C^1_\fin$, we let
\begin{equation}\label{eq:09jan2_2}
\iota_{X_P}(c)
\,=\,
\tint\sum_{i\in I}P_ic(u_i)\,\in\mc V/\partial\mc V=C^0\,.
\end{equation}
For $c\in C^k_\fin$ with $k\geq2$, we let
\begin{equation}\label{eq:09jan2_3}
\big\{{a_2}_{\lambda_2} \cdots {a_{k-1}}_{\lambda_{k-1}} a_k\big\}_{\iota_{X_P}(c)}
\,=\,
\sum_{i\in I}
\big\{{u_i}_{\partial}{a_2}_{\lambda_2} \cdots {a_{k-1}}_{\lambda_{k-1}} a_k\big\}_{c\,\,_\to} P_i\,,
\end{equation}
where, as usual, the arrow in the RHS means that $\partial$ is moved to the right.
The following proposition is a more precise form of the second part of Theorem 4.6
from \cite{DSK}.
\begin{proposition}\label{prop:09jan02}
The contraction operators $\iota_{X_P}$ defined by \eqref{eq:09jan2_2} and \eqref{eq:09jan2_3}
endow the cohomology complex $C_\fin$ with a $\mf g^\partial$-structure,
and the isomorphism of complexes $\rho:\,\Omega\to C_\fin$ (defined in Corollary \ref{cor:09jan02})
is an isomorphism of $\mf g^\partial$-complexes,
namely $\rho^{k-1}(\iota_{X_P}(\omega))=\iota_{X_P}(\rho^k(\omega))$
for all $\omega\in\Omega^k$ and $X_P\in\mf g^\partial$.
\end{proposition}
\begin{proof}
If $\tilde\omega\in\tilde\Omega^k$ is as in \eqref{eq:apr24_1}, with $f^{m_1,\cdots,m_k}_{i_1,\cdots,i_k}$
skew-symmetric with respect to simultaneous permutations of lower and upper indices,
we have, 
$$
\iota_{X_P}(\tilde\omega)
=
k\sum_{\substack{
i_2,\dots,i_k\in I\\m_2,\dots,m_k\in\mb Z_+
}}
\Big(
\sum_{i_1\in I,m_1\in\mb Z_+}
f^{m_1,\cdots,m_k}_{i_1,\cdots,i_k}
\big(\partial^{m_1}P_{i_1}\big)
\Big)
\delta u_{i_2}^{(m_2)}\wedge\cdots\wedge\delta u_{i_k}^{(m_k)}\,.
$$
Hence, applying 
equation \eqref{eq:09jan02_1}, we get
\begin{eqnarray*}
&& \{
{u_{i_2}}_{\lambda_2}\cdots{u_{i_{k-1}}}_{\lambda_{k-1}}u_{i_k}
\}_{\rho^{k-1}(\iota_{X_P}(\tilde\omega))} \\
&& =
k! 
\sum_{
\substack{
m_1,\cdots,m_k\in\mb Z_+
\\
i_1\in I
}}
\lambda_2^{m_2}\cdots\lambda_{k-1}^{m_{k-1}}(-\lambda_2-\cdots-\lambda_{k-1}-\partial)^{m_k}
\Big(
f^{m_1\cdots m_k}_{i_1\cdots i_k}
\big(\partial^{m_1}P_{i_1}\big)
\Big)\,.
\end{eqnarray*}
On the other hand, by \eqref{eq:09jan02_1} and \eqref{eq:09jan2_3}, we have
\begin{eqnarray*}
&\displaystyle{
\{
{u_{i_2}}_{\lambda_2}\cdots{u_{i_{k-1}}}_{\lambda_{k-1}}u_{i_k}
\}_{\iota_{X_P}(\rho^{k}(\tilde\omega))} 
=
\sum_{i_1\in I}
\{
{u_{i_1}}_{\partial}{u_{i_2}}_{\lambda_2}\cdots{u_{i_{k-1}}}_{\lambda_{k-1}}u_{i_k}
\}_{\rho^{k}(\tilde\omega)\,\,_\to}  P_{i_1}
}\\
&\displaystyle{
=
k! \sum_{
\substack{
m_1,\cdots,m_k\in\mb Z_+
\\
i_1\in I
}}
\lambda_2^{m_2}\cdots\lambda_{k-1}^{m_{k-1}}(-\lambda_2-\cdots-\lambda_{k-1}-\partial)^{m_k}
\Big(
f^{m_1\cdots m_k}_{i_1\cdots i_k}
\big(\partial^{m_1}P_{i_1}\big)
\Big)\,.
}
\end{eqnarray*}
\end{proof}

The space $C^k$ of $k$-$\lambda$-brackets, for $k\geq1$,
can be equivalently described
in terms of maps $S:\,(\mc V^{\ell})^{k-1}\to\mc V^{\oplus\ell}$
of \emph{differential type}, namely of the form:
\begin{equation}\label{eq:dic15_7}
S(P^1,\cdots,P^{k-1})_{i_k}
\,=\, 
\sum_{\substack{m_1,\cdots,m_{k-1}\in\mb Z_+ \\ i_1,\cdots,i_{k-1}\in I}}
f_{i_1,\cdots,i_{k-1},i_k}^{m_1,\cdots,m_{k-1}}
(\partial^{m_1}P^1_{i_1})\cdots(\partial^{m_{k-1}}P^{k-1}_{i_{k-1}})\,,
\end{equation}
where $f_{i_1,\cdots,i_{k-1},i_k}^{m_1,\cdots,m_{k-1}}\in\mc V$ is non-zero only
for finitely many choices of the upper and lower indices.
Indeed, given a $k$-$\lambda$-bracket $c\in C^k$,
the corresponding map of differential type is \eqref{eq:dic15_7},
where $f_{i_1,\cdots,i_{k-1},i_k}^{m_1,\cdots,m_{k-1}}$
is the coefficient of $\lambda_1^{m_1}\cdots\lambda_{k-1}^{m_{k-1}}$ 
in $\{{u_{i_1}}_{\lambda_1}\cdots{u_{i_{k-1}}}_{\lambda_{k-1}}u_{i_k}\}_c$.
It is easy to see that this gives a bijective correspondence between
the space $C^k$ of $k$-$\lambda$-brackets
and the space of maps  $S:\,(\mc V^{\ell})^{k-1}\to\mc V^{\oplus\ell}$
of differential type which are skew-symmetric, namely such that
\begin{equation}\label{a}
\tint S(P^{\sigma(1)},\dots,F^{\sigma(k-1)})P^{\sigma(k)} 
= \text{sign}(\sigma)\tint S(P^1,\dots,P^{k-1})P^k \,,
\end{equation}
for every $P^1,\cdots,P^k\in\mc V^\ell$ and every permutation $\sigma$.

An equivalent language is that of skew-symmetric poly-differential operators.
Recall from \cite{DSK} that a $k$-\emph{differential operator}
is a $\mb C$-linear map $S:\,(\mc V^{\ell})^{k}\to\mc V/\partial\mc V$, of the form
\begin{equation}\label{eq:dic15_1}
S(P^1,\cdots,P^k)
\,=\, 
\int \sum_{\substack{n_1,\cdots,n_{k}\in\mb Z_+ \\ i_1,\cdots,i_{k}\in I}}
f_{i_1,\cdots,i_k}^{n_1,\cdots,n_{k}}
(\partial^{n_1}P^1_{i_1})\cdots(\partial^{n_{k}}P^{k}_{i_{k}})\,.
\end{equation}
The operator $S$ is called skew-symmetric if
$$
\int S(P^1,\cdots,P^k)=\text{sign}(\sigma)\int S(P^{\sigma(1)},\cdots,P^{\sigma(k)})\,,
$$
for every $P^1,\cdots,P^k\in\mc V^\ell$ and every permutation $\sigma$.
Given a $k$-$\lambda$-bracket $c\in C^k$
we associate to it the following $k$-differential operator:
\begin{equation}\label{eq:vsept28_6}
S(P^1,\cdots,P^k)
\,=\, 
\int \sum_{\substack{n_1,\cdots,n_{k-1}\in\mb Z_+ \\ i_1,\cdots,i_{k}\in I}}
f_{i_1,\cdots,i_{k-1},i_k}^{n_1,\cdots,n_{k-1}}
(\partial^{n_1}P^1_{i_1})\cdots(\partial^{n_{k-1}}P^{k-1}_{i_{k-1}})P^k_{i_k}\,,
\end{equation}
where $f_{i_1,\cdots,i_{k-1},i_k}^{m_1,\cdots,m_{k-1}}$ is as in \eqref{eq:dic15_7}.
It is easy to see that the skew-symmetry property of the $k$-$\lambda$-bracket
is translated to the skew-symmetry of the $k$-differential operator.
Conversely, integrating by parts, any $k$-differential operator
can be written in the form \eqref{eq:vsept28_6}.
Thus we have a surjective map from the space of $k$-symbols
to the space $\Sigma^k$ of skew-symmetric $k$-differential operators.
By Proposition \ref{prop:20081222}, the $k$-differential operator $S$ can be written uniquely
in the form \eqref{eq:vsept28_6},
hence this map is an isomorphism.

The differential, the contraction operators, and the Lie derivatives for the variational complex 
$\Omega$ have an especially nice form when written in terms 
of poly-differential operators.
The following result was obtained in \cite[Section 4.5]{DSK}.
\begin{theorem}\label{th:09jan4}
The variational complex $(\Omega=\bigoplus_{k\in\mb Z_+}\Omega^k,\delta)$ is isomorphic to the complex
of poly-differential operators $(\Sigma=\bigoplus_{k\in\mb Z_+}\Sigma^k,d)$, where
\begin{equation}\label{eq:dic15_6}
(dS)(P^1,\cdots,P^{k+1})
\,=\,
\sum_{s=1}^{k+1}(-1)^{s+1}
\big(X_{P^s}S\big)(P^1,\stackrel{s}{\check{\cdots}},P^{k+1})\,.
\end{equation}
In this formula, if $S$ is as in \eqref{eq:dic15_1},
$X_PS$ denotes the $k$-differential operator obtained from $S$
by replacing the coefficients $f_{i_1,\cdots,i_k}^{n_1,\cdots,n_{k}}$
by $X_P(f_{i_1,\cdots,i_k}^{n_1,\cdots,n_{k}})$.

Furthermore, the $\mf g^\partial$-structure on $\Omega$ is translated to the following
$\mf g^\partial$-structure on $\Sigma$:
\begin{eqnarray}
(\iota_{X_{P^1}}S)(P^2,\cdots,P^k)
&=& 
S(P^1,P^2,\cdots,P^k)\,, \label{eq:dic15_12_a}\\
(L_{X_Q}S)(P^1,\cdots,P^k)
&=&
(X_QS)(P^1,\cdots,P^k)
+\sum_{s=1}^k S(P^1,\cdots,X_{P^s}(Q),\cdots,P^k)\,. \label{eq:dic15_12_b}
\end{eqnarray}
\end{theorem}
Equation \eqref{eq:dic15_6} is similar to formula (4.9) in \cite{D}.

We can write explicitly the bijection between the spaces $\Omega$ and $\Sigma$,
which gives the isomorphism of $\mf g^\partial$-complexes.
If $\omega\in\Omega^k$, the corresponding $k$-differential operator in $S_\omega\in\Sigma^k$
is given by
\begin{equation}\label{eq:09jan14_2}
S_\omega(P^1,\cdots,P^k)
\,=\,
\iota_{X_{P^k}}\cdots\iota_{X_{P^1}}\omega\,\in\Omega^0\,.
\end{equation}
This is clear by applying \eqref{eq:dic15_12_a} iteratively.
Due to the identification \eqref{eq:09jan14_2} of $\Omega$ with $\Sigma$,
we can think of an element $\omega\in\Omega^k$ as a map 
$\omega:\,\Lambda^k\mf g^\partial\to\Omega^0$,
given by
\begin{equation}\label{eq:09jan14_3}
\omega(X_1,\cdots,X_k)
\,=\,
\iota_{X_k}\cdots\iota_{X_1}\omega\,\in\Omega^0\,.
\end{equation}

Clearly, $\Sigma^0=\mc V/\partial\mc V$. 
Moreover, for $k\geq1$, using the bijections between the space $\Sigma^k$ of $k$-differential operators 
and the space of maps $(\mc V^\ell)^{k-1}\to\mc V^{\oplus\ell}$ of differential type,
we get that $\Sigma^1\simeq\mc V^{\oplus\ell}$,
and $\Sigma^2$ is naturally identified with the space of skew-adjoint 
differential operators $S:\,\mc V^\ell\to\mc V^{\oplus\ell}$.
With these identifications, $d,\,\iota_{X_P}$ and $L_{X_P}$, for $P\in\mc V^\ell$, are as follows.
For $\tint f\in\Sigma^0=\mc V/\partial\mc V$ we have 
\begin{equation}\label{eq:09jan4_1}
d\big(\tint f\big)=\frac{\delta f}{\delta u}\in\mc V^{\oplus\ell}
\,\,,\,\,\,\,
\iota_{X_P}\big(\tint f\big)=0
\,\,,\,\,\,\,
L_{X_P}\big(\tint f\big)=\tint X_P(f)\in\mc V/\partial\mc V
\,.
\end{equation}
For $F\in\Sigma^1=\mc V^{\oplus\ell}$, we have
\begin{eqnarray}
& d(F)=D_F(\partial)-D^*_F(\partial)\in\Sigma^2
\,\,,\,\,\,\,
\iota_{X_P}(F)=\tint\sum_{i\in I}P_iF_i\,\in\mc V/\partial\mc V
\,\,,
\label{eq:09jan4_2_a}\\
\vphantom{\Bigg\{}
& \iota_{X_P}(dF)=D_F(\partial)P-D^*_F(\partial)P\in\mc V^{\oplus\ell}
\,\,,\,\,\,\,
d(\iota_{X_P}F)=D^*_F(\partial)P+D^*_P(\partial)F\in\mc V^{\oplus\ell}
\,\,,
\label{eq:09jan4_2_b}\\
& L_{X_P}(F)
=
D_F(\partial)P+D^*_P(\partial)F\in\mc V^{\oplus\ell}
\,,
\label{eq:09jan4_2_c}
\end{eqnarray}
where 
$D_F(\partial)$ is the Fr\'echet derivative \eqref{eq:july18_5}.
Finally, a skew-adjoint operator $S(\partial)=(S_{ij}(\partial))_{i,j\in I}$ $\in\Sigma^2$ is closed 
if and only if the following equation holds (cf.\  \eqref{v024}):
\begin{equation}\label{v024_sec}
\big\{u_i\, _\mu\, S_{kj}(\lambda)\big\}_B
- \big\{u_j\, _\lambda\, S_{ki}(\mu)\big\}_B
=
\big\{S_{ij}(\lambda)\,_{\lambda+\mu}\,u_k\big\}_B\,.
\end{equation}
\begin{definition}\label{def:6feb09}
A matrix valued differential operator $S(\partial):\,\mc V^\ell\to\mc V^{\oplus\ell}$
is called \emph{symplectic} if it is skew-adjoint and satisfies 
equation \eqref{v024_sec}.
\end{definition}

Propositions \ref{prop:july19_1} and \ref{prop:july19_2} provide
a description of the kernel and the image of the variational derivative for normal $\mc V$,
based on Theorem \ref{th:july23}, namely the exactness of the variational complex
at $k=0$ and $1$.
The next corollary of this theorem provides a description of symplectic operators,
and it is based on the exactness of the variational complex
at $k=2$.
\begin{proposition}\label{prop:09jan7}
If $\mc V$ is a normal algebra of differential functions,
then any symplectic differential operator is of the form:
$S_F(\partial)=D_F(\partial)-D^*_F(\partial)$, for some $F\in\mc V^{\oplus\ell}$.
Moreover, $S_F=S_G$ if and only if $F-G=\frac{\delta f}{\delta u}$ for some $f\in\mc V$.
\end{proposition}

\begin{example}\label{ex:09feb16}
The simplest symplectic differential operators in the case $\ell=1$ are:
$$
S_{\frac{1}{2}u^{(n)}}(\partial) = \partial^n \,\text{ for } n \text{ odd },\,
S_{uu^{\prime}}(\partial) = u^{\prime}
+2u\partial,\, S_{\frac{1}{2}{u^{\prime}}^2}(\partial) = u^{\prime\prime}+ 2u^{\prime}\partial.  
$$
\end{example}

\begin{example}\label{ex:09jan7_1}
Let $\mc V$ be the localization of 
the algebra of differential polynomials 
$\mb C[u,u^\prime,\cdots]$ at $u^\prime$.
Then, for  $F=-\frac1{2u^\prime}\in\mc V$, the corresponding symplectic
operator
$$
S_F(\partial) = \frac1{u^{\prime}} \partial\circ \frac1{u^{\prime}} 
$$
is the \emph{Sokolov symplectic operator}.
For 
$G =- \partial^2\Big(\frac1{2 u^\prime}\Big)\in\mc V$,
the corresponding symplectic
operator
$$
S_G(\partial) = 
\partial \circ \frac1{u^\prime} \partial \circ \frac1{u^\prime} \partial
$$
is the \emph{Dorfman symplectic operator}.
As we will see in Section \ref{2006_sec-dirac}, these operators play an important role in the integrability
of the Krichever-Novikov equation.
\end{example}

\vspace{3ex}
\subsection{Projection operators.}~~
\label{sec:projop}
Another way to understand the variational complex uses
projection operators $\mc P_k,\, k\geq1$, of $\tilde\Omega^k$ onto its 
subspace 
complementary to $\partial\tilde\Omega^k$.
We can then identify the space of local $k$-forms $\Omega^k$ 
with the image of $\mc P_k$ in $\tilde\Omega^k$.

For $i\in I$ we define the operators $I_i:\,\tilde\Omega^k\to\tilde\Omega^{k-1},\,k\in\mb Z_+$, as
\begin{equation}\label{eq:apr25_1}
I_i\,=\,\sum_{n\in\mb Z_+}(-\partial)^n\circ\iota_{\partial/\partial u_i^{(n)}}\,,
\end{equation}
and, for $k\geq1$, we let $\mc P_k:\,\tilde\Omega^k\to\tilde\Omega^k$ be given by
\begin{equation}\label{eq:apr25_2}
\mc P_k(\tilde\omega)\,=\,\frac{(-1)^{k+1}}k \sum_{i\in I} I_i(\tilde\omega)\wedge\delta u_i
\,.
\end{equation}
This formula expresses the fact that, integrating by parts, one can bring any differential form
$\tilde\omega\in\tilde\Omega^k$ uniquely in the image of 
$\mc P^k$.
The projection operators $\mc P^k$ on the space of differential forms $\tilde\Omega$ are analogous to the projection operators on the space of polyvectors,
introduced in \cite{B}.

The following theorem shows that $\mc{P_k}$ is a projection operator onto a subspace
$\im(\mc P_k)\subset\tilde\Omega^k$ complementary to $\partial\tilde\Omega^k$, for $k\geq1$.
\begin{theorem}\label{2006_Tproj}
Let $k\geq1$.
\begin{enumerate}
\romanparenlist
\item $I_i\circ\partial=0$ for all $i\in I$. In particular $\mc{P_k}\circ\partial=0$.
\item If $\tilde\omega\in\tilde\Omega^k$, then
$\mc{P_k}(\tilde\omega)-\tilde\omega\in\partial\tilde\Omega^k$.
\item $\Ker\mc P_k=\partial\tilde\Omega^k$ and $\mc{P_k}^2=\mc{P_k}$.
\end{enumerate}
\end{theorem}
\begin{proof}
By \eqref{eq:comm_frac} and \eqref{eq:apr24_9} we have that
$\big[\iota_{\partial/\partial u_i^{(n)}} , \partial\big] = \iota_{\partial/\partial u_i^{(n-1)}}$,
where the RHS is stipulated to be zero for $n=0$.
It follows that
$$
I_i\circ\partial = \sum_{n\in\mb Z_+}(-\partial)^n\circ\iota_{\partial/\partial u_i^{(n)}}\circ\partial
= \sum_{n\geq1}(-\partial)^n\circ\iota_{\partial/\partial u_i^{(n-1)}}
 - \sum_{n\in\mb Z_+}(-\partial)^{n+1}\circ\iota_{\partial/\partial u_i^{(n)}} = 0\,,
$$
thus proving (i).
Notice that, if $\tilde\omega\in\tilde\Omega^k$, we have
\begin{equation}\label{eq:apr25_3}
\sum_{i\in I,n\in\mb Z_+}\delta u_i^{(n)}\wedge\iota_{\partial/\partial u_i^{(n)}}(\tilde\omega) 
= k\tilde\omega\,.
\end{equation}
On the other hand, since $\partial$ commutes with $\delta$
and it is a derivation of the wedge product in $\tilde\Omega$, we have that
\begin{equation}\label{eq:apr25_4}
\sum_{i\in I,n\in\mb Z_+}\delta u_i^{(n)}\wedge\iota_{\partial/\partial u_i^{(n)}}(\tilde\omega) 
\equiv
\sum_{i\in I,n\in\mb Z_+}\delta u_i\wedge(-\partial)^n \iota_{\partial/\partial u_i^{(n)}}(\tilde\omega) 
\,\,\,\, \mod\partial\tilde\Omega^k\,.
\end{equation}
Part (ii) is then an immediate consequence of equations \eqref{eq:apr25_3} and \eqref{eq:apr25_4}.
Finally, (iii) follows from (i) and (ii).
\end{proof}

Let
$\tilde \omega = \sum_{\substack{ i_1,\dots,i_k\in I \\ m_1,\dots,m_k\in \mb Z_+}} 
f_{i_1\dots i_k}^{m_1\dots m_k} 
\delta u_{i_1}^{(m_1)}\wedge\cdots\wedge \delta u_{i_k}^{(m_k)} \in \tilde \Omega^k$, 
where the coefficients $f_{i_1\dots i_k}^{m_1\dots m_k}\in\mc V$ are skew-symmetric with
respect to exchanging simultaneously the lower and the upper indices.
We have
\begin{equation}\label{eq:2006_main}
{\mc P}_k(\tilde \omega) =
\sum_{\substack{i_1\dots i_k\in I \\ m_1\dots m_k\in \mb Z_+}}
(-\partial)^{m_k} \Big(f_{i_1\dots i_k}^{m_1\dots m_k} 
\delta u_{i_1}^{(m_1)}\wedge\dots\wedge\delta u_{i_{k-1}}^{(m_{k-1})}\Big) 
\wedge \delta u_{i_k}
\,.
\end{equation}
We associate to the RHS of \eqref{eq:2006_main} the following
map $S:\,(\mc V^\ell)^{k-1}\to\mc V^{\oplus\ell}$ of differential type
(defined in Section \ref{sec:09jan2}): 
\begin{equation}\label{b}
S(P^1,\dots,P^{k-1})_{i_k} = 
\sum_{\substack{i_1\dots i_{k-1}\in I \\ m_1\dots m_k\in \mb Z_+}}(-\partial)^{m_k}
\Big(f_{i_1\dots i_k}^{m_1\dots m_k}  (\partial^{m_1}P^1_{i_1})\cdots 
(\partial^{m_{k-1}}P^{k-1}_{i_{k-1}})\Big)\,.
\end{equation}
It is immediate to check that $S$ is skew-symmetric in the sense of \eqref{a}.
Hence, using the projection operators,
we get a more direct identification of the spaces $\Omega^k$ and $\Sigma^k$.

For example, if $\tilde\omega=\sum_{i\in I,n\in\mb Z_+}f_i^n \delta u_i^{(n)} \in \tilde\Omega^1$, 
the corresponding element in $\Omega^1$ is
$\mc{P}_1(\tilde\omega)
= \sum_{i\in I} \Bigl(\sum_{n\in \mb{Z}_+} (-\partial)^n f_i^n\Bigr) \delta u_i$.
Therefore the projection operator gives an explicit identification 
of $\mc V^{\oplus\ell}$  with $\Omega^1$, which, to $F\in\mc V^{\oplus\ell}$, associates
\begin{equation}\label{eq:09jan15_1}
\omega_F=\sum_{i\in I}F_i\delta u_i
\,\in \mc P^1(\tilde\Omega^1)\simeq\Omega^1\,.
\end{equation}

As pointed out in the previous section,
the space of skew-symmetric maps of differential type
$S:\,(\mc V^\ell)^{k-1}\to\mc V^{\oplus\ell}$
is canonically identified with the space $\Sigma^k$
of skew-symmetric $k$-differential operators
$S:\,(\mc V^\ell)^k\to\mc V\big/\partial\mc V$.
In view of Remark \ref{rem:apr24},
the differential $\delta$ on $\Sigma^k$ is given by formula \eqref{2006_diff},
where $\tilde\omega\in\tilde\Omega^k$ is replaced by $S\in\Sigma^k$
and $X_i\in\mf g^\partial$.
Using formula \eqref{eq:09jan21_1} for the commutator of evolutionary vector fields,
it is straightforward to rewrite $\delta$ on $\Sigma^k$ in the form \eqref{eq:dic15_6}
(cf.\ \cite{D}, p.61).
Likewise, it is immediate to derive \eqref{eq:dic15_12_a} and \eqref{eq:dic15_12_b}
from \eqref{2006_contr} and \eqref{2006_lieder} respectively.
\begin{remark}\label{rem:2006}
We can use the isomorphism ${\mf g}^\partial\simeq \mc V^\ell$
to conclude that the space of local $k$-forms $\Omega^k$ can be 
identified with 
the subspace of $\Hom_{\mb C}(\Lambda^{k-1}{\mf g}^\partial,\Omega^1)$ 
consisting of
skew-symmetric (in the sense of \eqref{a}) linear maps
$S:\,\Lambda^{k-1}{\mf g}^\partial\to\Omega^1$ acting by differential operators
(in the sense of \eqref{eq:dic15_7}).
In order to find the expression of the map $S:\,\Lambda^{k-1}{\mf g}^\partial\to\Omega^1$
corresponding to $\omega\in\Omega^k$, we notice that equation  \eqref{b}
can be equivalently written as
$\sum_{i\in I}S(P^1,\dots,P^{k-1})_i\delta u_i
=\frac1{k!}\iota_{X_{P^1}}\cdots \iota_{X_{P^{k-1}}}\tilde\omega$.
We thus conclude that the map $S:\,\Lambda^{k_1}{\mf g}^\partial\to\Omega^1$ 
corresponding to $\omega\in\Omega^k$ is given by
\begin{equation}\label{c}
S\,:\,\,X_1\wedge\cdots\wedge X_{k-1}\,\mapsto\,\frac1{k!}\iota_{X_1}\cdots \iota_{X_{k-1}}\omega\,.
\end{equation}
\end{remark}
\begin{remark}\label{2006_l_nondeg}
Via the identifications $\mf g^\partial\simeq\mc V^\ell$ and $\Omega^1\simeq\mc V^{\oplus\ell}$,
the pairing \eqref{eq:may4_3} between $\mf g^\partial$ and $\Omega^1$ becomes \eqref{eq:may4_4}.
Since, by Proposition \ref{prop:20081222}, this pairing is non-degenerate,
we get the embedding 
$\Omega^1\hookrightarrow\Hom_\mb{C}(\mf{g}^\partial,\mc V/\partial\mc V)$,
discussed in Section \ref{sec:09jan2},
which associates to $\omega\in\Omega^1$ the corresponding map of
differential type $S:\,\mf g^\partial\to\mc V/\partial\mc V$.
However, this embedding is \emph{not} an isomorphism.
For example the linear map $\phi\in\Hom_\mb{C}(\mf{g}^\partial,\mc V/\partial\mc V)$
given by 
$\phi(X_P)=\sum_{i\in I}\int\frac{\partial P_i}{\partial u_i}$,
is not in the image of $\Omega^1$.
This is the same as saying that there is no element $F\in\mc V^{\oplus\ell}$ such that
$\tint F\cdot P=\tint\sum_i\partial P_i/\partial u_i$, for all $P\in\mc V^\ell$.
Indeed, by taking $P_j=\delta_{i,j}$, we get that $F_i\in \partial\mc V$ for all $i$,
and by taking $P_j=\delta_{i,j}u_k^{(N)},\,N\geq1$,
we get that $\tint F_i u_k^{(N)}=0$ for all $k\in I,N\geq1$.
By Lemma \ref{lem:feb18} this is possible only for $F=0$, which is a contradiction.
\end{remark}


\section{Dirac structures and the corresponding Hamiltonian equations.}\label{2006_sec-dirac}

\subsection{Derived brackets and the Courant-Dorfman product.}~~
\label{sec:derived}
\begin{definition}\label{def:july18}
Let $L$ be a Lie superalgebra with the bracket $[\cdot\,,\,\cdot]$.
Given an odd derivation $d:\,L\to L$ such that $d^2=0$,
the corresponding \emph{derived bracket} on $L$
is defined by the following formula:
\begin{equation}\label{eq:july18_7}
[a,b]_d\,=\, (-1)^{1-p(a)}[d(a),b]\,,
\end{equation}
where $p(a)\in\mb Z/2\mb Z$ denotes the parity of $a\in L$.
\end{definition}

This construction goes back to Koszul \cite{KS}.
It is not hard to check that the derived bracket satisfies the Jacobi identity
with respect to the opposite parity on $L$, namely
\begin{equation}\label{eq:july18_8}
[a,[b,c]_d]_d-(-1)^{(1-p(a))(1-p(b))}[b,[a,c]_d]_d= [[a,b]_d,c]_d\,.
\end{equation}
However, the derived bracket on $L$ with the opposite parity does not define 
a Lie superalgebra structure since, in general, skew-commutativity fails.

The relevant special case for us of the above construction is the following.
Consider the space $\tilde\Omega=\tilde\Omega(\mc V)$ 
with the parity induced by its $\mb Z_+$-grading
and let $L=\End_{\mb C}(\tilde\Omega)$ be the Lie superalgebra of endomorphisms
of the superspace $\tilde\Omega$.
Then $d=\ad\,\delta$ is an odd derivation of $L$ satisfying $d^2=0$,
and we can consider the corresponding derived bracket $[\cdot\,,\,\cdot]_d$ on $L$.

Note that we have a natural embedding of $\tilde\Omega^1\oplus\mf g\to L$
obtained by
$\tilde\omega\,\mapsto\,\tilde\omega\wedge\,\,,\,\,\,\, X\,\mapsto\,\iota_X$.
\begin{proposition}\label{prop:july18}
\begin{enumerate}
\alphaparenlist
\item
The image of $\tilde\Omega^1\oplus\mf g$ in $L$ is closed under the derived bracket 
$[\cdot\,,\,\cdot]_d$, and we have
\begin{equation}\label{2006_dp}
(\tilde\omega\oplus X)\circ(\tilde\eta\oplus Y)
=
(L_X(\tilde\eta)-\iota_Y(\delta\tilde\omega))\oplus[X,Y]\,,
\end{equation}
for every $X,Y\in\mf g$ and $\tilde\omega,\tilde\eta\in\tilde\Omega^1$.
\item
The bracket \eqref{2006_dp} induces a well-defined bracket on $\Omega^1\oplus\mf g^\partial$.
\end{enumerate}
\end{proposition}
\begin{proof}
First notice that, since $\delta$ is an odd derivation of the wedge product, we have
$d(\tilde\omega\wedge)=[\delta,\tilde\omega\wedge]=\delta(\tilde\omega)\wedge$ 
for every $\tilde\omega\in\tilde\Omega^1$,
and, by equation \eqref{eq:apr24_6},
we have $d(\iota_x)=[\delta,\iota_X]=L_X$ for every $X\in\mf g$.
Hence we have, by the definition \eqref{eq:july18_7} of the derived bracket, that
\begin{eqnarray*}
{[\tilde\omega\wedge,\tilde\eta\wedge]}_d &=& [\delta(\tilde\omega)\wedge,\tilde\eta\wedge]\,=\,0\,\,\,\,,
\,\,\,\,\,\,\,\,\,\,\,\,\,
{[\tilde\omega\wedge,\iota_X]}_d \,=\, [\delta(\tilde\omega)\wedge,\iota_X]
\,=\,-\iota_X(\delta \tilde\omega)\wedge\,,\\
{[\iota_X,\tilde\omega\wedge]}_d &=& [L_X,\tilde\omega\wedge]\,=\,L_X(\tilde\omega)\wedge\,\,,
\,\,\,\,
{[\iota_X,\tilde\iota_Y]}_d \,=\, [L_X,\iota_Y]\,=\,\iota_{[X,Y]}\,.
\end{eqnarray*}
In the last identity we used equation \eqref{eq:apr24_9}.
Equation \eqref{2006_dp} follows from the above relations, proving (a).
For part (b),
we just notice that $\tilde\Omega^1\oplus\mf g^\partial\subset\tilde\Omega^1\oplus\mf g$
is a subalgebra with respect to the bracket \eqref{2006_dp},
and $\partial\tilde\Omega^1\oplus\mf g^\partial\subset \tilde\Omega^1\oplus\mf g^\partial$
is an ideal in this subalgebra.
\end{proof}

\vspace{3ex}
\subsection{Definition of a Dirac structure.}~~
\label{sec:def-dirac}
Dirac structures were introduced independently by Courant and Dorfman (see \cite{D}),
as a generalization of the notions of Hamiltonian and symplectic operators. 
We  will discuss their relation to these operators 
in Sections \ref{sec:dirac-hamilt} and \ref{sec:dirac-sympl}.

Let $\mc V$ be an algebra of differential functions.
Recall from Proposition \ref{prop:20081222} and equation \eqref{eq:may4_3}
that we have a non-degenerate pairing 
$\mf g^\partial\times\Omega^1\to\mc V/\partial\mc V$, 
given by
$$
(X_P,\omega_F)=(\omega_F,X_P)=\iota_{X_P}(\omega_F)=\sum_{i\in I}\tint P_iF_i\,.
$$
We extend it to a symmetric $\Omega^0$-valued bilinear form
$(\,,\,) :\,(\Omega^1\oplus\mf{g}^\partial)\times(\Omega^1\oplus\mf{g}^\partial)
\to\Omega^0=\mc V/\partial\mc V$
by letting $\Omega^1$ and $\mf{g}^\partial$ be isotropic subspaces, i.e.\
$(\omega,\eta) =(X,Y) =0$,
for all $\omega,\eta\in\Omega^1,\,X,Y\in\mf g^\partial$.
In view of Proposition \ref{prop:july18}(b), 
we define the {\it Courant-Dorfman product} on $\Omega^1\oplus\mf{g}^\partial$ 
(with even parity) by 
\begin{equation}\label{2006_dp_copy}
(\omega\oplus X)\circ(\eta\oplus Y)=(L_X(\eta)-\iota_Y(\delta\omega))\oplus[X,Y]\,.
\end{equation}
By \eqref{eq:july18_8} and Proposition \ref{prop:july18}, 
this product satisfies the left Jacobi identity
$[A,[B,C]]-[B,[A,C]]$ $=[[A,B],C]$, for all $A,B,C\in\Omega^1\oplus\mf g^\partial$,
but it is not skew-commutative.
\begin{definition}\label{2006_def-dirac}
A {\it Dirac structure} is a subspace $\mc{L}\subset\Omega^1\oplus\mf{g}^\partial$
which is maximal isotropic with respect to the bilinear form $( \,,\,) $,
\begin{equation}\label{2006_28oct_1}
\mc{L}=\mc{L}^\bot=\big\{A\in\Omega^1\oplus\mf{g}^\partial\,\big|\,( A,B) =0\,,\,\,\text{for all\,\,} B\in\mc{L}\big\}\,,
\end{equation}
and which is closed with respect to the Courant-Dorfman product,
\begin{equation}\label{2006_28oct_2}
A,B\in\mc{L}\,\implies\,A\circ B\in\mc{L}\,.
\end{equation}
\end{definition}
\begin{remark}\label{rem:a}
It is clear that condition \eqref{2006_28oct_2} can be equivalently expressed,
using the first condition \eqref{2006_28oct_1}, by the relation
\begin{equation}\label{2006_28oct_3}
( A\circ B,C) =0\,,\,\,\text{for all\,\,} A,B,C\in\mc{L}\,.
\end{equation}
\end{remark}
\begin{proposition}\label{prop:finalissimo}
Assuming \eqref{2006_28oct_1},
condition \eqref{2006_28oct_3} is equivalent to the following
``integrability condition'' for a Dirac structure:
\begin{equation}\label{2006_28oct_4}
(L_{X_1}\omega_2,X_3)+(L_{X_2}\omega_3,X_1)+(L_{X_3}\omega_1,X_2)=0
\,\,,\,\,\,\,
\text{ for all }
\omega_1\oplus X_1,\omega_2\oplus X_2,\omega_3\oplus X_3\in\mc{L}\,.
\end{equation}
\end{proposition}
\begin{proof}
In order to derive \eqref{2006_28oct_4} from \eqref{2006_28oct_3} we notice that, for 
$A=\omega_1\oplus X_1,\,B=\omega_2\oplus X_2,\,C=\omega_3\oplus X_3\in\Omega^1\oplus\mf{g}^\partial$,
we have
\begin{equation}\label{2006_28oct_5}
( A\circ B,C) 
=(L_{X_1}(\omega_2),X_3)-(\iota_{X_2}(\delta\omega_1),X_3)+(\omega_3,[X_1,X_2])\,.
\end{equation}
The second term in the right hand side of \eqref{2006_28oct_5} is
$$
-(\iota_{X_2}(\delta\omega_1),X_3)
= (\iota_{X_3}(\delta\omega_1),X_2)
= (L_{X_3}\omega_1,X_2)
-
L_{X_2}(\iota_{X_3}(\omega_1))\,.
$$
Similarly, the last term in the right hand side of \eqref{2006_28oct_5} can be rewritten as
$$
(\omega_3,[X_1,X_2])=-\iota_{[X_2,X_1]}\omega_3
=
\iota_{X_1}(L_{X_2}(\omega_3))-L_{X_2}(\iota_{X_1}(\omega_3))
=
(L_{X_2}\omega_3,X_1) 
- 
L_{X_2}(\iota_{X_1}(\omega_3))\,.
$$
We can thus put the above equations together to get
\begin{equation*}
\begin{array}{c}
( A\circ B,C) 
= (L_{X_1}\omega_2,X_3) + (L_{X_2}\omega_3,X_1)  + (L_{X_3}\omega_1,X_2) \\
- L_{X_2}\big((\omega_1,X_3)+(\omega_3,X_1)\big)\,.
\end{array}
\end{equation*}
On  the other hand, by the assumption $A,B,C\in\mc{L}=\mc{L}^\bot$, 
we have $( A,C) =0$,
which precisely means that $(\omega_1,X_3)+(\omega_3,X_1)=0$.
This proves the equivalence of \eqref{2006_28oct_3} and \eqref{2006_28oct_4},
assuming \eqref{2006_28oct_1}.
\end{proof}
\begin{remark}\label{rem:vicapr09_2}
Define the following important subalgebras of a Dirac structure $\mc L$:
\begin{eqnarray*}
& \text{Ker}_1\mc L =\big\{\omega\oplus 0\in\mc L
\,\big|\, \omega\in\Omega^1\big\}\,,\quad
 \text{Ker}_2\mc L =\big\{0\oplus X\in\mc L
\,\big|\, X\in\mf g^\partial\big\}\,,\\
& \delta\mc L = \big\{\delta (\tint f)\oplus X\in\mc L
\,\big|\, \tint f\in\Omega^0,\,X\in\mf g^\partial\big\}\,.
\end{eqnarray*}
Note that the Courant-Dorfman product on $\text{Ker}_1\mc L$ is zero,
and on $\delta\mc L\supset\text{Ker}_2\mc L$ is as follows:
\begin{equation}\label{eq:lasteq}
\big(\delta(\tint f)\oplus X\big)\circ\big(\delta(\tint g)\oplus Y\big)
=
\delta\tint X(g)\oplus [X,Y]\,.
\end{equation}
Consequently, 
$\big(\delta(\tint f)\oplus X\big)\circ\big(\delta(\tint g)\oplus Y\big)\in\text{Ker}_2\mc L$
if $\tint X(g)=0$.
\end{remark}

\vspace{3ex}
\subsection{Hamiltonian functionals and Hamiltonian vector fields.}~~
\label{sec:equations-dirac}
Let $\mc{L}\subset\Omega^1\oplus\mf{g}^\partial$ be a Dirac structure.
We say that $\tint h\in\Omega^0$ is a {\it Hamiltonian functional},
associated to a {\it Hamiltonian vector field} $X$, if
$$
\delta (\tint h)\oplus X\in\mc{L}\,.
$$
We denote by $\mc{F}(\mc{L})\subset\Omega^0$ the subspace of all Hamiltonian functionals,
and by $\mc{H}(\mc{L})\subset\mf{g}^\partial$ the subspace of all Hamiltonian vector fields, i.e.\
\begin{eqnarray*}
\mc{F}(\mc{L}) &=& \Big\{\tint h \in\Omega^0\,\Big|\,\delta (\tint h)\oplus X\in\mc{L} 
\text{ for some } X\in\mf{g}^\partial\Big\}\,,\\
\mc{H}(\mc{L}) &=& \Big\{X\in\mf{g}^\partial\,\Big|\,\delta (\tint h)\oplus X\in\mc{L} 
\text{ for some } \tint h\in \mc V/\partial\mc V\Big\}\,.
\end{eqnarray*}
\begin{lemma}\label{La:4.8}
$\mc{H}(\mc{L})\subset\mf{g}^\partial$ is a Lie subalgebra,
and $\mc{F}(\mc{L})\subset\Omega^0$ is an $\mc{H}(\mc{L})$-submodule 
of the $\mf g^\partial$-module $\Omega^0$ with action given by Lie derivatives.
\end{lemma}
\begin{proof}
It is immediate from \eqref{eq:lasteq} and the fact that, by definition,
a Dirac structure $\mc{L}$ is closed under the Courant-Dorfman product.
\end{proof}
\begin{lemma}\label{2006_lpb}
Let $X,Y\in\mc{H}(\mc{L})$ be two Hamiltonian vector fields associated to the same Hamiltonian
functional $\tint f\in\mc{F}(\mc{L})$, i.e.\ $\delta (\tint f)\oplus X,\,\delta (\tint f)\oplus Y\in\mc{L}$.
Then
$$
\tint X(g)=\tint Y(g)\,,
$$
for every Hamiltonian functional $\int g\in\mc{F}(\mc{L})$.
\end{lemma}
\begin{proof}
By assumption $\delta (\tint f)\oplus X,\,\delta (\tint f)\oplus Y\in\mc{L}$,
and $\delta(\tint g)\oplus Z\in\mc{L}$ for some $Z\in\mf g^\partial$.
In particular, they are pairwise orthogonal with respect to the bilinear form $( \,,\,) $:
$$
( \delta (\tint f)\oplus X,\delta(\tint g)\oplus Z) 
=( \delta(\tint f)\oplus Y,\delta(\tint g)\oplus Z) =0\,,
$$
which means that
$\tint X(g) = -\tint Z(f) = \tint Y(g)$.
\end{proof}
Lemma \ref{2006_lpb} guarantees that the action of the Lie algebra $\mc{H}(\mc{L})$ 
on $\mc{F}(\mc{L})$, given by Lemma \ref{La:4.8}, induces a well-defined bracket
$$
\{\,,\,\}_\mc{L}\,:\,\,\mc{F}(\mc{L})\times\mc{F}(\mc{L})\,\to\,\mc{F}(\mc{L})\,,
$$
called the {\it Lie bracket} of Hamiltonian functionals, given by
\begin{equation}\label{2006_dic15_2}
\Big\{\tint f,\tint g\Big\}_\mc{L}=\tint X(g)\,,\,\,\text{ if } \delta(\tint f)\oplus X\in\mc{L}\,.
\end{equation}
\begin{proposition}\label{prop:oldlemma}
The bracket $\{\,,\,\}_\mc{L}$ defines a structure of a Lie algebra on $\mc{F}(\mc{L})$.
\end{proposition}
\begin{proof}
Skew-commutativity 
of $\{\,,\,\}_\mc{L}$ follows by the fact that $\mc{L}\subset\Omega^1\oplus\mf{g}^\partial$
is an isotropic subspace.
Indeed, if $\delta(\tint f)\oplus X,\,\delta(\tint g)\oplus Y\in\mc{L}$, we have
$$
\begin{array}{l}
\Big\{\tint f,\tint g\Big\}_\mc{L}+\Big\{\tint g,\tint f\Big\}_\mc{L} \\
= \tint X(g)+\tint Y(f) = ( \delta(\tint f)\oplus X,\delta(\tint g)\oplus Y)  = 0\,.
\end{array}
$$
Suppose now $\delta(\tint f_i)\oplus X_i\in\mc{L},\,i=1,2,3$.
We have
$$
\Big\{\tint f_1,\Big\{\tint f_2,\tint f_3\Big\}_\mc{L}\Big\}_\mc{L}
= \tint X_1(X_2(f_3)) = (L_{X_2}(\delta (\tint f_3)),X_1)\,.
$$
Hence Jacobi identity for $\{\,,\,\}_\mc{L}$ immediately follows from the integrability
condition \eqref{2006_28oct_4}.
\end{proof}

Let $\mc{L}\subset\Omega^1\oplus\mf{g}^\partial$ be a Dirac structure,
and let $\tint h\in\Omega^0$ be a Hamiltonian functional associated 
to the Hamiltonian vector field $X\in\mf{g}^\partial$, i.e.\ $\delta(\tint h)\oplus X\in\mc{L}$.
The corresponding {\it Hamiltonian evolution equation} is, 
by definition,
\begin{equation}\label{2006_evol}
\frac{du}{dt} = X(u)\,.
\end{equation}
An \emph{integral of motion} for the evolution equation \eqref{2006_evol}
is, by definition, a Hamiltonian functional
$\int f\in\mc F(\mc L)$ such that 
$\Big\{\tint h,\tint f\Big\}_\mc{L}\,\Big(=\tint X(f)\Big) = 0$.
\begin{remark}
One can think about the RHS of \eqref{2006_evol} as 
$\Big\{\tint h,f\Big\}_\mc{L}$,
but for a general Dirac structure such a notation can be misleading 
since, for $f\in \mc V$, the ``bracket"
$\{\tint h,f\}_\mc{L}$ depends on the particular choice of the Hamiltonian vector field $X$ 
associated to $\tint h$.
In other words, the analogue of Lemma \ref{2006_lpb} fails (it only holds under the sign of integration).
On the other hand, as we will see in Section \ref{sec:dirac-hamilt},
if $\mc L$ is the graph of a Hamiltonian map $H:\,\Omega^1\to\mf{g}^\partial$,
then to every Hamiltonian functional $\tint h\in\mc F(\mc L)$
there is a 
unique Hamiltonian vector field $X=H\big(\delta (\tint h)\big)$ associated to it, 
hence the notation $\{\tint h,f\}_\mc{L}$ makes sense.
\end{remark}

\begin{definition}\label{def:dirac-integr}
Given an element $\delta(\tint h)\oplus X\in\mc L$, the evolution equation \eqref{2006_evol}
is called \emph{integrable} (with respect to the Dirac structure $\mc L$)
if there exists an infinite sequence $\delta(\tint h_n)\oplus X_n\in\mc L$,
including the element $\delta(\tint h)\oplus X$,
such that all $\tint h_n$ are integrals of motion for all evolution equations
\begin{equation}\label{eq:vicapr09}
\frac{du}{dt_n}=X_n(u)\,,
\end{equation}
all evolutionary vector fields commute, i.e.\ $[X_m,X_n]=0$ for all $m,n\in\mb Z_+$,
and the integrals of motion $\tint h_n$, respectively the vector fields $X_n$,
span an infinite-dimensional subspace in $\Omega^0$, respectively in $\mf g^\partial$.
In this case the hierarchy \eqref{eq:vicapr09} of evolution equations
is also called integrable.
It follows from the definition of a Dirac structure that the above sequence 
$\delta\big(\tint h_n\big)\oplus X_n,\,n\geq0$, spans an isotropic subspace
of $\mc L$ with zero Courant-Dorfman product.
\end{definition}

\vspace{3ex}
\subsection{Pairs of compatible Dirac structures. }~~
\label{sec:bi-dirac}
The notion of compatibility of Dirac structures was introduced by Gelfand and Dorfman 
\cite{GD1},\cite{D}.

Given two Dirac structures $\mc L$ and $\mc L^\prime$,
we define the relation 
$\mc N_{\mc L,\mc L^\prime}\subset\mf g^\partial\oplus\mf g^\partial$ by
\begin{equation}\label{eq:20090322_1}
\mc N_{\mc L,\mc L^\prime}
\,=\,
\big\{X\oplus X^\prime\in\mf g^\partial\oplus\mf g^\partial\,\big|\,
\eta\oplus X\in\mc L,\,\eta\oplus X^\prime\in\mc L^\prime\,\text{ for some } \eta\in\Omega^1\big\}\,,
\end{equation}
and the adjoint relation 
$\mc N^*_{\mc L,\mc L^\prime}\subset\Omega^1\oplus\Omega^1$ by
\begin{equation}\label{eq:20090322_2}
\mc N^*_{\mc L,\mc L^\prime}
\,=\,
\big\{\omega\oplus\omega^\prime\in\Omega^1\oplus\Omega^1
\,\big|\,
(\omega,X)=(\omega^\prime,X^\prime)\,\text{ for all } X\oplus X^\prime
\in\mc N_{\mc L,\mc L^\prime}\big\}\,.
\end{equation}
\begin{definition}\label{2006_NRel}
Two Dirac structures $\mc{L},\,\mc{L}^\prime\,\subset \Omega^1\oplus\mf g^\partial$ 
are said to be \emph{compatible} if 
for all $X,X^\prime,Y,Y^\prime\in\mf g^\partial,\,
\omega,\omega^\prime,\omega^{\prime\prime}\in\Omega^1$
such that
$X\oplus X^\prime,\,Y\oplus Y^\prime\,\in\mc N_{\mc L,\mc L^\prime},\,
\omega\oplus\omega^\prime,\,\omega^\prime\oplus\omega^{\prime\prime}
\in\mc N^*_{\mc L,\mc L^\prime}$, 
we have
\begin{equation}\label{eq:20090320_1}
(\omega,[X,Y])-(\omega^\prime,[X,Y^\prime])-(\omega^\prime,[X^\prime,Y])
+(\omega^{\prime\prime},[X^\prime,Y^\prime])\,=\,0\,.
\end{equation}
\end{definition}
Note, of course, that 
$X\oplus X^\prime\in \mc N_{\mc L,\mc L^\prime}$ if and only if 
$X^\prime\oplus X\in \mc N_{\mc L^\prime,\mc L}$,
hence $(\mc{L},\mc{L}^\prime)$ is a pair of compatible Dirac structures
if and only if $(\mc L^\prime,\mc L)$ are compatible.

\vspace{3ex}
\subsection{Lenard scheme of integrability for a pair of Dirac structures. }~~
\label{2006_LenardScheme}
Recall that we think of $\omega\in\Omega^k$ as a $k$-linear skew-symmetric
function $\omega(X_1,\cdots,X_k)$, defined by \eqref{eq:09jan14_3}.
The following theorem is due to Dorfman \cite{D}.
\begin{theorem}\label{2006_LS}
Let $(\mc{L},\mc{L}^{\prime})$ be a pair of compatible Dirac structures.  Suppose that 
$X_0$ is a Hamiltonian vector field with respect to both 
$\mc{L}$ and $\mc{L}^{\prime}$, that is, there exists functionals  
$\int h_0, \int h_1  \in \Omega^0$ such that $\delta(\tint h_0) \oplus X_0 \in \mc{L}$
and $\delta(\tint h_1) \oplus X_0 \in\mc{L}^\prime$. 
Assume moreover that:
\begin{enumerate}
\item[(i)] if $\omega\in\Omega^1$ is such that 
$\delta\omega(X^\prime,Y^\prime)=0$ 
for all $X^\prime,Y^\prime\in\pi_2(\mc N_{\mc L,\mc L^\prime})$
(where $\pi_2:\,\mf g^\partial\oplus\mf g^\partial\to\mf g^\partial$
denotes projection on the second factor),
then $\delta\omega=0$;
\item[(ii)] there exists a sequence of 1-forms 
$\omega_n\in\Omega^1,\,n=0,1,\dots,N+1$ ($N\geq0$, and it can be infinite),
and of evolutionary vector fields $X_n\in\mf{g}^\partial,\,n=0,1,\dots,N$
(starting with the given $X_0$),
such that $\omega_0=\delta(\tint h_0),\,\omega_1=\delta(\tint h_1)$ and
$$
\omega_n\oplus X_n\in\mc{L}\,,\,\,\omega_{n+1}\oplus X_n\in\mc{L}^\prime\,,\,\,0\leq n\leq N\,.
$$
\end{enumerate}
Then:
\begin{enumerate}
\alphaparenlist
\item $\omega_n\oplus\omega_{n+1}\in\mc{N}^*_{\mc{L},\mc{L}^\prime}$ 
for all $n=0,\dots,N+1$;
\item $\omega_n$ is closed for all $n=0,\dots,N+1$, i.e.\ $\delta\omega_n=0$.
\end{enumerate}
\end{theorem}
\begin{proof}
By definition of $\mc{N}_{\mc{L},\mc{L}^\prime}$ and $\mc{N}_{\mc{L},\mc{L}^\prime}^*$,
(a) is equivalent to saying that, for $0\leq n\leq N$,
$$
(\omega_n,X) = (\omega_{n+1},X^\prime)\,,
$$
for all $X,X^\prime\in\mf{g}^\partial$ 
such that $\eta\oplus X\in\mc{L},\,\eta\oplus X^\prime\in\mc{L}^\prime$,
for some $\eta\in\Omega^1$.
By assumption (ii) we have
$\omega_n\oplus X_n\in\mc{L},\,\omega_{n+1}\oplus X_n\in\mc{L}^\prime$,
hence, since $\mc{L}$ and $\mc{L}^\prime$ are isotropic,
we get
$$
(\omega_n,X) = - (\eta,X_n) = (\omega_{n+1},X^\prime)\,,
$$
as we wanted.
We next prove (b) by induction on $n\geq0$.
By assumption $\omega_n=\delta(\tint f_n)$ for $n=0,1$.
Furthermore $\mc{L},\mc{L}^\prime$ is a pair of compatible Dirac structures,
namely \eqref{eq:20090320_1} holds.
In particular, since by (a) 
$\omega_n\oplus\omega_{n+1},\,
\omega_{n+1}\oplus\omega_{n+2}\in\mc{N}_{\mc{L},\mc{L}^\prime}^*$,
we have, by Definition \ref{2006_NRel}, that
\begin{equation}\label{2006_15gen_1}
\big(\omega_n,[X,Y]\big)
- \big(\omega_{n+1},[X,Y^\prime]\big)
- \big(\omega_{n+1},[X^\prime,Y]\big)
+ \big(\omega_{n+2},[X^\prime,Y^\prime]\big) = 0\,,
\end{equation}
for all $X\oplus X^\prime,\, Y\oplus Y^\prime\in\mc{N}_{\mc{L},\mc{L}^\prime}$.
Using equation \eqref{eq:09jan14_1} and recalling the notation \eqref{eq:09jan14_3},
equation \eqref{2006_15gen_1} can be written as follows:
\begin{equation}\label{2006_15gen_2}
\begin{array}{c}
\vphantom{\Big(}
- (\delta\omega_n)(X,Y)
+ (\delta\omega_{n+1})(X,Y^\prime)
+ (\delta\omega_{n+1})(X^\prime,Y)
- (\delta\omega_{n+2})(X^\prime,Y^\prime) \\
\vphantom{\Big(}
+L_{X}(\omega_n,Y)
- L_{Y}(\omega_n,X)
- L_{X}(\omega_{n+1},Y^\prime)
+ L_{Y^\prime}(\omega_{n+1},X) \\
\vphantom{\Big(}
- L_{X^\prime}(\omega_{n+1},Y)
+ L_{Y}(\omega_{n+1},X^\prime)
+ L_{X^\prime}(\omega_{n+2},Y^\prime)
- L_{Y^\prime}(\omega_{n+2},X^\prime) = 0\,.
\end{array}
\end{equation}
Since $\omega_{n}\oplus\omega_{n+1},\,
\omega_{n+1}\oplus\omega_{n+2}\in\mc{N}_{\mc{L},\mc{L}^\prime}^*$
and $X\oplus X^\prime,\,Y\oplus Y^\prime\in\mc{N}_{\mc{L},\mc{L}^\prime}$, we have
\begin{equation*}
\begin{array}{c}
(\omega_n,X) = (\omega_{n+1},X^\prime) \,,\,\, (\omega_{n+1},X) = (\omega_{n+2},X^\prime)\,, \\
(\omega_n,Y) = (\omega_{n+1},Y^\prime) \,,\,\, (\omega_{n+1},Y) = (\omega_{n+2},Y^\prime)\,,
\end{array}
\end{equation*}
hence the last eight terms in the LHS of \eqref{2006_15gen_2} cancel.
Moreover, by the inductive assumption $\omega_n$ and $\omega_{n+1}$ are closed,
so that equation \eqref{2006_15gen_2} reduces to
$(\delta\omega_{n+2})(X^\prime,Y^\prime)=0$,
which, by assumption (i), gives $\delta\omega_{n+2}=0$.
\end{proof}
\begin{remark}\label{rem:09feb9}
Let $(\mc L,\mc L^\prime)$ be a pair of compatible Dirac structures.
Suppose that $\pi_1(\mc L)=\Omega^1$.
In this case $\pi_2(\mc N_{\mc L,\mc L^\prime})=\pi_2(\mc L^\prime)$.
Hence the non-degeneracy condition (i) of Theorem \ref{2006_LS} reads:
if $\delta\omega(X^\prime,Y^\prime)=0$ for every $X^\prime,Y^\prime\in\pi_2(\mc L^\prime)$,
then $\delta\omega=0$.
\end{remark}

The following definition and proposition are a generalization, in the context of Dirac structures,
of Definition \ref{def:feb28}, Lemma \ref{lem:feb28} and Proposition \ref{prop:feb28_2}.
\begin{definition}
Let $\mc L$ and $\mc L^\prime$ be two subspaces of 
$\Omega^1\oplus\mf g^\partial$.
Two sequences $\{\omega_n\}_{n=0,\cdots,N+1}\subset\Omega^1$ 
and $\{X_n\}_{n=0,\cdots,N}\subset\mf g^\partial$
form an $(\mc L,\mc L^\prime)$-\emph{sequence} if
$$
\omega_n\oplus X_n\in\mc L
\,\,\text{ and }\,\,
\omega_{n+1}\oplus X_n\in\mc L^\prime\,\,\text{ for } n=0,\cdots,N\,.
$$
\end{definition}
\begin{proposition}\label{prop:09jan22}
Let $\mc L$ and $\mc L^\prime$ be isotropic subspaces of $\Omega^1\oplus\mf g^\partial$,
and suppose that $\{\omega_n\}_{n=0,\cdots,N+1}$ $\subset\Omega^1$ 
and $\{X_n\}_{n=0,\cdots,N}\subset\mf g^\partial$
form an $(\mc L,\mc L^\prime)$-\emph{sequence}.
\begin{enumerate}
\alphaparenlist
\item Then we have: $(\omega_m,X_n)=0$ for all $m=0,\cdots,N+1$ and $n=0,\cdots,N$.
\item Suppose in addition that
\begin{equation}\label{eq:09jan26}
\big(\Span_{\mb C}\{\omega_n\}_{0\leq n\leq N+1}\big)^\perp\subset \pi_2(\mc L^\prime)
\,\,,\,\,\,\,
\big(\Span_{\mb C}\{X_n\}_{0\leq n\leq N}\big)^\perp\subset \pi_1(\mc L)\,,
\end{equation}
where $\pi_1$ and $\pi_2$ denote projections on $\Omega^1$ and $\mf g^\partial$ respectively,
and the orthogonal complement is with respect to the pairing of $\Omega^1$ and $\mf g^\partial$.
Then we can extend this sequence to an infinite $(\mc L,\mc L^\prime)$-sequence
$\{\omega_n\}_{n\in\mb Z_+},\,\{X_n\}_{n\in\mb Z_+}$.
\end{enumerate}
\end{proposition}
\begin{proof}
Since $\mc L$ and $\mc L^\prime$ are isotropic, we have
\begin{eqnarray*}
&& 
(\omega_n,X_n)=0\,\,\text{for all\,\,} n=0,\cdots,N
\,\,,\,\,\,\,
(\omega_m,X_n)=-(\omega_n,X_m)\,\,\text{for all\,\,} m,n=0,\cdots,N
\,,\\
&& 
(\omega_{n+1},X_n)=0\,\,\text{for all\,\,} n=0,\cdots,N
\,\,,\,\,\,\,
(\omega_{m+1},X_n)=-(\omega_{n+1},X_m)\,\,\text{for all\,\,} m,n=0,\cdots,N
\,.
\end{eqnarray*}
From the above relations we get
$$
(\omega_m,X_n)=(\omega_{m+1},X_{n-1})\,,
$$
hence part (a) follows by an easy induction argument, as in the proof of Lemma \ref{lem:feb28}.
We then have, by (a),
$$
\omega_{N+1}\in\big(\Span_{\mb C}\{X_n\}_{0\leq n\leq N}\big)^\perp\,,
$$
so that, using the second inclusion in \eqref{eq:09jan26}, 
we find $X_{N+1}\in\mf g^\partial$
such that $\omega_{N+1}\oplus X_{N+1}\in\mc L$.
On the other hand, we also have
$$
X_{N+1}\in\big(\Span_{\mb C}\{\omega_n\}_{0\leq n\leq N+1}\big)^\perp\,,
$$
and using the first inclusion in \eqref{eq:09jan26} 
we find an element $\omega_{N+2}\in\Omega^1$
such that $\omega_{N+2}\oplus X_{N+1}\in\mc L^\prime$,
and (b) follows by induction.
\end{proof}
\begin{corollary}\label{09jan26}
Let $\mc V$ be a normal algebra of differential functions,
and let $(\mc{L},\mc{L}^{\prime})$ be a pair of compatible Dirac structures,
which is non-degenerate in the sense of condition (i) of Theorem \ref{2006_LS}.
Let, for some $N\geq0$,
$\{\omega_n\}_{n=0,\cdots,N+1}\subset\Omega^1$
and $\{X_n\}_{n=0,\cdots,N}\subset\mf g^\partial$
be such that
$\omega_0=\delta(\tint h_0),\,\omega_1=\delta(\tint h_1)\in\delta\Omega^0$,
$\omega_n\oplus X_n\in\mc L,\,\omega_{n+1}\oplus X_n\in\mc L^\prime$ for all $n=0,\cdots,N$,
and conditions \eqref{eq:09jan26} hold.
Then
there exist a sequence of Hamiltonian functionals $\{\tint h_n\}_{n\in\mb Z_+}$
such that $\delta(\tint h_n)=\omega_n$ for $n=0,\cdots,N+1$,
and a sequence of Hamiltonian vector fields $\{X_n\}_{n\in\mb Z_+}$
extending the given one for $n\leq N$,
such that 
$\delta(\tint h_n)\oplus X_n\in\mc L,\,
\delta(\tint h_{n+1})\oplus X_n\in\mc L^\prime$.
Furthermore,
\begin{equation}\label{eq:09jan26_1}
\Big\{\tint h_m, \tint h_n\Big\}_{\mc{L}} = \Big\{\tint h_m, \tint h_n\Big\}_{\mc{L}^{\prime}} = 0
\,\,,\,\,\,\,
\text{for all\,\,} m,n\in\mb Z_+\,.
\end{equation}
Namely the Hamiltonian functionals $\int h_n\,,n\in\mb Z_+$, are integrals of motion 
for all the vector fields $X_m \in\mf{g}^\partial,\, m\in\mb Z_+$,
and these vector fields are Hamiltonian with respect to both $\mc{L}$ and $\mc{L}^{\prime}$.
Finally, all vector fields $X_m,\,m\in\mb Z_+$, commute, provided 
that $\text{Ker}_2\mc L\cap\text{Ker}_2\mc L^\prime=0$.
\end{corollary}
\begin{proof}
Proposition \ref{prop:09jan22} guarantees the existence of 
an infinite $(\mc L,\mc L^\prime)$-sequence $\{\omega_n,X_n\}_{n\in\mb Z_+}$, 
extending the given one.
By Theorem \ref{2006_LS}, all elements $\omega_n\in\Omega^1$ are closed,
hence, by Theorem \ref{th:july23}, they are exact, namely $\omega_n=\delta(\tint f_n)$.
Finally, equation \eqref{eq:09jan26_1} follows by Proposition \ref{prop:09jan22}.
The last claim of the Corollary follows from Remark \ref{rem:vicapr09_2}.
\end{proof}

\vspace{3ex}
\subsection{The non-linear Schr\"odinger (NLS) hierarchy.}~~
\label{sec:09febmai}
Let $\mc V=\mb C[u^{(n)},v^{(n)}\,|\,n\in\mb Z_+]$.
Consider the following pair of subspaces of 
$\Omega^1\oplus\mf g^\partial\simeq\mc V^{\oplus2}\oplus\mc V^2$:
$$
\mc L=\left\{
\left(\begin{array}{c}
\frac{f}{v} \\ \frac{f+\partial g}{u}
\end{array}\right)
\oplus
\left(\begin{array}{c}
\partial\Big(\frac{f}{v}\Big)-4vg \\ \partial\Big(\frac{f+\partial g}{u}\Big)+4ug
\end{array}\right)
\right\}
\quad,\qquad
\mc L^\prime=\left\{
\left(\begin{array}{c}
f \\ g
\end{array}\right)
\oplus
\left(\begin{array}{c}
- g \\ f
\end{array}\right)
\right\}\,,
$$
where $f,g$ are arbitrary elements of $\mc V$ for $\mc L^\prime$,
while for $\mc L$ they are such that $\frac{f}{v},\,\frac{f+\partial g}{u}\in\mc V$.
\begin{proposition}\label{prop:20090320_nls}
\begin{enumerate}
\alphaparenlist
\item $(\mc L,\mc L^\prime)$ is a compatible pair of Dirac structures.
\item Condition (i) of Theorem \ref{2006_LS} holds.
\item Let $h_0=0,\,h_1=\frac12(u^2+v^2)\,\in\mc V$,
let $F^0,F^1\in\mc V^{\oplus2}$ be their variational derivatives,
and let 
$P^0=
\left(\begin{array}{c} -v \\ u\end{array}\right)\in\mc V^2\simeq\mf g^\partial$.
Then 
$$
F^0\oplus P^0\in\mc L\,\,,\,\,\,\,
F^1\oplus P^0\in\mc L^\prime\,,
$$
namely
$\Big\{F^0,F^1\Big\}$ and $\{P^0\}$ form an $(\mc L,\mc L^\prime)$-sequence,
i.e.\ condition (ii) of Theorem \ref{2006_LS} holds (for $N=0$).
\item The orthogonality conditions \eqref{eq:09jan26} hold (for $N=0$),
namely
$$
\big(\mb CF^0\oplus\mb CF^1\big)^\perp\subset\pi_2(\mc L^\prime)\,\,,\,\,\,\,
\big(\mb CP^0\big)^\perp\subset\pi_1(\mc L)\,.
$$
\end{enumerate}
\end{proposition}
\begin{proof}
Clearly, $\mc L^\prime$ is a Dirac structure, since it is the graph of the
operator $H=\left(\begin{array}{cc} 0&-1 \\ 1&0 \end{array}\right):\,\mc V^{\oplus2}\to\mc V^2$,
which is a Hamiltonian operator.
We next prove that $\mc L$ is a Dirac structure.
In particular, we start by showing that $\mc L\subset\mc V^{\oplus2}\oplus\mc V^2$
is a maximal isotropic subspace.
Namely,
given $B\in\mc V^{\oplus2}\oplus\mc V^2$,
we want to prove that
$(A,B)=0$ for all $A\in\mc L$
if and only if $B\in\mc L$.
An arbitrary element of $\mc L$ has the form
$$
A\,=\,
\left(\begin{array}{c}
\frac{f}{v} \\ \frac{f+\partial g}{u}
\end{array}\right)
\oplus
\left(\begin{array}{c}
\partial\Big(\frac{f}{v}\Big)-4vg \\ \partial\Big(\frac{f+\partial g}{u}\Big)+4ug
\end{array}\right)\,,
$$
and we can always write $B\in\mc V^{\oplus2}\oplus\mc V^2$ in the form
$$
B\,=\,
\left(\begin{array}{c} \frac{a}{v} \\ \frac{b}{u} \end{array}\right)
\oplus
\left(\begin{array}{c} c \\ d \end{array}\right)\,.
$$
Clearly, $B\in\mc L$ if and only if the following conditions hold:
\begin{equation}\label{eq:20090322_3}
b-a=\partial r\in\partial\mc V
\,\,,\,\,\,\,
c=\partial\left(\frac{a}{v}\right)-4vr
\,\,,\,\,\,\,
d=\partial\left(\frac{a+\partial r}{u}\right)+4ur
\,.
\end{equation}
On the other hand, the equation $(A,B)=0$ gives
$$
\int \left(
\frac{f}{v}c+\frac{f+\partial g}{u}d
+\left(\partial\left(\frac{f}{v}\right)-4vg\right)\frac{a}{v}
+\left(\partial\left(\frac{f+\partial g}{u}\right)+4ug\right)\frac{b}{u}
\right)=0\,,
$$
which, after integration by parts, becomes
$$
\int 
f\left(
\frac{c}{v}+\frac{d}{u}
-\frac1v\partial\left(\frac{a}{v}\right)
-\frac1u\partial\left(\frac{b}{u}\right)
\right)
+g\left(
4(b-a)
-\partial\left(\frac{d}{u}\right)
+\partial\left(\frac1u\partial\left(\frac{b}{u}\right)\right)
\right)
\,=\,0\,.
$$
Since the above equation holds for every $f,g$, we get the two identities:
\begin{eqnarray}
&&\displaystyle{
\frac{c}{v}+\frac{d}{u}
-\frac1v\partial\left(\frac{a}{v}\right)
-\frac1u\partial\left(\frac{b}{u}\right)
\,=\,0\,,
}\label{eq:20090322_4}\\
&&\displaystyle{
4(b-a)
-\partial\left(\frac{d}{u}\right)
+\partial\left(\frac1u\partial\left(\frac{b}{u}\right)\right)
\,=\,0\,.
}\label{eq:20090322_5}
\end{eqnarray}
The second equation, in particular, implies that $b-a=\partial r\in\partial\mc V$,
which is the first condition in \eqref{eq:20090322_3}.
Equations \eqref{eq:20090322_3} and \eqref{eq:20090322_4} then become
\begin{eqnarray}
&&\displaystyle{
\frac{c}{v}
-\frac1v\partial\left(\frac{a}{v}\right)
+4r
\,=\,0\,,
}\label{eq:20090322_6}\\
&&\displaystyle{
4r-\frac{d}{u}+\frac1u\partial\left(\frac{b}{u}\right)
\,=\,0\,,
}\label{eq:20090322_7}
\end{eqnarray}
which are the same as the second and third conditions in \eqref{eq:20090322_3}.
This proves that $\mc L$ is a maximal isotropic subspace of $\mc V^{\oplus2}\oplus\mc V^2$.

In order to show that $\mc L$ is a Dirac structure, we are left to prove that
it is closed under the Courant-Dorfman product \eqref{2006_dp_copy}.
Let $A$ and $B$ be two elements in $\mc L$; they can be written in the form
\begin{eqnarray}
A &=& F\oplus P
\,\,,\,\,\,\,
F=
\left(\begin{array}{c}
\!\!\!
\frac{a}{v}
\!\!\!
\\
\!\!\!
\frac{a+\partial b}{u}
\!\!\!
\end{array}\right)
\,\,,\,\,\,\,
P=
\left(\begin{array}{c}
\!\!\!
\partial\Big(\frac{a}{v}\Big)-4vb 
\!\!\!
\\ 
\!\!\!
\partial\Big(\frac{a+\partial b}{u}\Big)+4ub
\!\!\!
\end{array}\right)
=\partial F+4\left(\begin{array}{c}\!\!\! -v \!\!\!\\\!\!\! u \!\!\!\end{array}\right)b
\,,\label{eq:20090323_1}\\
B &=& G\oplus Q
\,\,,\,\,\,\,
G=
\left(\begin{array}{c}
\!\!\!
\frac{c}{v} 
\!\!\!
\\ 
\!\!\!
\frac{c+\partial d}{u}
\!\!\!
\end{array}\right)
\,\,,\,\,\,\,
Q=
\left(\begin{array}{c}
\!\!\!
\partial\Big(\frac{c}{v}\Big)-4vd 
\!\!\!
\\ 
\!\!\!
\partial\Big(\frac{c+\partial d}{u}\Big)+4ud
\!\!\!
\end{array}\right)
=\partial G+4\left(\begin{array}{c}\!\!\! -v \!\!\!\\\!\!\! u \!\!\!\end{array}\right)d
\,.
\label{eq:20090323_2}
\end{eqnarray}
Let $A\circ B=H\oplus R$, 
with $H=\left(\begin{array}{c} z \\ w \end{array}\right)\in\mc V^{\oplus 2}$
and $R\in\mc V^2$.
In order to prove that $A\circ B\in\mc L$, we need to show that
\begin{equation}\label{eq:20090323_3}
uw-vz=\partial r\in\partial\mc V
\,\,,\,\,\,\,
R=\partial H+4\left(\begin{array}{c} -v \\ u \end{array}\right)r\,,
\end{equation}
for some $r\in\mc V$.
By formula \eqref{2006_dp_copy} for the Courant-Dorfman product
and using equations \eqref{eq:09jan21_1}, \eqref{eq:09jan4_2_b} and \eqref{eq:09jan4_2_c},
we have:
\begin{eqnarray}
\label{eq:20090323_4}
H &=&
D_G(\partial)P+D_P^*(\partial)G-D_F(\partial)Q+D_F^*(\partial)Q
\,,\\
\label{eq:20090323_5}
R &=&
D_Q(\partial)P-D_P(\partial)Q
\,.
\end{eqnarray}
Using the formula for $P$ in \eqref{eq:20090323_1}
we get, using Lemma \ref{lem:20090323},
\begin{equation}\label{eq:20090323_6}
D_P^*(\partial)G+D_F^*(\partial)Q
=
4\left(\begin{array}{c} 
\frac{bc-ad+b\partial d-d\partial b}{u} \\ 
\frac{ad-bc}{v}
\end{array}\right)\,.
\end{equation}
Hence, the vector $H\in\mc V^{\oplus2}$ in \eqref{eq:20090323_4} becomes
\begin{equation}\label{eq:20090323_10}
H = \left(\begin{array}{c} z \\ w \end{array}\right)
= D_G(\partial)P-D_F(\partial)Q
+4\left(\begin{array}{c} 
\frac{bc-ad+b\partial d-d\partial b}{u} \\ 
\frac{ad-bc}{v}
\end{array}\right)\,.
\end{equation}
With similar computations involving Lemma \ref{lem:20090323}
we also get
\begin{equation}\label{eq:20090323_7}
D_G(\partial)P
=
\left(\begin{array}{c} 
\frac{D_c(\partial)P}{v} - \frac{c}{v^2}\left(\partial\frac{a+\partial b}{u}+4ub\right) \\ 
\frac{D_c(\partial)P+\partial D_d(\partial)P}{u} 
- \frac{c+\partial d}{u^2}\left(\partial\frac{a}{v}-4vb\right)
\end{array}\right)\,,
\end{equation}
and
\begin{equation}\label{eq:20090323_8}
D_F(\partial)Q
=
\left(\begin{array}{c} 
\frac{D_a(\partial)Q}{v} - \frac{a}{v^2}\left(\partial\frac{c+\partial d}{u}+4ud\right) \\ 
\frac{D_a(\partial)Q+\partial D_b(\partial)Q}{u} 
- \frac{a+\partial b}{u^2}\left(\partial\frac{c}{v}-4vd\right)
\end{array}\right)\,.
\end{equation}
Putting together equations 
\eqref{eq:20090323_10}, \eqref{eq:20090323_7} and \eqref{eq:20090323_8},
we get, after a straightforward computation, that
$$
uw-vz
=
\partial\left(D_d(\partial)P-D_b(\partial)Q
+\frac{c\partial b-a\partial d}{uv}\right)\,,
$$
namely the first condition in \eqref{eq:20090323_3} holds with
\begin{equation}\label{eq:20090323_9}
r=D_d(\partial)P-D_b(\partial)Q+\frac{c\partial b-a\partial d}{uv}\,.
\end{equation}
Next, let us compute the vector $R\in\mc V^2$ defined in \eqref{eq:20090323_5}.
Using the last formulas in \eqref{eq:20090323_1} and \eqref{eq:20090323_2},
and Lemma \ref{lem:20090323}, we get, after a straightforward computation,
\begin{equation}\label{eq:20090323_11}
R=
\partial\big(D_G(\partial)P-D_F(\partial)Q\big)
+4\left(\begin{array}{c} -v \\ u \end{array}\right)\big(D_d(\partial)P-D_b(\partial)Q\big)
+4
\left(\begin{array}{c} 
b\partial\frac{c+\partial d}{u}
-d\partial\frac{a+\partial b}{u} \\ 
d\partial\frac{a}{v}
-b\partial\frac{c}{v}
\end{array}\right)
\,.
\end{equation}
Puttin together \eqref{eq:20090323_10}, \eqref{eq:20090323_9} 
and \eqref{eq:20090323_11}, we get
$$
R-\partial H-4\left(\begin{array}{c} -v \\ u \end{array}\right)r
=
4\left(\begin{array}{c} 
b\partial\frac{c+\partial d}{u}
-d\partial\frac{a+\partial b}{u} \\ 
d\partial\frac{a}{v}
-b\partial\frac{c}{v}
\end{array}\right)
-4\partial
\left(\begin{array}{c} 
\frac{bc-ad+b\partial d-d\partial b}{u} \\ 
\frac{ad-bc}{v}
\end{array}\right)
-4\left(\begin{array}{c} 
\frac{a\partial d-c\partial b}{u} \\ 
\frac{c\partial b-a\partial d}{v} 
\end{array}\right)
\,.
$$
It is immediate to check that the RHS above is zero, thus proving
the second condition in \eqref{eq:20090323_3}.

In order to complete the proof of part (a), we are left to show that $\mc L$ and $\mc L^\prime$
are compatible,
namely that they satisfy the integrability condition \eqref{eq:20090320_1}.
By the definition \eqref{eq:20090322_1}, the relation 
$\mc N_{\mc L,\mc L^\prime}\subset\mc V^2\oplus\mc V^2$
consists of elements of type $P\oplus P^\prime$,
where $P$ is as in \eqref{eq:20090323_1} and $P^\prime=JF$,
with $J=\left(\begin{array}{cc} 0&-1 \\ 1&0 \end{array}\right)$.
Moreover, it is not hard to check that the adjoint relation 
$\mc N^*_{\mc L,\mc L^\prime}\subset\mc V^{\oplus 2}\oplus\mc V^{\oplus 2}$
consists of elements of type
$F\oplus F^\prime$, 
where again $F$ is as in \eqref{eq:20090323_1}
and
$F^\prime\,=\,-J\partial F+4\left(\begin{array}{c} u \\ v \end{array}\right) b$.
Therefore, the integrability condition \eqref{eq:20090320_1}
reads as follows:
\begin{equation}\label{eq:20090323_n1}
(H,[P,Q])-(H^\prime,[P,Q^\prime])-(H^\prime,[P^\prime,Q])
+(H^{\prime\prime},[P^\prime,Q^\prime])\,=\,0\,,
\end{equation}
where
\begin{eqnarray}
\label{eq:20090323_n2}
&& P=\partial F+4J \left(\begin{array}{c} u \\ v \end{array}\right) b
\quad,\qquad
P^\prime=JF
\quad,\qquad
F= \left(\begin{array}{c}
\frac{a}{v} \\ \frac{a+\partial b}{u}
\end{array}\right)\,, \\
\label{eq:20090323_n3}
&& Q=\partial G+4J \left(\begin{array}{c} u \\ v \end{array}\right) d
\quad,\qquad
Q^\prime=JG
\quad,\qquad
G= \left(\begin{array}{c}
\frac{c}{v} \\ \frac{c+\partial d}{u}
\end{array}\right)\,, \\
\label{eq:20090323_n4}
&& 
H= \left(\begin{array}{c}
\frac{f}{v} \\ \frac{f+\partial g}{u}
\end{array}\right)
\,\,,\,\,\,\,
H^\prime=-J\partial H+4 \left(\begin{array}{c} u \\ v \end{array}\right) g
\,\,,\,\,\,\,
H^{\prime\prime}=-J\partial H^\prime+4 \left(\begin{array}{c} u \\ v \end{array}\right) h\,,
\end{eqnarray}
and where $f,g,h\in\mc V$ are related by the following identity
\begin{equation}
\label{eq:20090323_n5}
\frac1{uv}\partial h+\frac1v\partial\frac{f}{v}+\frac1u\partial\frac{f+\partial g}{u}=0\,.
\end{equation}
In \eqref{eq:20090323_n1} $[\,\cdot,\cdot\,]$ denotes the bracket
of $\mc V^2\simeq\mf g^\partial$, namely (cf.\ \eqref{eq:09jan21_1}),
$[P,Q]=D_Q(\partial)P-D_P(\partial)Q$,
and $(\,\cdot,\cdot\,)$ denotes the usual pairing 
$\mc V^{\oplus2}\times\mc V^2\to\mc V/\partial\mc V$.
Notice that, by construction, we have
\begin{equation}
\label{eq:20090323_n6}
F\oplus P\,,\,\,G\oplus Q\,,\,\,H\oplus JH^\prime\,,\,\,H^\prime\oplus JH^{\prime\prime}\,\in\,\mc L\,.
\end{equation}
Since $\mc L$ is closed under the Courant Dorfman product, we have that
$$
(F\oplus P)\circ(G\oplus Q)
=
\big(D_G(\partial)P-D_F(\partial)Q+D_P^*(\partial)G+D_F^*(\partial)Q\big)\oplus[P,Q]
\,\in\,\mc L\,.
$$
Hence, since $H\oplus JH^\prime\in\mc L$ and 
$\mc L$ is a maximal isotropic subspace of $\mc V^{\oplus2}\oplus\mc V^2$,
it follows that
\begin{eqnarray}\label{eq:20090324_2}
(H,[P,Q])
&=&
-(D_G(\partial)P-D_F(\partial)Q+D_P^*(\partial)G+D_F^*(\partial)Q,JH^\prime)
\nonumber \\
&=&
(H^\prime,JD_G(\partial)P-JD_F(\partial)Q+JD_P^*(\partial)G+JD_F^*(\partial)Q)\,.
\end{eqnarray}
In the last identity we used the fact that $J:\,\mc V^{\oplus 2}\to\mc V^2$ 
is a skew-adjoint operator.
Recalling the identities \eqref{eq:20090323_n2} and \eqref{eq:20090323_n3},
and using Lemma \ref{lem:20090323}, we also have
\begin{eqnarray}
\label{eq:20090324_3}
(H^\prime,[P^\prime,Q])
&=&
(H^\prime,D_Q(\partial)JF-JD_F(\partial)Q)\,,\\
\label{eq:20090324_4}
(H^\prime,[P,Q^\prime])
&=&
(H^\prime,JD_G(\partial)P-D_P(\partial)JG)\,.
\end{eqnarray}
Furthermore, using the fact that $J$ is skew-adjoint, we can easily get
\begin{equation}
\label{eq:20090324_5}
(H^{\prime\prime},[P^\prime,Q^\prime])
=
(D_F(\partial)JG-D_G(\partial)JF,JH^{\prime\prime})\,.
\end{equation}
We can then put together equations \eqref{eq:20090324_2}, \eqref{eq:20090324_3}, 
\eqref{eq:20090324_4} and \eqref{eq:20090324_5},
to rewrite the integrability condition \eqref{eq:20090323_n1}
as follows
\begin{equation}\label{eq:20090324_6}
(H^\prime,
D_P(\partial)JG-D_Q(\partial)JF
+JD_P^*(\partial)G+JD_F^*(\partial)Q)
+(D_F(\partial)JG-D_G(\partial)JF,JH^{\prime\prime})
\,=\,0\,.
\end{equation}
Since $H^\prime\oplus JH^{\prime\prime}\in\mc L$ and $\mc L$ is maximal isotropic,
in order to prove equation \eqref{eq:20090324_6}
it suffices to show that
\begin{equation}\label{eq:20090324_7}
\big(D_F(\partial)JG-D_G(\partial)JF\big)
\oplus
\big(D_P(\partial)JG-D_Q(\partial)JF
+JD_P^*(\partial)G+JD_F^*(\partial)Q\big)\,\in\,\mc L\,,
\end{equation}
whenever $F\oplus P,\,G\oplus Q\in\mc L$.
We start by proving that $D_F(\partial)JG-D_G(\partial)JF\in\pi_1(\mc L)$.
Indeed, it follows by a straightforward computation
involving Lemma \ref{lem:20090323}, that
$$
D_F(\partial)JG-D_G(\partial)JF
=
\left(\begin{array}{c} 
\frac{D_a(\partial)JG-D_c(\partial)JF}{v} \\ 
\frac{D_a(\partial)JG-D_c(\partial)JF+\partial\big(D_b(\partial)JG-D_d(\partial)JF\big)}{u}
\end{array}\right)\,,
$$
namely the RHS has the form 
$\left(\begin{array}{c} \frac{z}{v} \\ \frac{z+\partial w}{u} \end{array}\right)\in\pi_1(\mc L)$,
with
\begin{equation}\label{eq:20090324_8}
z=D_a(\partial)JG-D_c(\partial)JF
\quad,\qquad
w=D_b(\partial)JG-D_d(\partial)JF\,.
\end{equation}
In order to prove \eqref{eq:20090324_7} we are thus left to prove the following identity:
\begin{eqnarray}\label{eq:20090324_9}
& D_P(\partial)JG-D_Q(\partial)JF
+JD_P^*(\partial)G+JD_F^*(\partial)Q
=
\partial
\big(D_F(\partial)JG-D_G(\partial)JF\big) \nonumber\\
& +4J\left(\begin{array}{c} u \\ v \end{array}\right)
\big(D_b(\partial)JG-D_d(\partial)JF\big)\,.
\end{eqnarray}
Using equations \eqref{eq:20090323_n2}-\eqref{eq:20090323_n3} 
and Lemma \ref{lem:20090323}, it is not hard to check that
\begin{eqnarray}
\label{eq:20090324_10}
D_P(\partial)JG
&=&
\partial\big(D_F(\partial)JG\big)
+4J\left(\begin{array}{c} u \\ v \end{array}\right)D_b(\partial)JG
-4Gb\,,\\
\label{eq:20090324_11}
D_Q(\partial)JF
&=&
\partial\big(D_G(\partial)JF\big)
+4J\left(\begin{array}{c} u \\ v \end{array}\right)D_d(\partial)JF
-4Fd\,.
\end{eqnarray}
Hence, using \eqref{eq:20090324_10} and \eqref{eq:20090324_11},
equation \eqref{eq:20090324_9} becomes
\begin{equation}\label{eq:20090324_12}
JD_P^*(\partial)G+JD_F^*(\partial)Q
+4Fd-4Gb
\,=\,0\,.
\end{equation}
Equation \eqref{eq:20090324_12} can be now checked directly
using equation \eqref{eq:20090323_6}.
This completes the proof of part (a).

Let us next prove part (b).
From the expression of $\mc L$ and $\mc L^\prime$,
and from the definition \eqref{eq:20090322_1} of $\mc N_{\mc L,\mc L^\prime}$,
it follows that
$\pi_2(\mc N_{\mc L,\mc L^\prime})$ consists of elements of the form
\begin{equation}\label{eq:star}
\left(\begin{array}{c} 
-\frac{f+\partial g}{u} \\
\frac{f}{v}
\end{array}\right)
\,\in\mc V^2\,,
\end{equation}
where $f,g$ are such that $\frac{f}{v},\frac{f+\partial g}{u}\,\in\mc V$.
Suppose now that $M(\partial)$ is a $2\times2$ matrix valued differential operator
such that
\begin{equation}\label{eq:star2}
\tint P\cdot M(\partial)Q\,=\,0\,,
\end{equation}
for all $P,Q\in\pi_2(\mc N_{\mc L,\mc L^\prime})$.
Recall that $P\in\pi_2(\mc N_{\mc L,\mc L^\prime})$
is an arbitrary element of the form \eqref{eq:star}.
Therefore, if \eqref{eq:star2} holds for every such $P$,
we deduce, by Proposition \ref{prop:20081222},
that $M(\partial)Q=\text{const.}
\left(\begin{array}{c} u \\ v \end{array}\right)$.
Since this has to be true for every $Q$ of the form \eqref{eq:star},
we conclude, by a simple differential order consideration, that $M(\partial)=0$,
thus proving (b).

For part (c), we just notice that 
$F^0\oplus P^0=
\left(\begin{array}{c} 0 \\ 0\end{array}\right)\oplus \left(\begin{array}{c} -v \\ u\end{array}\right)$
is the element of $\mc L$ corresponding $f=0$ and $g=\frac14$,
while
$F^1\oplus P^0=
\left(\begin{array}{c} u \\ v\end{array}\right)\oplus \left(\begin{array}{c} -v \\ u\end{array}\right)$
if the element of $\mc L^\prime$ corresponding to $f=u$ and $g=v$.

We are left to prove part (d). The first orthogonality condition is trivial, since 
$\pi_2(\mc L^\prime)=\mc V^2$.
For the second one we have that, if
$F=\left(\begin{array}{c} z/v \\ w/u\end{array}\right)\in(\mb CP^0)^\perp$, then
$$
\tint F\cdot P^0
=\tint (w-z)=0\,,
$$
namely $w-z\in\partial\mc V$, which exactly means that $F\in\pi_1(\mc L)$.
\end{proof}

Corollary \ref{09jan26} allows us to extend the $(\mc L,\mc L^\prime)$-sequence
$\{F^0,F^1\}\,,\,\,\{P^0\}$
to an infinite $(\mc L,\mc L^\prime)$-sequence 
$\{F^n\}_{n\in\mb Z_+},\,\{P^n\}_{n\in\mb Z_+}$.
Moreover, by Proposition \ref{prop:20090320_nls}, all elements 
$F^n\in\mc V^{\oplus\ell}\simeq\Omega^1,\,n\in\mb Z_+$,
are closed,
hence, since $\mc V$ is normal and by Theorem \ref{th:july23},
they are exact:
$F^n=
\left(\begin{array}{c} \frac{\delta h_n}{\delta u} \\ \frac{\delta h_n}{\delta v}\end{array}\right)$
for some $h_n\in\mc V$, for every $n\in\mb Z_+$.
This gives us an infinite hierarchy of Hamiltonian equations ($n\in\mb Z_+$):
\begin{equation}\label{eq:20090320_2}
\left(\begin{array}{c} 
\frac{du}{dt} \\ 
\frac{dv}{dt} 
\end{array}\right)
= P^n\,,
\end{equation}
for which all $\tint h_m,\,m\in\mb Z_+$, are integrals of motion.
It is easy to compute the first few integrals of motion and
evolutionary vector fields of the hierarchy:
\begin{eqnarray*}
&&\displaystyle{
h_0=0
\,\,,\,\,\,\,
F^0=\left(\begin{array}{c} 0 \\ 0 \end{array}\right)
\,\,,\,\,\,\,
P^0=\left(\begin{array}{c} -v \\ u \end{array}\right)
\,,
} \\
&&\displaystyle{
h_1=\frac12(u^2+v^2)
\,\,,\,\,\,\,
F^1=\left(\begin{array}{c} u \\ v \end{array}\right)
\,\,,\,\,\,\,
P^1=\left(\begin{array}{c} u^\prime \\ v^\prime\end{array}\right)
\,,
} \\
&&\displaystyle{
h_2=\frac12(uv^\prime-u^\prime v)
\,\,,\,\,\,\,
F^2=\left(\begin{array}{c} v^\prime \\ -u^\prime \end{array}\right)
\,\,,\,\,\,\,
P^2
=\left(\begin{array}{c} v^{\prime\prime}+2v(u^2+v^2) \\ 
-u^{\prime\prime}-2u(u^2+v^2) \end{array}\right)
\,,
} \\
&&\displaystyle{
h_3=\frac12\big({u^\prime}^2+{v^\prime}^2-(u^2+v^2)^2\big)
\,\,,\,\,\,\,
F^3
=\left(\begin{array}{c} -u^{\prime\prime}-2u(u^2+v^2) \\ 
-v^{\prime\prime}-2v(u^2+v^2) \end{array}\right)
\,,
} \\
&&\displaystyle{
\qquad\qquad\qquad\qquad\qquad\qquad\qquad\quad
P^3
=\left(\begin{array}{c} 
-u^{\prime\prime\prime}-2\partial\big(u(u^2+v^2)\big)-4v(vu^\prime-uv^\prime) \\ 
-v^{\prime\prime\prime}-2\partial\big(v(u^2+v^2)\big)+4u(vu^\prime-uv^\prime)
\end{array}\right)
\,,
} \\
&&\displaystyle{
h_4\!=\!
\frac12(u^\prime v^{\prime\prime}-v^\prime u^{\prime\prime})+2(v^3u^\prime-u^3v^\prime)
\,,\,\,
F^4
\!=\!
\left(\begin{array}{c}
\!\!\!\!
-v^{\prime\prime\prime}-2\partial\big(v(u^2+v^2)\big)+4u(vu^\prime-uv^\prime) 
\!\!\!\!\\
\!\!\!\!
u^{\prime\prime\prime}+2\partial\big(u(u^2+v^2)\big)+4v(vu^\prime-uv^\prime) 
\!\!\!\!
\end{array}\right)
\dots
}
\end{eqnarray*}
Therefore the first three equations of the hierarchy \eqref{eq:20090320_2} are:
$$
\frac{d}{dt_0}
\left(\begin{array}{c} u \\ 
v \end{array}\right)
=
\left(\begin{array}{c} -v \\ 
u \end{array}\right)
\,\,,\,\,\,\,
\frac{d}{dt_1}
\left(\begin{array}{c} u \\ 
v \end{array}\right)
=
\left(\begin{array}{c} u^\prime \\ 
v^\prime \end{array}\right)
\,\,,\,\,\,\,
\frac{d}{dt_2}
\left(\begin{array}{c} u \\ 
v \end{array}\right)
=
\left(\begin{array}{c} v^{\prime\prime}+2v(u^2+v^2) \\ 
-u^{\prime\prime}-2u(u^2+v^2) \end{array}\right)\,.
$$
The third equation is called the NLS system.
The first four integrals of motion are $h_n,\,n=1,2,3,4$, written above.
It is not hard to see by induction on $n$ that the 1-forms $F^n,\,n\geq1$,
have differential order $n-1$,
hence they, and consequently the $h_n$,
are linearly independent.
Finally, $\text{Ker}_2\mc L^\prime=0$, hence all evolutionary vector fields $X_{P^n}$ commute,
and they are linearly independent for $n\geq1$ since the $F^n$'s are.
Thus the NLS system (and the whole NLS hierarchy) is integrable.

In fact, using the conditions $F^n\oplus P^n\in\mc L$ 
and $F^{n+1}\oplus P^n\in\mc L^\prime$,
it is not hard to find a recursive formula for $F^n$ and $P^n,\,n\in\mb Z_+$.
We have
$$
F^n
=
\left(\begin{array}{c} 
\frac{f_n}{v} \\
\frac{f_n+\partial g_n}{u}
\end{array}\right)
\quad,\qquad
P^n
=
\left(\begin{array}{c} 
-\frac{f_{n+1}+\partial g_{n+1}}{u} \\
\frac{f_{n+1}}{v} 
\end{array}\right)\,,\,\,n\geq0\,,
$$
and the elements $f_n,\,g_n\in\mc V$ are given by the recursive equations:
$f_0=0,\,g_0=\frac14$, and
\begin{eqnarray*}
f_{n+1}
&=&
v\partial\left(\frac{f_n+\partial g_n}{u}\right)+4uvg_n\,,\\
\partial g_{n+1}
&=&
-u\partial\left(\frac{f_n}{v}\right)-v\partial\left(\frac{f_n+\partial g_n}{u}\right)
\,.
\end{eqnarray*}


\vspace{3ex}
\subsection{The Hamiltonian case.}~~
\label{sec:dirac-hamilt}
In Section \ref{sec:1.1} we gave a definition of a Hamiltonian operator
in relation to Poisson vertex algebras (see Definition \ref{def:09jan19}).
As we shall see in Theorem \ref{2006_th-may30}, the following definition 
is equivalent to it in the case of differential operators.
\begin{definition}\label{2006_def-hs}
A {\it Hamiltonian operator} is a linear map $H:\,\Omega^1\to\mf{g}^\partial$ such that its graph
\begin{equation*}
  \mc{G}(H) = \{ \omega\oplus H(\omega) \,|\, \omega \in \Omega^1\} \subset \Omega^1\oplus \mf{g}^\partial\,,
\end{equation*}
is a Dirac structure.
\end{definition}
In the case of the Dirac structure given by the graph $\mc G(H)$ of the Hamiltonian operator $H$,
the corresponding space of Hamiltonian functionals $\mc F(\mc G(H))$
is the whole space $\Omega^0$,
while the space of Hamiltonian vector fields $\mc H(\mc G(H))$
is the Lie subalgebra $H(\delta\Omega^0)\subset\mf g^\partial$.
Hence, by the discussion in Section \ref{sec:equations-dirac},
the space $\Omega^0=\mc V/\partial\mc V$ acquires a structure of a Lie algebra
with the bracket
\begin{equation}\label{eq:09jan20_1}
\big\{\tint f,\tint g\big\}_H
\,=\,
\tint H\big(\delta\big(\tint f\big)\big) (g)
\,=\,
\Big(H\big(\delta\big(\tint f\big)\big),\delta\big(\tint g\big)\Big)\,\in\Omega^0\,,
\end{equation}
and the evolution equation associated to a Hamiltonian functional $\tint h\in\Omega^0$
takes the form
\begin{equation}\label{eq:09jan20_2}
\frac d{dt} f
\,=\,
H\big(\delta\big(\tint h\big)\big) (f)\,.
\end{equation}
Notice that,
in the special case of a Hamiltonian operator 
$H:\,\Omega^1=\mc V^{\oplus\ell}\to\mf g^\partial=\mc V^\ell$
given by a matrix valued differential operator
$H(\partial)=(H_{ij}(\partial))_{i,j\in I}$, 
the above formula \eqref{eq:09jan20_1} for the Lie bracket reduces to equation \eqref{k18},
and the evolution equation \eqref{eq:09jan20_2} reduces to \eqref{k17}.
This coincidence is not surprising, as will be clear from Theorem \ref{2006_th-may30}.
\begin{proposition}\label{2006_def-ham2}
Let $H:\, \Omega^1\to\mf{g}^\partial$ be a linear map,
and let $\mc{G}(H)=\big\{\omega\oplus H(\omega)\,|\,\omega\in\Omega^1\big\}
\subset\Omega^1\oplus\mf{g}^\partial$ be its graph.
\begin{enumerate}
\alphaparenlist
\item The space $\mc{G}(H)\subset\Omega^1\oplus\mf{g}^\partial$
is isotropic with respect to the bilinear form $(\,,\,)$, 
i.e.\ $\mc{G}(H)\subset\mc{G}(H)^\perp$,
if and only if 
it is maximal isotropic,
i.e.\ $\mc{G}(H)^\perp=\mc{G}(H)$,
which in turn is equivalent to saying that
$H$ is skew-adjoint:
\begin{equation}\label{2006_28oct_6}
(\omega,H(\eta))=-(\eta,H(\omega))\,,\,\,\text{for all\,\,}\omega,\eta\in\Omega^1\,.
\end{equation}
\item A skew-adjoint operator $H:\,\Omega^1\to\mf g^\partial$
is a Hamiltonian operator if and only if one of 
the following equivalent conditions holds:
\begin{enumerate}
\item[(i)]
for every $\omega_1,\omega_2,\omega_3\in\Omega^1$, we have
\begin{equation}\label{2006_28oct_4_ham}
(L_{H(\omega_1)}\omega_2,H(\omega_3))+
(L_{H(\omega_2)}\omega_3,H(\omega_1))+
(L_{H(\omega_3)}\omega_1,H(\omega_2))
\,=\,0\,,
\end{equation}
\item[(ii)] 
for every $\omega,\eta\in\Omega^1$ we have
\begin{equation}\label{2006_may30-1}
H\Big(L_{H(\omega)}(\eta) - \iota_{H(\eta)}(\delta \omega)\Big)
\,=\, \big[H(\omega),H(\eta)\big]\,.
\end{equation}
\end{enumerate}
\end{enumerate}
\end{proposition}
\begin{proof}
Equation \eqref{2006_28oct_6}
is equivalent to 
$(\omega\oplus H(\omega),\eta\oplus H(\eta))=0$
for every $\omega,\eta\in\Omega^1$,
which in turn is equivalent to $\mc{G}(H)\subset\mc{G}(H)^\perp$.
On the other hand, if $H$ is skew-adjoint,
we have
$(\omega\oplus X,\eta\oplus H(\eta))=(\eta,X-H(\omega))$,
and this is zero for every $\eta\in\Omega^1$ if and only if $X=H(\omega)$.
Hence $\mc{G}(H)^\perp\subset\mc{G}(H)$, thus proving (a).
For part (b), 
we notice that equation \eqref{2006_28oct_4_ham} is the same as the integrability condition \eqref{2006_28oct_4}
for the Dirac structure $\mc G(H)$.
Hence, by Remark \ref{rem:a}, $H$ is a Hamiltonian operator if and only if (i) holds.
Moreover, by definition of the Courant-Dorfman product, we have
$$
\big(\omega\oplus H(\omega)\big)\circ \big(\eta\oplus H(\eta)\big)
\,=\, \Big( L_{H(\omega)}(\eta)-\iota_{H(\eta)}(\delta\omega)\Big)\oplus\big[H(\omega),H(\eta)\big]\,,
$$
and this element lies in $\mc{G}(H)$ if and only if \eqref{2006_may30-1} holds.
\end{proof}
\begin{theorem}\label{2006_th-may30}
Let $H:\,\Omega^1=\mc V^{\oplus\ell}\to\mf{g}^\partial=\mc V^\ell$ 
be a matrix valued differential operator, 
i.e.\ $H=\big(H_{ij}(\partial)\big)_{i,j\in I}$.
Then Definitions \ref{def:09jan19} and \ref{2006_def-hs} of a Hamiltonian operator are equivalent.
\end{theorem}
\begin{proof}
Due to Propositions \ref{prop:09jan16} and \ref{2006_def-ham2},
the only thing to check is that equation \eqref{2006_may30-1} reduces,
if $H$ is a differential operator, to equation \eqref{eq:09jan16}.
Let $\omega=\omega_F,\,\eta=\omega_G$ for $F,G\in \mc V^{\oplus\ell}$, 
be as in \eqref{eq:09jan15_1}.
Equation \eqref{2006_may30-1} then reads
\begin{equation}\label{2006_may31-1}
H\Big(L_{H(\omega_F)}(\omega_G) - \iota_{H(\omega_G)}(\delta \omega_F)\Big)
\,=\, \big[H(\omega_F),H(\omega_G)\big]\,.
\end{equation}
Using equation \eqref{eq:09jan4_2_c} for $L_{X_P}(\omega_F)$,
the element in $\mc V^{\oplus\ell}$ corresponding to $L_{H(\omega_F)}(\omega_G)\in\Omega^1$ is
$D_G(\partial)H(\partial)F+D^*_{H(\partial)F}(\partial)G$.
Similarly, using equation \eqref{eq:09jan4_2_b},
the element in $\mc V^{\oplus\ell}$ corresponding 
to $\iota_{H(\omega_G)}(\delta\omega_F)\in\Omega^1$ is
$D_F(\partial)H(\partial)G-D^*_F(\partial)H(\partial)G$.
Furthermore, recalling the correspondence \eqref{2006_iso2}
and formula \eqref{eq:09jan21_1},
the element in $\mc V^\ell$ corresponding 
to $\big[H(\omega_F),H(\omega_G)\big]\in\mf g^\partial$ is
$D_{H(\partial)G}(\partial)H(\partial)F-D_{H(\partial)F}(\partial)H(\partial)G$.
Putting together the above observations,
we can rewrite equation \eqref{2006_may31-1} as \eqref{eq:09jan16},
and vice versa.
\end{proof}

Next, we discuss how the notion of compatibility of Dirac structures
is translated in the case of Hamiltonian operators.
First
we recall the definition of the Schouten bracket \cite{D}.
\begin{definition}\label{def:09jan20_1}
If $H,\,K:\,\Omega^1\to\mf{g}^\partial$ are skew-adjoint (with respect to $(\,,\,)$) operators,
their {\it Schouten bracket} is a tri-linear map
$[H,K]_{SB}\,:\,\Omega^1\times\Omega^1\times\Omega^1\,\to\,\Omega^0$,
given by (cf.\ equation \eqref{2006_28oct_4_ham})
\begin{equation}\label{2006_dic14_3}
[H,K]_{SB}(\omega_1,\omega_2,\omega_3)
= (\omega_1, H(L_{K(\omega_2)} \omega_3)) + (\omega_1, K(L_{H(\omega_2)} \omega_3)) 
+ (\text{cycl. perm.})\,.
\end{equation}
\end{definition}
We have the following corollary of Proposition \ref{2006_def-ham2}(b):
\begin{corollary}\label{cor:09jan20}
A linear map  $H:\,\Omega^1\to\mf g^\partial$
is a Hamiltonian operator if and only if 
it is skew-adjoint and $[H,H]_{SB}=0$.
\end{corollary}
\begin{proposition}\label{2006_BiHamiltonian}
Let $H,\, K:\,\Omega^1\to\mf{g}^\partial$ be Hamiltonian operators.
Their graphs $\mc G(H)$ and $\mc G(K)$ are compatible Dirac structures 
(see Definition \ref{2006_NRel}) if and only if we have
$$
[H,K]_{SB}(\omega_1,\omega_2,\omega_3)
\,=\,0\,,
$$
for all $\omega_1,\omega_2,\omega_3\in\Omega^1$,
such that $K(\omega_2)\in \im(H),\,H(\omega_2)\in\im(K)$.
\end{proposition}
\begin{proof}
For the Dirac structures $\mc{G}(H)$ and $\mc{G}(K)$, we have
\begin{equation}\label{eq:09jan21_2}
\mc{N}_{\mc{G}(H),\mc{G}(K)} = \big\{H(\eta)\oplus K(\eta)\,\big|\, 
\eta\in\Omega^1\big\} \subset\mf{g}^\partial\oplus\mf{g}^\partial\,.
\end{equation}
The adjoint relation is, by the non-degeneracy of $(\,,\,)$,
\begin{equation*}
\mc{N}^*_{\mc{G}(H),\mc{G}(K)} = \big\{\omega\oplus \omega^\prime\,\big|\,
H(\omega)=K(\omega^\prime)\big\} \subset\Omega^1\oplus\Omega^1.
\end{equation*}
Then by 
Definition \ref{def:1.22} $\mc G(H)$ and $\mc G(K)$ are compatible
if for all $\omega,\omega^\prime,\omega^{\prime\prime}\in\Omega^1$ such that
\begin{equation}\label{2006_5}
H(\omega)=K(\omega^\prime), \quad H(\omega^\prime)=K(\omega^{\prime\prime}),
\end{equation}
and for all $\eta,\theta\in\Omega^1$, we have 
\begin{equation}\label{2006_6}
\begin{array}{l}
\vphantom{\Big(}
\big(\omega,[H(\eta),H(\theta)]\big) 
- \big(\omega^\prime,[H(\eta),K(\theta)]\big) \\
\vphantom{\Big(}
- \big(\omega^\prime,[K(\eta),H(\theta)]\big)
+ \big(\omega^{\prime\prime},[K(\eta),K(\theta)]\big) = 0\,.
\end{array}
\end{equation}
We next use equation \eqref{eq:09jan14_1_new} to rewrite 
\eqref{2006_6} as follows:
\begin{equation}\label{2006_dic14_2}	
\begin{array}{c}
\vphantom{\Big(}
\big(L_{H(\eta)}\omega,H(\theta)\big) 
- \big(L_{H(\eta)}\omega^\prime,K(\theta)\big) 
- \big(L_{K(\eta)}\omega^\prime,H(\theta)\big) 
+ \big(L_{K(\eta)}\omega^{\prime\prime},K(\theta)\big) \\
\vphantom{\Big(}
- L_{H(\eta)}\big(\omega,H(\theta)\big) 
+ L_{H(\eta)}\big(\omega^\prime,K(\theta)\big) 
+ L_{K(\eta)}\big(\omega^\prime,H(\theta)\big)
- L_{K(\eta)}\big(\omega^{\prime\prime},K(\theta)\big) = 0\,.
\end{array}
\end{equation}
By \eqref{2006_5} and skew-adjointness of $H$ and $K$, the last four terms in the LHS of \eqref{2006_dic14_2}
cancel.
Moreover, by assumption, $[H,H]_{SB}=[K,K]_{SB}=0$, so that equation \eqref{2006_dic14_2} becomes
\begin{equation*}	
\begin{array}{l}
\vphantom{\Big(}
\big(\eta,H(L_{H(\omega)}\theta)\big)
+ \big(\omega,H(L_{H(\theta)}\eta)\big)
+ \big(\theta,K(L_{H(\eta)}\omega^\prime)\big) \\
\vphantom{\Big(}
+ \big(\theta,H(L_{K(\eta)}\omega^\prime)\big)
+ \big(\eta,K(L_{K(\omega^{\prime\prime})}\theta)\big)
+ \big(\omega^{\prime\prime},K(L_{K(\theta)}\eta)\big) = 0\,.
\end{array}
\end{equation*}
If we then use again \eqref{2006_5}, the above identity becomes
$[H,K]_{SB}(\eta,\omega^\prime,\theta) = 0$.
\end{proof}
\begin{remark}\label{rem:09jan20}
Let $H,K:\,\Omega^1\to\mf{g}^\partial$ be Hamiltonian operators. 
It is clear from Definition \ref{def:09jan20_1}  that the Schouten bracket
$[H,K]_{SB}$ is linear in each factor.
Hence the following conditions are equivalent:
\begin{enumerate}  
\item[(i)] any linear combination $\alpha H+\beta K$ is a Hamiltonian operator,
\item[(ii)] the sum $H+K$ is a Hamiltonian operator,
\item[(iii)] $[H,K]_{SB} = 0$.
\end{enumerate}
When the above equivalent conditions hold we say that $(H,K)$
form a \emph{bi-Hamiltonian pair} (cf.\ Definition \ref{def:1.22}).
It follows from Proposition \ref{2006_BiHamiltonian} that if $(H,K)$
form a bi-Hamiltonian pair, then the graphs $\mc G(H)$ and $\mc G(K)$
are compatible Dirac structures.
\end{remark}

\begin{remark}\label{rem:09jan21}
Theorems \ref{2006_LS} and \ref{2006_th-may30}
and Remark \ref{rem:09jan20} imply Theorem \ref{th:1.24}.
Indeed, due to Theorem \ref{2006_th-may30} and Remark \ref{rem:09jan20},
if $(H,K)$ form a bi-Hamiltonian pair (as in Definition \ref{def:1.22}),
the corresponding graphs $\mc G(H)$ and $\mc G(K)$ are compatible Dirac structures.
In order to apply Theorem \ref{2006_LS}, we need to check that assumptions (i) and (ii) 
of this theorem hold.
For condition (i),
by formula \eqref{eq:09jan21_2} we have that 
$\pi_2\big(\mc N_{\mc G(H),\mc G(K)}\big)=\im(K)\subset\mf g^\partial$.
Hence condition (i) of Theorem \ref{2006_LS} becomes the following:
if $F\in\mc V^{\oplus\ell}$ is such that 
\begin{equation}\label{eq:09jan21_3}
\tint \big(K(\partial)G^2\big)\cdot\big(D_F(\partial)-D^*_F(\partial)\big)\big(K(\partial)G^1\big)
\,=\,0
\,\,,\,\,\,\,
\text{ for all } G^1,G^2\in\mc V^{\oplus\ell}\,,
\end{equation}
then $D_F(\partial)-D^*_F(\partial)=0$.
For this we used the identifications $\Omega^1\simeq\mc V^{\oplus\ell}$ 
and $\mf g^\partial\simeq\mc V^\ell$, and formula \eqref{eq:09jan4_2_a}.
Using the fact that $K$ is skew-adjoint, equation \eqref{eq:09jan21_3} becomes
$$
\tint G^2\cdot\big(K(\partial) \big(D_F(\partial)-D^*_F(\partial)\big) K(\partial)\big)G^1
\,=\,0\,.
$$
Hence, recalling that the pairing $(\cdot\,,\,\cdot)$ is non-degenerate,
we see that condition (i) of Theorem \ref{2006_LS}
is exactly the non-degeneracy assumption for $K$ in Theorem \ref{th:1.24}.
Next, suppose, as in Theorem \ref{th:1.24}, that 
$\{F^n\}_{n=0,\cdots,N}\subset\mc V^{\oplus\ell}$
is an $(H,K)$-sequence,
and that $F^0=\frac{\delta h_0}{\delta u},\,F^1=\frac{\delta h_1}{\delta u}$.
Then we have $\omega_{F^n}\oplus X_{H(\partial)F^n}\in\mc G(H)$
and, since $H(\partial)F^n=K(\partial)F^{n+1}$, we also have
$\omega_{F^{n+1}}\oplus X_{H(\partial)F^n}\in\mc G(K)$,
which is condition (ii) of Theorem \ref{2006_LS}.
Then, by Theorem \ref{2006_LS}(b), all the elements $\omega_{F^n},\,n=0,\cdots,N$,
are closed, thus proving Theorem \ref{th:1.24}.
\end{remark}
\begin{remark}
Let $H(\partial)$ be a Hamiltonian differential operator, 
and let $\tint h\in\mc V/\partial\mc V$ be a local functional,
and consider the corresponding Hamiltonian evolution equation \eqref{2006_evol},
where $X=X_{H(\partial)\frac{\delta h}{\delta u}}$.
This equation is integrable with respect to the Dirac structure $\mc G(H)$
if and only if it is integrable in the sense of Definition \ref{def:vicapr09}.
\end{remark}

\vspace{3ex}
\subsection{The symplectic case.}~~
\label{sec:dirac-sympl}
In Section \ref{sec:09jan2} we gave a definition of a symplectic operator
as a closed 2-form in the variational complex.
As we shall see in Theorem \ref{2006_th-may30_sym}, the following definition 
is equivalent to it in the case of differential operators.
\begin{definition}\label{2006_def-hs_sym}
A {\it symplectic operator} is a linear map $S:\,\mf{g}^\partial\to\Omega^1$ such that its graph
\begin{equation*}
  \mc{G}(S) = \{ S(X)\oplus X \,|\, X \in \mf g^\partial\} \subset \Omega^1\oplus \mf{g}^\partial\,,
\end{equation*}
is a Dirac structure.
\end{definition}
In the case of a Dirac structure given by the graph $\mc G(S)$ of a symplectic operator $S$,
the corresponding space of Hamiltonian functionals is $\mc F(\mc G(S))=\delta^{-1}(S(\mf g^\partial))$,
while the space of Hamiltonian vector fields is $\mc H(\mc G(S))=S^{-1}(\delta\Omega^0)$.
Hence, by the discussion in Section \ref{sec:equations-dirac},
the space $\delta^{-1}(S(\mf g^\partial))\subset \mc V/\partial\mc V$ 
acquires a structure of a Lie algebra with the bracket
\begin{equation}\label{eq:09jan20_1_sym}
\big\{\tint f,\tint g\big\}_S
\,=\,
\tint X (g)
\,\,,\,\,\,\,
\text{ where } X\in\mf g^\partial \text{ is s.t. } S(X)=\delta(\tint f)\,.
\end{equation}
Moreover, 
given a Hamiltonian vector field  $X\in\mc H(\mc G(S))$ associated
to a Hamiltonian functional $\tint h\in\mc F(\mc G(S))$,
i.e.\  such that $S(X)=\delta(\tint h)$,
the corresponding evolution equation is \eqref{2006_evol}.
\begin{proposition}\label{2006_def-ham2_sym}
Let $S:\, \mf{g}^\partial\to\Omega^1$ be a linear map,
and let $\mc{G}(S)=\big\{S(X)\oplus X\,|\,X\in\mf g^\partial\big\}
\subset\Omega^1\oplus\mf{g}^\partial$ be its graph.
\begin{enumerate}
\alphaparenlist
\item The space $\mc{G}(S)\subset\Omega^1\oplus\mf{g}^\partial$
is isotropic with respect to the bilinear form $(\,,\,)$, 
i.e.\ $\mc{G}(S)\subset\mc{G}(S)^\perp$,
if and only if 
it is maximal isotropic,
i.e.\ $\mc{G}(S)^\perp=\mc{G}(S)$,
which in turn is equivalent to saying that
$S$ is skew-adjoint:
\begin{equation}\label{2006_28oct_6_sym}
(S(X),Y)=-(S(Y),Y)\,,\,\,\text{for all\,\,} X,Y\in\mf g^\partial\,.
\end{equation}
\item A skew-adjoint operator $S:\,\mf g^\partial\to\Omega^1$
is a symplectic operator if and only if one of 
the following equivalent conditions holds:
\begin{enumerate}
\item[(i)]
for every $X_1,X_2,X_3\in\mf g^\partial$, we have
\begin{equation}\label{2006_28oct_4_ham_sym}
(L_{X_1}S(X_2),X_3)+
(L_{X_2}S(X_3),X_1)+
(L_{X_3}S(X_1),X_2)\,=\,0\,,
\end{equation}
\item[(ii)] 
for every $X,Y\in\mf g^\partial$ we have
\begin{equation}\label{2006_may30-1_sym}
L_{X}(S(Y)) - \iota_{Y}(\delta S(X))
\,=\, S\big([X,Y]\big)\,.
\end{equation}
\end{enumerate}
\end{enumerate}
\end{proposition}
\begin{proof}
The proof is the same as that of Proposition \ref{2006_def-ham2}.
\end{proof}
\begin{theorem}\label{2006_th-may30_sym}
Let $S:\,\mf{g}^\partial=\mc V^\ell\to\Omega^1=\mc V^{\oplus\ell}$ 
be a differential operator, 
$S=\big(S_{ij}(\partial)\big)_{i,j\in I}$.
Then Definitions \ref{def:6feb09} and \ref{2006_def-hs_sym} of a symplectic operator 
are equivalent.
\end{theorem}
\begin{proof}
In view of Proposition \ref{2006_def-ham2_sym}, we only have to check that,
if $S=(S_{ij}(\partial))_{i,j\in I}:\,\mc V^\ell\to\mc V^{\oplus\ell}$
is a skew-adjoint matrix valued differential operator,
then equation \eqref{2006_may30-1_sym} reduces to equation \eqref{v024_sec}.
Let $X=X_P,\,Y=X_Q$, with $P,Q\in \mc V^\ell$. 
We can use equations \eqref{eq:09jan21_1}, \eqref{eq:09jan4_2_b} and \eqref{eq:09jan4_2_c}
to rewrite equation \eqref{2006_may30-1_sym} as follows:
$$
D_{S(\partial)Q}(\partial)P+D_P^*(\partial)\big(S(\partial)Q\big)
-D_{S(\partial)P}(\partial)Q+D_{S(\partial)P}^*(\partial)Q
=S(\partial)\big(D_Q(\partial)P-D_P(\partial)Q\big)\,.
$$
Written out explicitly, the above equation reads (for $k\in I$):
\begin{equation}\label{eq:09feb10_1}
\begin{array}{c}
\displaystyle{
\sum_{i,j\in I, n\in\mb{Z}_+}
\Bigg(
\frac{\partial \big(S_{ki}(\partial) Q_i\big)}{\partial u_j^{(n)}} \Big(\partial^n P_j\Big)
+  (-\partial)^n\bigg(\frac{\partial P_j}{\partial u_k^{(n)}} \Big(S_{ji}(\partial) Q_i\Big)\bigg)
}\\
\displaystyle{
- \frac{\partial \big(S_{kj}(\partial) P_j\big) }{\partial u_i^{(n)}} \Big(\partial^n Q_i \Big)
+ (-\partial)^n\bigg(\frac{\partial \big(S_{ij}(\partial)P_j\big)}{\partial u_k^{(n)}} Q_i\bigg)\Bigg)
}\\
\displaystyle{
\,=\,
\sum_{i,j\in I, n\in\mb{Z}_+} 
\Bigg(
S_{ki}(\partial)
\bigg(\frac{\partial Q_i}{\partial u_j^{(n)}} \partial^n P_j\bigg)
-
S_{kj}(\partial)
\bigg(\frac{\partial P_j}{\partial u_i^{(n)}}\partial^n Q_i\bigg)
\Bigg)\,.
}
\end{array}
\end{equation}
We next use Lemma \ref{2006_l-may31} to rewrite, in the LHS of \eqref{eq:09feb10_1},
the first term as
\begin{equation}\label{eq:09feb10_2}
\sum_{i,j\in I, n\in\mb{Z}_+}
\Bigg(
\bigg(\frac{\partial S_{ki}(\partial)}{\partial u_j^{(n)}}Q_i\bigg) \Big(\partial^n P_j\Big)
+S_{ki}(\partial) \bigg(\frac{\partial Q_i}{\partial u_j^{(n)}} \Big(\partial^n P_j\Big)\bigg)
\Bigg)\,,
\end{equation}
the third term as
\begin{equation}\label{eq:09feb10_3}
- \sum_{i,j\in I, n\in\mb{Z}_+}
\Bigg(
\bigg(\frac{\partial S_{kj}(\partial)}{\partial u_i^{(n)}} P_j\bigg) \Big(\partial^n Q_i \Big)
+
S_{kj}(\partial) \bigg(\frac{\partial P_j }{\partial u_i^{(n)}} \Big(\partial^n Q_i \Big)\bigg)
\Bigg)\,,
\end{equation}
and the last term as
\begin{equation}\label{eq:09feb10_4}
\sum_{i,j\in I, n\in\mb{Z}_+}
(-\partial)^n\Bigg(
Q_i \bigg(\frac{\partial S_{ij}(\partial)}{\partial u_k^{(n)}}P_j\bigg)
+
\frac{\partial P_j}{\partial u_k^{(n)}} \big(S^*_{ji}(\partial) Q_i\big)
\Bigg)\,.
\end{equation}
The first term in the RHS of \eqref{eq:09feb10_1} cancels with the second term in \eqref{eq:09feb10_2},
and similarly 
the second term in the RHS of \eqref{eq:09feb10_1} 
cancels with the second term in \eqref{eq:09feb10_3}.
Furthermore, by the skew-adjointness assumption on $S$, 
the second term in the LHS of \eqref{eq:09feb10_1} 
cancels with the second term in \eqref{eq:09feb10_4}.
Hence, equation \eqref{eq:09feb10_1} can be rewritten as
$$
\sum_{i,j\in I, n\in\mb{Z}_+}
\Bigg(
\bigg(\frac{\partial S_{ki}(\partial)}{\partial u_j^{(n)}}Q_i\bigg) \Big(\partial^n P_j\Big)
- 
\bigg(\frac{\partial S_{kj}(\partial)}{\partial u_i^{(n)}} P_j\bigg) \Big(\partial^n Q_i \Big)
+
(-\partial)^n
Q_i \bigg(\frac{\partial S_{ij}(\partial)}{\partial u_k^{(n)}}P_j\bigg)
\Bigg)
\,=\,
0\,.
$$
Since the above identity holds for every $P,Q\in\mc V^{\ell}$,
we can replace $\partial$ acting on $P$ by $\lambda$ and $\partial$ acting on $Q$ by $\mu$.
We thus get that \eqref{2006_may30-1_sym} is equivalent
to the following equation (for $i,j,k\in I$):
\begin{equation}\label{eq:09feb10_5}
\sum_{n\in\mb{Z}_+}
\Bigg(
\frac{\partial S_{ki}(\mu)}{\partial u_j^{(n)}} \lambda^n
- 
\frac{\partial S_{kj}(\lambda)}{\partial u_i^{(n)}} \mu^n 
+
(-\lambda-\mu-\partial)^n
\frac{\partial S_{ij}(\lambda)}{\partial u_k^{(n)}}
\Bigg)
\,=\,
0\,.
\end{equation}
To conclude, we just notice that, by the Definition \ref{ex:1.6} of the Beltrami $\lambda$-bracket,
equation \eqref{eq:09feb10_5} is the same as equation \eqref{v024_sec}.
\end{proof}

Next, we study how the notion of compatibility of Dirac structures is translated in the case
of symplectic operators.
Let $S:\,\mc V^\ell\to\mc V^{\oplus\ell}$ and $T:\,\mc V^\ell\to\mc V^{\oplus\ell}$
be symplectic operators,
and consider the corresponding Dirac structures $\mc G(S)$ and $\mc G(T)$.
Recalling the definition \eqref{eq:20090322_1}, we have
\begin{equation}\label{eq:20090324_n1}
\mc N_{\mc G(S),\mc G(T)}
=
\big\{X\oplus X^\prime\in\mf g^\partial\oplus\mf g^\partial\,\big|\,S(X)=T(X^\prime)\big\}\,,
\end{equation}
and, similarly, recalling \eqref{eq:20090322_2}, we have
\begin{equation}\label{eq:20090324_n2}
\mc N^*_{\mc G(S),\mc G(T)}
=
\big\{\omega\oplus \omega^\prime\in\Omega^1\oplus\Omega^1\,\big|\,
(\omega,X)=(\omega^\prime,X^\prime)\,,\,\,\text{for all\,\,} X\oplus X^\prime\in\mc N_{\mc G(S),\mc G(T)}
\big\}\,.
\end{equation}
We also consider the space
\begin{equation}\label{eq:20090324_n3}
\mc M_{\mc G(S),\mc G(T)}
=
\big\{
S(X)\oplus T(X)\,\big|\,X\in\mf g^\partial
\big\}
\,\subset\Omega^1\oplus\Omega^1\,.
\end{equation}
It is immediate to check that 
$\mc M_{\mc G(S),\mc G(T)}\subset\mc N^*_{\mc G(S),\mc G(T)}$
for every pair of symplectic operators $S$ and $T$,
while the opposite inclusion is not necessarily true.
\begin{definition}\label{def:20090324_1}
A pair $(S,T)$ of symplectic operators is called 
\emph{strong}, if 
\begin{equation}\label{eq:20090324_n5}
\mc N^*_{\mc G(S),\mc G(T)}=\mc M_{\mc G(S),\mc G(T)}\,.
\end{equation}
\end{definition}
\begin{proposition}\label{prop:20090324}(cf.\ \cite{D})
Let $(S,T)$ be a strong pair of symplectic operators.
The following three conditions are equivalent:
\begin{enumerate}
\romanparenlist
\item $\mc G(S),\mc G(T)$ are compatible Dirac structures,
\item for every $X,X^\prime,Y,Y^\prime\in\mf g^\partial$
satisfying $S(X)=T(X^\prime),\,S(Y)=T(Y^\prime)$,
there exists $Z\in\mf g^\partial$ such that
\begin{equation}\label{eq:20090324_n6}
L_{X^\prime}S(Y)-L_XS(Y^\prime) = S(Z)
\quad,\qquad
L_{X^\prime}T(Y)-L_XT(Y^\prime) = T(Z)\,.
\end{equation}
\item for $X_1,X_1^\prime,X_2,X_2^\prime,X_3,X_3^\prime\,\in\mf g^\partial$ 
such that $S(X_i)=T(X_i^\prime),\,i=1,2,3$, we have
\begin{equation}\label{eq:20090324_n4}
\big(L_{X_1}S(X_2),X_3^\prime\big)
+\big(L_{X_1^\prime}S(X_2),X_3\big)
+ (\text{cycl. perm.})\,=\,0\,.
\end{equation}
\end{enumerate}
\end{proposition}
\begin{proof}
Recall Definition \ref{2006_NRel}.
By \eqref{eq:20090324_n1} 
and assumption \eqref{eq:20090324_n5},
$\mc G(S)$ and $\mc G(T)$ are compatible Dirac structures
if and only if we have
\begin{equation}\label{eq:20090324_n7}
(S(Z^\prime),[X,Y])-(S(Z),[X,Y^\prime])-(S(Z),[X^\prime,Y])
+(T(Z),[X^\prime,Y^\prime])\,=\,0\,,
\end{equation}
for every $X,X^\prime,Y,Y^\prime,Z,Z^\prime\in\mf g^\partial$
such that 
\begin{equation}\label{eq:20090324_n8}
S(X)=T(X^\prime)
\quad,\qquad
S(Y)=T(Y^\prime)
\quad,\qquad
S(Z)=T(Z^\prime)\,.
\end{equation}
Using identity \eqref{eq:09jan14_1_new},
we can rewrite equation \eqref{eq:20090324_n7} as follows:
\begin{eqnarray}\label{eq:20090324_n9}
& -\big(L_XS(Z^\prime),Y\big)
+\big(L_XS(Z),Y^\prime\big)
+\big(L_{X^\prime}S(Z),Y\big)
-\big(L_{X^\prime}T(Z),Y^\prime\big) \\
& L_X\big(S(Z^\prime),Y\big)
-L_X\big(S(Z),Y^\prime\big)
-L_{X^\prime}\big(S(Z),Y\big)
+L_{X^\prime}\big(T(Z),Y^\prime\big)
\,=\,0\,.\nonumber
\end{eqnarray}
By equations \eqref{eq:20090324_n8} and the skew-adjointness 
of the operators $S$ and $T$,
we have 
$(S(Z^\prime),Y)=(S(Z),Y^\prime)$ and $(S(Z),Y)=(T(Z),Y^\prime)$,
hence the last four terms in the LHS of \eqref{eq:20090324_n9} cancel.
Equation \eqref{eq:20090324_n9} then reads
\begin{equation}\label{eq:20090324_n10}
\big(L_{X^\prime}S(Z)-L_XS(Z^\prime),Y\big)
\,=\,
\big(L_{X^\prime}T(Z)-L_XT(Z^\prime),Y^\prime\big) \,.
\end{equation}
Since the above equation holds for every $Y,Y^\prime\in\mf g^\partial$
such that $S(Y)=T(Y^\prime)$,
by assumption \eqref{eq:20090324_n5}, it is equivalent
to saying that there exists an element $W\in\mf g^\partial$ such that
$L_{X^\prime}S(Z)-L_XS(Z^\prime)=S(W),\,
L_{X^\prime}T(Z)-L_XT(Z^\prime)=T(W)$.
If we replace $Z$ by $Y$ and $W$ by $Z$, these equations are the same as
\eqref{eq:20090324_n6}.
This proves the equivalence of conditions (i) and (ii).
Next, we prove that equation \eqref{eq:20090324_n10} is equivalent to condition (iii).
Since $S$ and $T$ are symplectic operators,
we have, by the integrability condition \eqref{2006_28oct_4_ham_sym}, that
\begin{eqnarray*}
\big(L_XS(Z^\prime),Y\big)
&=&
-\big(L_{Z^\prime} S(Y),X\big)
-\big(L_Y S(X),Z^\prime\big)\,,
\\
\big(L_{X^\prime}T(Z),Y^\prime\big)
&=&
-\big(L_{Z}T(Y^\prime),X^\prime\big)
-\big(L_{Y^\prime}T(X^\prime),Z\big)\,.
\end{eqnarray*}
Using the above identities, equation \eqref{eq:20090324_n10} becomes
\begin{eqnarray*}
& \big(L_{Z^\prime} S(Y),X\big)
+\big(L_Y S(X),Z^\prime\big)
+\big(L_{X^\prime}S(Z),Y\big)
+\big(L_XT(Z^\prime),Y^\prime\big) \\
& +\big(L_{Z}T(Y^\prime),X^\prime\big)
+\big(L_{Y^\prime}T(X^\prime),Z\big)
\,=\,
0\,,
\end{eqnarray*}
which is the same as equation \eqref{eq:20090324_n4} for $X_1=X,\,X_2=Z,\,X_3=Y$,
and similarly for $X_i^\prime,\,i=1,2,3$.
\end{proof}
\begin{definition}\label{def:bisympl}
A strong pair $(S,T)$ of symplectic operators which satisfies 
one of the equivalent conditions (i)--(iii) of Proposition \ref{prop:20090324}
is called a \emph{bi-symplectic pair}.
\end{definition}

In view of Proposition \ref{prop:20090324}, we get the following corollaries
of Theorem \ref{2006_LS}, Proposition \ref{prop:09jan22}
and Corollary \ref{09jan26}, which translate the Lenard scheme of integrability
in the symplectic case.
\begin{theorem}\label{th:20090324}
Let $\mc V$ be an algebra of differential functions in the variables $\{u_i\}_{i\in I}$,
and let $S(\partial),\,T(\partial)$ be two symplectic operators 
given by differential operators: $\mc V^\ell\to\mc V^{\oplus\ell}$.
Suppose moreover that $\{P^n\}_{0\leq n\leq N}\subset\mc V^\ell$
is an $(S,T)$-sequence, namely
$$
S(\partial)P^{n+1}=T(\partial)P^{n},\,n=0,\dots,N-1\,.
$$
\begin{enumerate}
\alphaparenlist
\item If the orthogonality condition
$$
\big(\Span_{\mb C}\{P^n\}_{0\leq n\leq N}\big)^\perp\subset \im S(\partial)
$$
hods, then we can extend the sequence 
$\{P^n\}_{0\leq n\leq N}$
to an infinite $(S,T)$-sequence $\{P^n\}_{n\in\mb Z_+}$.
\item Suppose in addition that $(S,T)$ is a bi-symplectic pair, 
and that it satisfies the following non-degeneracy condition:
if, for some $F\in\mc V^{\oplus\ell}$,
$\tint Q\cdot\big(D_F(\partial)-D_F^*(\partial)\big)P=0$
for all $P,Q\in\mc V^\ell$ such that $T(\partial)P,\,T(\partial)Q\,\in\im(S(\partial))$,
then $D_F(\partial)-D_F^*(\partial)=0$.
Suppose, moreover, that 
$$
S(\partial)P^0=\frac{\delta \tint h_0}{\delta u}
\quad,\qquad
T(\partial)P^0=\frac{\delta \tint h_1}{\delta u}\,,
$$
for some local functionals $\tint h_0,\,\tint h_1\in\mc V/\partial\mc V$.
Then all elements $F^n=S(\partial)P^n,\,n\in\mb Z_+$, are closed, 
namely $D_{F^n}(\partial)=D^*_{F^n}(\partial)$.
In particular, if the algebra $\mc V$ is normal,
then they are all exact,
namely there exist local functionals $\tint h_n\in\mc V/\partial\mc V$ such that
$$
F^n=S(\partial)P^n=\frac{\delta \tint h_n}{\delta u}
\,\,,\,\,\,\,n\in\mb Z_+\,.
$$
These local functionals are in involution with respect to both Lie brackets
associated to $S$ and $T$ (cf.\ \eqref{2006_dic15_2}):
$$
\big\{\tint h_m,\tint h_n\big\}_S=\big\{\tint h_m,\tint h_n\big\}_T=0
\,\,,\,\,\,\,\text{for all\,\,} m,n\in\mb Z_+\,.
$$
\item
Suppose, in addition, 
that $\Ker(S)\cap\Ker(T)=0$. 
Then the evolutionary vector fields $X_{P^n},\,n\in\mb Z_+$, commute, 
hence the equations of the hierarchy
$$
\frac{du}{dt_n}=P^n,\,n\in\mb Z_+\,,
$$
are compatible, and the local functionals in involution $\tint h_n$
are their integrals of motion.
Thus, this hierarchy is integrable,
provided that the set $\{F^n\}_{n\in\mb Z_+}$ spans an infinite-dimensional space.
\end{enumerate}
\end{theorem}

\vspace{3ex}
\subsection{The potential KdV hierarchy.}~~
\label{sec:pkdv}
Let $\mc V=\mb C[u,u^\prime,u^{\prime\prime},\dots]$,
and consider the following pair of differential operators on $\mc V$ ($c\in\mb C$):
\begin{equation}\label{eq:09may5_1}
S(\partial)=\partial
\quad,\qquad
T(\partial)=u^{\prime\prime}+2u^\prime\partial+c\partial^3
=2(u^\prime)^{1/2}\partial\circ(u^\prime)^{1/2}+c\partial^3\,.
\end{equation}
\begin{proposition}\label{prop:09may5_pkdv}
\begin{enumerate}
\alphaparenlist
\item $(S,T)$ form a bi-symplectic pair (cf.\ Definition \ref{def:bisympl}).
\item The pair $(S,T)$ satisfies the following non-degeneracy condition 
(cf.\ Theorem \ref{th:20090324}(b)): if 
\begin{equation}\label{eq:20090505_11}
\tint Q\cdot M(\partial) P=0\,\,,
\end{equation}
where $M(\partial)$ is a differential operator,
for all $P,Q\in\mc V$ such that $T(\partial)P,\,T(\partial)Q\in\im S(\partial)$,
then $M(\partial)=0$.
\end{enumerate}
\end{proposition}
\begin{proof}
We already know, from Example \ref{ex:09feb16}, that $S(\partial)$ and $T(\partial)$
are symplectic operators.
%
We want to prove that the pair $(S,T)$ is strong 
(cf.\ Definition \ref{def:20090324_1}).
Let $F\oplus G\in\mc N^*_{\mc G(S),\mc G(T)}$, namely 
\begin{equation}\label{eq:20090505_8}
\tint FP=\tint GQ\,,
\end{equation}
whenever 
\begin{equation}\label{eq:20090505_4}
\partial P
=
c\partial^3 Q+2\partial(u^\prime Q)-u^{\prime\prime}Q\,.
\end{equation}
We need to prove that $F\oplus G\in\mc M_{\mc G(S),\mc G(T)}$,
namely that there exists $R\in\mc V$ such that 
\begin{equation}\label{eq:20090505_5}
F=\partial R
\quad,\qquad
G=c\partial^3 R+2u^\prime\partial R+u^{\prime\prime}R\,.
\end{equation}
Equation \eqref{eq:20090505_4} 
implies that $u^{\prime\prime}Q\in\partial\mc V$,
namely $Q=\frac1{u^{\prime\prime}}\partial f$ for some $f\in\mc V$.
Hence, all pairs $(P,Q)$ solving \eqref{eq:20090505_4} are of the form
\begin{equation}\label{eq:20090505_6}
P = c\partial^2\Big(\frac{\partial f}{u^{\prime\prime}}\Big)
+2\frac{u^\prime}{u^{\prime\prime}}\partial f-f
\quad,\qquad
Q = \frac{\partial f}{u^{\prime\prime}}\,,
\end{equation}
for some $f\in\mc V$ such that 
\begin{equation}\label{eq:20090505_7}
\partial f\in u^{\prime\prime}\mc V\,.
\end{equation}
Equation \eqref{eq:20090505_8}, combined with \eqref{eq:20090505_6}, gives,
after integration by parts, that
\begin{equation}\label{eq:20090505_9}
\int f\Big\{
c\partial\Big(\frac{\partial^2 F}{u^{\prime\prime}}\Big)
+\partial\Big(\frac{u^\prime}{u^{\prime\prime}}F\Big)
+F
-\partial\Big(\frac{G}{u^{\prime\prime}}\Big)
\Big\}\,=\,0\,.
\end{equation}
Note that the space of elements $f\in\mc V$ satisfying condition \eqref{eq:20090505_7}
contains the ideal generated by $(u^{\prime\prime})^2$.
Hence, we can use Proposition \ref{prop:20081222}(b) to conclude,
from \eqref{eq:20090505_9}, that
\begin{equation}\label{eq:20090505_10}
c\partial\Big(\frac{\partial^2 F}{u^{\prime\prime}}\Big)
+\partial\Big(\frac{u^\prime}{u^{\prime\prime}}F\Big)
+F
=
\partial\Big(\frac{G}{u^{\prime\prime}}\Big)\,.
\end{equation}
It immediately follows from \eqref{eq:20090505_10} that $F=\partial R$ for some $R\in\mc V$.
If we combine this fact with equation \eqref{eq:20090505_10},
we easily get that $F$ and $G$ have the form \eqref{eq:20090505_5}.

In order to prove part (a) we are left to prove that the pair $(S,T)$
satisfies condition (ii) in Proposition \ref{prop:20090324}.
Let $P_0,\,P_1,\,Q_0,\,Q_1\in\mc V$ be such that
\begin{equation}\label{eq:final1}
S(\partial)P_0=T(\partial)P_1
\quad,\qquad
S(\partial)Q_0=T(\partial)Q_1\,.
\end{equation}
According to condition (ii), we need to find $R\in\mc V$ such that
$$
\begin{array}{l}
L_{X_{P_1}}\big(S(\partial)Q_0\big)
-
L_{X_{P_0}}\big(S(\partial)Q_1\big)
= S(\partial)R\,, \\
L_{X_{P_1}}\big(T(\partial)Q_0\big)
-
L_{X_{P_0}}\big(T(\partial)Q_1\big)
= T(\partial)R\,.
\end{array}
$$
Using \eqref{eq:09jan4_2_c} the above equations can be rewritten as follows
\begin{equation}\label{eq:final2}
\begin{array}{l}
D_{S(\partial)Q_0}(\partial)P_1
-
D_{S(\partial)Q_1}(\partial)P_0
+
D_{P_1}^*(\partial)S(\partial)Q_0
-
D_{P_0}^*(\partial)S(\partial)Q_1
=
S(\partial) R\,, \\
D_{T(\partial)Q_0}(\partial)P_1
-
D_{T(\partial)Q_1}(\partial)P_0
+
D_{P_1}^*(\partial)T(\partial)Q_0
-
D_{P_0}^*(\partial)T(\partial)Q_1
=
T(\partial) R\,.
\end{array}
\end{equation}
In order to check these equations
we use the fact that the pairs $(P_0,P_1)$ and $(Q_0,Q_1)$
are of the form \eqref{eq:20090505_6},
and Lemma \ref{lem:20090323}.
A long but straightforward computation, which is left to the reader,
shows that equations \eqref{eq:final2} hold for
$$
R=D_{Q_0}(\partial)P_1
-
D_{Q_1}(\partial)P_0
+P_1\partial Q_1
-Q_1\partial P_1\,.
$$

Let us next prove part (b).
Recall from the above discussion that $Q\in\mc V$ is such that
$T(\partial)Q\in\im S(\partial)$
if and only if $Q=\frac1{u^{\prime\prime}}\partial f$
for some $f\in\mc V$ satisfying condition \eqref{eq:20090505_7}
(in particular, for every element of the ideal of $\mc V$ generated by $(u^{\prime\prime})^2$).
By Proposition \ref{prop:20081222}(b), condition \eqref{eq:20090505_11} implies that
$M(\partial)\frac{\partial f}{u^{\prime\prime}}=0$,
for every $f\in\mc V$ satisfying \eqref{eq:20090505_7},
and this of course implies that $M(\partial)=0$, as we wanted.
\end{proof}

Let $P^0=1,\,P^1=u^\prime$.
We clearly have $S(\partial)P^1=T(\partial)P^0$,
namely $\{P^0,P^1\}$ is an $(S,T)$-sequence.
Moreover, we have 
$\big(\Span_\mb C\{P^0,P^1\}\big)^\perp\subset (P^0)^\perp=\partial\mc V=\im S(\partial)$,
namely the orthogonality condition in Theorem \ref{th:20090324}(a) holds.
We also have
$S(\partial)P^0=\frac{\delta h_0}{\delta u},\,S(\partial)P^1=\frac{\delta h_1}{\delta u}$,
for $\tint h_0=0$ and $\tint h_1=-\frac12\tint (u^\prime)^2$.
Hence, all the assumptions 
of Theorem \ref{th:20090324} hold,
and we can extend $\{P^0,P^1\}$ to an infinite sequence 
$\{P^n\}_{n\in\mb Z_+}$,
such that
$S(\partial)P^n=T(\partial)P^{n-1}=\frac{\delta h_n}{\delta u}$,
for some local functionals $\tint h_n\in\mc V/\partial\mc V$.
This gives us an infinite hierarchy of Hamiltonian equations associated to
both symplectic operators $S(\partial)$ and $T(\partial)$ ($n\in\mb Z_+$):
\begin{equation}\label{eq:20090505_1}
\frac{du}{dt}
= P^n\,,
\end{equation}
for which all local functionals $\tint h_m,\,m\in\mb Z_+$ are integrals of motion.

It is easy to compute
the first few terms of the sequence $\{P^n,\tint h_n\}_{n\in\mb Z_+}$.
In fact the whole sequence $P^n,\,n\in\mb Z_+$, 
can be obtained inductively by $P^0=1$ and the recursive equation
\begin{equation}\label{eq:20090505_3}
\partial P^{n+1}
=
c\partial^3 P^n+2u^\prime\partial P^n+u^{\prime\prime} P^n\,,
\end{equation}
while the local functionals $\tint h_n,\,n\in\mb Z_+$,
are obtained by solving the variational problem 
$$
\frac{\delta h_n}{\delta u}
=
\partial P^n\,.
$$
We then get
\begin{eqnarray*}
&&\displaystyle{
P^1=u^\prime
\,\,,\,\,\,\,
\tint h_1=-\frac12\tint(u^\prime)^2
\quad,\qquad
P^2=
cu^{\prime\prime\prime}+\frac32 (u^\prime)^2
\,\,,\,\,\,\,
\tint h_2
=\int \Big(\frac c2(u^{\prime\prime})^2-\frac12(u^\prime)^3\Big)
\,,
} \\
&&\displaystyle{
P^3=
c^2u^{(5)}+5cu^\prime u^{\prime\prime\prime}\!+\!\frac52 c (u^{\prime\prime})^2
\!+\!\frac52 (u^\prime)^3
\,,\,\,
\tint h_3
=\int\!\!
\Big(
-\frac12 c^2 (u^{\prime\prime\prime})^2+\frac52 cu^\prime(u^{\prime\prime})^2-\frac58(u^\prime)^4
\Big)
\,\cdots
} 
\end{eqnarray*}
The corresponding Hamiltonian equations are as follows:
$$
\frac{du}{dt_1}=u^\prime
\,\,,\,\,\,\,
\frac{du}{dt_2}=
cu^{\prime\prime\prime}+\frac32 (u^\prime)^2
\,\,,\,\,\,\,
\frac{du}{dt_3}=
c^2u^{(5)}+5cu^\prime u^{\prime\prime\prime}+\frac52 c 
(u^{\prime\prime})^2+\frac52 (u^\prime)^3,
\,\dots
$$
The second evolution equation above is known as the potential KdV-equation.

We claim that \eqref{eq:20090505_1} is an integrable hierarchy of evolution equations
and hence, in particular, the potential KdV equation is integrable
(see Definition \ref{def:dirac-integr}).
First, the elements $F^n=\partial P^n,\,n\geq1$,
are linearly independent, since, 
by the recurrence formula \eqref{eq:20090505_3},
the highest degree term (in $u,u^\prime,\dots$) in $P^n$ is obtained
by putting $c=0$.
In this case the recurrence equation \eqref{eq:20090505_3} can be solved explicitly:
$$
P^n_{c=0}=\frac{(2n-1)!!}{n!}(u^\prime)^n\,,
$$ 
and the corresponding integrable hierarchy 
$$
\frac{du}{dt_n}=(u^\prime)^n
$$
may be called the dispersionless pKdV hierarchy.
We thus get
$P^n=P^n_{c=0}+$ terms of lower degree.
The linear independence of the $F^n$'s follows immediately.
Finally the evolutionary vector fields $P^n$ commute
since $\Ker\big(S(\partial)\big)\cap\Ker\big(T(\partial)\big)=0$.

\vspace{3ex}
\subsection{The KN hierarchy.}~~
\label{sec:kn}
Let $\mc V$ be an algebra of differential functions in one variable
in which $u^\prime$ is invertible,
for example $\mc V=\mb C[u,(u^\prime)^{\pm1},u^{\prime\prime},\dots]$.
Consider the following pair of differential operators on $\mc V$ ($c\in\mb C$):
$$ 
S(\partial)
=
(u^\prime)^{-1}\partial\circ(u^\prime)^{-1}
\quad,\qquad
T(\partial)
=
\partial\circ(u^\prime)^{-1}\partial\circ(u^\prime)^{-1}\partial\,.
$$ 
Recall from Example \ref{ex:09jan7_1} that $S(\partial)$ and $T(\partial)$ are symplectic operators.
\begin{proposition}\label{prop:09may5_kn}
$(S,T)$ form a bi-symplectic pair,
and the non-degeneracy condition of Proposition \ref{prop:09may5_pkdv}(b) holds.
\end{proposition}
\begin{proof}
The proof is similar to the proof of Proposition \ref{prop:09may5_pkdv}
and it is left to the reader.
\end{proof}

As usual, in order to find an integrable hierarchy, 
it is convenient to start from the kernel of $S(\partial)$. 
Let then $P^0=u^\prime$, so that $S(\partial)P^0=0$.
We have the following identities, which can be checked directly,
$$ 
\partial\Big(\frac1{u^\prime}\partial\Big(\frac1{u^\prime}\partial u^\prime\Big)\Big) 
=
\frac{\delta}{\delta u}\Big(\frac12\frac{(u^{\prime\prime})^2}{(u^\prime)^2}\Big)
=
\frac1{u^\prime}\partial\Big(\frac{u^{\prime\prime\prime}}{u^\prime}
-\frac{3}{2}\frac{(u^{\prime\prime})^2}{(u^\prime)^2}
\Big)
=
\frac{u^{(4)}}{(u^\prime)^2}-4\frac{u^{\prime\prime}u^{\prime\prime\prime}}{(u^\prime)^3}
+3\frac{(u^{\prime\prime})^3}{(u^\prime)^4}\,.
$$ 
These identities can be rewritten in the form
$T(\partial)P^0=S(\partial)P^1=\frac{\delta h_1}{\delta u}$,
where
$P^1=
u^{\prime\prime\prime}
-\frac{3}{2}\frac{(u^{\prime\prime})^2}{u^\prime}$
and
$\tint h_1=
\frac12\int\frac{(u^{\prime\prime})^2}{(u^\prime)^2}$.
In particular, $\{P^0,P^1\}$ is an $(S,T)$-sequence.
Moreover, we have 
$\big(\Span_\mb C\{P^0,P^1\}\big)^\perp\subset (P^0)^\perp
=\frac1{u^\prime}\partial\mc V=\im S(\partial)$,
hence the orthogonality condition in Theorem \ref{th:20090324}(a) holds.
Therefore all the assumptions 
of Theorem \ref{th:20090324} hold,
and we can extend $\{P^0,P^1\}$ to an infinite sequence 
$\{P^n\}_{n\in\mb Z_+}$,
such that
$S(\partial)P^n=T(\partial)P^{n-1}=\frac{\delta h_n}{\delta u}$,
for some local functionals $\tint h_n\in\mc V/\partial\mc V$.
This gives us an infinite hierarchy of Hamiltonian equations associated to
both symplectic operators $S(\partial)$ and $T(\partial)$
for which all local functionals $\tint h_m,\,m\in\mb Z_+$ are integrals of motion.
The 1-st equation of the hierarchy, associated to $P^1$, is
$$
\frac{du}{dt_1}
=
u^{\prime\prime\prime}-\frac{3}{2}\frac{(u^{\prime\prime})^2}{u^\prime}\,,
$$
which is known as the Krichever-Novikov (KN) equation.

To prove integrability of the KN equation (and of the whole hierarchy $\frac{du}{dt_n}=P^n$),
it suffices to show that the elements $F^n=S(\partial) P^n,\,n\geq1$,
are linearly independent.
For this, we notice that,
by the recurrence relation $S(\partial)P^{n+1}=T(\partial)P^n$,
the elements $P^n$ have the form
$P^n=u^{(2n+1)}+$ terms of lower differential order.
The linear independence of $P^n,\,n\geq0$, 
and hence of $F^n,\,n\geq1$, follows immediately.
Finally the evolutionary vector fields $P^n$ commute
since $\Ker\big(S(\partial)\big)=\mb Cu^\prime$,
and therefore $\Ker\big(S(\partial)\big)\cap\Ker\big(T(\partial)\big)=0$.

\end{document}